\documentclass[twocolumn]{emulateapj}

\usepackage[usenames,dvipsnames,svgnames,table]{xcolor}
\usepackage{graphicx,color,soul}
\usepackage{latexsym}
\usepackage{amsmath,amssymb}      
\usepackage[draft=false]{hyperref}
\hypersetup{
     colorlinks   = true,
     citecolor    = blue
}

\def\apj{ApJ}
\def\apjs{ApJS}
\def\aap{A\&A}
\def\mnras{MNRAS}
\def\prd{Phys.~Rev.~D}

\begin{document}

\title{CAFE: A New Relativistic MHD Code}

\author{F. D. Lora-Clavijo}
\affiliation{Instituto de Astronom\'{\i}a, Universidad Nacional Aut\'{o}noma de M\'{e}xico, AP 70-264, Distrito Federal 04510, M\'{e}xico.}

\author{A. Cruz-Osorio}
\affiliation{Instituto de F\'{\i}sica y Matem\'{a}ticas, Universidad Michoacana de San Nicol\'as de Hidalgo. Edificio C-3, Cd. Universitaria, 58040 Morelia, Michoac\'{a}n, M\'{e}xico.}
\affiliation{Instituto de Astronom\'{\i}a, Universidad Nacional Aut\'{o}noma de M\'{e}xico, AP 70-264, Distrito Federal 04510, M\'{e}xico.}

\author{F. S. Guzm\'an}
\affiliation{Instituto de F\'{\i}sica y Matem\'{a}ticas, Universidad Michoacana de San Nicol\'as de Hidalgo. Edificio C-3, Cd. Universitaria, 58040 Morelia, Michoac\'{a}n, M\'{e}xico.}

\email{FDLC: fdlora@astro.unam.mx}
\email{ACO: aosorio@astro.unam.mx}
\email{FSG: guzman@ifm.umich.mx}

\date{\today}

\begin{abstract}
We introduce CAFE, a new independent code designed to solve the equations of Relativistic ideal Magnetohydrodynamics (RMHD) in 3D. We present the standard tests for a RMHD code and for the Relativistic Hydrodynamics (RHD) regime since we have not reported them before. The tests include the 1D Riemann problems related to blast waves, head-on collision of streams and states with transverse velocities, with and without magnetic field, which is aligned or transverse, constant or discontinuous across the initial discontinuity. Among the 2D and 3D tests, without magnetic field we include the 2D Riemann problem, a one dimensional shock tube along a diagonal, the high speed Emery wind tunnel, the Kelvin-Helmholtz instability, a set of jets and a 3D spherical blast wave, whereas in the presence of a magnetic field we show the magnetic rotor, the cylindrical explosion, a case of Kelvin-Helmholtz instability and a 3D magnetic field advection loop. The code uses High Resolution Shock Capturing methods and we present the error analysis for a combination that uses  the HLLE flux formula combined with linear, PPM and fifth order WENO reconstructors. We use the flux-CT and the divergence cleaning methods to control the divergence free magnetic field constraint.
\end{abstract}


\maketitle

\section{Introduction}

Models of high energy astrophysics are closely related to relativistic fluid dynamics, because most of the sources are identified with the dynamics of gas or plasma associated to a high energy source. In the most complex cases, the models involve magnetic fields and cooling processes associated to various reactions taking place in the plasma. We know moreover, that  interesting sources of this sort involve also strong gravitational fields. In this sense, the most complete models approaching realistic scenarios within the field of high energy astrophysics involve three main ingredients: relativistic hydrodynamics (RHD), magnetic fields (RMHD) and strong gravitational fields (GRMHD), which are further complicated with the introduction of various cooling processes and realistic equations of state. In this way, the most modern relativistic models involve the solution of the coupled system of equations composed by the Einstein-Euler-Maxwell equations under very general conditions.

Given the complexity of this system of partial differential equations, these have been solved numerically for particular scenarios as a system of evolution equations, which requires the development of a code. Particular examples of high energy phenomena with some of these ingredients are the propagation of jets on flat space-times, accretion of tori around black holes, supernovae core collapse processes, mergers of binary neutron stars, etc. \citep{2013LRR....16....1A,2008LRR....11....7F}.

Astrophysical models of rapidly moving gas involve also a degree of idealization, in particular the gas equation of state, the conditions on magnetic fields and sometimes symmetries. The more powerful a code is, the less idealizations it assumes. Various codes have been presented that are distinguished in terms of what type of problems each one is able to solve. As examples we mention some of the currently most used codes. Cactus Einstein Toolkit, a multi usage package mounted on Cactus \citep{Goodale2002a}, capable of solving the general relativistic MHD \citep{2014CQGra..31a5005M}. Whisky, a code that in its most sophisticated version can evolve general relativistic resistive magnetohydrodynamics \citep{2013PhRvD..88d4020D}, also mounted Cactus. GENESIS, which is a code capable of solving the RMHD equations for relativistic and ultrarelativistic flows \citep{1999ApJS..122..151A,2005A&A...436..503L}. HARM, a general relativistic code for a fixed space-time \citep{2012ascl.soft09005G} and its latest version including radiation terms \citep{2014MNRAS.441.3177M}. HAD, that is capable of evolving binary compact stars in the presence of magnetic fields, in general relativity \citep{2013PhRvD..88d3011P}. 
There are also independent codes, dealing with general relativistic hydrodynamics, for instance the one in \citep{2012PhRvD..85l4010E}. 
CoCoNuT, evolves the General relativistic magneto-hydrodynamics to simulate the core collapse of massive stars and the evolution of neutron stars \citep{2013hsa7.conf..940C}. 
Specific purpose codes also include \citep{2011MNRAS.414.1467P}, designed to evolve the accretion of magnetized winds onto black holes. The PLUTO code, solves the RMHD equations \citep{2012ApJS..198....7M}, so as that in \citep{2011ApJS..193....6B}. And a certainly more complete list of codes designed for various purposes that can be found in for instance \citep{2008LRR....11....7F}.

Even though the codes are extremely advanced, including coupling to General Relativity and radiation cooling processes 
\citep[e.g.,][]{2011MNRAS.417.2899Z, 2012ApJS..201....9F}, and knowing that new state of the art numerical methods to handle the High Resolution Shock Capturing (HRSC) schemes in relativistic and General Relativistic hydrodynamics are now being studied and tested  \citep[e.g.,][]{2012A&A...547A..26R,2014CQGra..31g5012R}, it is our intention in this paper, to present and certify our code CAFE, which in its first version focuses on the solution of RMHD equations, with the intention of extending it to general background space-times and will be applied to the study of accretion processes on black holes. Additionally, considering we have not presented the tests of the RHD implementation before, we include also in this paper the tests in this regime.

The paper is organized as follows. In Section \ref{sec:mhd} we present the standard ideal RMHD equations that are solved. In Section \ref{sec:tests} we show the tests for the zero magnetic field case, which reduces the system to the pure RHD regime and also show how our implementation performs on the standard tests of the RMHD. In section \ref{sec:conclusions} we mention some final comments. Finally, in appendix~\ref{appendix} we provide details of the implementation in cylindrical coordinates, required in one of the tests.

\section{SRMHD Equations and Numerical Methods}
\label{sec:mhd}

\subsection{Ideal SRMHD Equations}

Special relativistic ideal magnetohydrodynamic equations (SRMHD) can be 
derived from the conservation of the rest mass, 
the local conservation of the stress-energy 
tensor $T^{\mu \nu}$ and Maxwell equations

\begin{equation}
 \partial_{\nu} (\rho_{0}u^{\nu}) = 0, ~~ \partial_{\nu} (T^{\mu \nu}) = 0, ~~
 \partial_{\nu} (^{*}F^{\mu \nu}) = 0 \label{eq:LC}, 
\end{equation}

\noindent  where $\rho_{0}$ is the rest mass density of a fluid, 
$u^{\mu}$ is the 4-velocity of the fluid elements  and 
$^{*}F^{\mu \nu} = \frac{1}{2}\epsilon^{\mu \nu \delta \lambda}F_{\delta \lambda}$ 
are the components of the Faraday dual tensor, 
where $\epsilon^{\mu \nu \delta \lambda}$ are the components 
of the Levi-Civita tensor. We use the rescaled Faraday tensor 
and its dual with  the factor $1/\sqrt{4\pi}$, in order to avoid 
the inclusion of the permittivity and permeability of free 
space in cgs-Gaussian units. Notice that our definitions 
assume geometric units where $c=1$.

In this work we consider the magnetized fluid is described 
by the sum of the stress-energy tensor of a perfect fluid $T^{\mu \nu}_{fluid}$ 
and Maxwell stress-energy tensor $T^{\mu \nu}_{EM}$

\begin{eqnarray}
T^{\mu \nu} &=& T^{\mu \nu}_{fluid} + T^{\mu \nu}_{EM},  \\
T^{\mu \nu}_{fluid} &=& \rho_0 h u^{\mu}u^{\nu} + p \eta^{\mu \nu}, 
\label{eq:TEMfluid} \nonumber \\ \nonumber
T^{\mu \nu}_{EM}  &=& \left( u^{\mu}u^{\nu} 
+ \frac{1}{2}\eta^{\mu\nu}\right) b^2 -b^{\mu}b^{\nu} \label{eq:EM_Maxw},
\end{eqnarray}

\noindent where $p$ is the fluid pressure, 
$b^{\mu}=u_{\nu} ^{*}F^{\mu \nu}$ is the 
component of the Maxwell tensor parallel to the 4-velocity 
of the fluid (magnetic 4-vector measured by an observer comoving with the fluid), 
$b^2 = b^{\mu}b_{\mu}$ is the magnitude of the magnetic 
field, $\eta_{\mu \nu}$ are the components of the Minkowski 
metric and $h=1 + \varepsilon + p/\rho_{0}$ is the specific enthalpy, 
where $\varepsilon$ is the rest frame specific internal energy density of the fluid.
Then, the total stress-energy tensor reads

\begin{equation}
T^{\mu \nu} = \rho_0 h^* u^{\mu}u^{\nu} + p^*\eta^{\mu \nu} - b^{\mu}b^{\nu} 
\label{eq:TotalEMT},
\end{equation}  

\noindent where $p^*=p+p_m$ is the addition of the gas pressure $p$ 
and the magnetic pressure $p_m = b^2/2$. In the same way, 
$h^* = h + h_m$ is the sum of the specific enthalpy  of the fluid $h$ and 
the specific magnetic enthalpy $h_m = b^2/\rho_0$.

In order to track the evolution of the fluid it is convenient to 
write down SRMHD as a system of flux balance laws 
\citep{2006ApJ...637..296A,2008LRR....11....7F} on the Minkowski 
space-time. Following \citep{2006ApJ...637..296A}, the evolution 
scheme is written as a first-order hyperbolic system of flux-conservative 
equations for the conservative variables ($D,M_j,\tau, B^k$), 
which are defined in terms of the primitive variables ($\rho_0,v_j,p,B^k$) as

\begin{eqnarray}
 D &=& \rho_0 W, \\
 M_j &=& \rho_0h^* W^2 v_j - b^0b_j, \\
 \tau &=&  \rho_0h^* W^2 - p^* - (b^0)^2 - D, \\
 B^k &=& B^k,
\end{eqnarray}

\noindent where $W = u^0$ is the Lorentz factor, $v^{i}$ 
is the 3-velocity measured  by an Eulerian observer and defined 
in terms of the spatial part of the 4-velocity $u^{i}$, 
as $v^{i}=\frac{u^{i}}{W}$ and 
$B^k$ is the spatial magnetic field measured by an Eulerian observer. 
Thus, the SRMHD Euler equations (\ref{eq:LC}) form a system of conservation laws,
which can be written, in cartesian coordinates, as

\begin{eqnarray}
&& \partial_0 {\bf f}^{0} 
+ \partial_i {\bf f}^{i} 
= {\bf 0}, \label{eq:valencia} \\ 
\nonumber \\
&& {\bf \nabla} \cdot {\bf B} = 
\partial_i B^{i}
= 0 \label{eq:constraint}, 
\end{eqnarray}

\noindent where ${\bf f}^{0}$ is the vector whose 
entries are the conservative variables and the vector ${\bf f}^{i}$ contains
the fluxes along the spatial directions. All these ingredients are explicitly

\begin{eqnarray}
&&{\bf f}^{0} = \left[ D,M_{j},\tau,B^k \right], \label{eq:cvars} \\ \nonumber
&&{\bf f}^{i} = [ D v^i, M_{j} v^i + p^* \delta^{i}_{j} - b_jB^i/W,  \\
&& ~~~~~~~~~ \tau v^i + p^*v^i - b^0 B^i/W,
v^i B^k - v^k B^i ] \label{eq:fluxes},
\end{eqnarray}

\noindent where $\delta^i_j$ is the Kronecker delta. 

In these definitions the components of the magnetic field measured by the comoving observer and an Eulerian observer are related as follows 

\begin{eqnarray}
b^0 &=& W B^iv_i, \\
b^i &=& \frac{B^i}{W} + b^0 v^i,
\end{eqnarray}

\noindent where the magnitude of the magnetic field can be written as

\begin{equation}
b^2 = \frac{B^2 + (b^0)^2}{W^2},
\end{equation}

\noindent with $B^2 = B^iB_i$. Finally, the RMHD system of equations (\ref{eq:valencia},\ref{eq:constraint}), is a set of eight equations for either, the primitive variables $\rho_0,v_j,\varepsilon,p,B^k$ or the conservative variables $D,S_j,\tau,B^k$. As usual, the system is closed with an equation of state relating $p=p(\rho_{0},\varepsilon)$. Specifically, we choose the fluid to obey an ideal gas equation of state $p = (\Gamma-1)\rho_{0}\varepsilon$, where $ \Gamma $ is the adiabatic index.

\subsection{Numerical Methods}

For a time-dependent PDE problem, a complete basic solver has several components: grid generation, initial conditions, boundary conditions, spatial discretization and time integration. In our case, the relativistic magnetized Euler evolution equations are solved on a single uniform cell centered grid. The integration in time uses the method of lines, with a third order total variation diminishing Runge-Kutta time integrator (RK3) \citep{Shu1988439}. The SRMHD equations are discretized using a finite volume approximation together with High Resolution Shock Capturing methods. Thus, the system of equations (\ref{eq:valencia}) can be expressed in a semi-discrete form as follows

\begin{eqnarray}
\nonumber \frac{d}{dt} {\bf f}^{0}_{(i,j,k)} =  
&-& \frac{ {\bf F}^{x}_{(i+1/2,j,k)} - {\bf F}^{x}_{(i-1/2,j,k)} }{\Delta x} \\  
&-& \frac{ {\bf F}^{y}_{(i,j+1/2,k)} - {\bf F}^{y}_{(i,j-1/2,k)} }{\Delta y} \\ \nonumber 
&-& \frac{ {\bf F}^{z}_{(i,j,k+1/2)} - {\bf F}^{z}_{(i,j,k-1/2)} }{\Delta z}, \label{eq:dis}
\end{eqnarray}

\noindent where ${\bf F}^{x}_{i\pm 1/2,j,k}$, ${\bf F}^{y}_{i,j\pm 1/2,k}$ and ${\bf F}^{z}_{i,j,k\pm 1/2}$ are the numerical fluxes at the respective cell interface. The rigth hand side of this expression is what we will call right hand side of the conservative variables from now on.

CAFE provides different types of spatial reconstruction schemes which are applied on the primitive variables $\{\rho, v^j, p, B^k\}$. Specifically, we use the MINMOD and MC linear piecewise reconstructors, which are second-order methods. For higher order reconstructions, we use the third-order piecewise parabolic method (PPM), which was developed by \citep{1984JCoPh..54..174C} and adapted to the relativistic case by \citep{1996JCoPh.123....1M}. We use this recipe with parameters $K_{0}=1.0$, $\eta^{1}=5.0$,  $\eta^{2}=0.5$, $\epsilon^{1}=0.1$,  $\epsilon^{2}=1.0$, $\omega^{(1)}=0.52$ and  $\omega^{(1)}=10.0$. In regions near contact discontinuities, the interpolation is modified in such a way that in the vicinity of local extrema, the scheme reduces to a piecewise constant approximation in order to avoid shock oscillations. We also use the fifth-order weighted Essentially Non Oscillatory (WENO5), which approaches the variables with high order of accuracy using polynomial interpolation, see \citep{Titarev2004238, Harten19973}, and is efficient at capturing the structure of turbulent fluxes \citep{2012A&A...547A..26R}. Our framework is such that other schemes can be incorporated easily.

\subsubsection{Approximate Riemann Solver}

In order to solve the system of equations \ref{eq:dis}, we have implemented the HLLE \citep{doi:10.1137/1025002,doi:10.1137/0725021} approximate Riemann solver formula 

\begin{equation}
{\bf F}^i = \frac{\xi^{+}{\bf f}^{i}_L - \xi^{-}{\bf f}^{i}_R + \xi^{+}\xi^{-}({\bf f}^{0}_R - {\bf f}^{0}_L)}{\xi^{+} - \xi^{-}},
\end{equation}

\noindent where ${\bf f}^i_L$ and ${\bf f}^i_R$ are the fluxes whereas ${\bf f}^0_L$ and ${\bf f}^0_R$ are the conservative variables at the left and right sides of the interface between cells;  $\xi^{+} = max(0,\xi^p_R,\xi^p_L)$, $\xi^{-} = min(0,\xi^p_R,\xi^p_L)$ and $\xi^p_{L,R}$ denote the $p$ eigenvalue of the Jacobian matrix for the left and right states respectively.

Different approximate Riemann solvers require different characteristic information from the Jacobian matrix 
$ {\cal \vec{A}}^i = \partial {\bf f}^{(i)} / \partial {\bf f}^{(0)}$, for instance the Marquina \citep{doi:10.1137/0915054} and Roe \citep{Roe1981357} approximate Riemann solvers, require the eigenvalues and eigenvectors. One of the appealing properties of the HLLE approximate Riemann solver is that it requires only the eigenvalues of the Jacobian matrix. Specifically, the HLLE solver uses a two-wave approximation to compute the fluxes across the discontinuity at the cell interfaces. One disadvantage of this flux formula is that it does not resolve properly the contact discontinuity and it is more dissipative than other methods like HLLC which actually does solve the contact discontinuity \citep{1994ShWav...4...25T,2005MNRAS.364..126M,2006MNRAS.368.1040M}. However, less dissipative formulas may produce undesirable effects, such as the carbuncle artifact along the axis of propagation of strong shocks \citep{2008ApJS..176..467W}.

Furthermore, in our HRSC scheme that solves the relativistic magnetized Euler equations (\ref{eq:valencia}), in order to calculate the eigenvalue structure, we follow the formalism introduced in  \citep{1989rfmw.book.....A} and further developed in \citep{2010ApJS..188....1A,2006ApJ...637..296A}. The expressions of the seven eigenvalues associated to entropic, Alfv\'en fast and slow magnetosonic waves are the following:

\begin{eqnarray}
&&\xi =  v^i ,  \\ \nonumber 
\\ 
&&\xi_{\pm} = \frac{b^i \pm \sqrt{{\cal E}} W v^i}{b^0 \pm \sqrt{{\cal E}} W }, 
\end{eqnarray}

\noindent where ${\cal E} = \rho_0 h + b^2$ is the total energy density measured by an observer comoving with the fluid. The fast and slow magnetosonic waves, which are required in the computation of the numerical fluxes, can be obtained by solving the following quartic equation in each direction $i$ for the unknown $\xi$
\begin{equation}
\rho_0 h \left(\frac{1}{c_s^2} - 1 \right) a^4 - \left(\rho_0h + \frac{b^2}{c_s^2} \right)a^2 {\cal G} + {\cal H}^2 {\cal G} = 0, \label{eq:quartic}
\end{equation} 
where 
\begin{eqnarray}
\nonumber a &=& W(-\xi + v^i), 
~~~~ {\cal H} = b^i - b^0 \xi, \\
\nonumber {\cal G} &=& -\xi^2 + 1,
\end{eqnarray}

\noindent and  $c_s$ is the sound speed.

In order to solve this equation, we use an approximate method for the calculation of the eigenvalues related to the fast magnetosonic waves. The method was introduced by \citep{2005A&A...436..503L} and basically consists in reducing the original quartic equation (\ref{eq:quartic}) to a quadratic equation, which can be solved analytically. Finally,  Due to the structure of the HLLE formula, which uses the upper and lower bounds of the eigenvalues, it is not necessary to incorporate the slow magnetosonic waves.

\subsubsection{The divergence-free magnetic field constraint}

Although the analytic solutions of Maxwell equations guarantee the constraint (\ref{eq:constraint}), the experience shows that it is not the case when calculating numerical solutions of these equations. Instead, the numerical evolution of initial data involving  Maxwell equations may eventually lead to a violation of the divergence free constraint (\ref{eq:constraint}), giving as a consequence the development of unphysical results like the presence of magnetic monopolar sources. \\ \\

{\it a. Flux Constraint Transport} \\ \\

\begin{figure}
\begin{center}
\includegraphics[width=7.0cm]{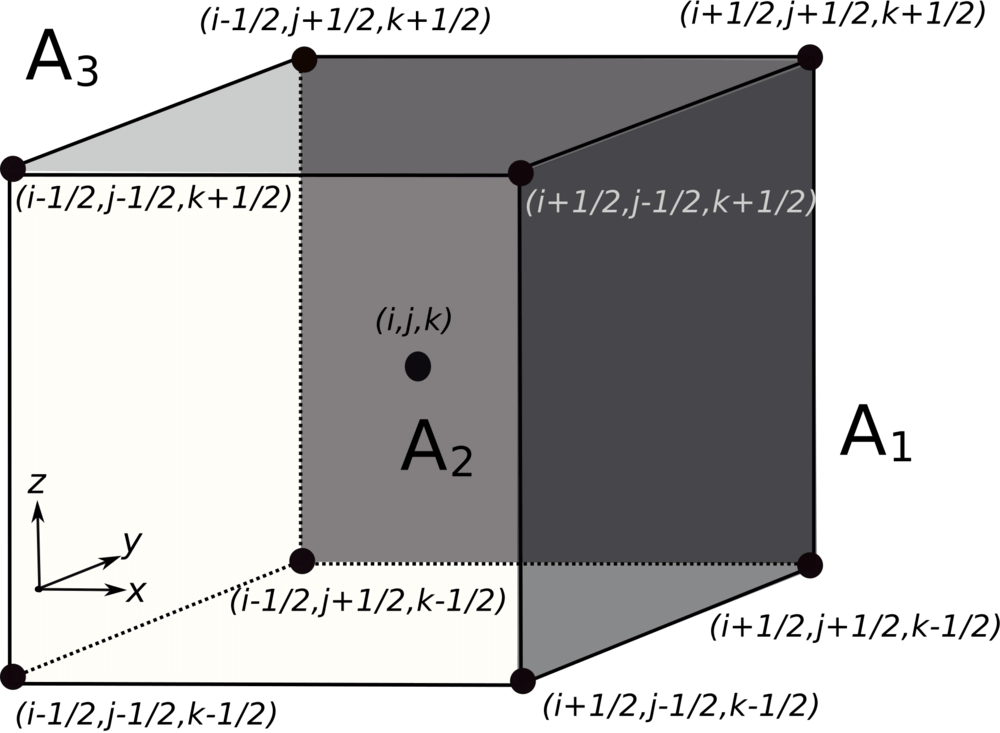}
\end{center}
\caption{\label{fig:cube} This figure shows a numerical cell centered at $(i,j,k)$, and contains the detailed labels of the corners required in the calculations.}
\end{figure}

There are various methods available to control the growth of the constraint violation \citep[e.g.,][]{2000JCoPh.161..605T}, in our code we use a version of the constrained transport method ``CT'' originally proposed in \citep{1988ApJ...332..659E}, which is based on the use of the fluxes computed with a conservative scheme. This algorithm is known as flux-CT and was proposed in \citep{2001ApJS..132...83B}. Following \citep{2007CQGra..24S.235G}, the resulting discretized cell-faced evolution equations for the magnetic field components are given in terms of ${\bf \Omega} \equiv {\bf v} \times {\bf B}$ as

\begin{eqnarray}
\frac{dB^x_{(i+\frac{1}{2},j,k)}}{dt}&=&  \frac{\Omega^{z}_{(i+\frac{1}{2},j+\frac{1}{2},k)}-\Omega^{z}_{(i+\frac{1}{2},j-\frac{1}{2},k)}}{\Delta y} \label{eq:dBxdt}\\
&-& \frac{\Omega^{y}_{(i+\frac{1}{2},j,k+\frac{1}{2})}-\Omega^{y}_{(i+\frac{1}{2},j,k-\frac{1}{2})}}{\Delta z} , \nonumber \\
\frac{dB^y_{(i,j+\frac{1}{2},k)}}{dt}&=&  \frac{\Omega^{x}_{(i,j+\frac{1}{2},k+\frac{1}{2})}-\Omega^{x}_{(i,j+\frac{1}{2},k-\frac{1}{2})}}{\Delta z} \label{eq:dBydt}\\
&-& \frac{	\Omega^{z}_{(i+\frac{1}{2},j+\frac{1}{2},k)}-\Omega^{z}_{(i-\frac{1}{2},j+\frac{1}{2},k)}}{\Delta x}, \nonumber \\
\frac{dB^z_{(i,j,k+\frac{1}{2})}}{dt}&=&  \frac{\Omega^{y}_{(i+\frac{1}{2},j,k+\frac{1}{2})}-\Omega^{y}_{(i-\frac{1}{2},j,k+\frac{1}{2})}}{\Delta x} \label{eq:dBzdt}  \\
&-& \frac{\Omega^{x}_{(i,j+\frac{1}{2},k+\frac{1}{2})}-\Omega^{x}_{(i,j-\frac{1}{2},k+\frac{1}{2})}}{\Delta y},\nonumber
\end{eqnarray}

\noindent where $B_{x(i+1/2,j,k)}$, $B_{y(i,j+1/2,k)}$ and 
$B_{z(i,j,k+1/2)}$ are the magnetic field components defined as the average on each surface $A_1$, $A_2$ and $A_3$ 
respectively, see Figure \ref{fig:cube}. Moreover, the $\Omega$ terms in the right hand side of equations (\ref{eq:dBxdt},\ref{eq:dBydt},\ref{eq:dBzdt}) are defined at the cell vertex. These terms are computed using the simple average of the neighboring values of the numerical fluxes at the intercells as follows

\begin{eqnarray}
\Omega^{x}_{(i,j+1/2,k+1/2)} &=& \frac{1}{4} (  F^{yz}_{(i,j+\frac{1}{2},k)} + F^{yz}_{(i,j+\frac{1}{2},k+1)} \\
                       &-& F^{zy}_{(i,j,k+\frac{1}{2})} - F^{zy}_{(i,j+1,k+\frac{1}{2})}  ),  \label{eq:omegafluxes} \nonumber \\
\Omega^{y}_{(i+1/2,j,k+1/2)} &=& \frac{1}{4} (  F^{zx}_{(i,j,k+\frac{1}{2})} + F^{zx}_{(i+1,j,k+\frac{1}{2})} \\
                       &-& F^{xz} _{(i+\frac{1}{2},j,k)} - F^{xz}_{(i+\frac{1}{2},j,k+1)}  ), \nonumber\\
\Omega^{z}_{(i+1/2,j+1/2,k)} &=& \frac{1}{4}  (  F^{xy}_{(i+\frac{1}{2},j,k)} + F^{xy}_{(i+\frac{1}{2},j+1,k)} \\
                       &-& F^{yx}_{(i,j+\frac{1}{2},k)} - F^{yx}_{(i+1,j+\frac{1}{2},k)}  ), \nonumber                      
\end{eqnarray}

\noindent where {\small $\Omega^{x} = -F^{zy} = F^{yz}$}, {\small $\Omega^{y} = F^{zx} =- F^{xz}$} and {\small $\Omega^{z} = -F^{yx} = F^{xy}$ }, are the appropriate flux components $F^{ij}$ computed with the HLLE flux formula along each spatial direction.\\ 

\begin{figure}
\begin{center}
\includegraphics[width=7.0cm]{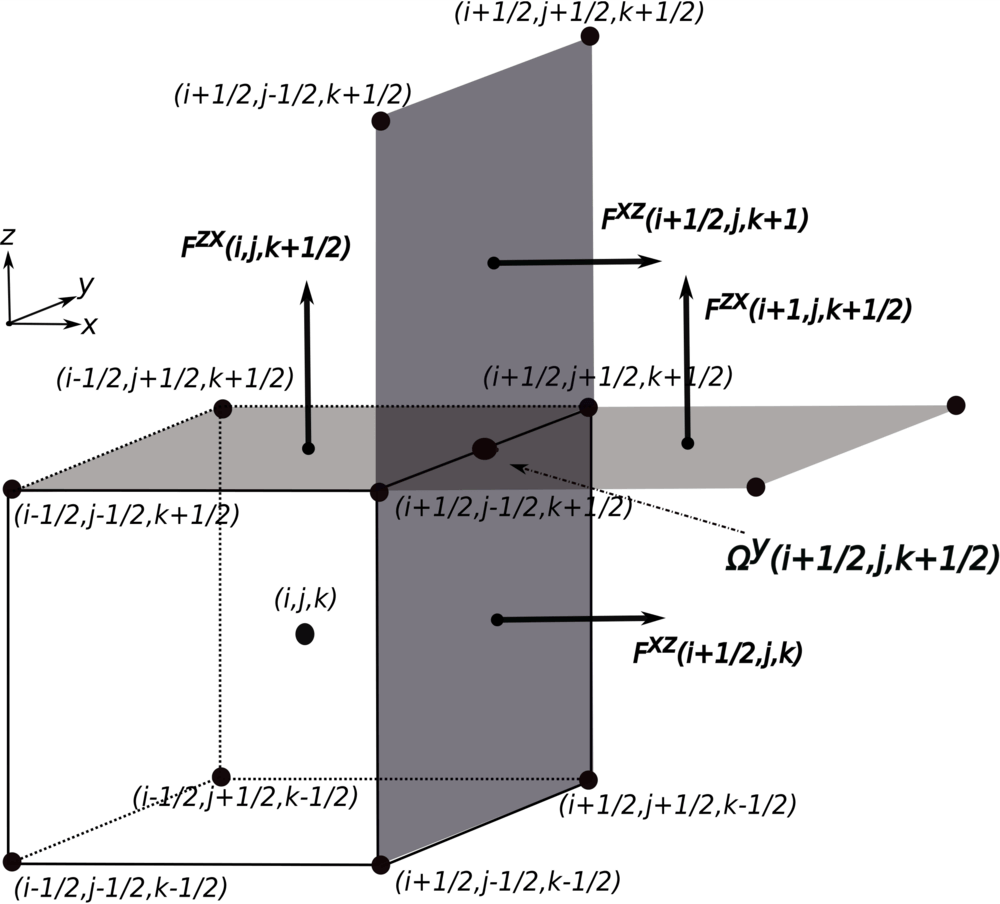}
\end{center}
\caption{\label{fig:cruz} Numerical cell centered at $(i,j,k)$ and two adjacent faces. We show an example to illustrate the cell corner interpolation of the numerical fluxes using the four values computed using the HLLE flux formula. We can see that the fluxes $F^{xz}$ and $F^{zx}$ are computed along $x$ and $z$ directions at points {\small $(i,j,k+\frac{1}{2})$, $(i+1,j,k+\frac{1}{2})$, $(i+\frac{1}{2},j,k)$} and {\small $(i+\frac{1}{2},j,k+1)$}. We use this values to compute {\small $\Omega^{y}_{(i+1/2,j,k+1/2)}$} using the formula given by the equation (\ref{eq:omegafluxes}). }
\end{figure}

In Figure \ref{fig:cruz} we show an example of the values used to compute the cell vertex flux $\Omega^{y}_{(i+1/2,j,k+1/2)}$. As we can see, for example in equation (\ref{eq:omegafluxes}), we need the four values $F^{zx}_{(i,j,k+\frac{1}{2})}$, $F^{zx}_{(i+1,j,k+\frac{1}{2})}$, $F^{xz} _{(i+\frac{1}{2},j,k)}$ and $F^{xz}_{(i+\frac{1}{2},j,k+1)}$, where two of these are computed along the $x$-direction whereas the other two along the $z$-direction. The procedure to calculate $\Omega^x$ and $\Omega^z$ is  similar.

Once the magnetic field is evolved at the interfaces, in order to compute the cell centered magnetic field  we use a simple average in each direction: {\small $B^x_{(i,j,k)}=\frac{1}{2}(B^x_{(i-\frac{1}{2},j,k)} + B^x_{(i+\frac{1}{2},j,k)} )$} and similarly {\small  $B^y_{(i,j,k)}=\frac{1}{2}( B^y_{(i,j-\frac{1}{2},k)}+ B^y_{(i,j+\frac{1}{2},k)})$} and {\small $B^z_{(i,j,k)}= \frac{1}{2}(B^z_{(i,j,k-\frac{1}{2})}+B^z_{(i,j,k+\frac{1}{2})})$}. Finally, the discretized magnetic field divergence, calculated at the corner of the grid, is given by

\begin{eqnarray}
&&\nabla \cdot {\bf B}(i+1/2,j+1/2,k+1/2) = \nonumber \\
&&\frac{1}{4 \Delta x}\sum_{jj=j}^{j+1}\sum_{kk=k}^{k+1}(B^{x}(i+1,jj,kk)-B^{x}(i,jj,kk) ) + \nonumber\\
&&  \frac{1}{4 \Delta y}\sum_{ii=i}^{i+1}\sum_{kk=k}^{k+1}(B^{y}(ii,j+1,kk)-B^{y}(ii,j,kk) ) +\nonumber\\
&& \frac{1}{4 \Delta z}\sum_{ii=i}^{i+1}\sum_{jj=j}^{j+1}(B^{z}(ii,jj,k+1)-B^{z}(ii,jj,k) ),\nonumber
\end{eqnarray}

\noindent which is the discretized expression the constraint has to be monitored with.

{\it b. Hyperbolic divergence cleaning} \\ \\

We also implemented the divergence cleaning method, which preserves the magnetic field constraint by solving a modified version of Maxwell equations $\partial_{\nu}\big(^{*}F^{\mu \nu}+ \eta^{\mu \nu}\psi\big)= \kappa n^{\mu}\psi$, where a new field variable $\psi$ and a diffusive term $\kappa n^{\mu}\psi$ are added. The parameter $\kappa$ may be adjusted in order to absorb errors in the constraint, and $\psi$ vanishes when ${\bf \nabla} \cdot {\bf B}=0$ is exactly satisfied. Here $n^{\mu}$ is the normal vector to the hypersurface. The hyperbolic divergence cleaning method used was proposed by \citep{2002JCoPh.175..645D}. To see how explicitly Maxwell equations are modified, we refer the reader to \citep{2011MNRAS.414.1467P,2010PhRvD..81l4023L}. The field variable $\psi$ satisfies the damped wave equation $\partial_{\nu} \partial^{\nu} \psi = -\kappa \partial_{\nu} (n^{\nu}\psi)$, this means that the amplitude of $\psi$ decreases in time during the evolution, recovering the unmodified Maxwell equations. Thus, the modified Maxwell equations in special relativity are written as

\begin{equation}
{\small \partial_{t} B^{j} + \partial_{i}\left(v^{i}B^{j}-v^{j}B^{i}+\eta^{ij}\psi \right)= 0},
\end{equation}
 
\noindent and aditionally, we have an evolution equation for the 
new variable $\psi$ given by

\begin{equation} 
\partial_{t}\psi + \partial_{i} B^{j}  = -\kappa \psi .
\end{equation}
 
\noindent which is incorporated to the set of evolution equations. 

\subsubsection{Recovery of primitive variables}

The code evolves the conservative $\{D, S_j,\tau, B^k\}$, but not the primitive variables $\{\rho, v^j, p, B^k\}$. However, the numerical fluxes depend on both sets of variables. Therefore, after each time step within the evolution scheme, one needs to recover the primitive variables out of the conservative ones. By definition, the conservative quantities can be written in terms of the primitives, however a solution to the inverse problem is not known and a numerical algorithm is required.

The method we use is based on \citep{2006MNRAS.368.1040M} and is as follows. As a starting point, the definitions of  $M^{i}$, $\tau$, and an auxiliary variable $Z = \rho h W^{2}$ are used. Then, it is necessary to compute $M^{2}=M_{i}M^{i}$ and $\tau$ in terms only of $Z,W$ and $B$ as follows

\begin{eqnarray}
\nonumber  M^{2} &=& (Z+B^{2})^{2} \left( 1-W^{-2}\right)-\left(\frac{{\bf B\cdot M}}{Z} \right)^{2}(2Z+B^{2}),\label{eq:ss} \\ \\
 \tau &=& Z+B^{2}-p-D-\frac{B^{2}}{2W^{2}} -\frac{1}{2}\left(\frac{{\bf B\cdot M}}{Z} \right)^{2}, \label{eq:tau}
\end{eqnarray}

\noindent where $B^{2}=B_{i}B^{i}$ and $p$ can be expressed in terms of $Z$ and $W$

\begin{equation}
p  = \frac{\Gamma-1}{\Gamma}\frac{Z-DW}{W^{2}}.\label{eq:press}
\end{equation}

\noindent From equation (\ref{eq:ss}) it is possible to express the Lorentz factor in terms of $Z$ , $B^2$ and ${\bf B \cdot M}$

\begin{equation}
W  = \frac{1}{\sqrt{1 - \frac{ ({\bf B\cdot M})^{2} (2Z+B^{2}) + M^{2} Z^{2}}{(Z+B^{2} )^{2} Z^{2} }}},\label{eq:Lorentz}
\end{equation}

\noindent and then substituting this into (\ref{eq:tau}) we obtain 

\begin{equation}
f(Z)= Z+B^{2}-(\tau+D)-p-\frac{B^{2}}{2W^{2}} -\frac{1}{2}\left(\frac{{\bf B\cdot M}}{Z} \right)^{2} = 0. \label{eq:fun}
\end{equation} 

\noindent In order to solve this algebraic equation, we use a numerical iterative algorithm, which is a combination of the Newton-Raphson and bisection methods \citep{1992nrfa.book.....P}. The Newton-Raphson method requires the derivative $df(Z)/dZ$

\begin{equation}
\frac{df(Z)}{dZ}= 1 -\frac{dp}{dZ}+\frac{B^{2}}{W^{3}}\frac{dW}{dZ} + \frac{({\bf B\cdot M})^{2}}{Z^{3}}, \label{eq:dfun}
\end{equation}

\noindent where 

\begin{eqnarray}
\nonumber \frac{dp}{dZ}  &=& \frac{\Gamma-1}{\Gamma}\frac{ W(1+DdW/dZ)- 2ZdW/dZ}{W^{3}},\label{eq:dpress_rec} \\ \nonumber \\
\nonumber \frac{dW}{dZ}  &=& -W^{3}\frac{ M^2 Z^3 + (B^4 + 3 B^2 Z + 3 Z^2) ({\bf B \cdot M})^2}{Z^3 (B^2 + Z)^3}.
\end{eqnarray}

Finally, after calculating $Z$ it is possible to find the other primitive variables as follows: once $Z$ is known, $W$ can be recovered from equation (\ref{eq:Lorentz}), then the pressure from equation (\ref{eq:press}), $\rho$ from the definiton of $D$ and the velocity components from the expresion 

\begin{equation}
v^{i} =\frac{S^{i}+({\bf B\cdot M}) B^{i}/Z}{Z+B^{2}},
\end{equation}

\noindent which is obtained from the definiton of $M_{i}$.

\section{Numerical Tests}
\label{sec:tests}

The first set of tests involve the evolution with the magnetic field switched off, which is the domain of the Relativistic Hydrodynamics (RHD), considered to be as important because we have not shown prior evidence of the ability of our code to handle this system. A second set of tests involves non-trivial magnetic fields, and includes a complete set of RMHD tests.

\subsection{RHD Tests}

In order to illustrate how our implementation handles the evolution of a relativistic gas, in this subsection we present the standard tests showing that our code works properly. The 1D tests are Riemann problems under various conditions and we compare the numerical results with the exact solution we implemented based on \citep{1994JFM...258..317M,2003LRR.....6....7M,2013arXiv1303.3999L}.

We calculate the numerical solution of these tests using various limiters, however, unless otherwise specified: all the results in the 1D test figures corresponding to Reimann problems use the HLLE formula and the MC limiter, the problem is solved in the domain $[-0.5,0.5]$ with $N = 400$ identical cells,  a Courant factor $CFL=0.25$, and with the initial discontinuity located at $x=0$. The resolutions we have used for the error estimates are $\Delta x_1=1/200$, $\Delta x_2=1/400$, $\Delta x_3=1/800$,  $\Delta x_4=1/1600$, $\Delta x_5=1/3200$ and $\Delta x_6=1/6400$. For these tests we are using the 3D code with five cells along the transverse directions. The various parameters of 1D tests are summarized in Table \ref{tab:RHD_tests}.

\begin{table}
\centering
\resizebox{8cm}{!}{
\begin{tabular}{c|c|c|c|c|c|c}\hline \hline
${\bf Test ~ type}$ & $\Gamma$ & $\rho_0$ & $p$ & $v^x$ & $v^y$ & $v^z$  \\ \hline \hline
${\bf Test ~ 1}$ &  &  &  &  &  &   \\
Left state  & 5/3 & 10.0 & 13.33 & 0.0 & 0.0  & 0.0   \\ 
Right state &  & 1.0 & $10^{-8}$ & 0.0  & 0.0  & 0.0   \\ \hline \hline
${\bf Test ~ 2}$ &  &  &  &  &  &   \\
Left state  & 5/3 & 1.0 & 1000.0 & 0.0 & 0.0  & 0.0   \\ 
Right state &  & 1.0 & 0.01 & 0.0  & 0.0  & 0.0    \\ \hline \hline
${\bf Test ~ 3}$ &  &  &  &  &  &   \\
Left state  & 4/3 & 1.0 & 0.001 & 0.999999995 & 0.0  & 0.0  \\ 
Right state &  & 1.0 & 0.001 & -0.999999995  & 0.0  & 0.0    \\  \hline \hline
${\bf Test ~ 4}$ &  &  &  &  &  &   \\
Left state  & 4/3 & 1.0 & 1.0 & 0.9 & 0.0  & 0.0   \\ 
Right state &  & 1.0 & 10.0 & 0.0  & 0.0  & 0.0    \\ \hline \hline
${\bf Test ~ 5}$ &  &  &  &  &  &   \\
Left state  & 5/3 & 1.0 & 1000.0 & 0.0 & 0.0  & 0.0   \\ 
Right state &  & 1.0 & 0.01 & 0.0  & 0.99  & 0.0    \\ \hline \hline
${\bf Test ~ 6}$ &  &  &  &  &  &   \\
Left state  & 5/3 & 1.0 & 1000.0 & 0.0 & 0.9  & 0.0   \\ 
Right state &  & 1.0 & 0.01 & 0.0  & 0.9  & 0.0    \\ \hline \hline
\end{tabular}
}
\caption{\label{tab:RHD_tests} Parameters for the various RHD 1D Riemann problems.} 
\end{table}

\subsubsection{Test 1: Relativitic Blast Wave (a)}

\begin{figure}
\begin{center}
\includegraphics[width=7.0cm]{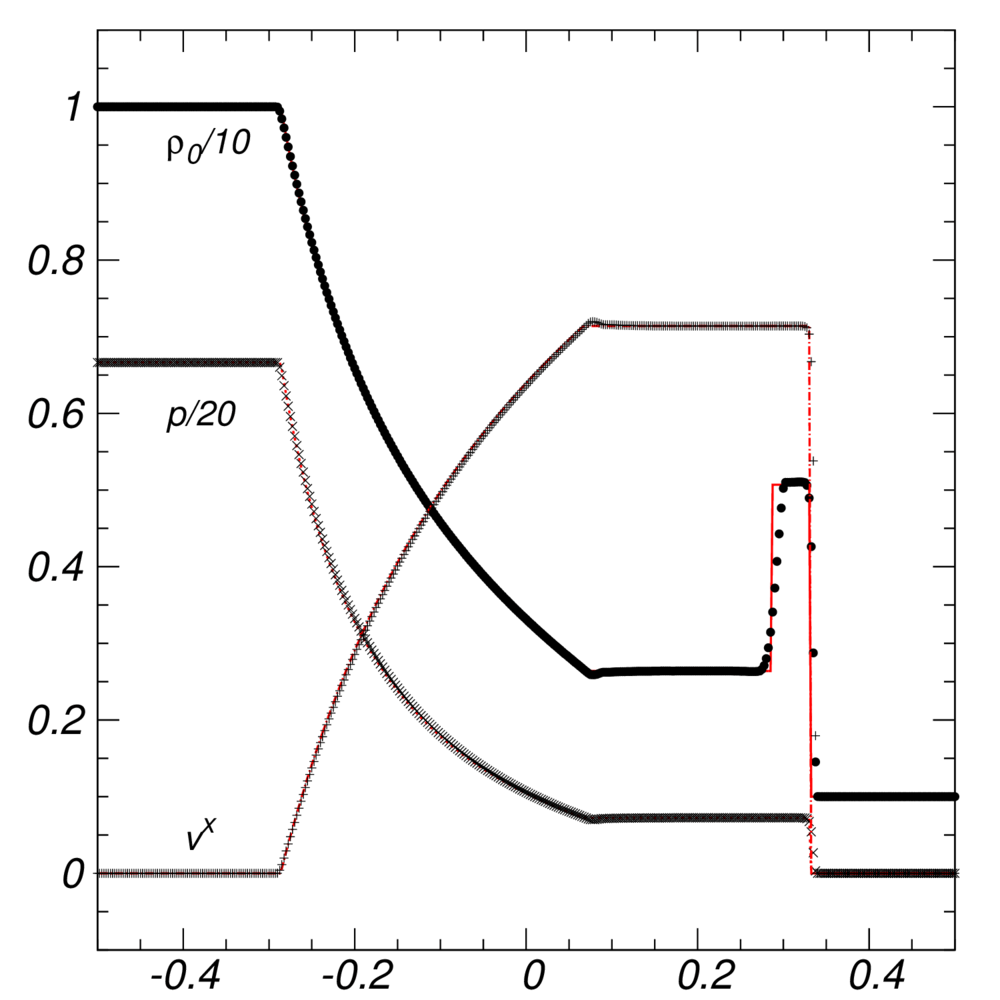}
\end{center}
\caption{\label{fig:Midst1D} Test 1: mildly relativistic blast wave explosion problem at $t=0.4$. We show proper rest mass density, pressure and velocity. }
\end{figure}

The first 1D Riemann problem test corresponds to a mildly relativistic blast wave explosion, characterized by an initial static state with higher pressure in the region on the left. The results can be seen in Figure \ref{fig:Midst1D}, where we compare the numerical solution (points) with the exact solution (lines). The comparison between the exact and numerical solutions is as good as that obtained by other codes \citep[e.g.,][]{2003LRR.....6....7M,2002A&A...390.1177D,2006ApJS..164..255Z}. In this test the most important feature is that with a relatively small number of cells the shock speed is pretty much the exact one.

\subsubsection{Test 2: Relativitic Blast Wave (b)}

\begin{figure}
\begin{center}
\includegraphics[width=7.0cm]{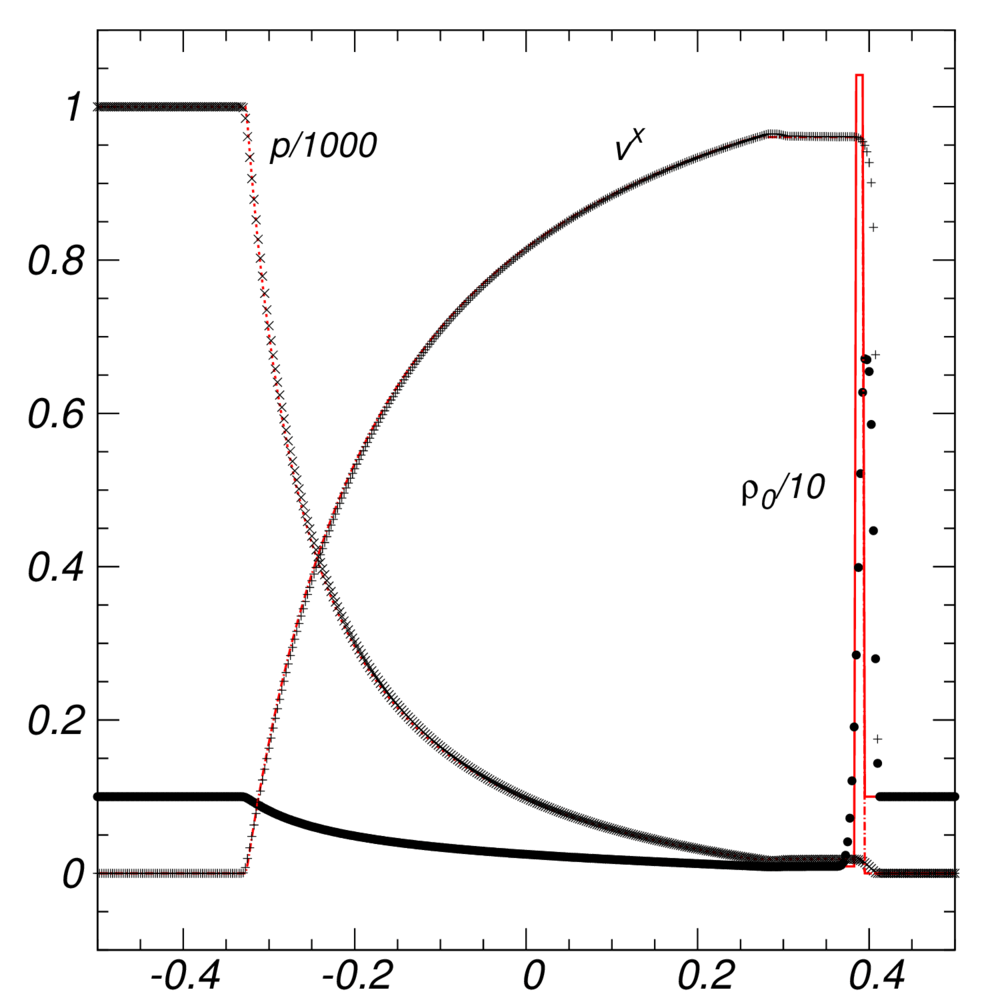}
\end{center}
\caption{\label{fig:Strong1D} Test 2: strong relativistic blast wave problem at time $t = 0.4$. As in the previous case, proper rest mass density, pressure and velocity are shown. The value of the numerical density peak of magnitude $7.3$, is similar to what is found with other schemes \citep[e.g.,][]{2003LRR.....6....7M,2002A&A...390.1177D}.}
\end{figure}

In this problem, unlike the previous one, the evolution of the initial discontinuity produces a sharper  blast moving to the right. The standard initial data are those in \citep{2003LRR.....6....7M}.  In Figure \ref{fig:Strong1D} we show our results and contrast them with the exact solution. Due to the important difference of pressure between the left and right states,  behind the shock there is an extremely thin dense shell, which is a feature expected to be controlled by a code. The fact that the thin shell is not well resolved is a matter of resolution, and in this particular case $\Delta x_1$ is not enough, however with $\Delta x_2$ the shell is well resolved and within the convergence regime.

\subsubsection{Test 3: Head-on Stream Collision}

\begin{figure}
\begin{center}
\includegraphics[width=7.0cm]{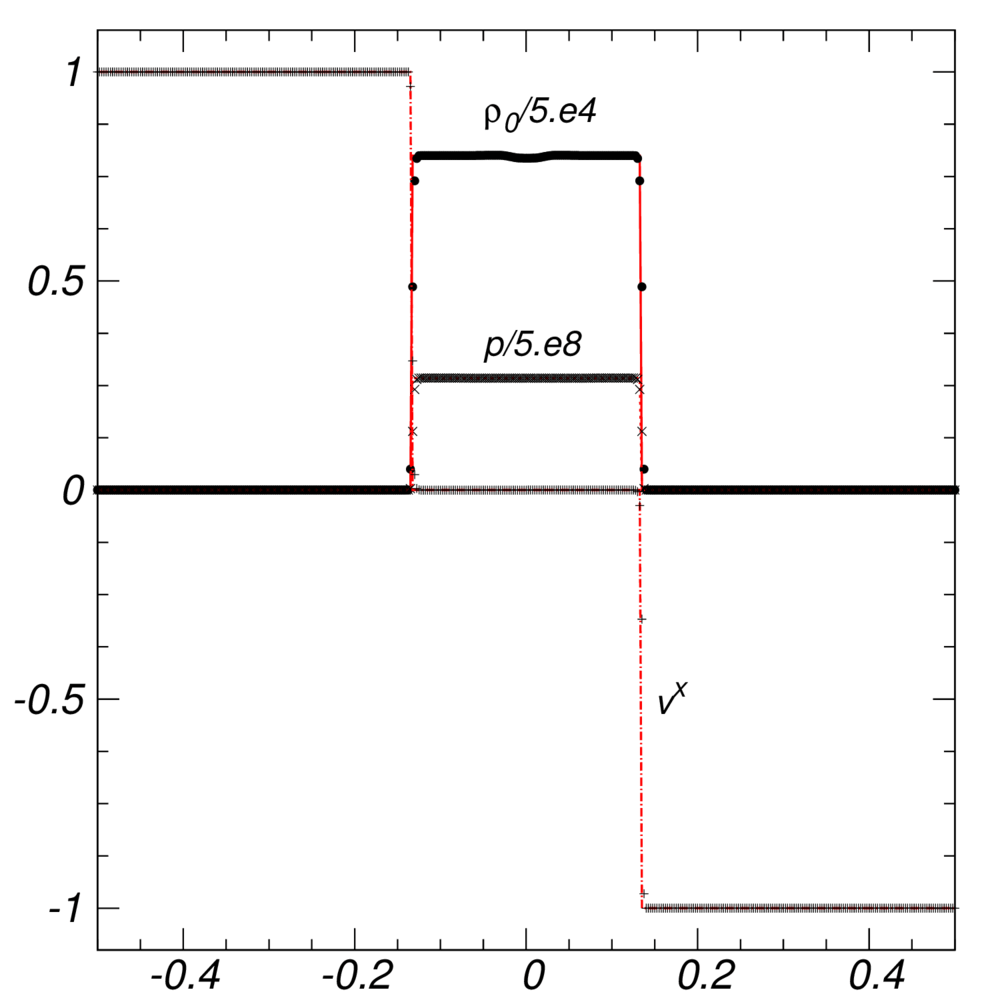}
\end{center}
\caption{\label{fig:Headon} Test 3: head-on Stream Collision. This test shows that our implementation is capable to track the evolution of the fluid with Lorentz factors of the order of $10000$. The snapshot is taken at $t = 0.4$. We show the proper rest mass density, pressure and velocity.}
\end{figure}

In this case, the initial velocity in the two chambers is high and with opposite direction, and consequently  two strong shocks form and propagate to the left and right decelerating the gas to a very low speed. The evolution produces Lorentz factors of the order of 10000, which tests the capability of an implementation to control extremely high fluid speeds. The numerical results compared with the exact solution are shown in Figure \ref{fig:Headon}. Due to the strength of the shocks,  unphysical oscillations may appear behind them, and in order to avoid these oscillations we use the MINMOD reconstructor, which is more dissipative than the MC.

\subsubsection{Test 4: Strong Reverse Shock}

\begin{figure}
\begin{center}
\includegraphics[width=7.0cm]{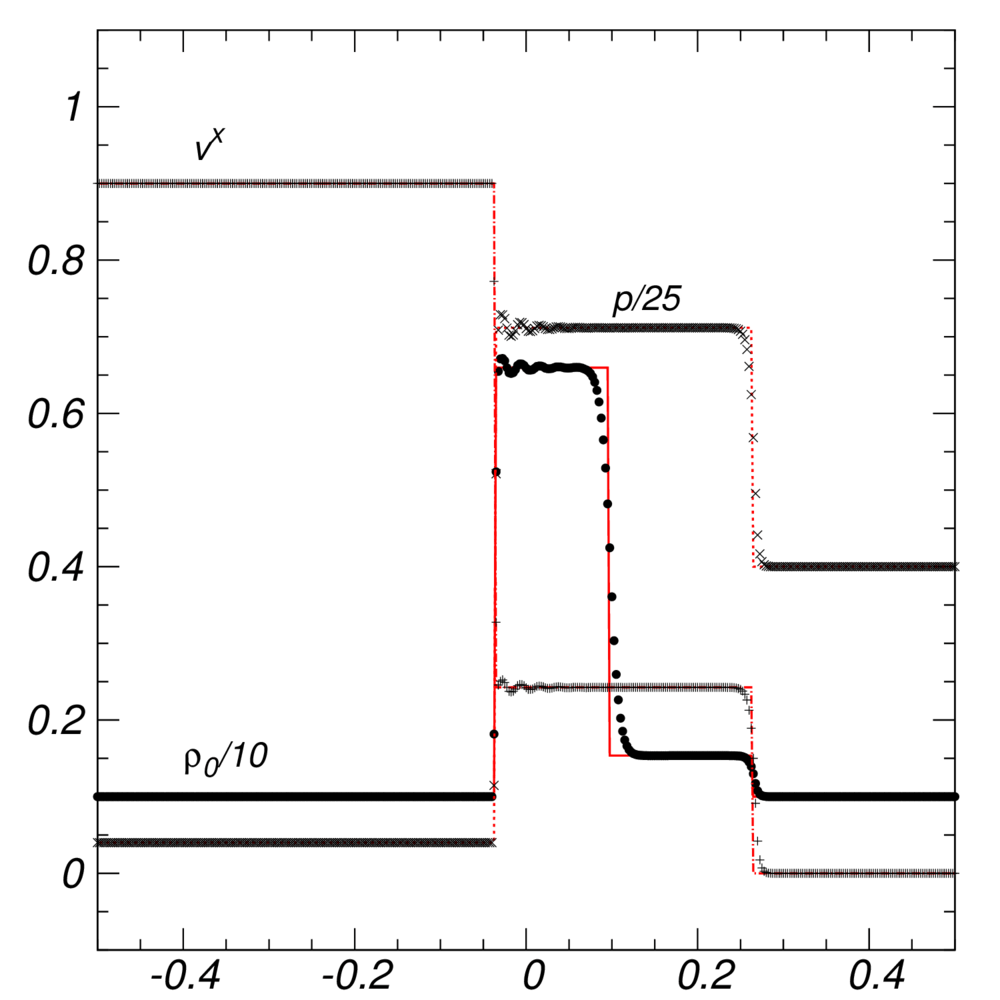}
\end{center}
\caption{\label{fig:Reverse} Test 4: strong reverse shock at $t=0.4$. 
We show the proper rest mass density, the pressure and the velocity.}
\end{figure}

In this problem a strong reverse shock forms, in which post-shock oscillations are visible for the numerical methods used in our simulations. Specifically these oscillations are more evident in the pressure and density. The numerical results compared with the exact solution are shown in Figure \ref{fig:Reverse}. Under these extreme conditions none of the reconstructors used here is capable of diminishing the oscillations, however the amplitude of the oscillations converges to zero with resolution.

\subsubsection{Test 5: Non-Zero Transverse Velocity: Easy Test}

\begin{figure}
\begin{center}
\includegraphics[width=7.0cm]{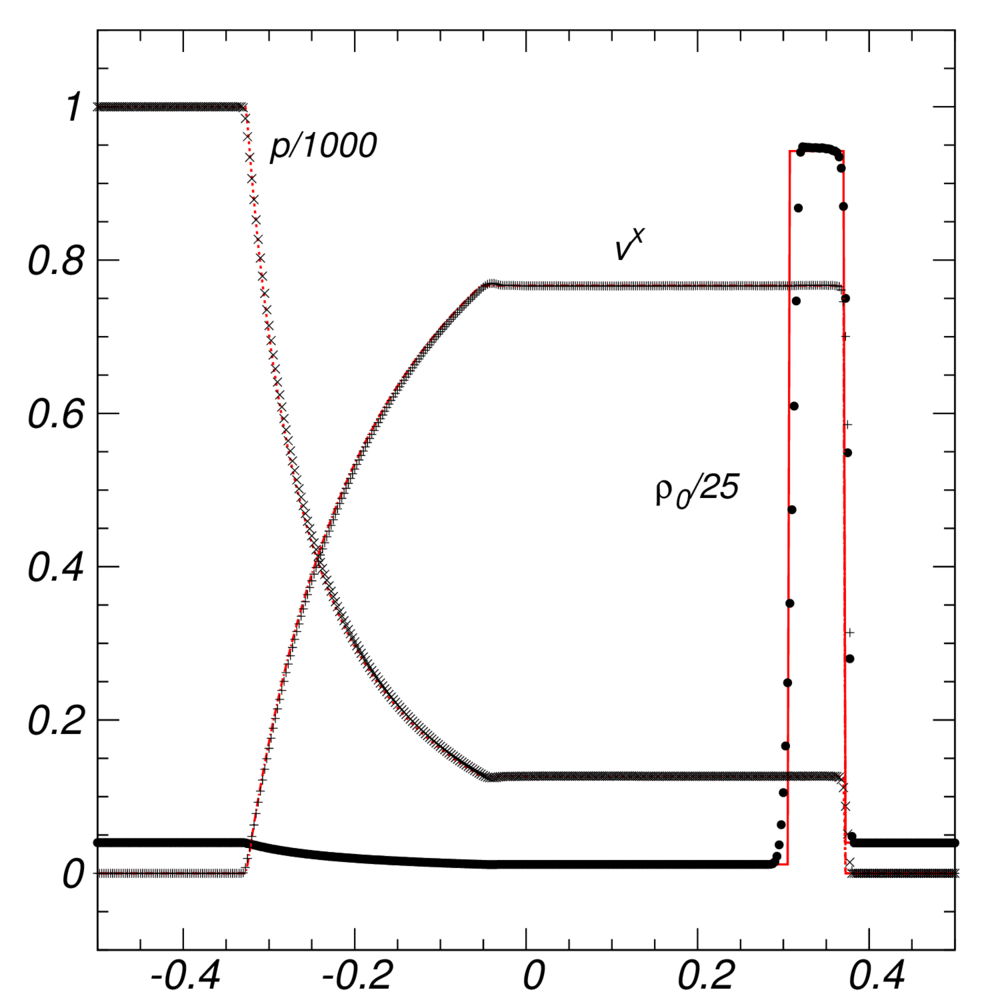}
\end{center}
\caption{\label{fig:NEasy} Test 5: non-zero transverse velocity Easy Test  at $t=0.4$. This test problem is relatively easy and can be resolved with $\Delta x_1$. Again we show the proper rest mass density, the pressure and the velocity.}
\end{figure}

Many problems of interest in hydrodynamics involve strong shear flows. For example, astrophysical jets include shearing layers of ambient material into the fast jet flow. It is therefore important to test the ability of numerical codes to handle Riemann problems with velocity components transverse to the direction of propagation of the main flow. In this first case, the problem is relatively easy because the transverse velocity is in the cold gas of the right state, not in the relativistically hot left state or in the rarefaction fan which subsequently propagates into it. The numerical results compared with the exact solution are shown in Figure \ref{fig:NEasy}, and there is no major difficulty to resolve the shock using resolution $\Delta x_1$. 

\subsubsection{Test 6: Non-Zero Transverse Velocity: Hard Test}

\begin{figure*}
\begin{center}
\includegraphics[width=5.5cm]{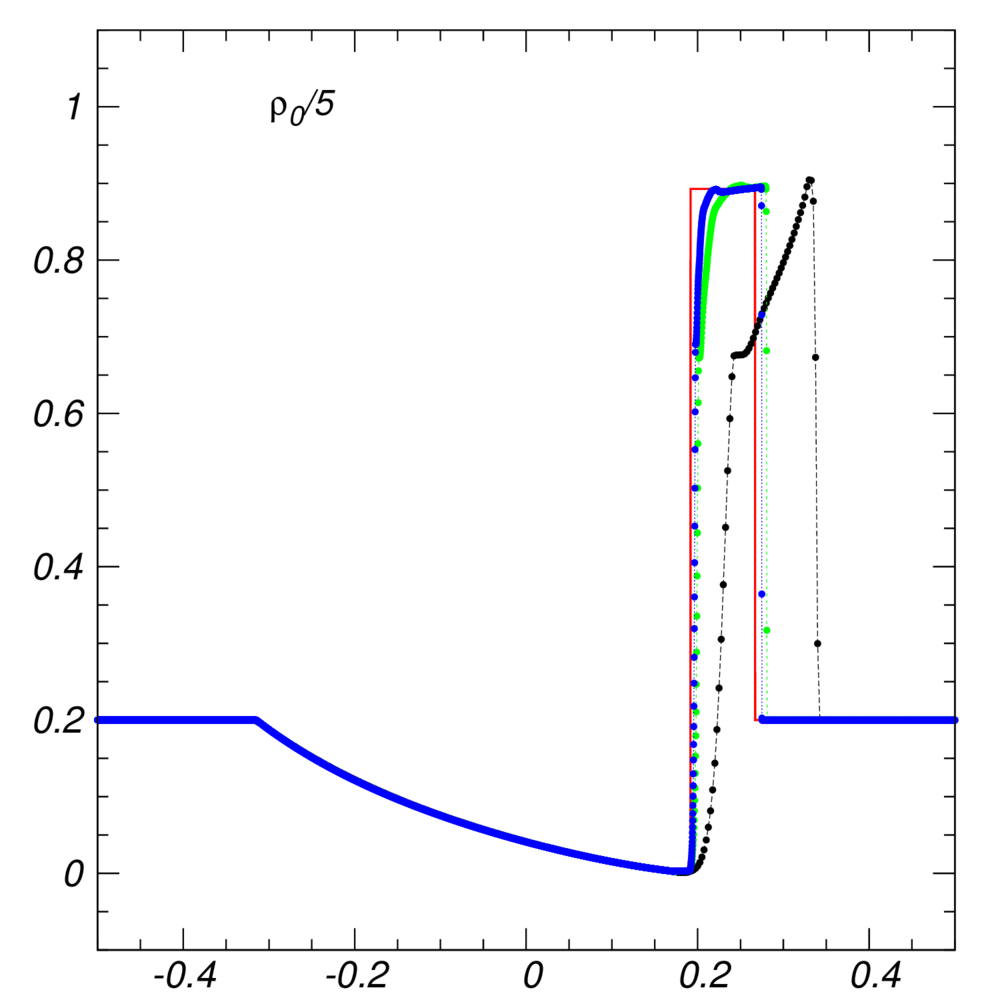}
\includegraphics[width=5.5cm]{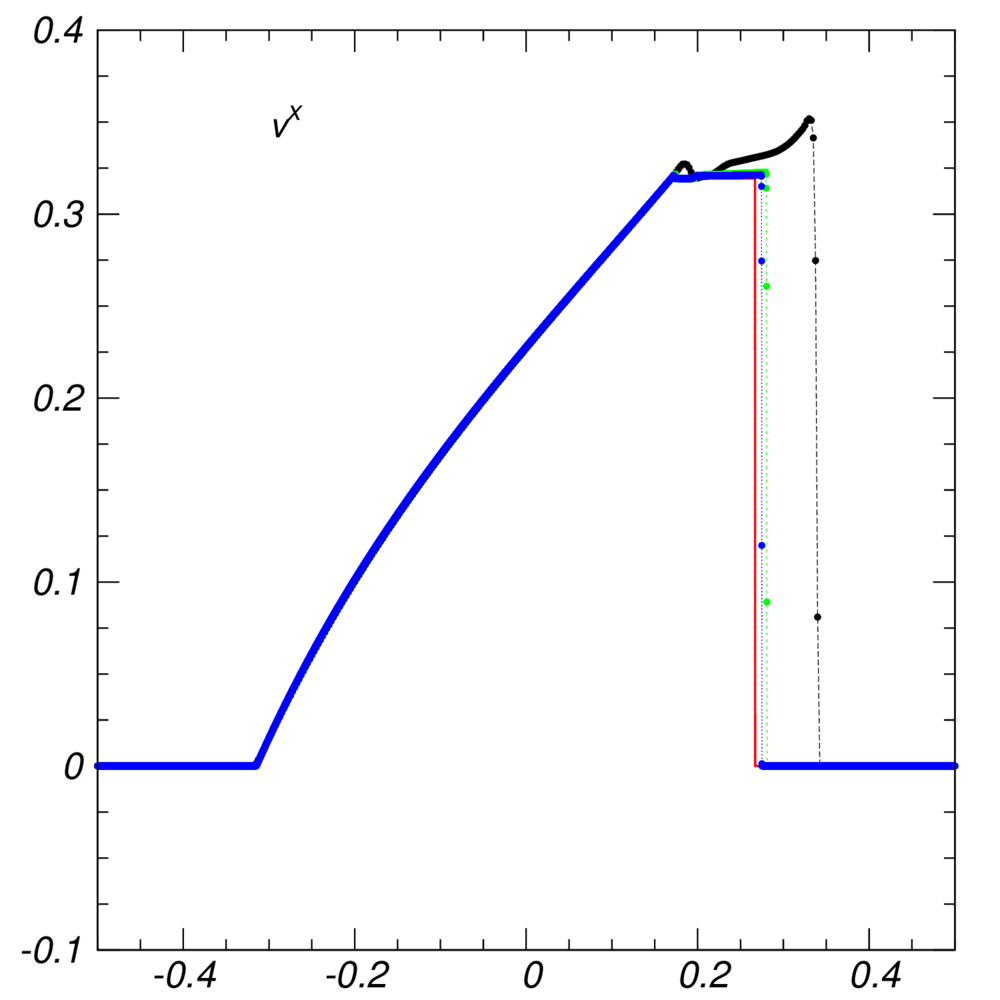}
\includegraphics[width=5.5cm]{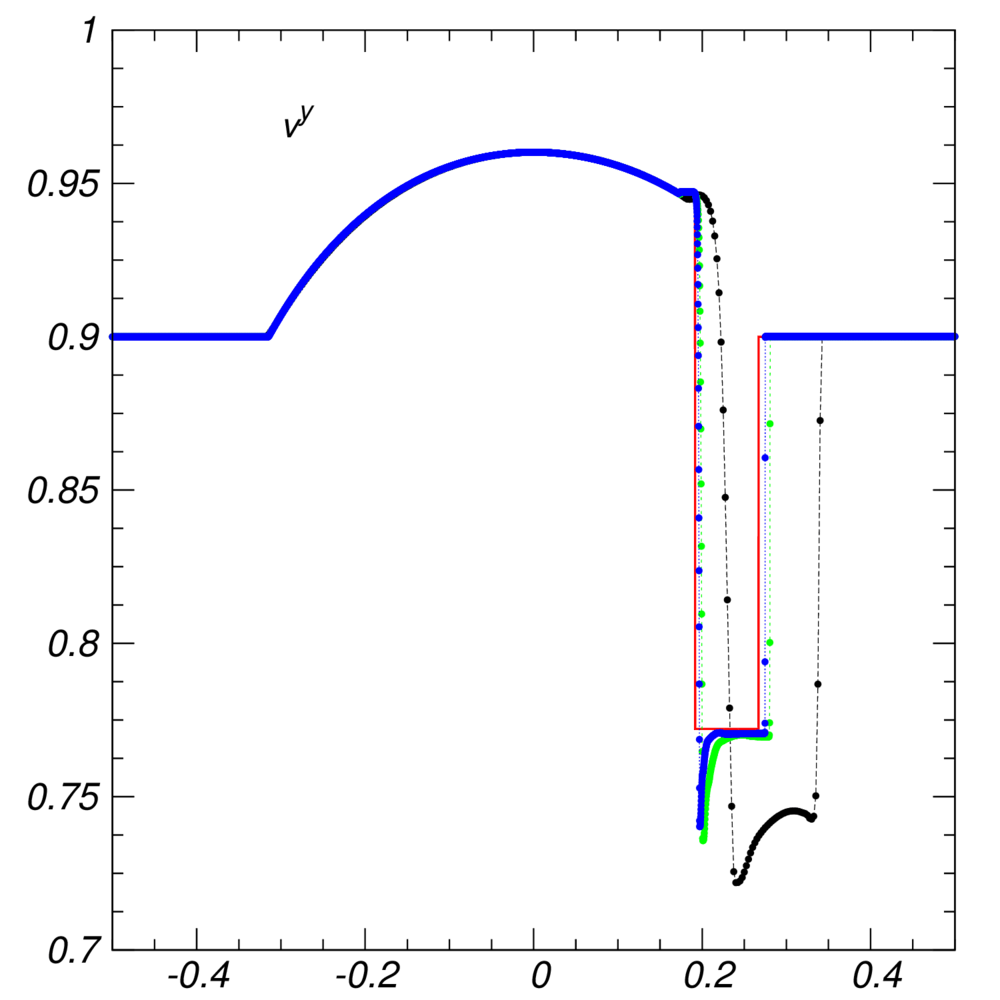}
\end{center}
\caption{\label{fig:NHard} Test 6: non-zero transverse velocity Hard Test at $t=0.6$. In this figure, we show the results for three different resolutions, with $\Delta x_2 = 1/400$ (stars), $\Delta x_5 = 1/3200$ (diamonds) and $\Delta x_6 = 1/6400$ (points). As a consistency check, and in agreement with \citep{2006ApJS..164..255Z}, we verified that when the resolution is increased, the numerical solution is closer to the exact solution (lines).}
\end{figure*}

This is a very severe test requiring very high resolution to resolve the complicated structure of the transverse velocity. This test is particularly hard because the transverse velocity is high not only where the gas is cold but also in the hot region. The numerical results compared with the exact solution are shown in Figure \ref{fig:NHard} and the initial set of parameters can be seen in Table \ref{tab:RHD_tests}.

\subsubsection{Error estimates for the 1D RHD tests}

In order to systematically compare the numerical solutions under different combinations of limiters implemented in our code, we have calculated the error for each 1D Riemann test, and the results are summarized in table \ref{tab:rhd_error}. In all the numerical solutions presented we found consistency, that is, the error decreases when resolution is increased. For this we calculate the $L_1$ norm of the error of these tests as compared with the exact solution. Since we always consider the resolution factors of two, the order of convergence is given by $log(error_i/error_{i-1})/log(2)$, where $i$ is the error calculated with the resolution $\Delta x_i$. We carried out the tests with five resolutions and for the methods used in our code, involving the use of only the RK3 time integrator, first order convergence is expected since the problems start with discontinuous initial data. For tests 1, 2,  3, 4 and 5 we spotted the desirable nearly first order convergence regime for at least a combination of resolution and reconstructor used. The strong test 6 on the other hand, needs more resolution than the previous ones, however we nearly approach first order convergence for the highest resolutions.

It is important to mention that with error estimates, we also locate the resolutions required to safely work on a convergence regime of resolution. In fact these error estimates could be important in AMR codes, in which the resolution in a certain region may be prescribed by the strength and type of local Riemann problems contained in the tests presented here.

\begin{table*}
\begin{center}
\begin{tabular}{c||c|c|c|c||c|c|c|c} \hline \hline
 ${\it Resolution}$ & MM & MC & WENO5 & PPM & MM & MC & WENO5 & PPM \\ \hline \hline
\multicolumn{1}{|c}{} & \multicolumn{4}{c|}{Error} & \multicolumn{4}{|c|}{Order of convergence} \\ \hline \hline
\multicolumn{1}{|c}{} & \multicolumn{8}{c|}{${\bf Test ~ 1}$} \\ \hline \hline
$\Delta x_1$   & 1.33e-1 & 8.09e-2 & 8.05e-2 & 3.28e-1 & .... & .... & .... & .... \\ \hline \hline
$\Delta x_2$   & 7.39e-2 & 4.22e-2 & 4.25e-2 & 1.98e-1 & 0.85 & 0.94 & 0.92 & 0.73 \\ \hline \hline
$\Delta x_3$   & 3.71e-2 & 2.25e-2 & 2.28e-2 & 1.22e-1 & 0.99 & 0.90 & 0.90 & 0.70 \\ \hline \hline
$\Delta x_4$   & 2.12e-2 & 1.19e-2 & 1.44e-2 & 7.49e-2 & 0.81 & 0.92 & 0.67 & 0.70 \\ \hline \hline
$\Delta x_5$   & 1.17e-2 & 7.01e-3 & 9.02e-3 & 4.45e-2 & 0.86 & 0.76 & 0.68 & 0.75 \\ \hline \hline
\multicolumn{1}{|c}{} & \multicolumn{8}{c|}{${\bf Test ~ 2}$} \\ \hline \hline
$\Delta x_1$   & 2.59e-1 & 1.95e-1 & 1.91e-1 & 2.58e-1 & .... & .... & .... & .... \\ \hline \hline
$\Delta x_2$   & 1.87e-1 & 1.39e-1 & 1.28e-1 & 1.95e-1 & 0.47 & 0.49 & 0.58 & 0.40 \\ \hline \hline
$\Delta x_3$   & 1.37e-1 & 8.86e-2 & 6.98e-2 & 1.46e-1 & 0.45 & 0.65 & 0.87 & 0.42 \\ \hline \hline
$\Delta x_4$   & 9.11e-2 & 5.07e-2 & 3.69e-2 & 9.70e-2 & 0.58 & 0.81 & 0.92 & 0.60 \\ \hline \hline
$\Delta x_5$   & 5.16e-2 & 2.72e-2 & 1.82e-2 & 6.03e-2 & 0.82 & 0.89 & 1.01 & 0.69 \\ \hline \hline
\multicolumn{1}{|c}{} & \multicolumn{8}{c|}{${\bf Test ~ 3}$} \\ \hline \hline
$\Delta x_1$   & 3.24e+2 & 3.71e+2 & - & 5.35e+2 & .... & .... & .... & .... \\ \hline \hline
$\Delta x_2$   & 1.42e+2 & 1.78e+2 & - & 2.35e+2 & 1.19 & 1.05 &   -  & 1.18 \\ \hline \hline
$\Delta x_3$   & 8.06e+1 & 9.55e+1 & - & 1.33e+2 & 0.82 & 0.90 &   -  & 0.82 \\ \hline \hline
$\Delta x_4$   & 3.49e+1 & 4.53e+1 & - & 5.88e+1 & 1.20 & 1.08 &   -  & 1.18 \\ \hline \hline
$\Delta x_5$   & 1.96e+1 & 2.34e+1 & - & 3.29e+1 & 0.83 & 0.95 &   -  & 0.84 \\ \hline \hline
\multicolumn{1}{|c}{} & \multicolumn{8}{c|}{${\bf Test ~ 4}$} \\ \hline \hline
$\Delta x_1$   & 7.87e-2 & 6.81e-2 & 6.06e-2 & 1.90e-1 & .... & .... & .... & .... \\ \hline \hline
$\Delta x_2$   & 4.85e-2 & 3.35e-2 & 3.52e-2 & 1.10e-1 & 0.70 & 1.02 & 0.78 & 0.79 \\ \hline \hline
$\Delta x_3$   & 2.91e-2 & 2.03e-2 & 2.01e-2 & 7.35e-2 & 0.74 & 0.72 & 0.81 & 0.58 \\ \hline \hline
$\Delta x_4$   & 1.77e-2 & 1.17e-2 & 1.19e-2 & 4.19e-2 & 0.71 & 0.79 & 0.76 & 0.81 \\ \hline \hline
$\Delta x_5$   & 1.02e-2 & 7.13e-3 & 7.16e-2 & 2.44e-2 & 0.80 & 0.71 & 0.73 & 0.78  \\ \hline \hline
\multicolumn{1}{|c}{} & \multicolumn{8}{c|}{${\bf Test ~ 5}$} \\ \hline \hline
$\Delta x_1$   & 7.14e-1 & 4.28e-1 & 3.73e-1 & 1.69e+0 & .... & .... & .... & .... \\ \hline \hline
$\Delta x_2$   & 4.06e-1 & 2.34e-1 & 2.01e-1 & 1.01e+0 & 0.81 & 0.87 & 0.89 & 0.74 \\ \hline \hline
$\Delta x_3$   & 2.37e-1 & 1.30e-1 & 1.12e-1 & 6.65e-1 & 0.77 & 0.85 & 0.84 & 0.60 \\ \hline \hline
$\Delta x_4$   & 1.30e-1 & 7.25e-2 & 6.12e-2 & 3.90e-1 & 0.87 & 0.84 & 0.87 & 0.77 \\ \hline \hline
$\Delta x_5$   & 7.75e-2 & 4.21e-2 & 3.56e-2 & 2.31e-1 & 0.74 & 0.78 & 0.78 & 0.76 \\ \hline \hline
\multicolumn{1}{|c}{} & \multicolumn{8}{c|}{${\bf Test ~ 6}$} \\ \hline \hline
$\Delta x_1$   & 6.34e-1 & 5.78e-1 & 5.77e-1 & 7.31e-1 & .... & .... & .... & .... \\ \hline \hline
$\Delta x_2$   & 5.01e-1 & 3.90e-1 & 4.04e-1 & 6.16e-1 & 0.34 & 0.57 & 0.51 & 0.24 \\ \hline \hline
$\Delta x_3$   & 3.45e-1 & 2.57e-1 & 2.59e-1 & 4.59e-1 & 0.54 & 0.60 & 0.64 & 0.42 \\ \hline \hline
$\Delta x_4$   & 2.19e-1 & 1.50e-1 & 1.51e-1 & 3.46e-1 & 0.66 & 0.78 & 0.78 & 0.40 \\ \hline \hline
$\Delta x_5$   & 1.30e-1 & 8.63e-2 & 8.73e-2 & 2.32e-1 & 0.75 & 0.80 & 0.79 & 0.57 \\ \hline \hline
$\Delta x_6$   & 7.78e-2 & 5.13e-2 & 5.33e-2 & 1.44e-1 & 0.74 & 0.75 & 0.71 & 0.69 \\ \hline \hline
\end{tabular}
\end{center}
\caption{\label{tab:rhd_error} $L_1$ norm of the error in density for four  schemes  with different numerical reconstructors. The $L_1$ norm is computed at $t=0.4$, except for test 6, which is presented at $t=0.6$. We also show the order of convergence between the different pairs of resolutions. The results achieve convergence as expected for problems with sharp discontinuities.}
\end{table*}

\subsubsection{Smooth initial profile}

In order to know how the code performs evolving initial smooth profiles, following \citep{2006ApJS..164..255Z} and \citep{2012A&A...547A..26R}, we set the smooth initial density profile immersed on a reference constant density ambient. Specifically the density profile is

\begin{equation}
\rho(x) = \left\{ 
	\begin{array}{ll}
	1 + \exp[-1/(1-x^2/L^2)], &~~\mbox{if}~~ |x|<L\\
         1, & ~~ \mbox{otherwise} .
         \end{array} \right. \nonumber\\
\end{equation}

\begin{figure}
\begin{center}
\includegraphics[width=8cm]{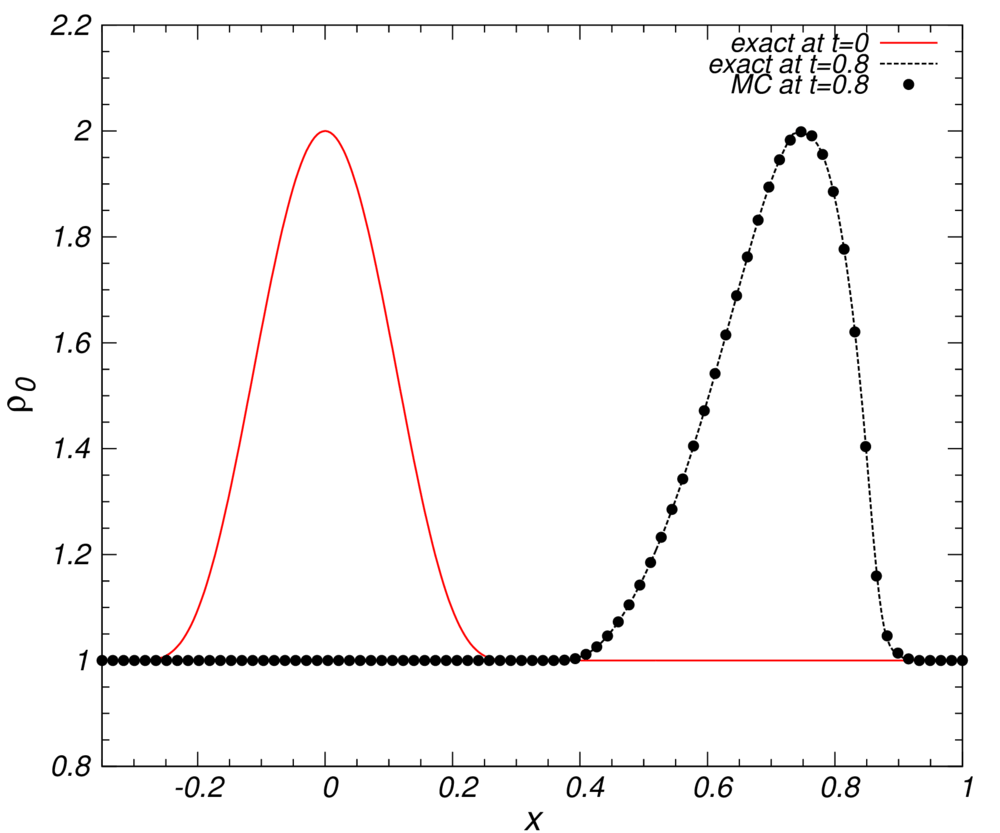}
\end{center}
\caption{\label{fig:smooth_test} Initial profile and a snapshot at $t=0.8$ of the density for the smooth profile. The continuous line corresponds to the initial profile, the dashed line to the exact solution at $t=0.8$ and the dots indicate the numerical solutions calculated using the MC reconstructor.}
\end{figure}

\noindent The fluid obeys the isentropic relation $p=K\rho^{\Gamma}$ whereas the initial velocity field is subject to the condition that the invariant

\begin{equation}
J_{-} = \frac{1}{2}\ln\left(\frac{1+v}{1-v}\right) - \frac{1}{\sqrt{\Gamma-1}}\ln\left(\frac{\sqrt{\Gamma-1}-c_s}{\sqrt{\Gamma-1}-c_s}\right)
\end{equation}

\noindent has to be constant in the whole domain. The construction of this invariant assumes that the velocity in the ambient region is set to zero. The parameters used for the test are $L=0.3$, $\Gamma=5/3$ and $K=100$ and the domain along the $x$ direction $[-0.35,1]$ is covered with a number of cells. In Fig. \ref{fig:smooth_test} we show the initial profile and a snapshot of the numerical solution using the MC reconstructor at $t=0.8$, as shown in 
\citep{2006ApJS..164..255Z} and \citep{2012A&A...547A..26R}. A convergence test was performed  comparing the numerical with the exact solution for various resolutions and reconstructors. The order of convergence of the different combinations are collected in Table \ref{tab:smooth_error}.

As expected, the various reconstructors approach the second order convergences unlike the first order convergence achieved for the tests with initial shocks. The convergence found is comparable with that used in previous studies (e.g. \cite{2006ApJS..164..255Z}) using RK3.

\begin{table}
\centering
\begin{tabular}{c|c|c|c|c} \hline \hline
 ${\it Cells}$  & MM       & MC & WENO5 & PPM \\ \hline \hline
\multicolumn{5}{|c|}{${\bf Smooth ~profile~ test}$} \\ \hline \hline
80   	& ....      &  ....  	& ....  & ....   \\ \hline \hline
160  	& 2.11  & 2.57  &  2.27 & 1.90  \\ \hline \hline
320  	& 2.09  & 2.37   & 2.30  & 1.86  \\ \hline \hline
640  	& 2.04  & 2.18   & 2.48  & 1.91  \\ \hline \hline
1280& 2.00  & 2.10   & 2.60  & 1.95  \\ \hline \hline
\end{tabular} 
\caption{\label{tab:smooth_error} Order of convergence of the $L_1$ norm of the error in the density for four  schemes  with different numerical reconstructors. The $L_1$ norm is computed at $t=0.8$ for a test with smooth initial data. The left column indicates the number of cells used to cover the domain $[-0.35,1]$ along the $x$ direction.  }
\end{table}

\subsubsection{The first 2D test: shock Tube on the Diagonal}

Multidimensional relativistic simulations are more difficult to carry out than the one-dimensional ones because the components of the velocity, which are spatially interpolated separately, eventually may cause the velocity to be greater than $v^2 >1$, especially in the ultrarelativistic regime, due to numerical errors in the reconstruction. For this reason, in some cases it is necessary to use more dissipative methods and in some regions low order reconstructors. The first 2D test consists in the evolution of a one dimensional shock-tube problem along a diagonal of a plane.

\begin{figure}
\begin{center}
\includegraphics[width=8.0cm]{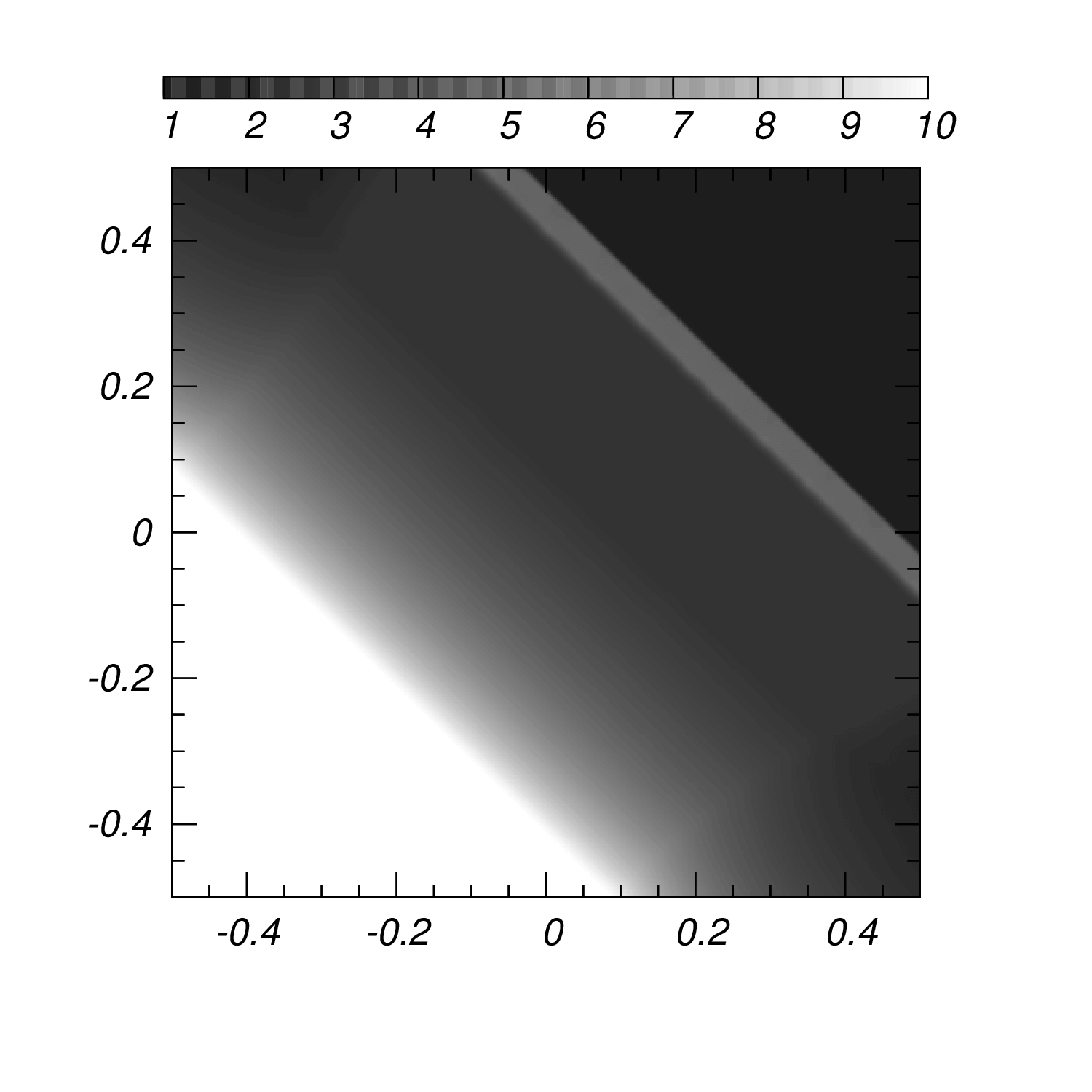}
\includegraphics[width=6.8cm]{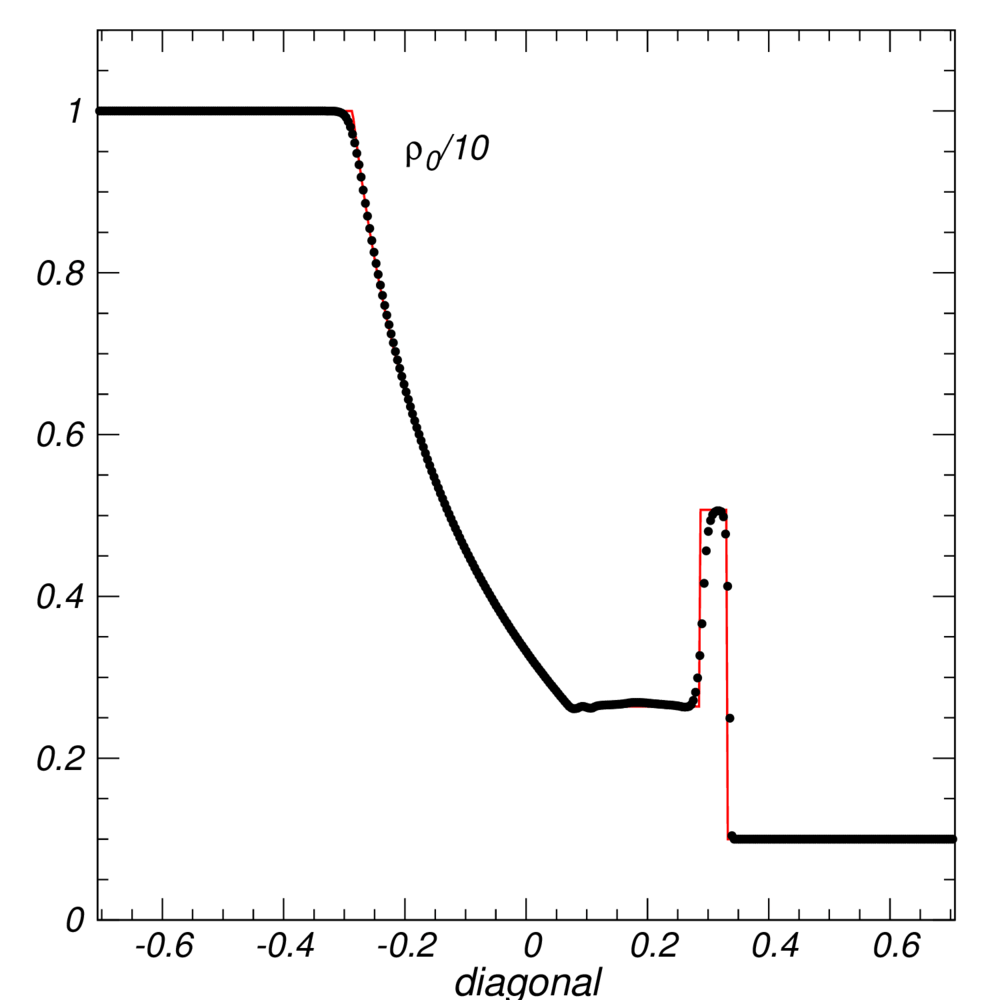}
\end{center}
\caption{\label{fig:D_ST} (Top) A 2D snapshot at $t=0.4$ of the rest mass density for test 1 with the shock propagating along a diagonal direction. (Bottom) Comparison with the exact solution as seen along the diagonal perpendicular to that of the propagation.}
\end{figure}

In order to check the code is able to handle the fluxes along two different directions simultaneously we implement the initial data of the 1D test 1, however with the initial shock propagating along the diagonal $\hat{x}+\hat{y}$ direction. The initial data are set on a 2D domain $[-0.5,0.5]\times[-0.5,0.5]$ which is covered with 400$\times$ 400 cells. In Fig. \ref{fig:D_ST} we show the 2D profile as seen from the $z$ axis and a snapshot of the density for the solution at $t=0.4$ compared with the exact solution. In this example we use the MC reconstructor.

\subsubsection{Two-dimensional Riemann problem}

\begin{figure*}
\begin{center}
\includegraphics[width=8.0cm]{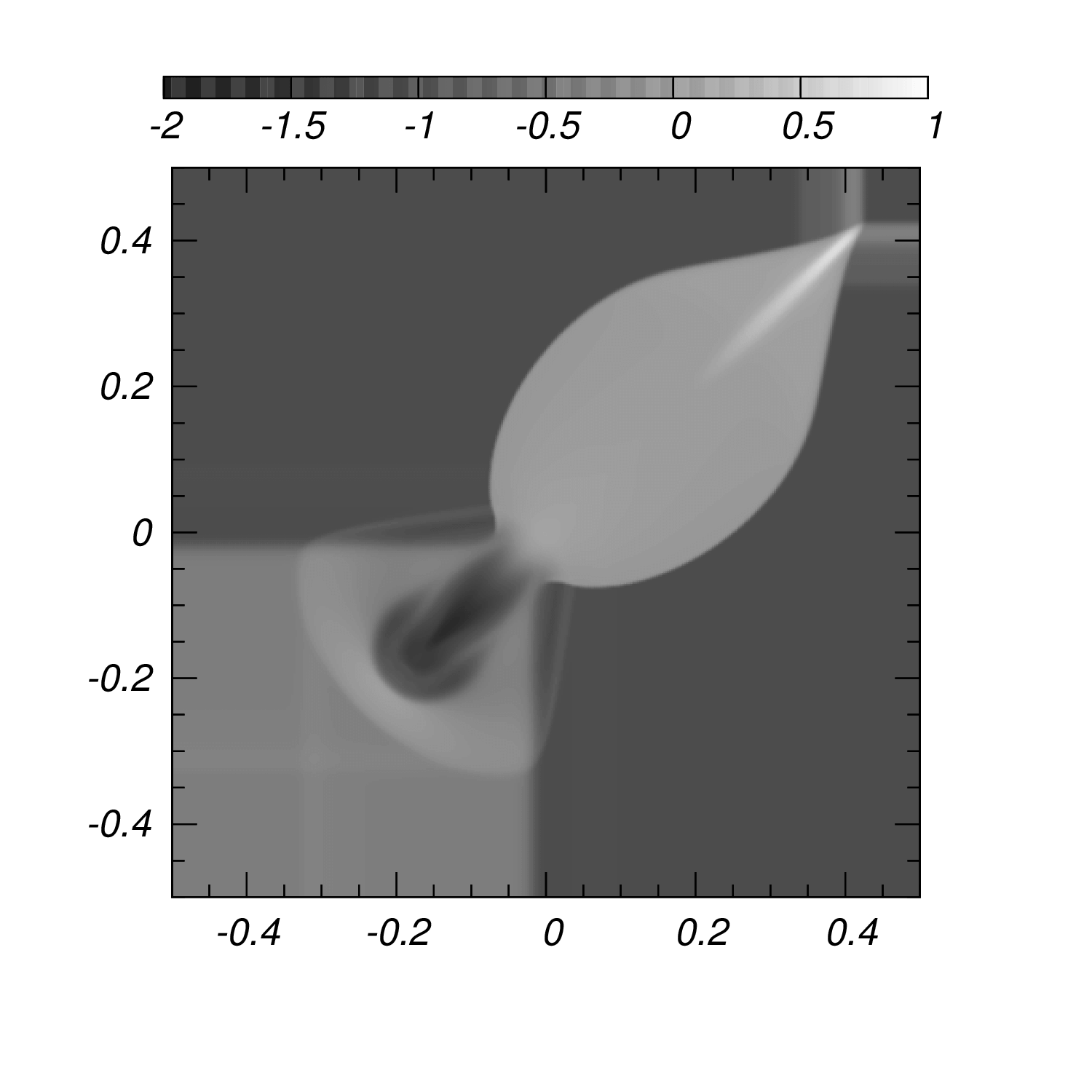}
\includegraphics[width=8.0cm]{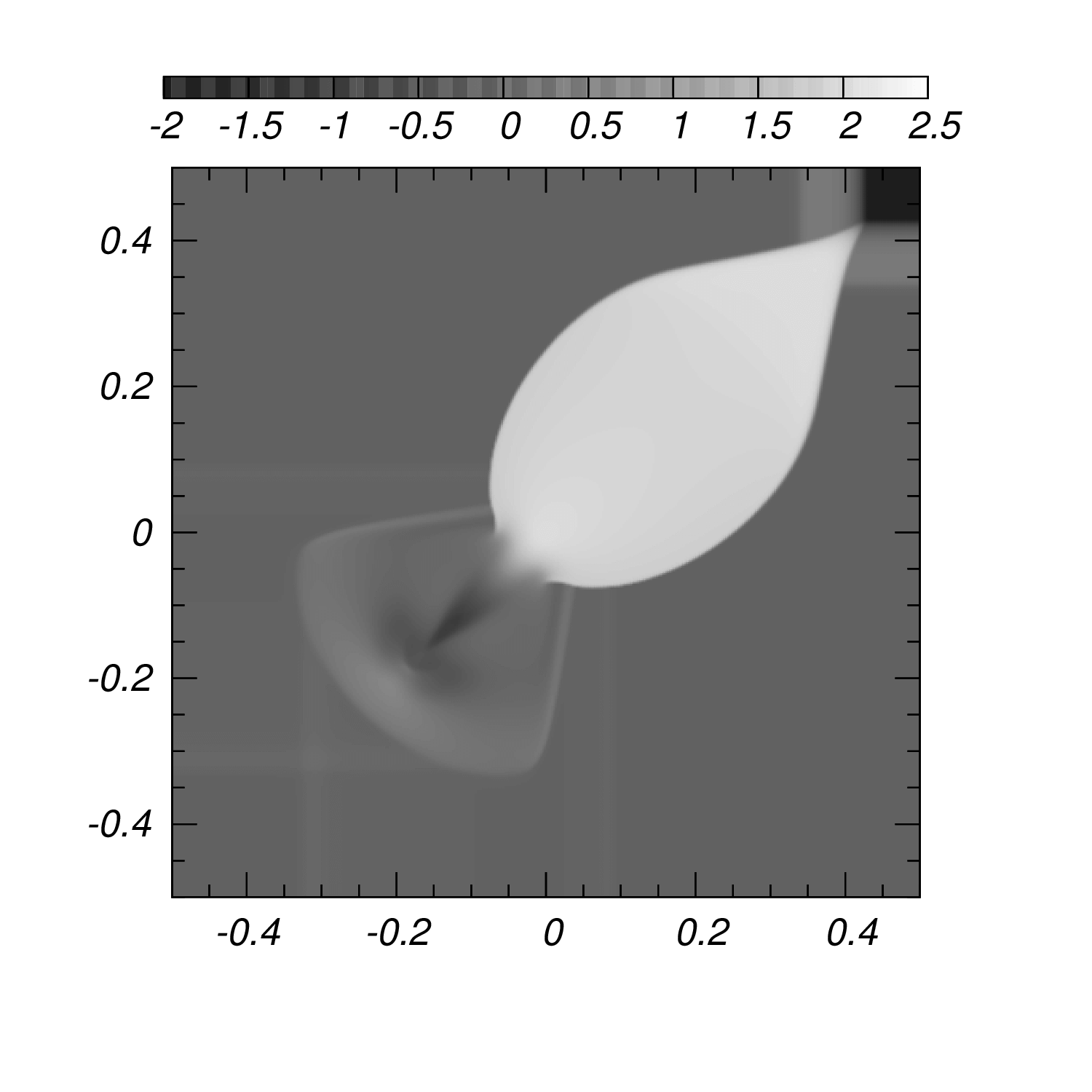}
\end{center}
\caption{\label{fig:2dshock-tube} Porper rest mass density (top) and pressure (bottom) in logarithmic scale for the 2D relativistic Riemann problem at $t=0.4$. The simulation was carried out on the numerical domain $x,y \in [-0.5,0.5]$, covered with $400\times 400$ cells with a courant factor of $CFL=0.25$, using the HLLE solver and the MINMOD limiter.}
\end{figure*}

The relativistic 2D Riemann problem basically involves the interaction of shock, rarefaction and contact waves initially separated by four quadrants of constant values at initial time. In the context of classical hydrodynamics this problem was formulated in \citep{doi:10.1137/S1064827595291819} and its extension to the relativistic case in \citep{2002A&A...390.1177D}, where the initial condition involve two shocks and two tangential discontinuities. In this simulation, we use the HLLE flux formula and the MINMOD limiter. The problem is defined in the domain $[-0.5,0.5]\times[-0.5,0.5]$, which is covered with $400 \times 400$ cells. The integration was carried out with Courant factor $CFL = 0.25$ and  we imposed outflow boundary conditions. Specifically, the initial state has the following parameters

\begin{equation}
\nonumber (\rho,p,v^x,v^y) =
\left \{
\begin{array}{cc}
 (0.1,1.0,0.99,0)^{TL} \\
 (0.1,0.01,0,0)^{TR}   \\
 (0.5,1.0,0,0)^{BL}    \\
 (0.1,1.0,0,0.99)^{BR}
\end{array}
\right. ,
\end{equation}

\noindent where the labels correspond to Top-Left (TL), Top-Right (TR), Bottom-Left (BL) and Bottom-Right (BR) quadrants of the $xy-$plane. In Fig. \ref{fig:2dshock-tube} we show the logarithm of the proper rest mass density and the pressure at $t=0.4$.

The morphology shows various features, including a bow shock and a small jet moving diagonally in the initially high density region (BL quadrant), and in the opposite direction the fluid moves towards the region of lower density in a filamentary form.

\subsubsection{Relativistic Emery's Wind tunnel}

\begin{figure}
\begin{center}
\includegraphics[width=10cm]{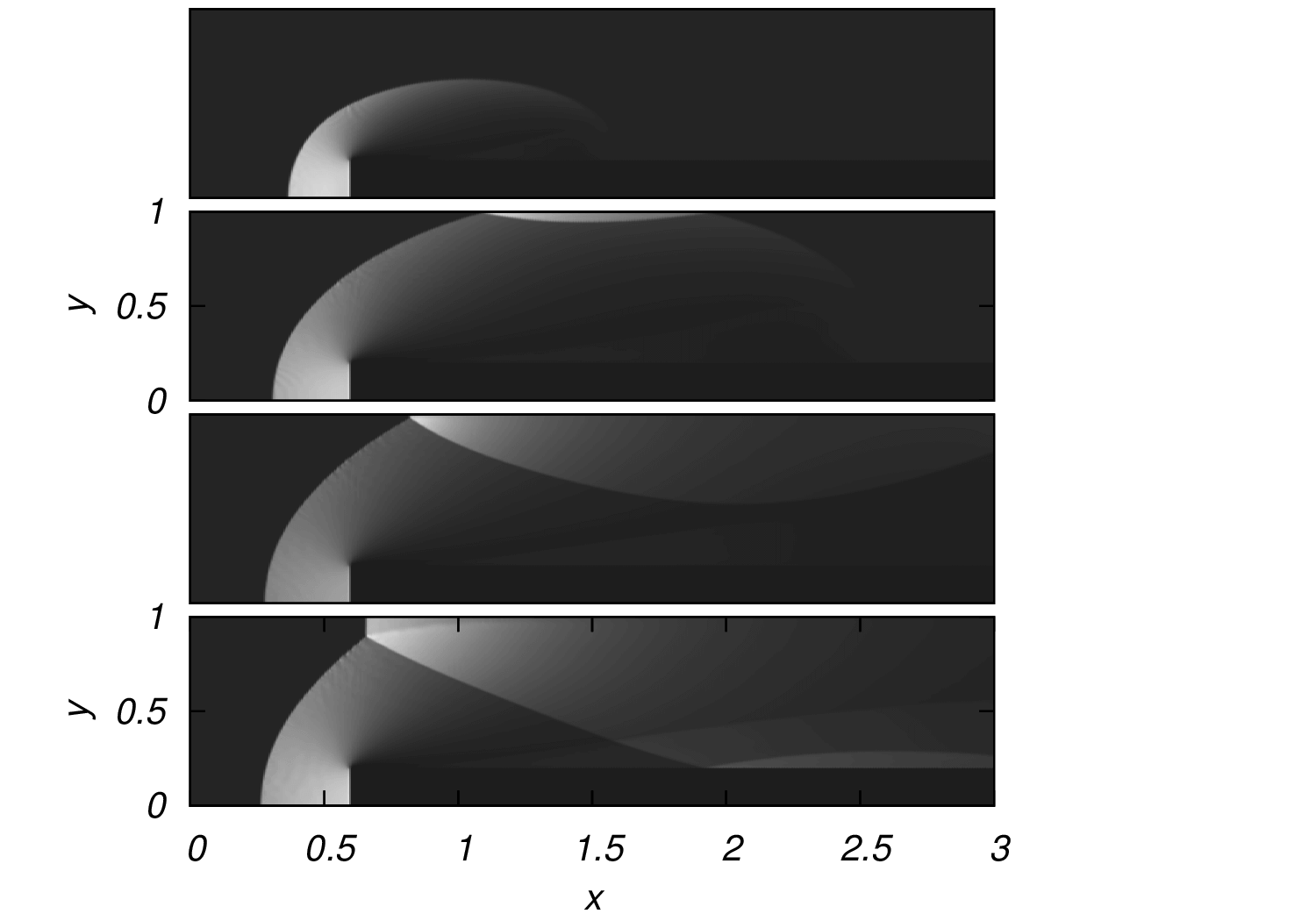}
\end{center}
\caption{\label{fig:Emery} Rest mass density of the gas for the Emery wind tunnel at $t=1,2,3,4$. The snapshots show the different artifacts expected to happen during the evolution.  The domain $[0,3]\times[0,1]$ is covered with 600$\times$200 cells and we use CFL=0.25.}
\end{figure}

This is a test proposed for classical hydrodynamics in \citep{1968JCoPh...2..306E,1984JCoPh..54..115W} that has been extended to the relativistic case \citep{2004A&A...428..703L,2006ApJS..164..255Z}. It consists in the flow entering from the left side of the domain and encounters a step. The standard initial conditions are $\rho = 1.4$, $p = 1.0$, $v^x = 0.999$, $v^y = 0$ with $\Gamma = 1.4$, with the step starting at one fifth from the horizontal and vertical domains \citep{2006ApJS..164..255Z}. The boundary conditions are inflow at the left boundary, outflow at the right, reflecting at the top and bottom and the step boundaries. The results using MC are shown in Fig. \ref{fig:Emery}. The global features of this test by $t=1$ are that a reverse shock to the left is formed, which subsequently faces the constant entrance of the fluid from the left to form a bow shock. This shock then expands and by $t=2$ it reaches the upper boundary and gets reflected as shown in the snapshot at $t=3$ and finally it bounces back again from the step upwards as seen in $t=4$. Also a Mach stem is formed  vertically at the top boundary. Unlike in the Newtonian case, the contact discontinuity caused by the corner of the step does not develop any Kelvin-Helmholtz instability near the Mach stem.

\subsubsection{Relativistic Kelvin Helmholtz Instability in 2D}

Another test in 2D is the response of the code to unstable initial conditions and the resolution of small structures with low resolution. The Kelvin-Helmoltz (KH) instability develops when the initial conditions of a gas in two different states separated by different membranes is perturbed. Specifically, a KH instability can occur when there is velocity shear in a continuous fluid, or when there is a velocity difference across the interface between two states of the fluid. In this test one assumes a chamber filled with gas in a given state and a strip in a different state. In our case we use the following set up for $\Gamma=5/3$:

\begin{equation}
\nonumber (\rho,p,v^x,v^y) =
\left \{
\begin{array}{cc}
 (2,2.5,0.5,0), ~~~ if ~|y| < 0.25 \\
 (1,2.5,-0.5,0), ~~ if~ |y|\ge 0.25.
\end{array}
\right. 
\end{equation}

\noindent Additionally the velocities are perturbed such that $v^x=v^x \times (1+0.01 \cos (10 \pi x) \cos (10\pi y))$ and $v^y = v^y \times (0.01 \cos(10 \pi  x) \cos(10\pi y))$. In these numerical simulations, we cover the domain $x,y \in [-0.5,0.5]$ with $400\times 400$ cells, and use a courant factor of $CFL = 0.25$ and periodic boundary conditions in all faces.

\begin{figure*}
\begin{center}
\includegraphics[width=8cm]{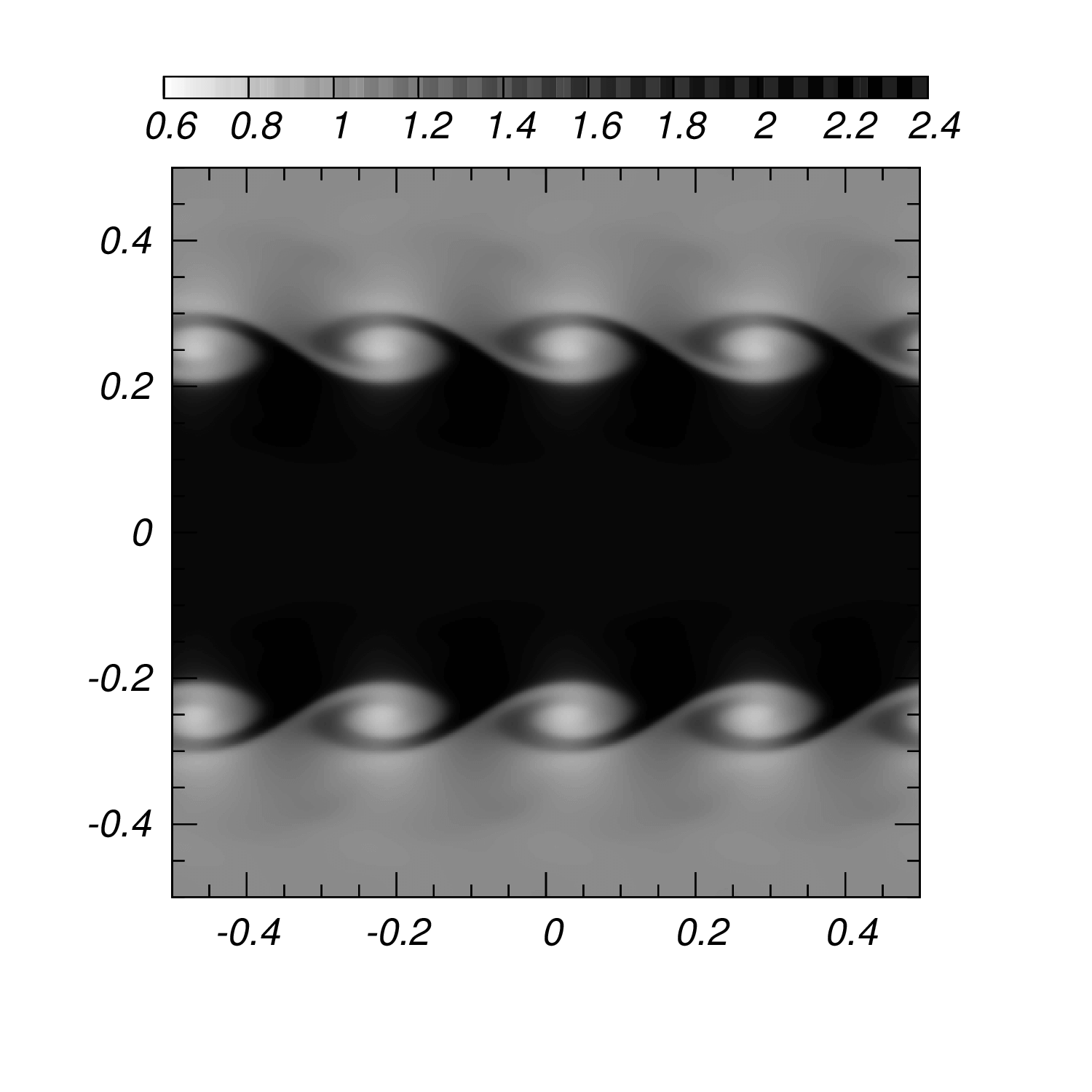}
\includegraphics[width=8cm]{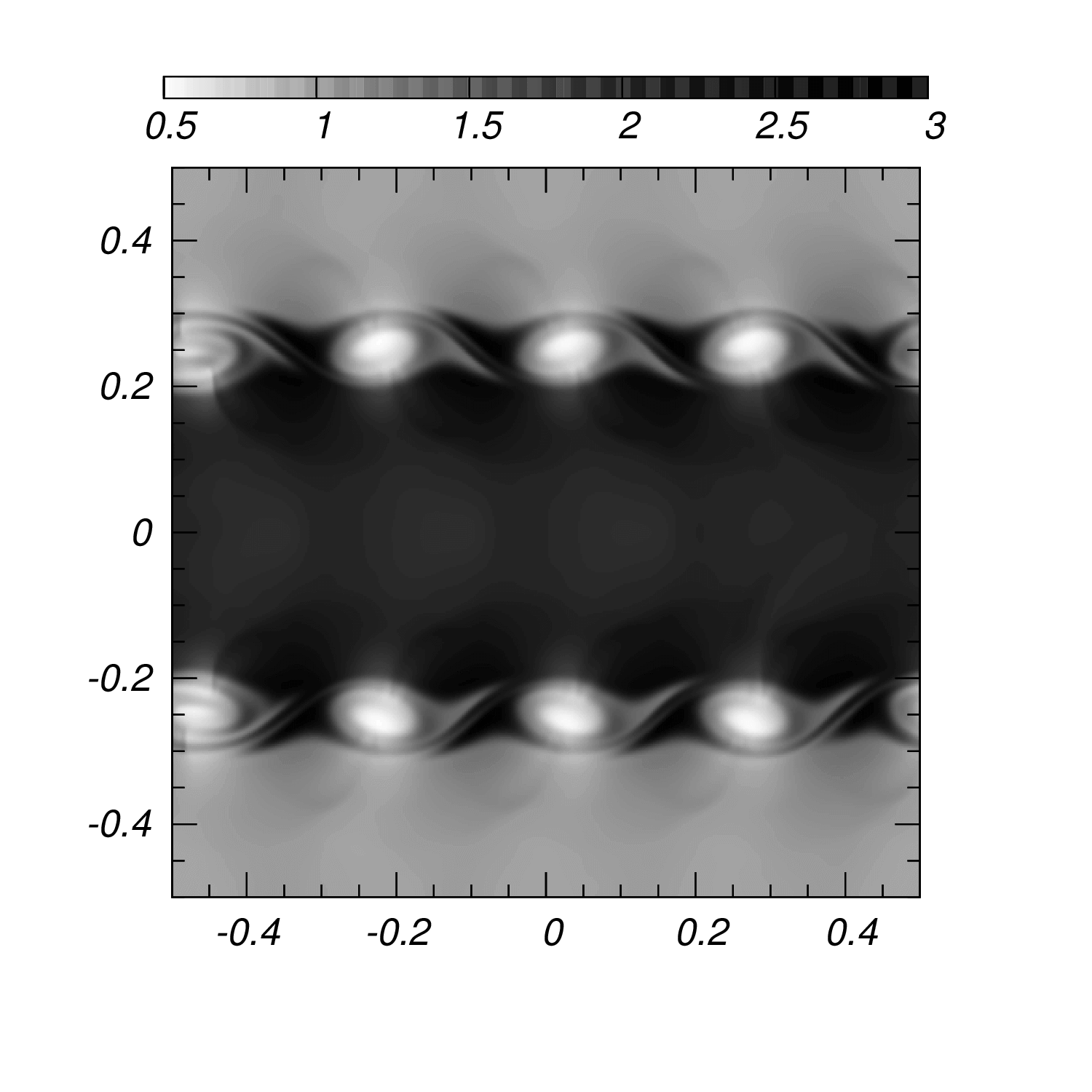}
\includegraphics[width=8cm]{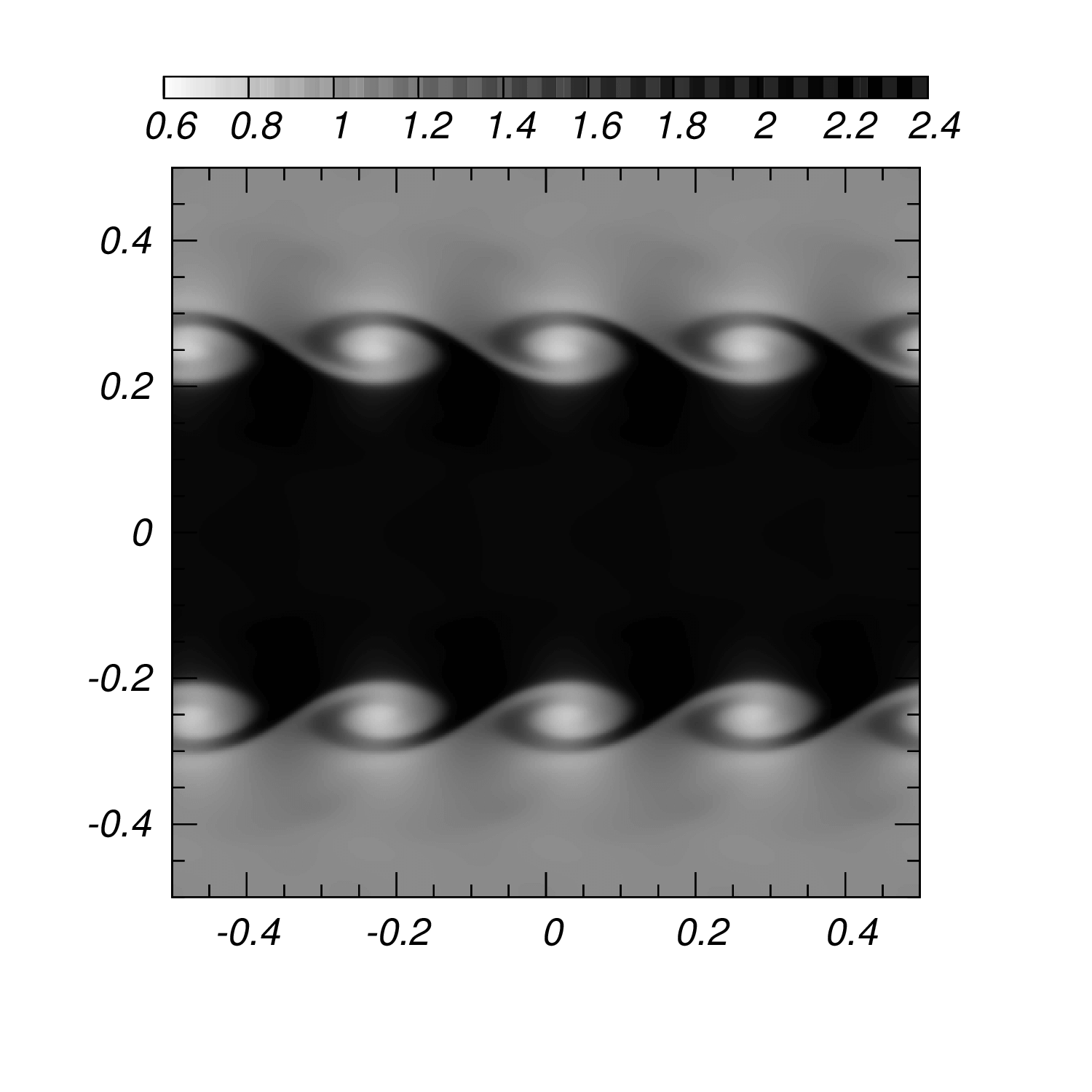}
\includegraphics[width=8cm]{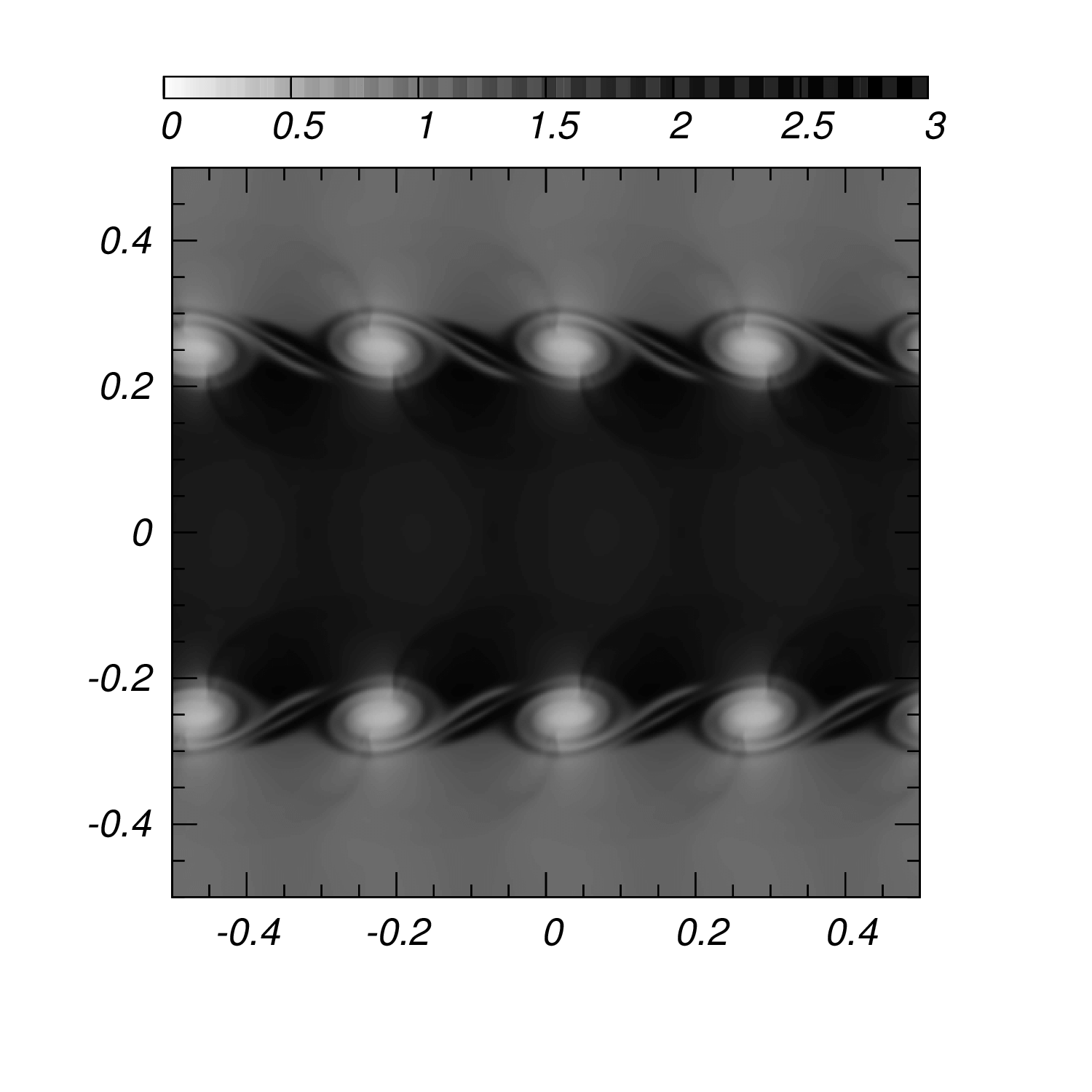}
\end{center}
\caption{\label{fig:KH1} Classical test of the relativistic Kelvin Helmholtz Instability. 
We cover the domain $[-0.5,0.5]\times[-0.5,0.5]$ with 400$\times$400 cells at $t=1.5$. We show, the proper rest mass density with four different limiters: MINMOD (top-left), MC (top-right), PPM (bottom-left) and WENO5 (bottom-right).}
\end{figure*}

\begin{figure*}
\begin{center}
\includegraphics[width=8cm]{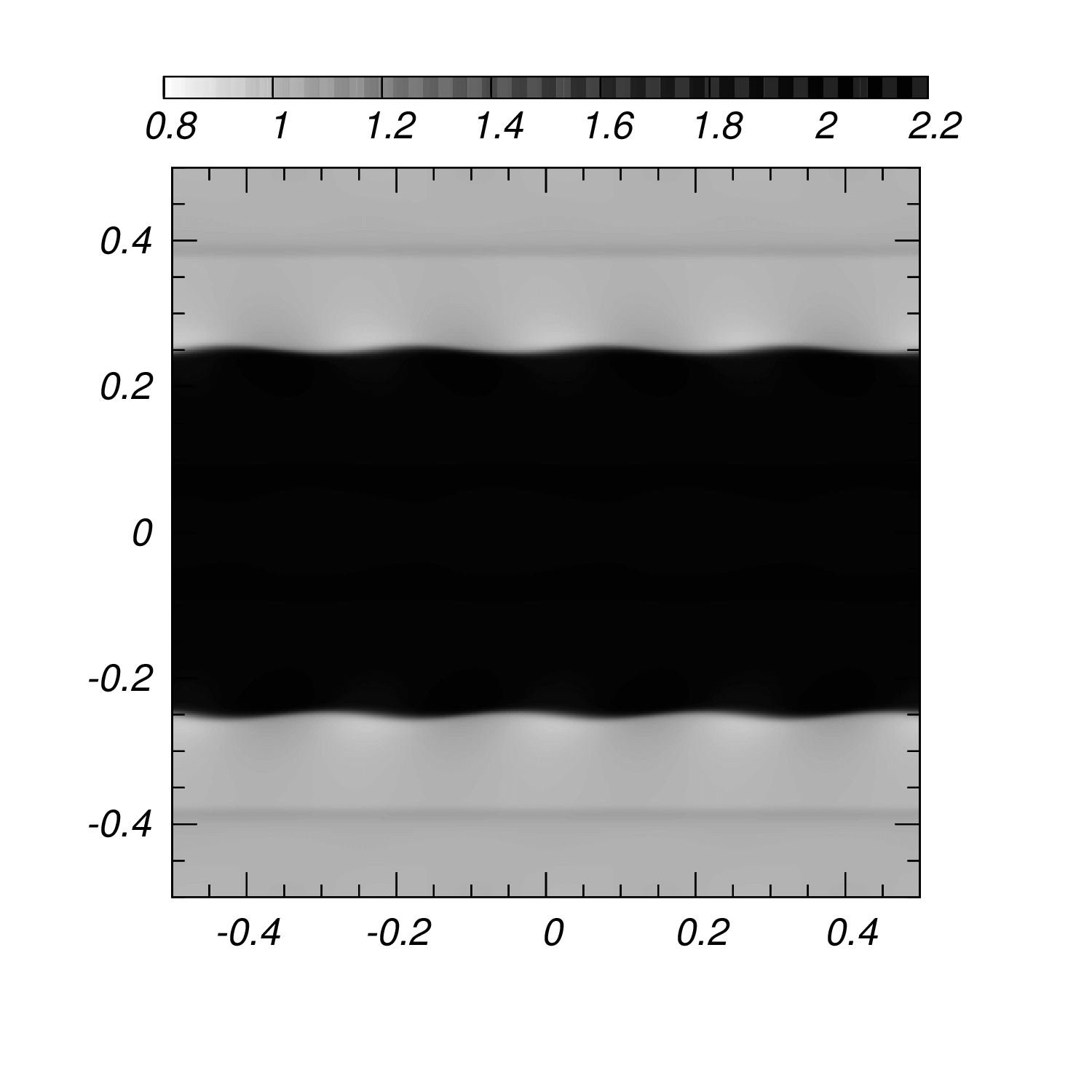}
\includegraphics[width=8cm]{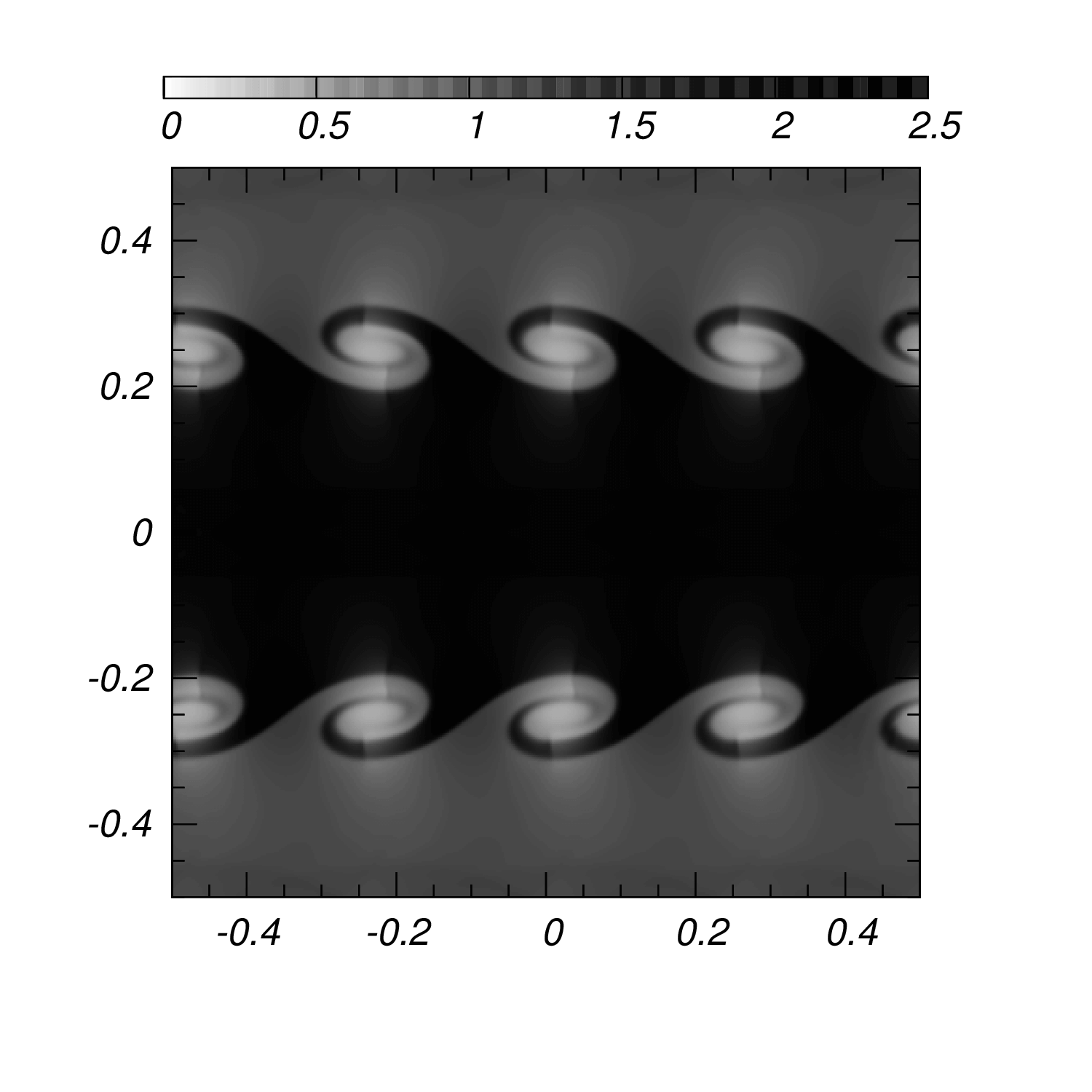}
\includegraphics[width=8cm]{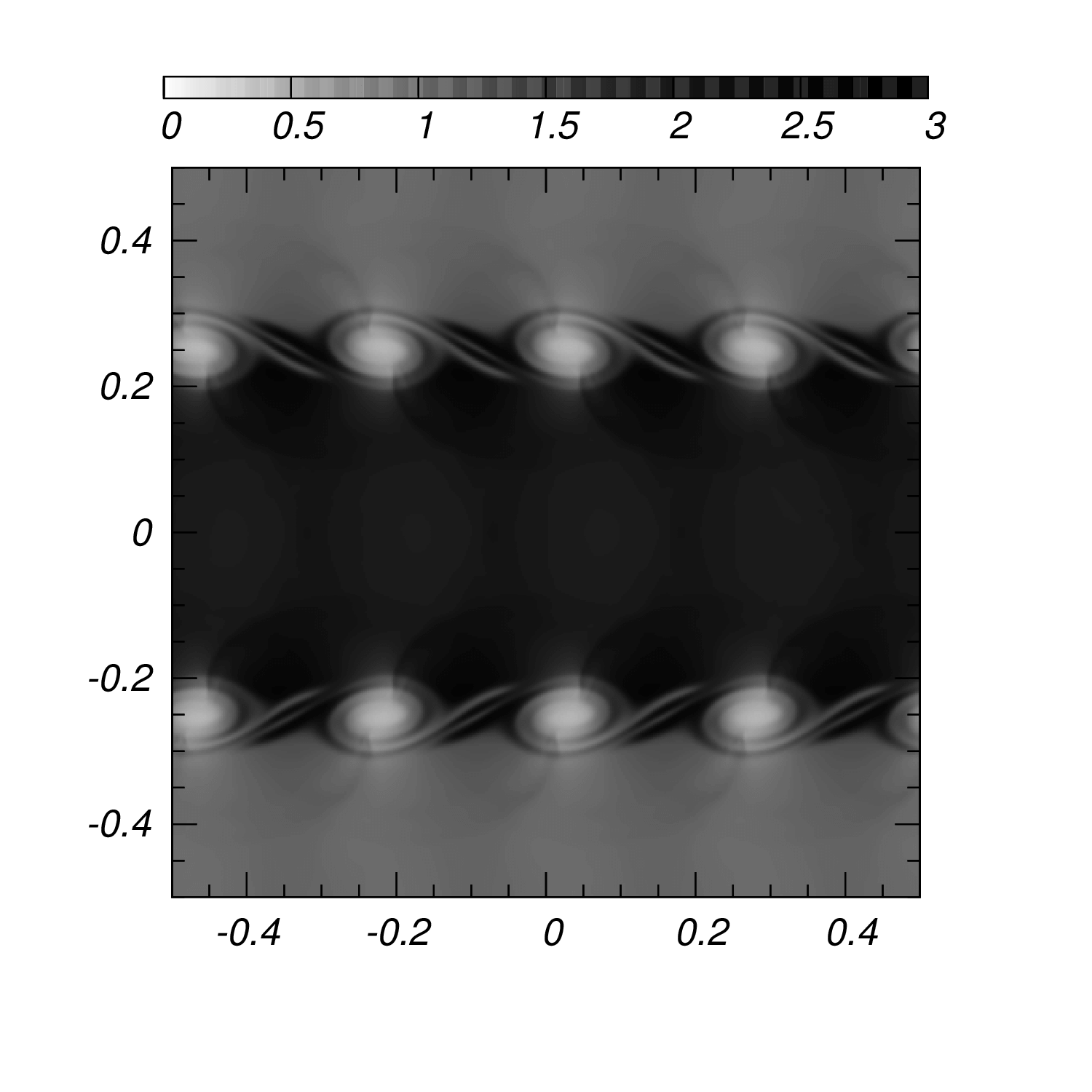}
\includegraphics[width=8cm]{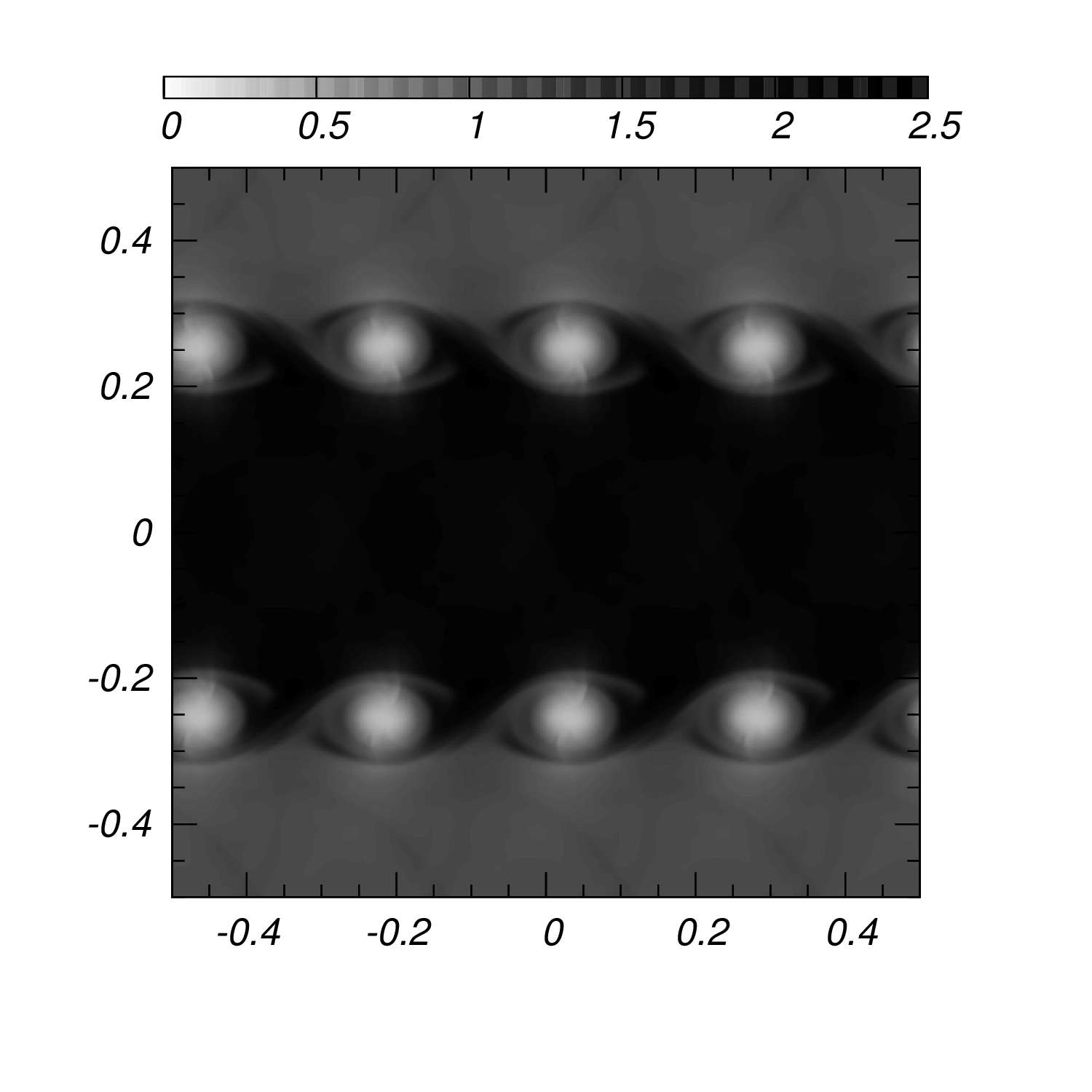}
\end{center}
\caption{\label{fig:KH2}  Relativistic Kelvin Helmholtz instability at different stages. Results obtained using WENO5 at different times are shown. The figures correspond to $t=0.5$ (top-left), $1$ (top-right), $1.5$ (bottom-left) and $2$ (bottom-right). }
\end{figure*}

In Fig. \ref{fig:KH1}, we show the KH instability test at $t=1.5$. We present the proper rest mass density using different reconstructors. The figure shows the density computed with MINMOD (top-left), MC (top-right), PPM (bottom-left) and WENO5 (bottom-right) limiters in combination with the HLLE approximate Riemann solver. As we can see, MC and WENO5 present more sub-structure than MINMOD and PPM, because the later ones  introduce more dissipation. However, the less dissipative a limiter is, more chances there are that unphysical oscillations appear especially when the gas velocity approaches the speed of light. Thus, in order to avoid these oscillations, when the condition $v^2<1-10^{-6}$ is violated, the code uses a constant piecewise reconstructor. In Fig. \ref{fig:KH2}, we show the morphology at various times for different stages of the instability using WENO5 at $t=0.5, 1, 1.5 , 2$.

Aside of the morphological tests, it would also be interesting to estimate the saturation time of the various initial perturbations and its comparison with the linear perturbation theory as in \citep{2004A&A...427..415}. Nevertheless, this task would involve a more systematic analysis, and here we only point out the different features produced by the use of different reconstructors.

\subsubsection{RHD Jets}

The last of the 2D numerical RHD tests corresponds to an axisymmetric relativistic jet, in cylindrical coordinates, injected towards an homogeneos medium. The details of the SRMHD evolution equations in  
cylindrical coordinates can be found in appendix~\ref{appendix}. The beam of the jet is injected with a velocity $v_{b}$ through a circular region of radius $r_{b}=1$. The density of the beam $\rho_b$ and ambient density $\rho_{m}$ are related by $\eta=\rho_{b}/\rho_{m}$, where usually $\eta$ is less than 1. The relativistic Mach number in the beam is defined as ${\cal M}_{b}= M_{b}W_{b}\sqrt{(1-c_{s}^{2})}$, where $M_{b}$ is the classical definition of the Mach number, $W_{b}$ is the Lorentz factor and $c_{s}$ is the sound speed of the fluid. Finally, the pressure of the fluid is constant everywhere at initial time. 
Outflow boundary conditions are used at the boundaries except inside the beam radius, where the values of the variables are kept constant.  In general, the resulting morphology of the relativistic jets shows a bow shock surrrounding a central cocoon, that contains jet gas mixed with shocked ambient gas at the contact discontinuity between them while the mixing is enhanced by turbulent motions.
Internal shocks are produced due to the lack of pressure equilibrium between the beam and the cocoon.

\begin{figure}
\begin{center}
\includegraphics[width=8cm]{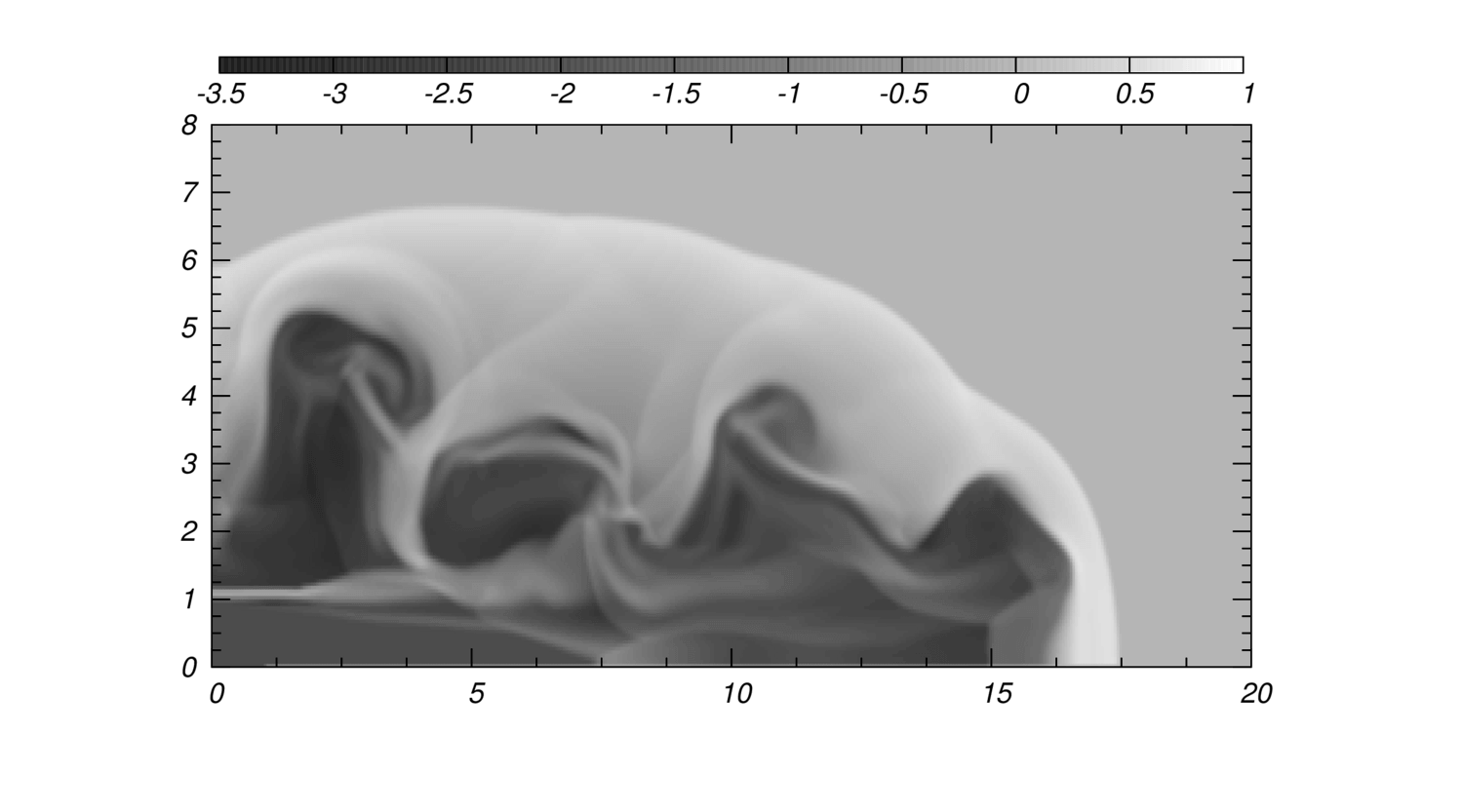}
\end{center}
\caption{\label{fig:jetZanna} Morphology in logarithmic scale of rest mas density for the supersonic jet in \citep{2002A&A...390.1177D} at time $t=40$ using MC. The injection is made with Lorentz factor $~7$ and relativistic Mach number $\sim$18.}
\end{figure}

In Figure \ref{fig:jetZanna} we show the propagation of one of the hydrodynamical jets presented in \citep{2002A&A...390.1177D}. The domain is $[0,8] \times [0,20]$ in $r$ and $z$ directions respectively, where we use a uniform resolution $\Delta r= \Delta z = 0.05$.  The static medium density is $\rho_{m}=10.0$, pressure $p_{m}=0.01$, adiabatic index $\Gamma=5/3$. The jet is injected in a circular region  with radius $r_{b}=1.0$, $v_{z}=0.99$, $\rho_{b}=0.1$. In the figure we show a snapshot at time $t=40$ using MC. The morphology is similar to that obtained in \citep{2002A&A...390.1177D} when using CENO3.

\begin{figure}
\begin{center}
\includegraphics[width=8cm]{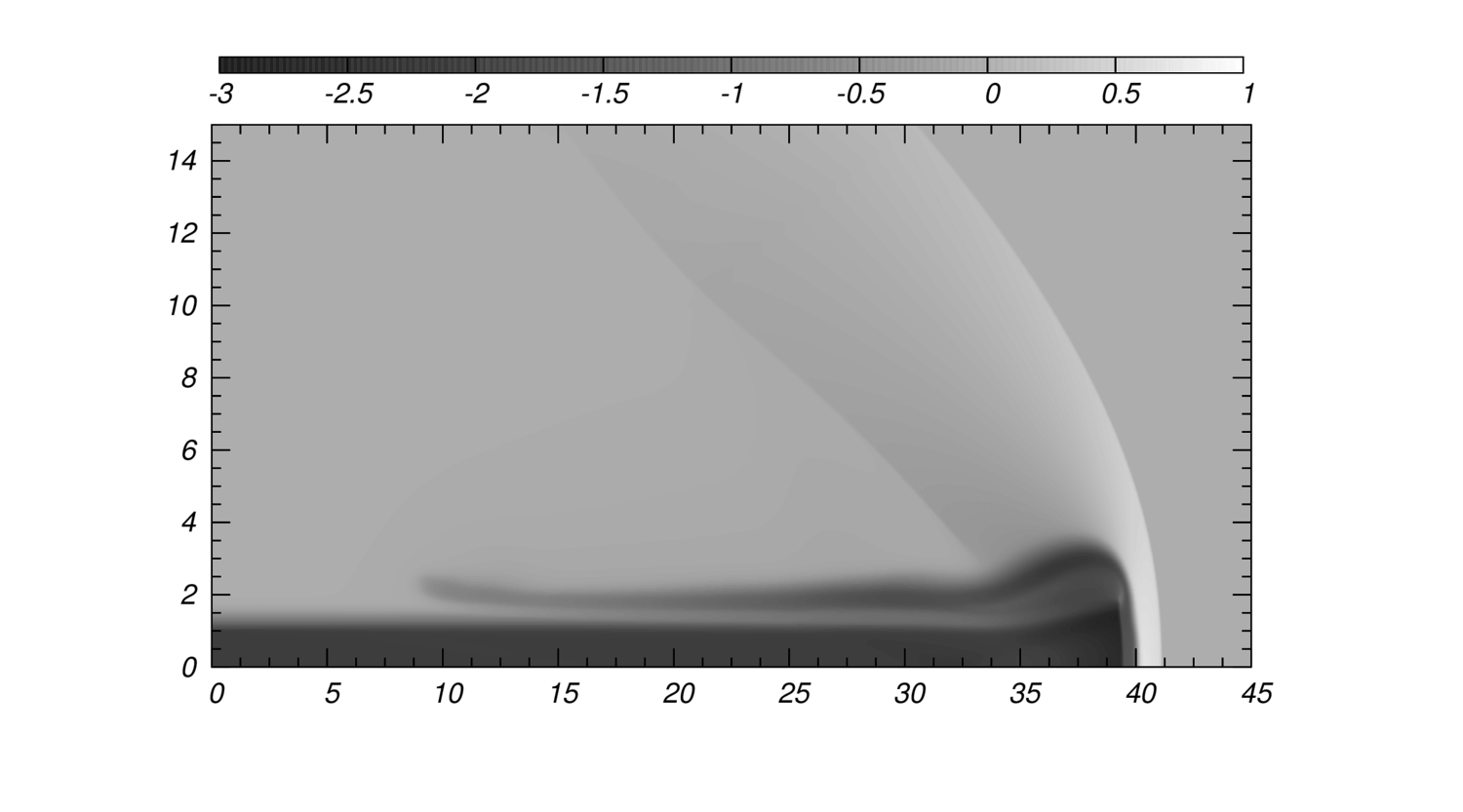}
\includegraphics[width=8cm]{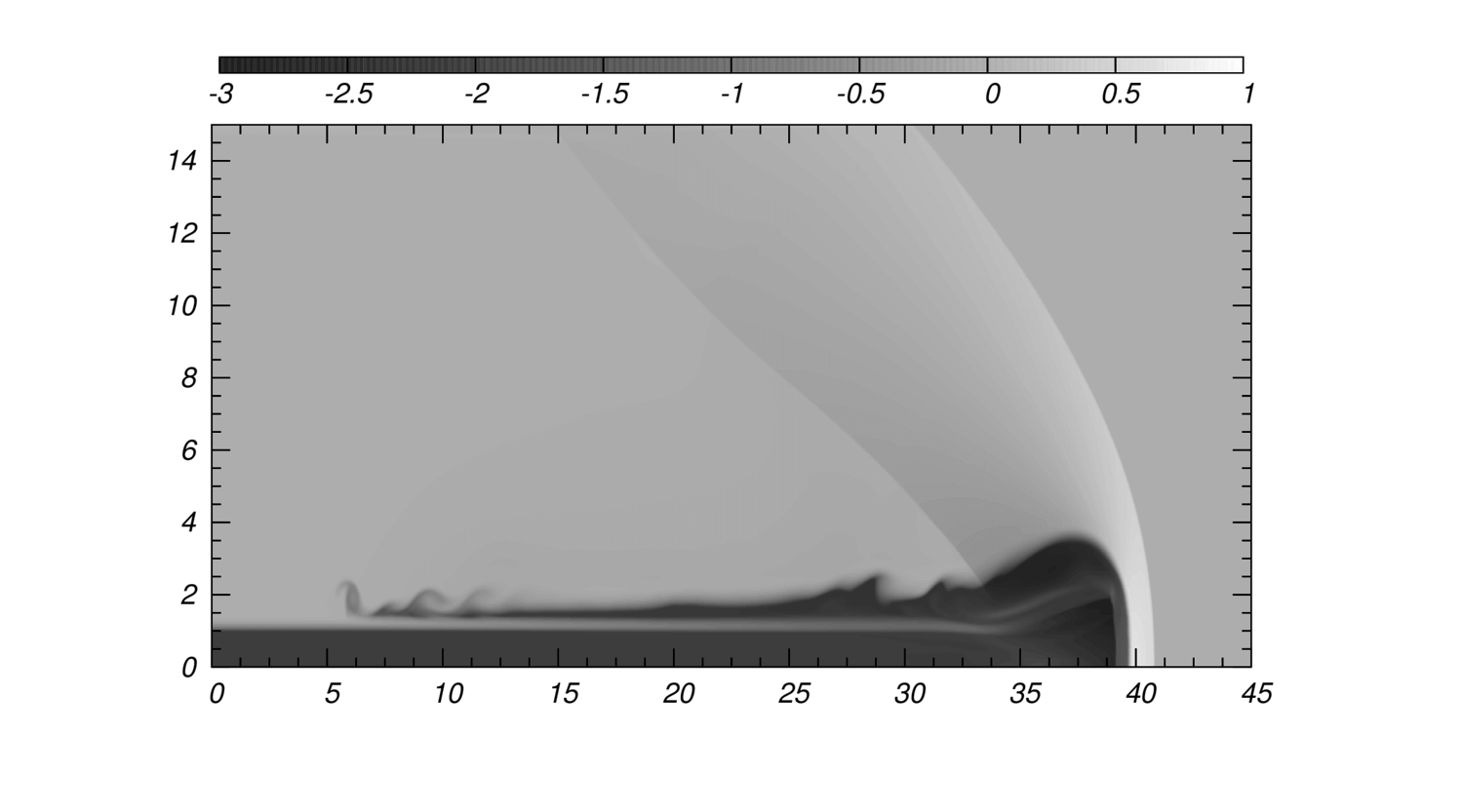}
\includegraphics[width=8cm]{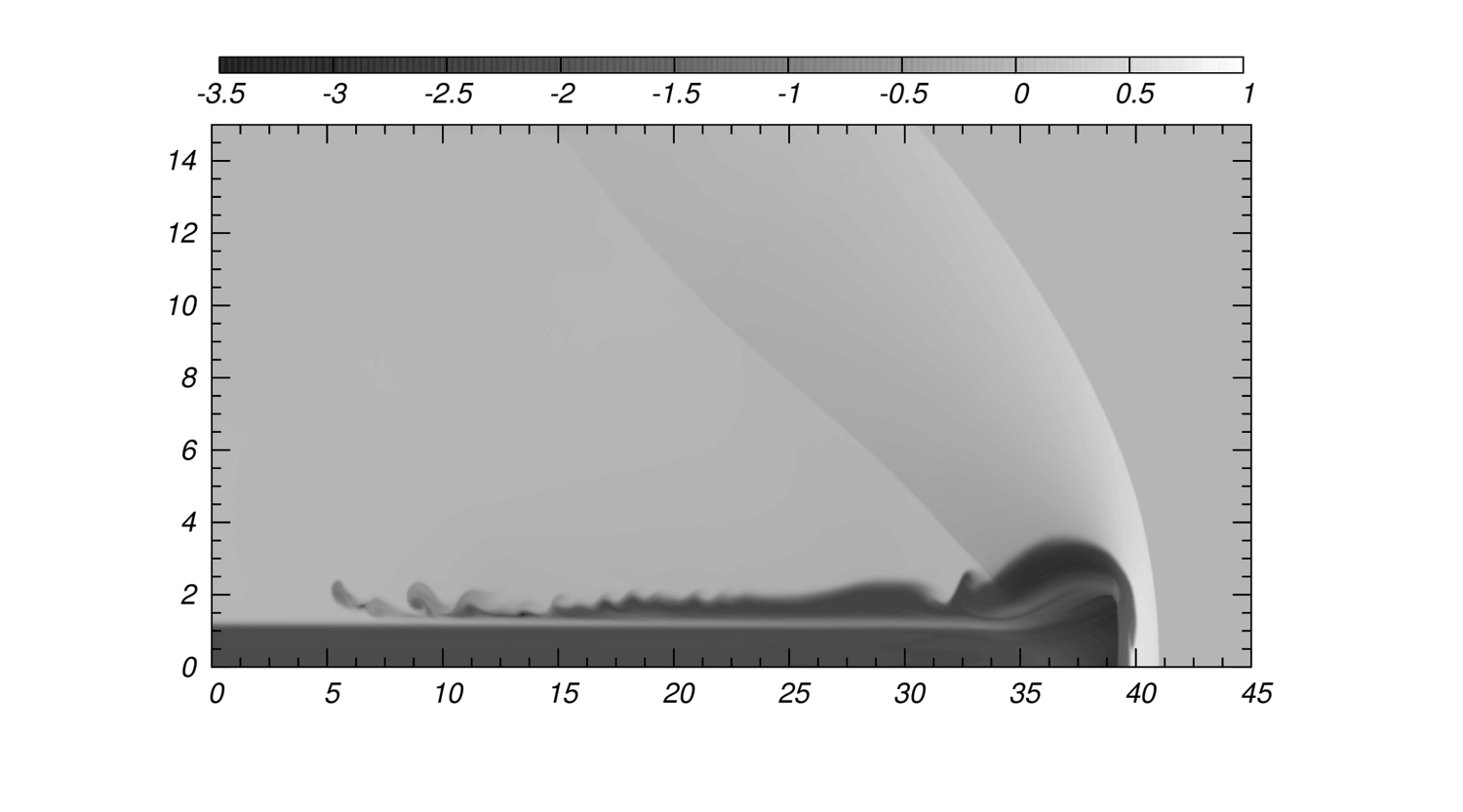}
\end{center}
\caption{\label{fig:jetA1} Logarithm of the rest mass density for the hot jet model A1 in \citep{1997ApJ...479..151M} at time $t=48.82$. In this case we use MINMOD (top), MC (middle) and WENO5 (bottom) reconstructors, in the domain  $[0,15]\times[0,60]$ covered with $576 \times 1920$ cells.}
\end{figure}

In Figure \ref{fig:jetA1} we present the hot model $A1$ in \citep{1997ApJ...479..151M} at time $t=48.82$. The parmeters of the injected jet are: adiabatic index $\Gamma = 4/3$, $v_{b}=0.99$, $\rho_{b}=0.01$, $\eta=0.01$ and the medium parameters are $\rho_{m}=1.0$ and $M_{b}=1.72$. The particular feature of this model is that the bow shock is extended, and has a very thin cocoon as can be seen in the Figure. We compare the morphology of the rest mass density using (from top to bottom) MINMOD, MC and WENO5 reconstructors, and we can see that WENO5 method captures the turbulent shocks in the cocoon region much better than the other reconstructors.

\begin{figure*}
\begin{center}
\includegraphics[width=8cm]{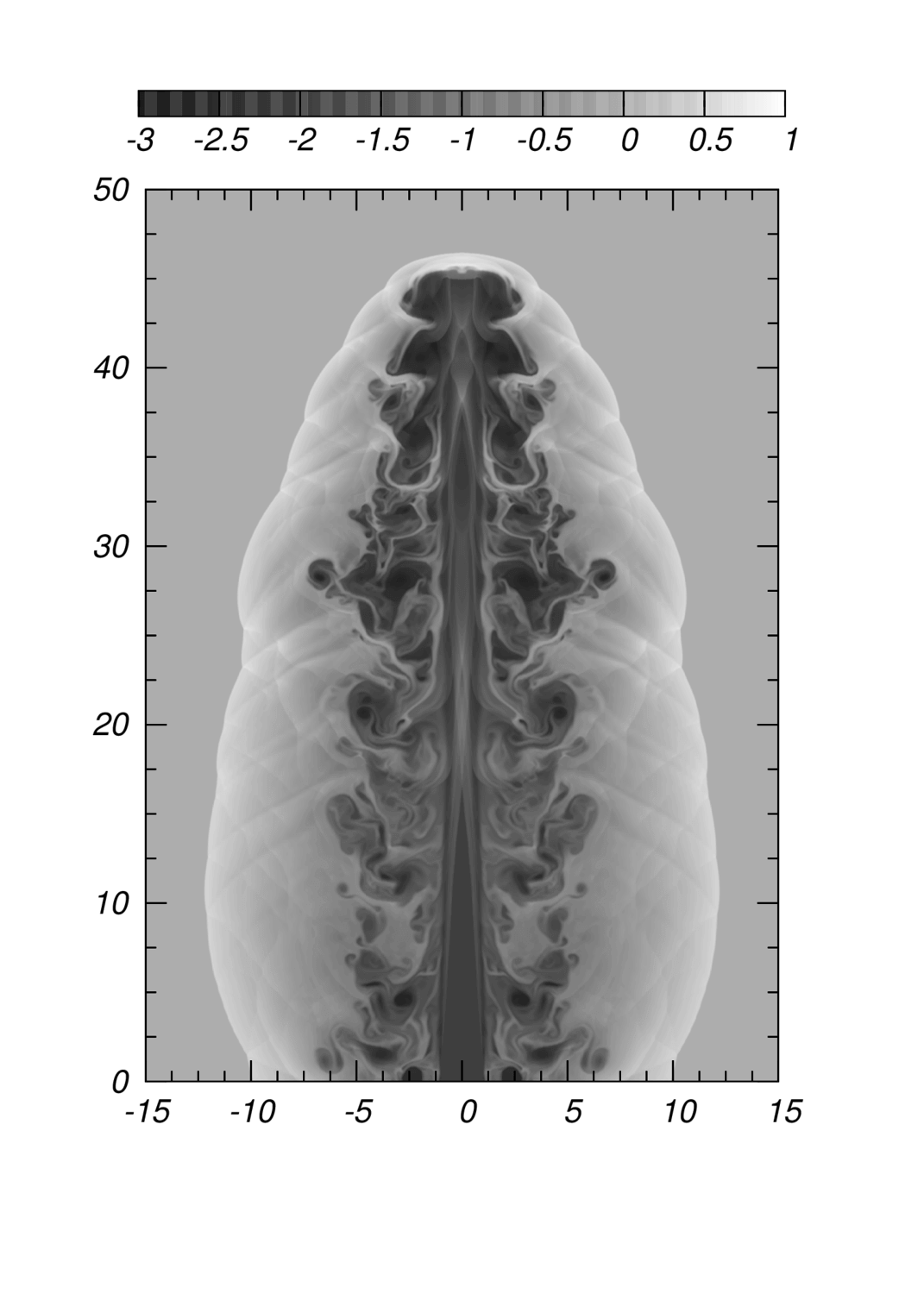}
\includegraphics[width=8cm]{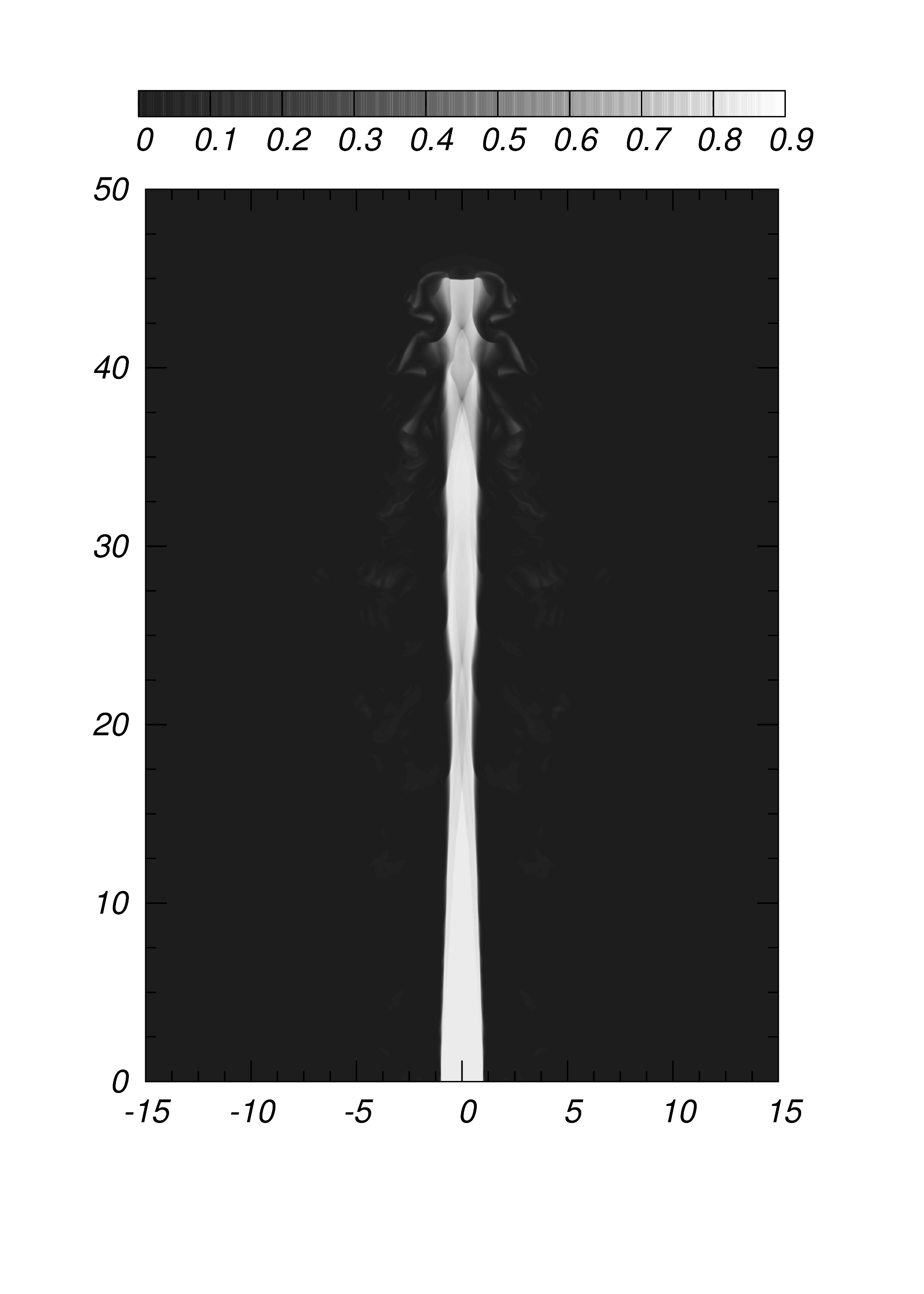}
\end{center}
\caption{\label{fig:jetC2}. Model C2 in \citep{1997ApJ...479..151M}. We show the logarithm of both the rest mass density and Lorentz factor at time $t=110.67$. In this case we use the MC reconstrutor. The domain $[0,15]\times[0,60]$ is covered with $576 \times 1920$ cells.}
\end{figure*}

In Figure \ref{fig:jetC2} we show the rest mass density and the Lorentz factor at time $t=110.67$ of the model C2 in \citep{1997ApJ...479..151M}. The jet parameters are $v_{b}=0.99$, adiabatic index $\Gamma = 5/3$, $\rho_{b}=0.01$, $\eta=0.01$ and the ambient parameters $\rho_{m}=1.0$, $M_{b}=6.0$. We use the MC reconstructor in this case.  This model has the extended bow shock surrounding the jet, and also has a larger cocoon containing the spots with structure.

\subsubsection{3D Spherical Blast Wave}

\begin{figure}
\begin{center}
\includegraphics[width=6.0cm]{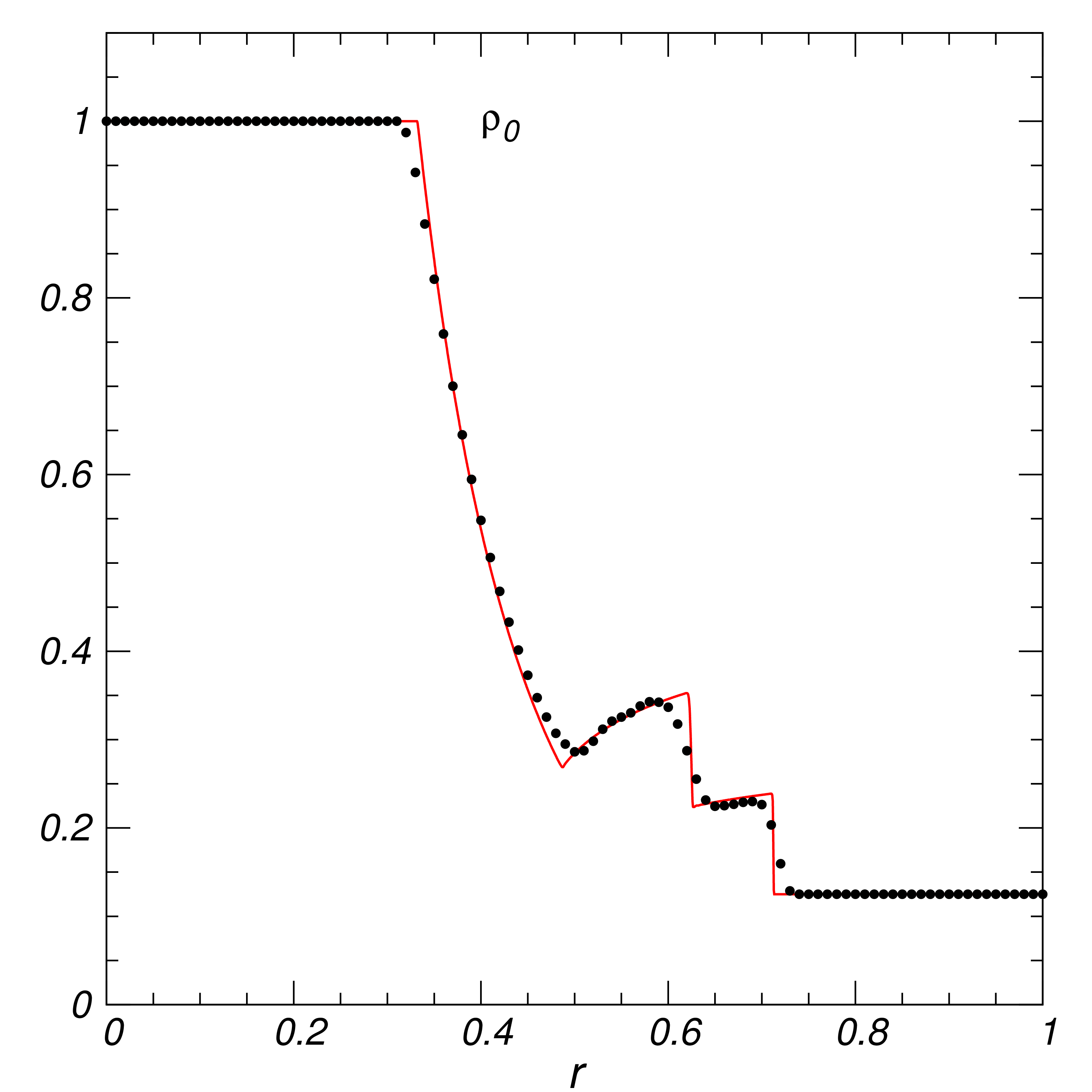}
\includegraphics[width=6.0cm]{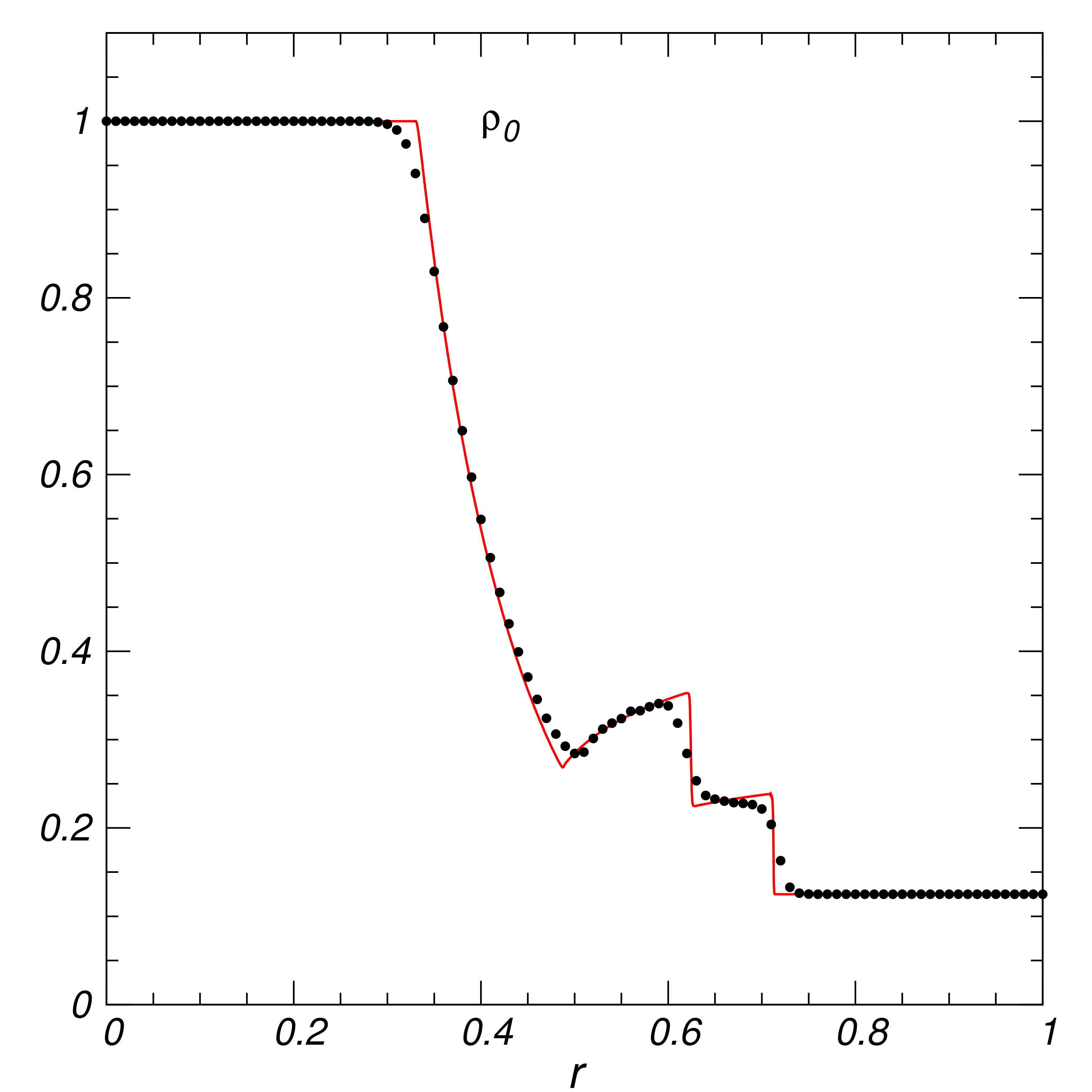}
\end{center}
\caption{\label{fig:BlastWave} Rest mass density of the fluid for the spherical blast wave at $t=0.3$. In these figures, we present the results obtained with two different schemes (points) as well as the reference solution (solid line) which was obtained with the 1D relativistic spherically simmetric code. The schemes used are: MC (top) and WENO5 (bottom).} 
\end{figure}

In order to observe the performance of the code in three dimensions, we consider the spherical blast wave test in relativistic hydrodynamics which involves non-grid-aligned shocks. Specifically, the initial data involves the following parameteres 

\begin{equation}
\nonumber (\rho,p,v^r) =
\left \{
\begin{array}{cc}
 (1.0,1.0,0), ~~~ if ~r < 0.5 \\
 (0.125,0.1,0), ~~ \mbox{elsewhere},
\end{array}
\right. 
\end{equation}

\noindent with adiabatic index $\Gamma=1.4$. Since the analytic solution for this problem is not known, we use as a reference solution, the one calculated with our one-dimensional spherically-symmetric numerical code. The simulation with the spherically symmetric code is done on the domain $r\in[0,1]$ covered with 2500 cells, whereas the 3D code uses the domain $x,y,z \in [0,1]$ covered with $100^3$ cells. Figure \ref{fig:BlastWave} shows  the rest mass density of the fluid computed with two different reconstructors, MC and WENO5 at $t=0.3$, and compared with the obtained with the spherically symmetric code. Even though the resolution used by the 3D code is considerably lower than that of the spherically symmetric code, the numerical solution reproduces all the features captured by the 1D code.

\subsubsection{Relativistic Kelvin Helmholtz Instability in 3D}

Tracking the development of turbulent zones depends highly on the dissipation of the numerical method used. This is specially important in 3D, because the memory required may be restrictive. We illustrate the performance of our code with a variation of the initial conditions shown for the Kelvin-Hemlholtz instabilities above. For this we follow \citep{2011ApJS..193....6B,2012A&A...547A..26R}, where the authors propose a density profile and velocity field with the following components.

\begin{eqnarray}
\rho(y) &=& \left\{ \begin{array}{ll}
	\rho_0 + \rho_1 \tanh [(y-0.5)/a], &~~\mbox{if}~~ y>0\\
         \rho_0 - \rho_1 \tanh [(y+0.5)/a], & ~~ \mbox{if}~~ y \le 0 .\end{array} \right. \nonumber\\
v^x(y) &=& \left\{ \begin{array}{ll}
	V_{\mbox{s}} \tanh [(y-0.5)/a], &~~\mbox{if}~~ y>0\\
        V_{\mbox{s}} \tanh [(y+0.5)/a], & ~~ \mbox{if}~~ y \le 0 .\end{array} \right. \label{eq:3dthc}\\
v^y(x,y) &=& \left\{ \begin{array}{ll}
	A_0 V_{\mbox{s}} \sin (2\pi x) \exp[-(y-0.5)^2/\sigma], &~~\mbox{if}~~ y>0\\
        -A_0 V_{\mbox{s}} \sin (2\pi x) \exp [-(y+0.5)/\sigma], & ~~ \mbox{if}~~ y \le 0 .\end{array} \right. \nonumber
\end{eqnarray}

\noindent where $\rho_0=0.505$, $\rho_1=0.495$, $a=0.1$, $V_{\mbox{s}}=0.5$ and $\Gamma=4/3$. The problem is solved in the domain 
$-0.5 \le x \le 0.5$
$-1 \le y \le 1$
$-0.5 \le z \le 0.5$

\noindent using $256 \times 512 \times 256$ cells.

\begin{figure}
\begin{center}
\includegraphics[width=6.0cm]{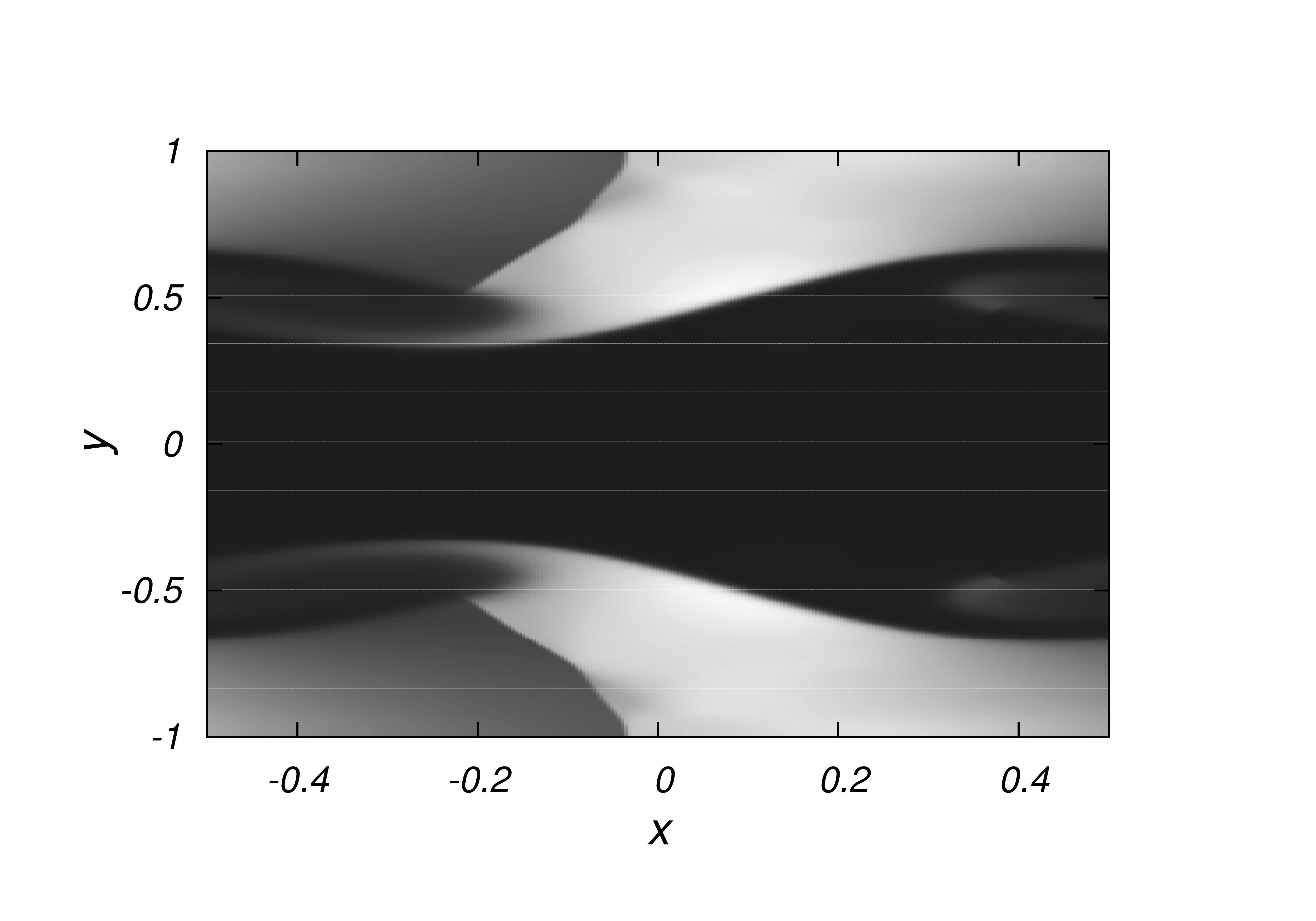}
\includegraphics[width=6.0cm]{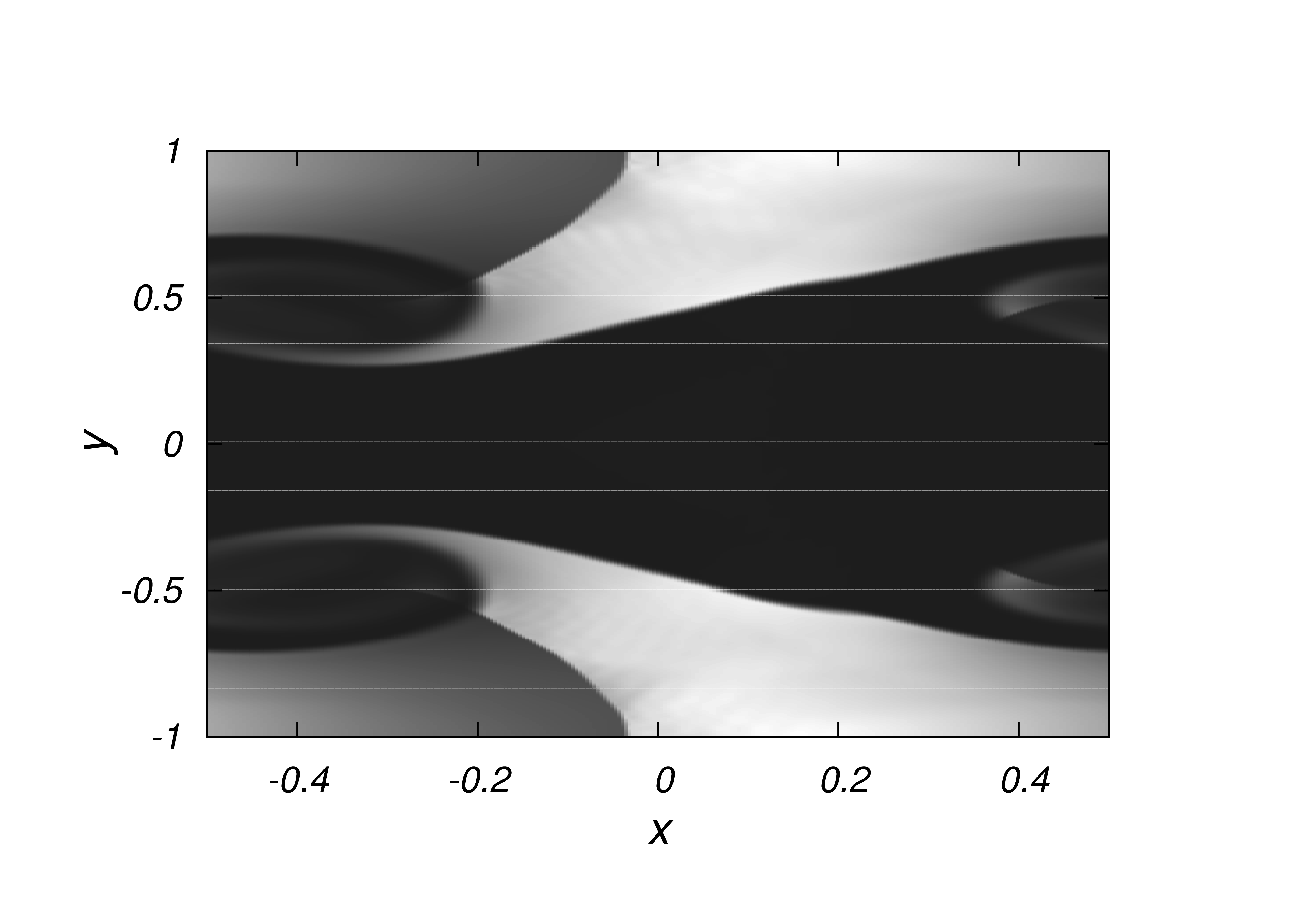}
\includegraphics[width=6.0cm]{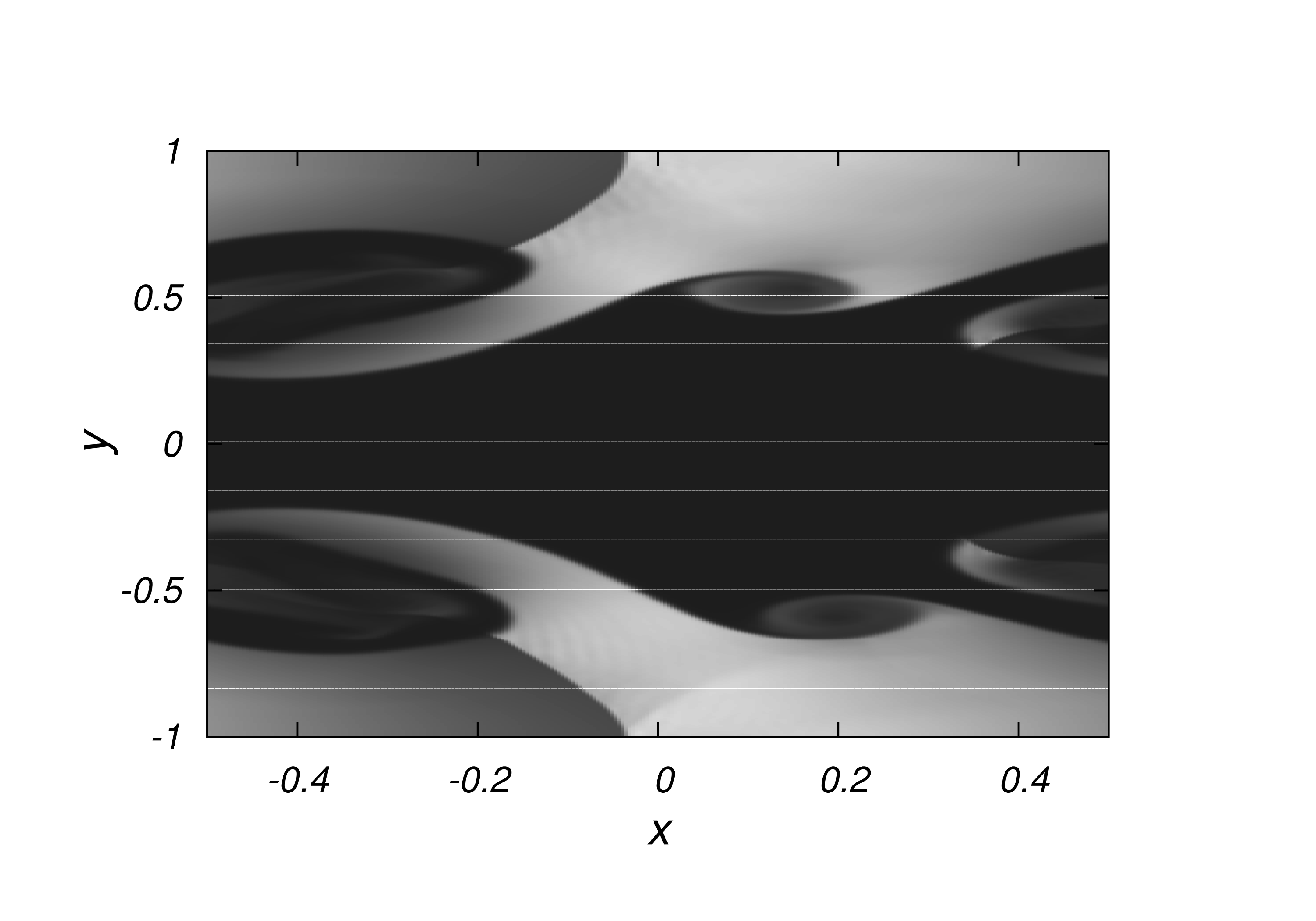}
\end{center}
\caption{\label{fig:THC3D} 3D development of Kelvin-Helmholtz instabilities starting from the initial data in (\ref{eq:3dthc}). We show the results at $t=3$ using MINMOD, MC and WENO.} 
\end{figure}

The 3D nature of the test is given by the addition of a non-trivial $v^z$ component that has a random amplitude between 0 and 0.01 added at initial time. This random perturbation triggers a small asymmetry during the evolution. In Fig. \ref{fig:THC3D} we show the results at time $t=3$ using three different reconstructors. In this situation the WENO limiter is the one that best captures the development of small structure, whereas the linear reconstructors add enough dissipation as to wash out the small structures. \\

\subsection{Relativistic Magnetohydrodynamic tests}

Now we present the standard 1D and 2D tests for the RMHD according to the standard literature \citep{2001ApJS..132...83B,1999MNRAS.303..343K,2003A&A...400..397D}. The 1D tests, which are summarized in Table \ref{tab:MHD_tests}, consist in 1D Riemann problems with the special feature that now there is at least a non-trivial component of the magnetic field. Again, for 1D tests we are using the 3D code with five cells along the transverse directions. We compare the numerical solution with the exact solution, which was computed using Giacomazzo's code \citep{2006JFM...562..223G,2007CQGra..24S.235G}. Unless otherwise stated, the results presented in the figures are obtained with the MINMOD limiter, HLLE Riemann solver and Flux-CT method to preserve the constraint (\ref{eq:constraint}). Furthermore, in order to compare the accuracy using different limiters, we have calculated an error for each 1D Riemann test using the limiters MINMOD, MC, PPM and WENO5. In each figure, we show the proper rest mass density $\rho_0$, the total pressure $p+p_{m}$, the magnetic field components $B^y$, $B^z$ and the velocity components $v^x$, $v^y$. 

We test our multidimensional scheme, with  the following 2D simulations in the presence of magnetic field: relativistic cylindrical explosion, relativistic rigid rotor, a relativistic Kelvin-Helmholtz instability and finally a 3D test regarding the magnetic field loop advection test.  

\begin{table}
\centering
\resizebox{8.5cm}{!}{
\begin{tabular}{c|c|c|c|c|c|c|c|c|c}\hline \hline
${\bf Test ~ type}$ & $\Gamma$ & $\rho_0$ & $p$ & $v^x$ & $v^y$ & $v^z$ & $B^x$ & $B^y$ & $B^z$ \\ \hline \hline
${\bf Test ~ 1}$ &  &  &  &  &  &  &  &  &  \\
Left state  & 4/3 & 1.0 & 1000.0 & 0.0 & 0.0  & 0.0  & 1.0  & 0.0  & 0.0  \\ 
Right state &  & 0.1 & 1.0 & 0.0  & 0.0  & 0.0  & 1.0 & 0.0 & 0.0  \\ \hline \hline
${\bf Test ~ 2}$ &  &  &  &  &  &2 &  &  &  \\
Left state  & 4/3 & 1.0 & 1.0 & $5/\sqrt{26}$ & 0.0  & 0.0  & 10.0  & 10.0  & 0.0  \\ 
Right state &  & 1.0 & 1.0 & $-5/\sqrt{26}$  & 0.0  & 0.0  & 10.0 & -10.0 & 0.0  \\ \hline \hline 
${\bf Test ~ 3}$   &  &  &  &  &  &  &  &  &  \\
Left state  & 2 & 1.0 & 1.0 & 0.0 & 0.0  & 0.0  & 0.5  & 1.0  & 0.0 \\ 
Right state &  & 0.125 & 0.1 & 0.0  & 0.0  & 0.0  & 0.5 & -1.0 & 0.0  \\ \hline \hline
${\bf Test ~ 4}$   &  &  &  &  &  &  &  &  &  \\
Left state  & 5/3 & 1.0 & 30.0 & 0.0 & 0.0  & 0.0  & 5.0  & 6.0  & 6.0 \\ 
Right state &  & 1.0 & 1.0 & 0.0  & 0.0  & 0.0  & 5.0 & 0.7 & 0.7  \\ \hline \hline
${\bf Test ~ 5}$   &  &  &  &  &  &  &  &  &  \\
Left state  & 5/3 & 1.0 & 1000.0 & 0.0 & 0.0  & 0.0  & 10.0  & 7.0  & 7.0  \\ 
Right state &  & 1.0 & 0.1 & 0.0  & 0.0  & 0.0  & 10.0 & 0.7 & 0.7  \\ \hline \hline 
${\bf Test ~ 6}$   &  &  &  &  &  &  &  &  &  \\
Left state  & 5/3 & 1.0 & 0.1 & 0.999 & 0.0  & 0.0  & 10.0  & 7.0  & 7.0  \\ 
Right state &  & 1.0 & 0.1 & -0.999  & 0.0  & 0.0  & 10.0 & -7.0 & -7.0  \\ \hline \hline
${\bf Test ~ 7}$   &  &  &  &  &  &  &  &  &  \\
Left state  & 5/3 & 1.08 & 0.95 & 0.4 & 0.3  & 0.2  & 2.0  & 0.3  & 0.3  \\ 
Right state &  & 1.0 & 1.0 & -0.45  & -0.2  & 0.2  & 2.0 & -0.7 & 0.5  \\ \hline \hline 
${\bf Test ~ 8}$   &  &  &  &  &  &  &  &  &  \\
Left state  & 5/3 & 1.0 & 5.0 & 0.0 & 0.3  & 0.4  & 1.0  & 6.0  & 2.0  \\ 
Right state &  & 0.9 & 5.3 & 0.0  & 0.0  & 0.0  & 1.0 & 5.0 & 2.0  \\ \hline \hline 
\end{tabular}
}
\caption{\label{tab:MHD_tests} Initial conditions for the magnetized 1D Riemann problems. $\Gamma$ corresponds to the adiabatic index and $(\rho_0,v^i,p,B^k)$ are the initial primitive variables of the left and right states.}
\end{table} 

\subsubsection{Test 1: Komissarov shock tube} 

\begin{figure}
\begin{center}
\includegraphics[width=4.0cm]{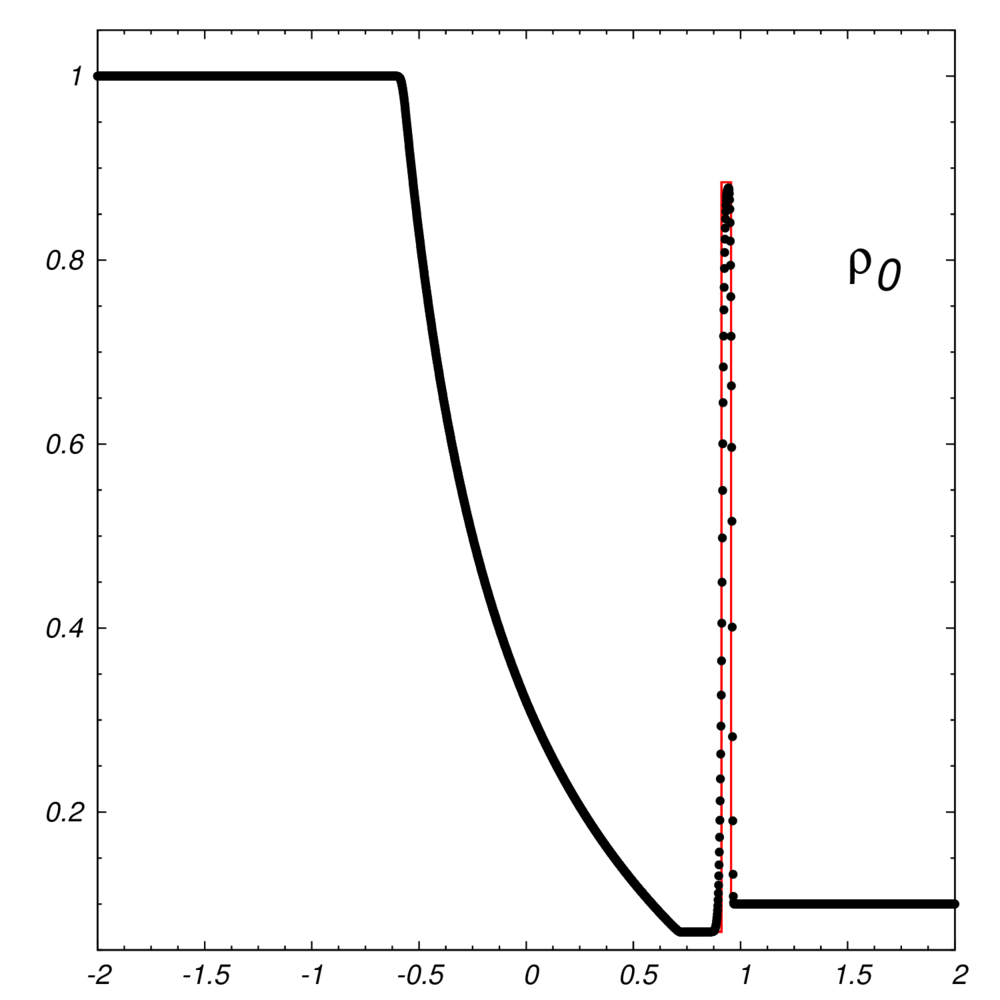}
\includegraphics[width=4.0cm]{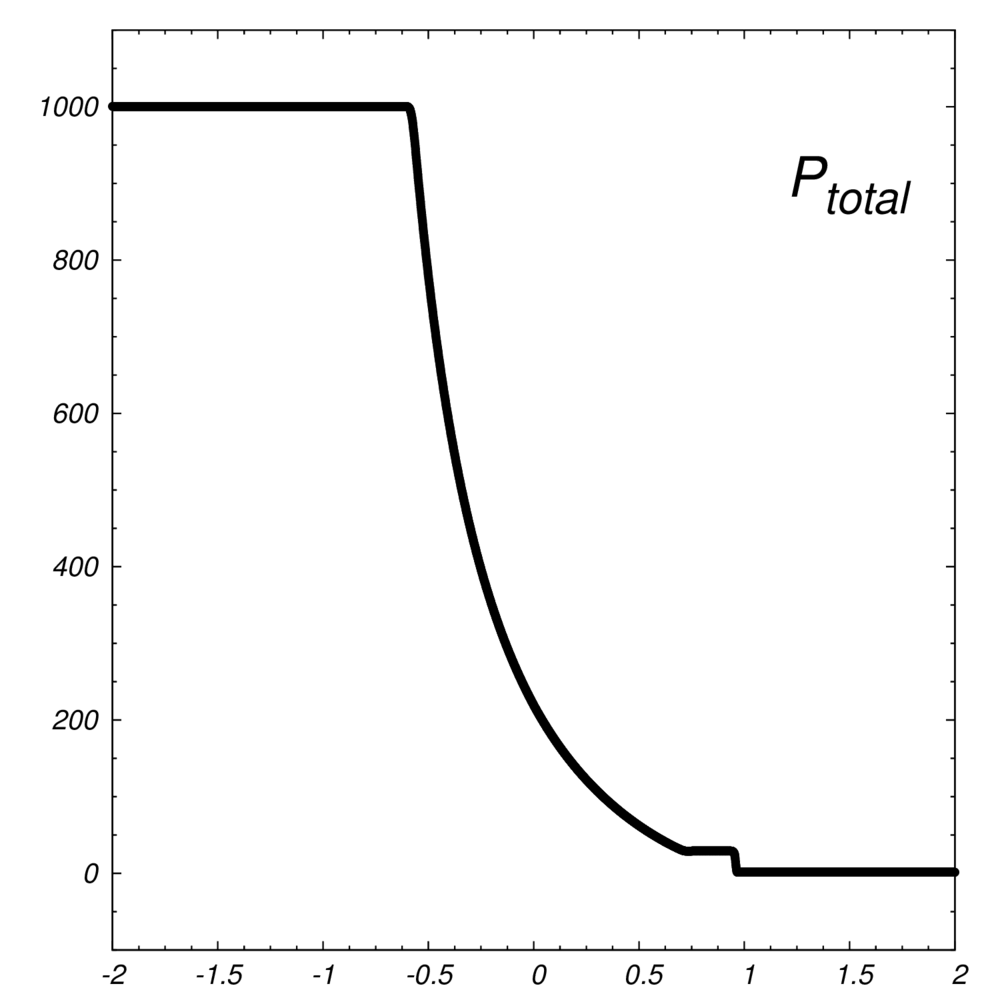}
\includegraphics[width=4.0cm]{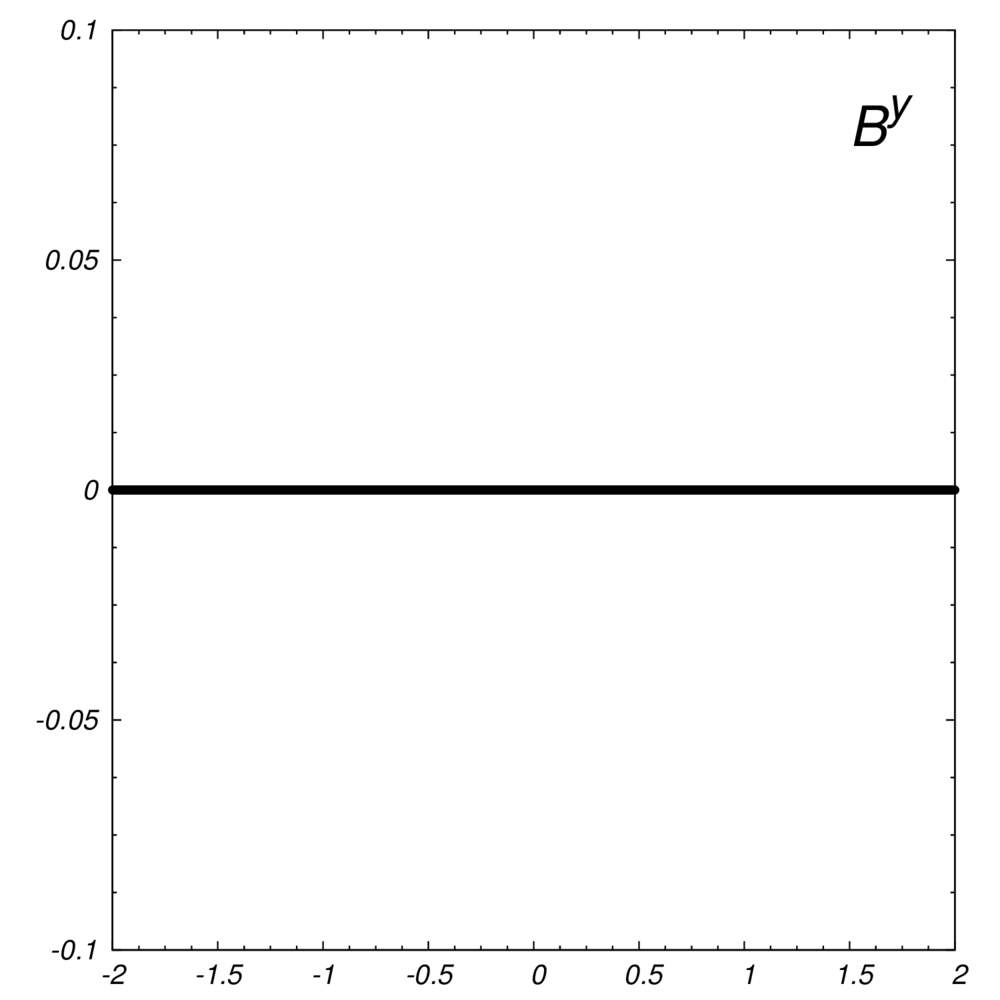}
\includegraphics[width=4.0cm]{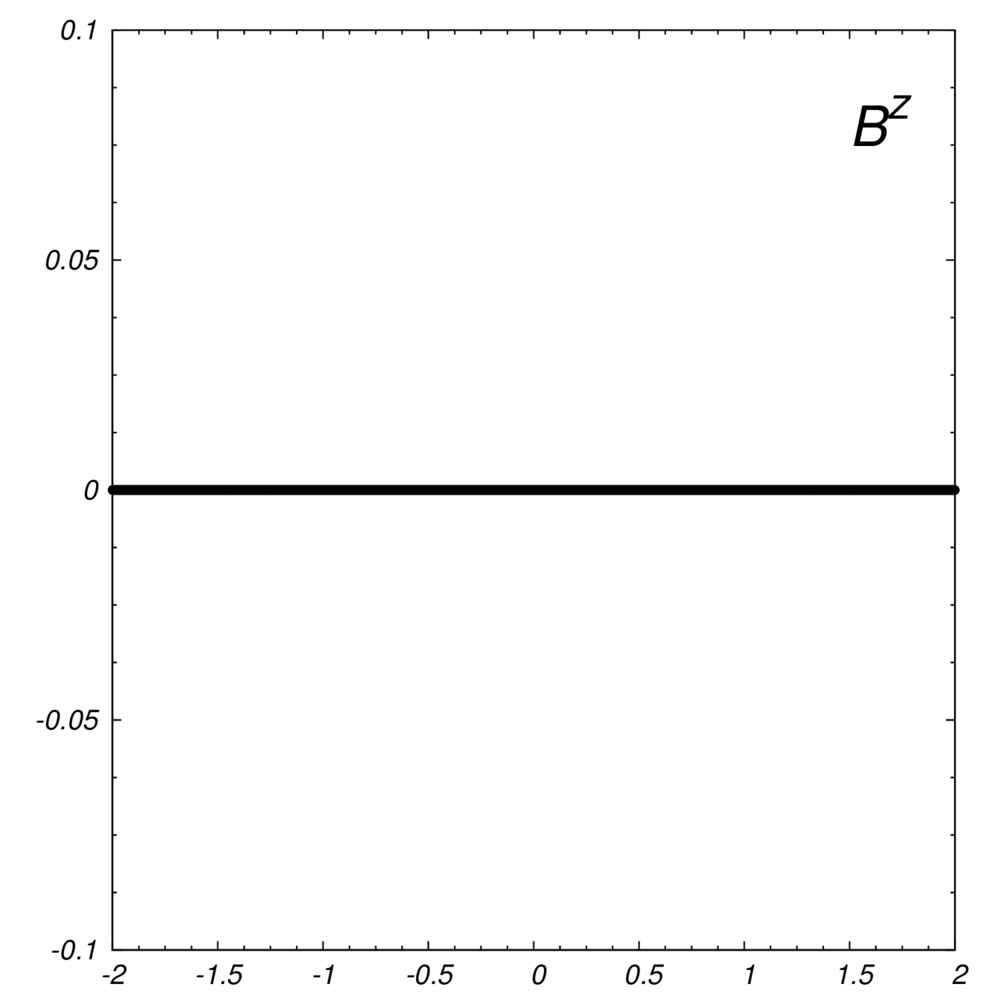}
\includegraphics[width=4.0cm]{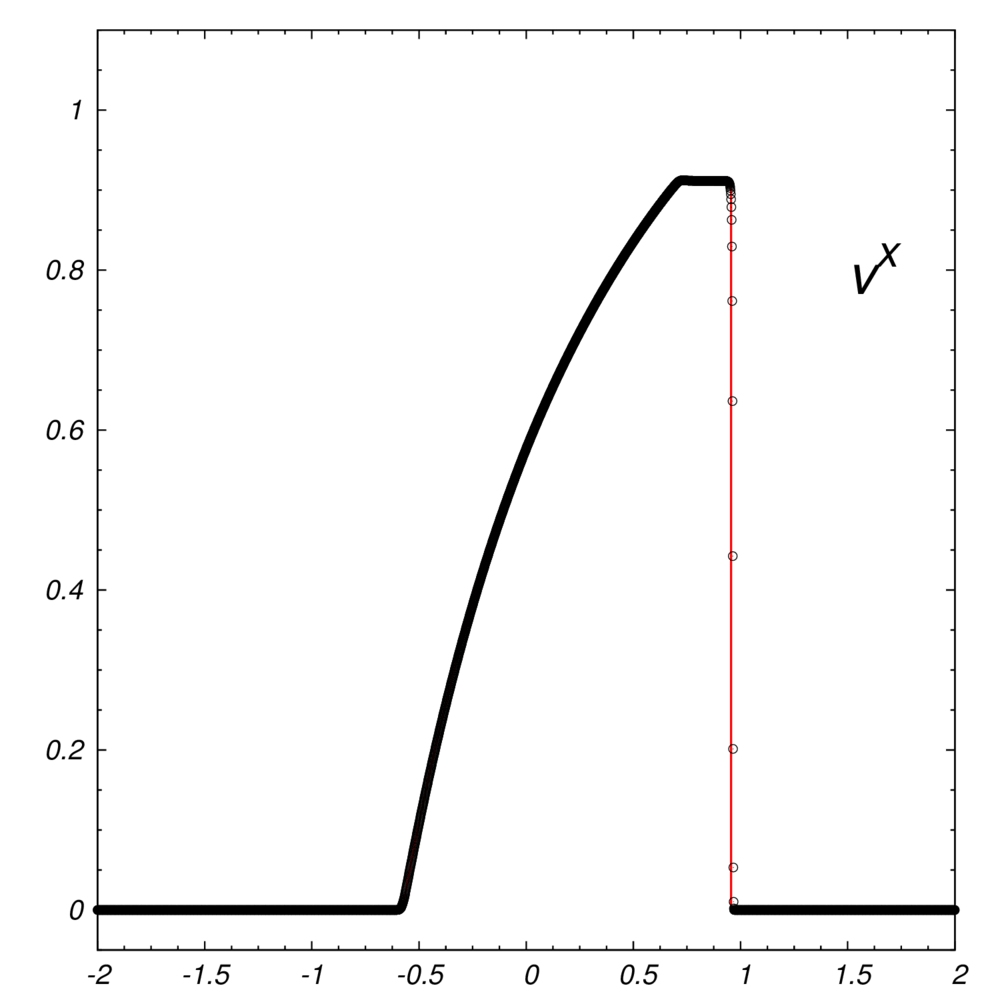}
\includegraphics[width=4.0cm]{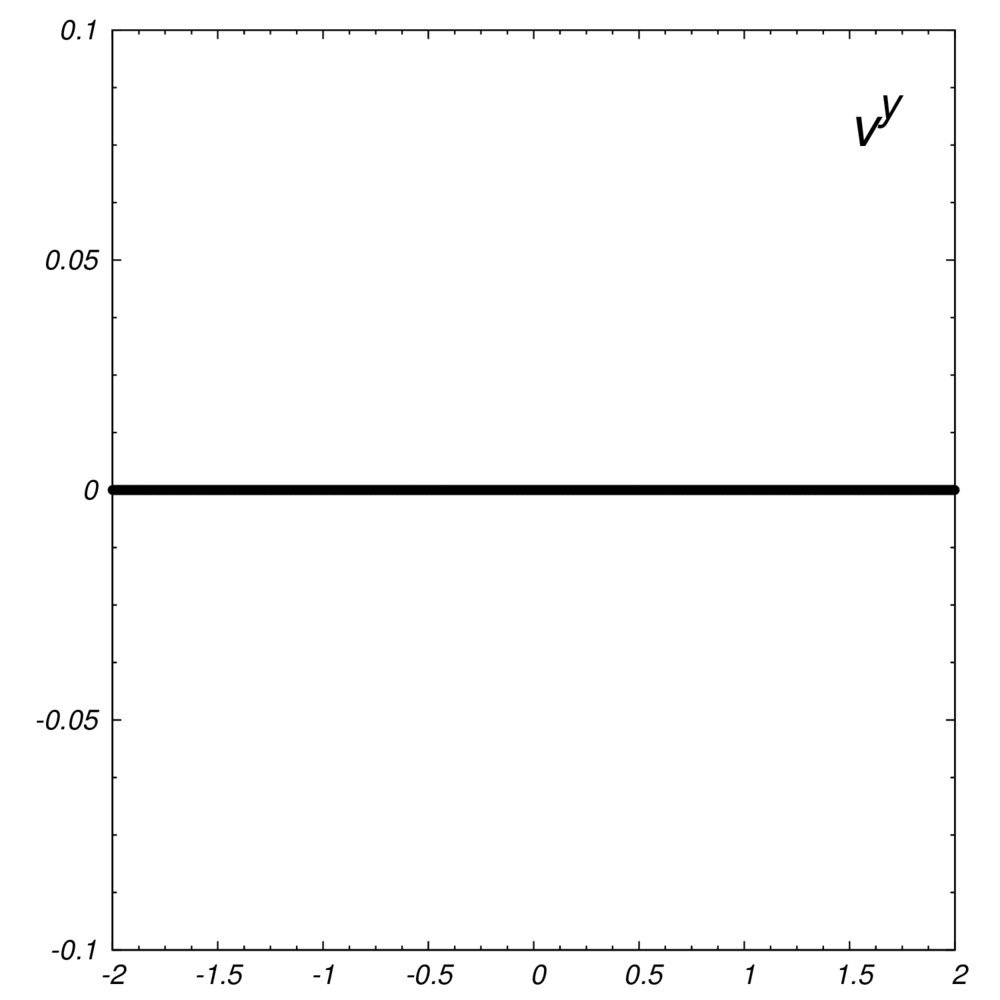}
\end{center}
\caption{\label{fig:KomissarovST} Test 1: Komissarov shock tube at $t=1.0$. We use $3200$ cells in the domain $[-2,2]$ and $CFL=0.1$.} 
\end{figure}

In the plots of 1D tests we again use lines to represent the exact solution and points to represent the numerical solutions. 
The first test is a shock tube test with pressure ratio of $p\sim 10^{3}$ and a constant  magnetic field along the shock direction. In Figure  \ref{fig:KomissarovST}, we show the numerical and exact solutions computed in the domain $x\in[-2,2]$ at $t=1.0$, using resolution $\Delta x=1/800$.  A fast rarefaction zone moves to the left and a fast shock to the right from the contact discontinuity. The high difference of the pressure produces a thin shell in the density that moves with relativistic velocities in the shock direction. We can also verify in the figure that the transverse components of the magnetic and velocity fields are zero at all times during the evolution as expected. We obtain similar results using different reconstructors and divergence control methods. 

\subsubsection{Test 2: Komissarov collision test}

\begin{figure}
\begin{center}
\includegraphics[width=4.0cm]{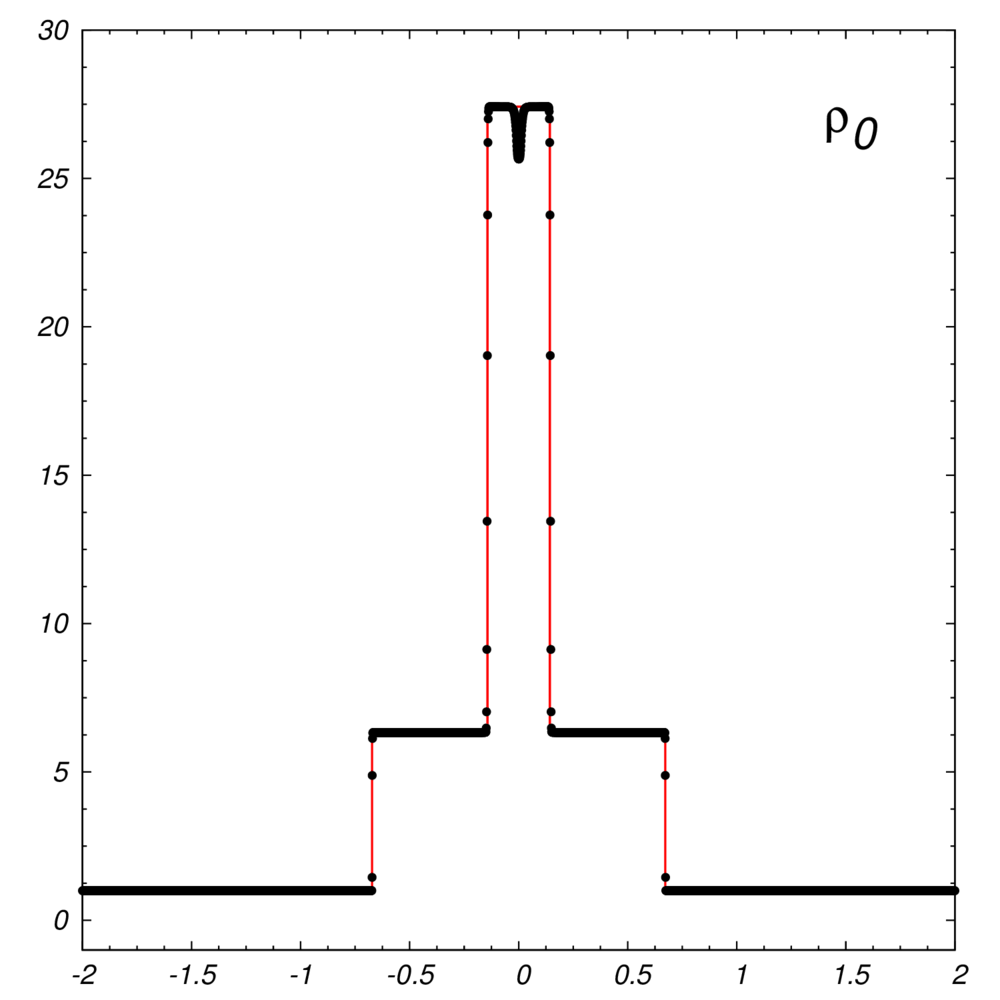}
\includegraphics[width=4.0cm]{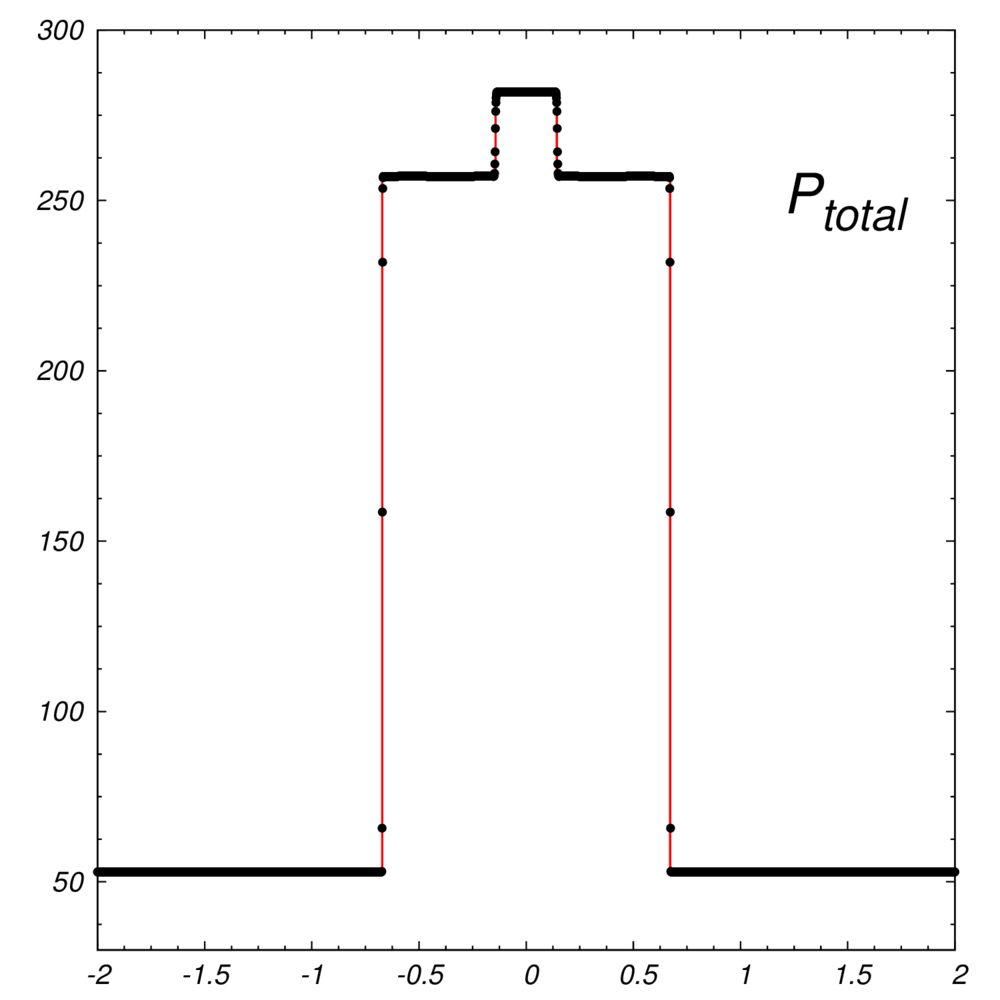}
\includegraphics[width=4.0cm]{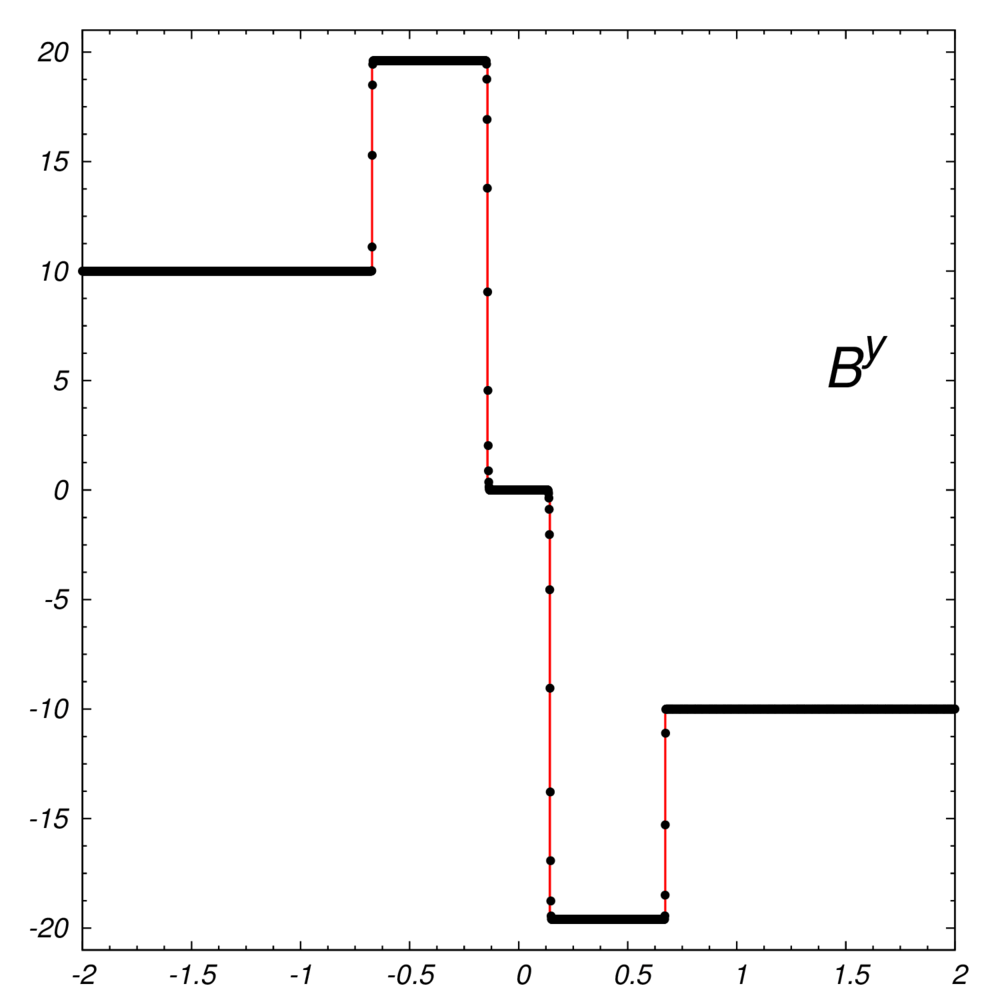}
\includegraphics[width=4.0cm]{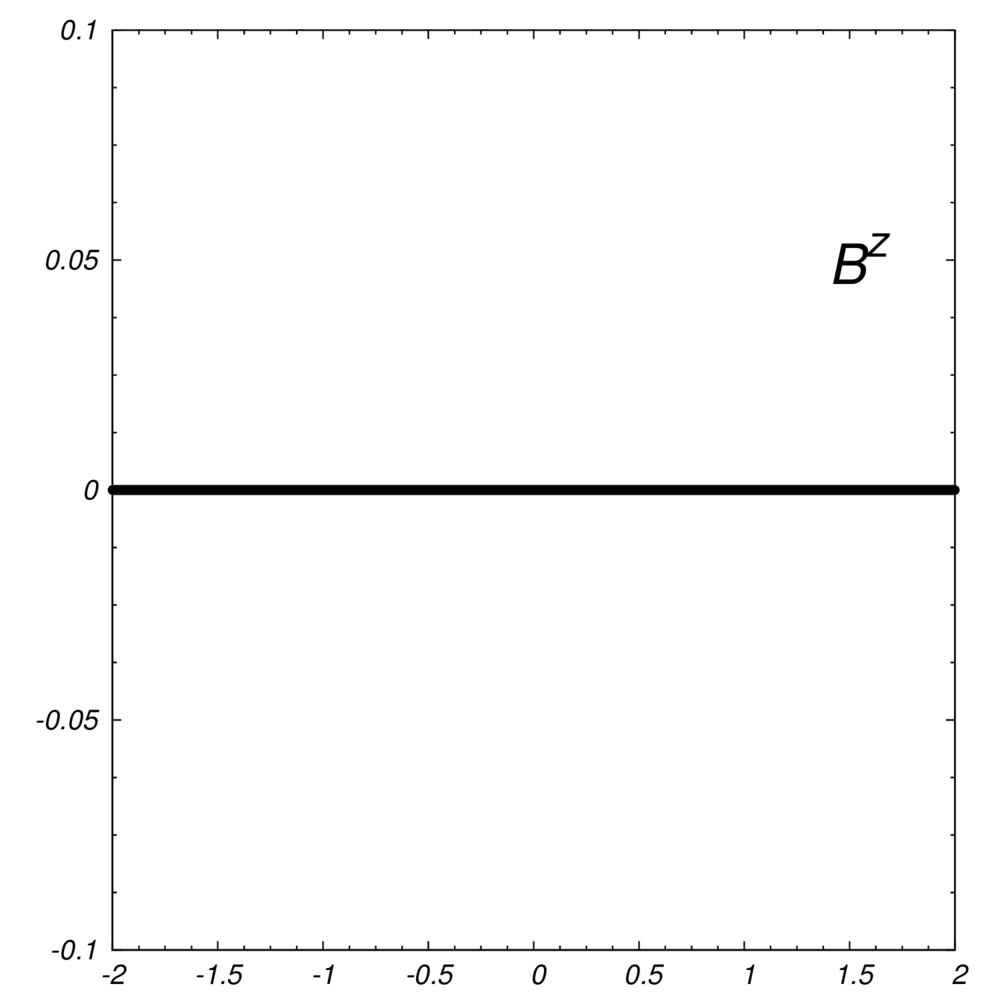}
\includegraphics[width=4.0cm]{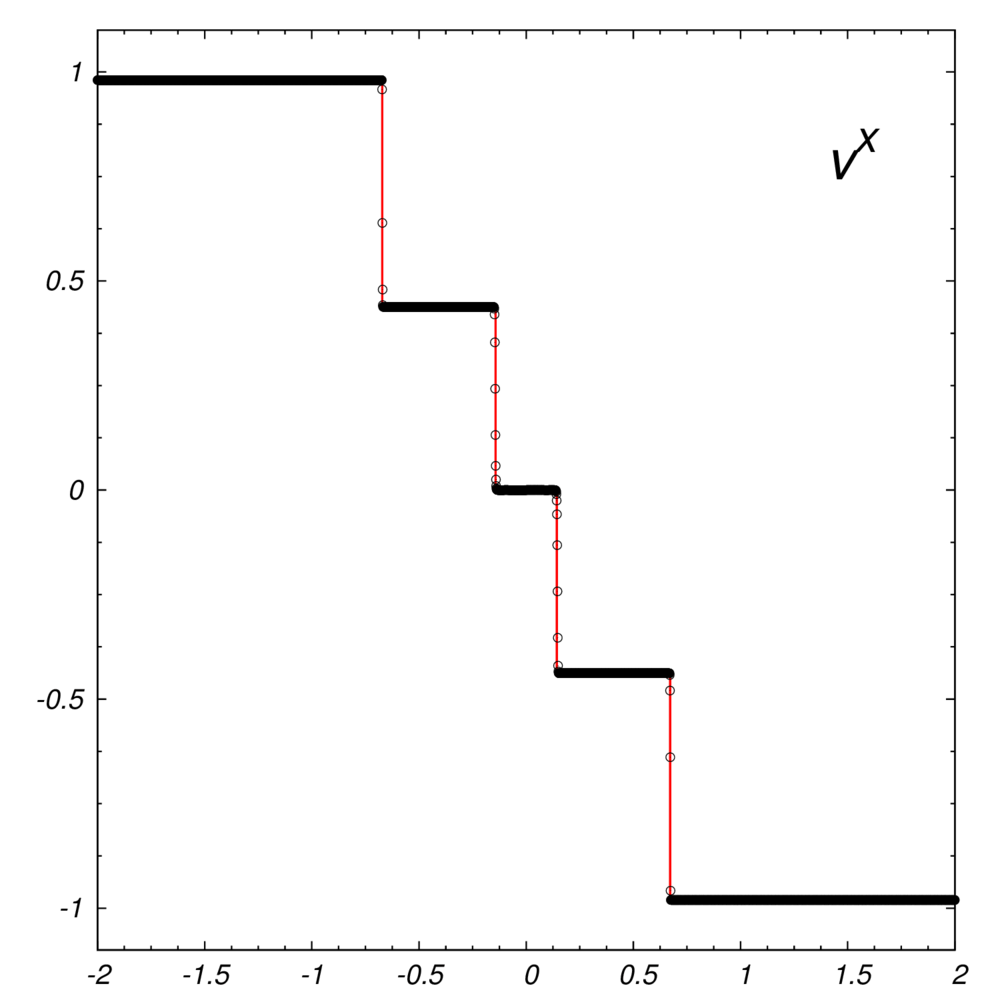}
\includegraphics[width=4.0cm]{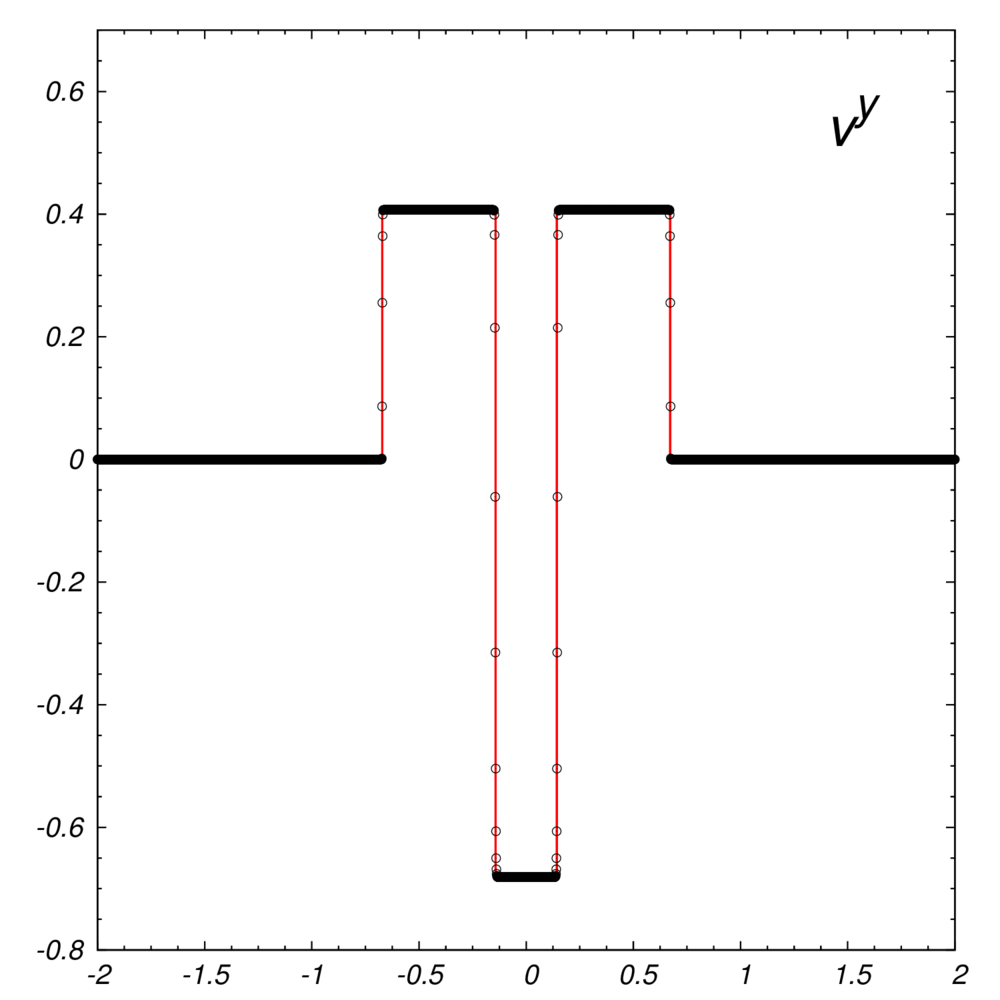}
\end{center}
\caption{\label{fig:KomissarovColl} Test 2: Komissarov collision test at $t=1.22$. We cover the domain $[-2,2]$ with 3200 cells and use $CFL=0.1$.}
\end{figure}

This is the collision of streams moving in opposite direction with initial head-on velocity $v^{x}= 0.98058$, in this test the dynamics of the fluid is immersed in a magnetic field with constant $x$ component and discontinuous $y$ component. We compare the results with the exact solution in Figure \ref{fig:KomissarovColl}. As in the Komissarov shock tube test, we use a numerical domain $x\in[-2,2]$ at $t=1.22$, using resolution $\Delta x=1/800$.  In this case two slow and two fast shocks move to the left and to the right as expected. We experimented with different reconstructors and found that the MC limiter, unlike the other limiters, develops numerical oscillations. 

\subsubsection{Test 3:  Balsara 1 test}

\begin{figure}
\begin{center}
\includegraphics[width=4.0cm]{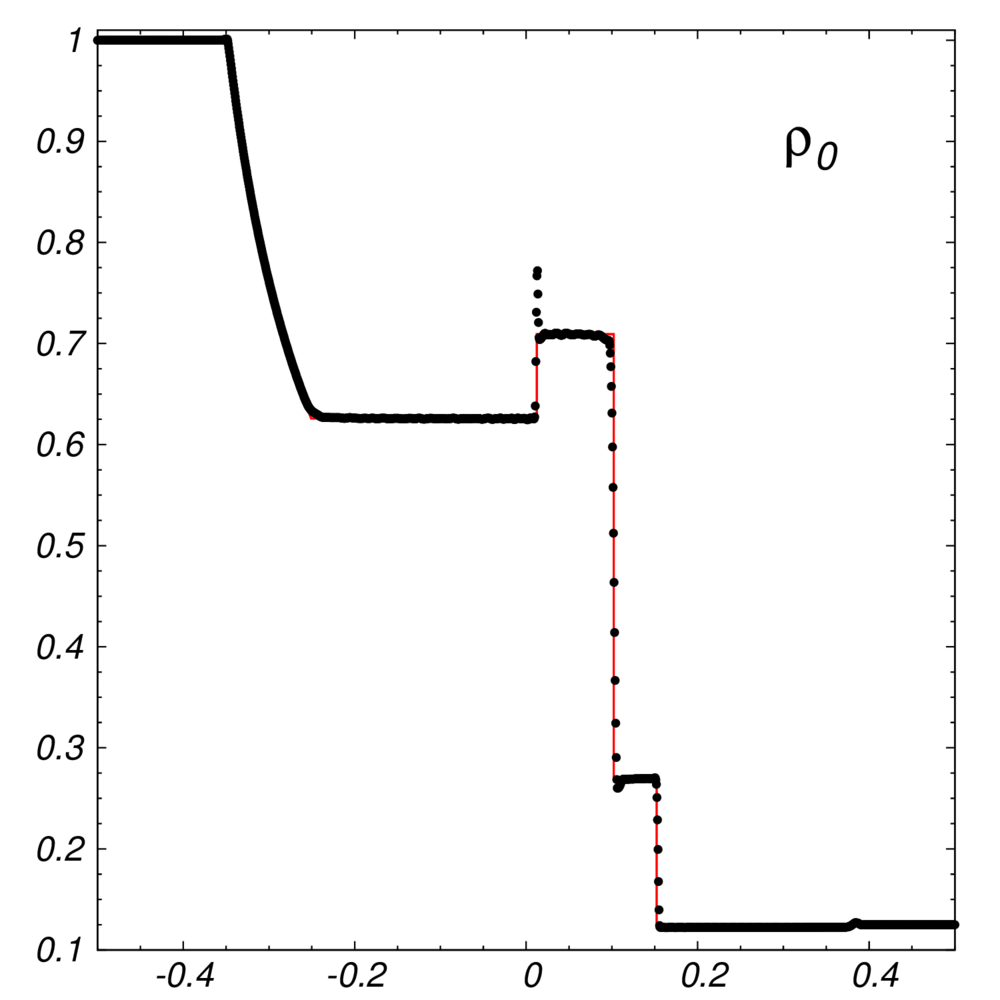}
\includegraphics[width=4.0cm]{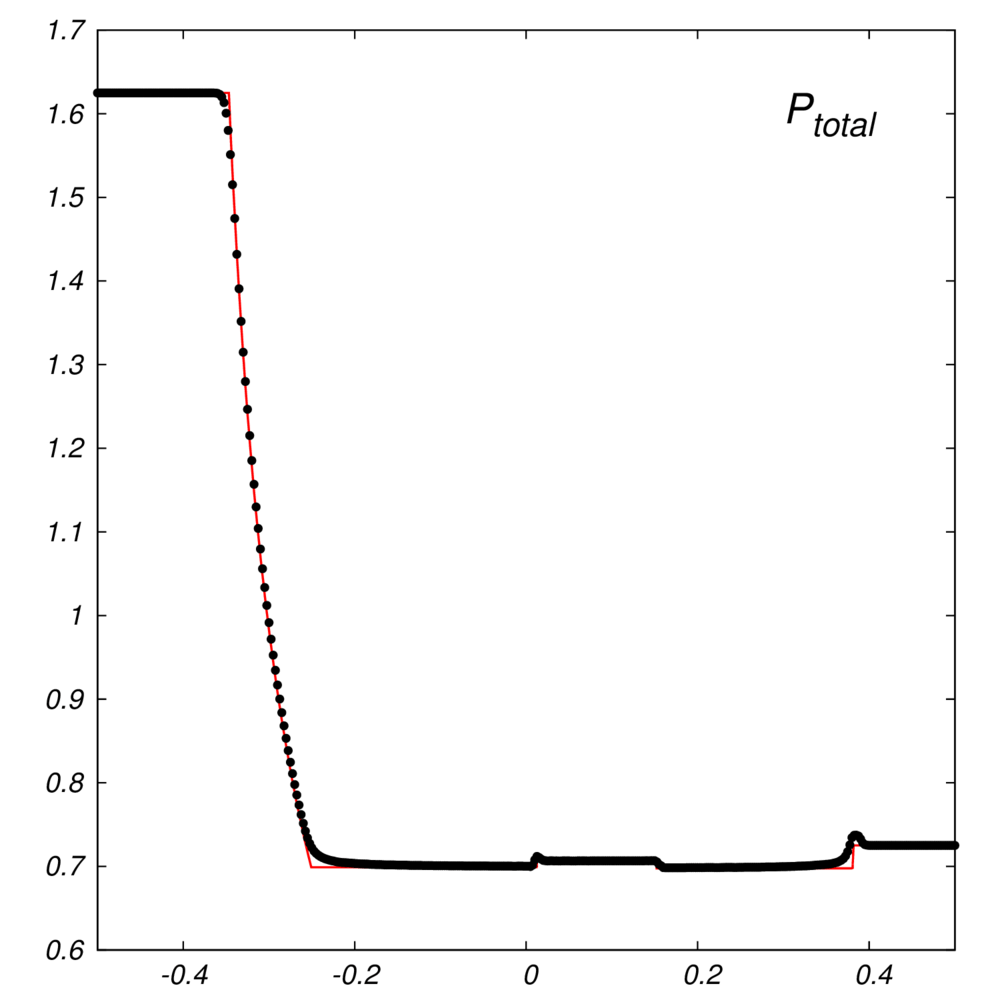}
\includegraphics[width=4.0cm]{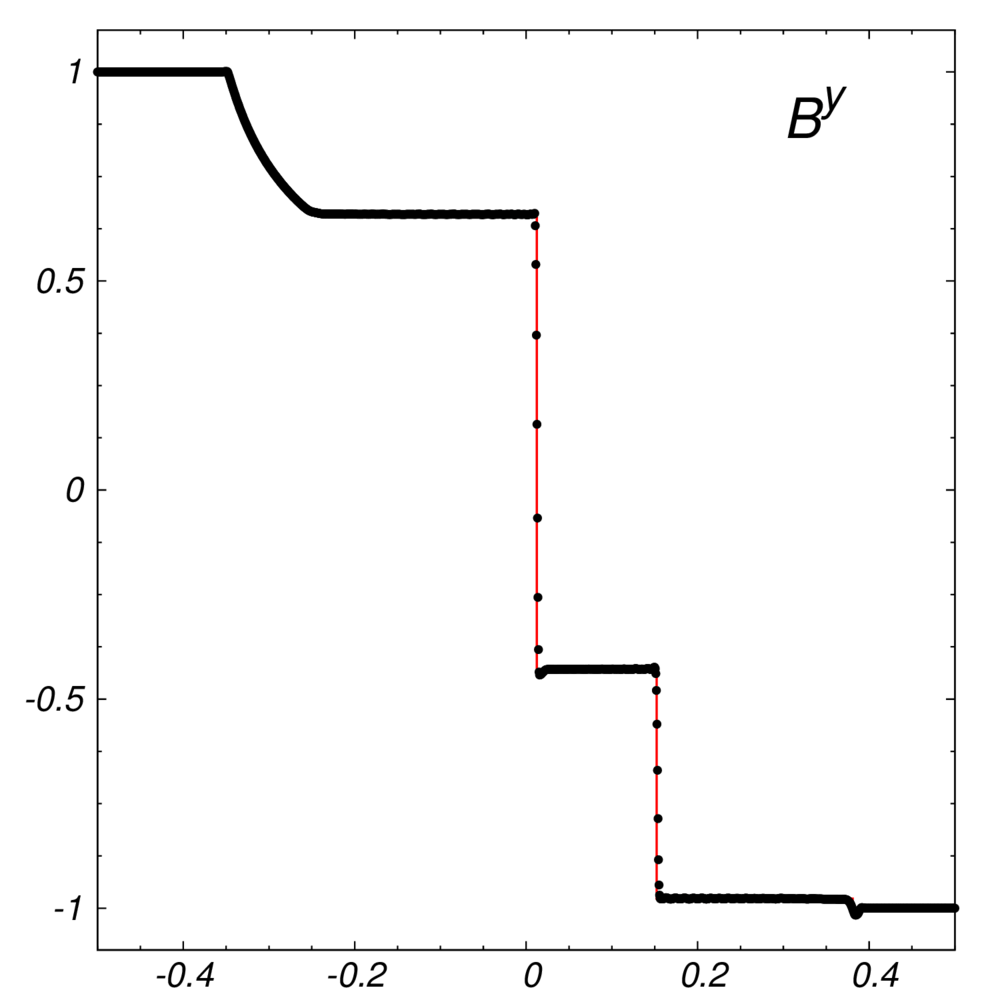}
\includegraphics[width=4.0cm]{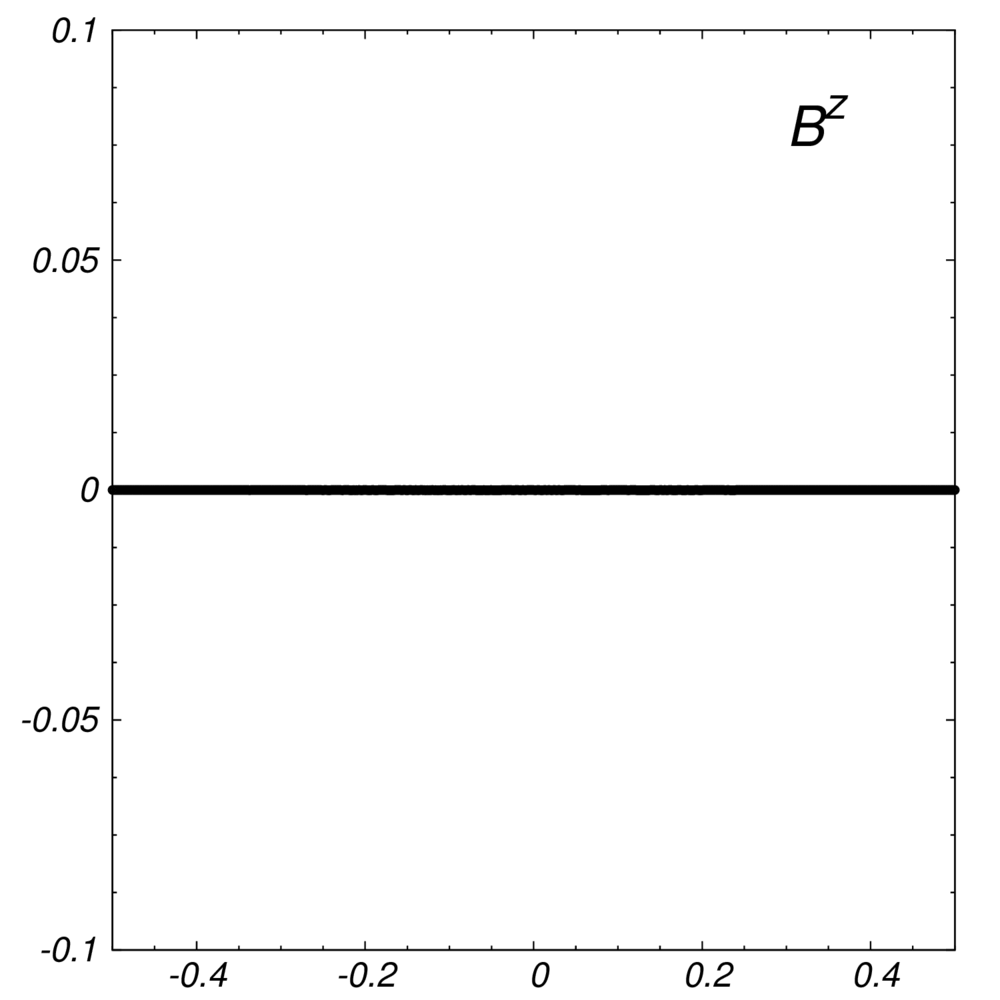}
\includegraphics[width=4.0cm]{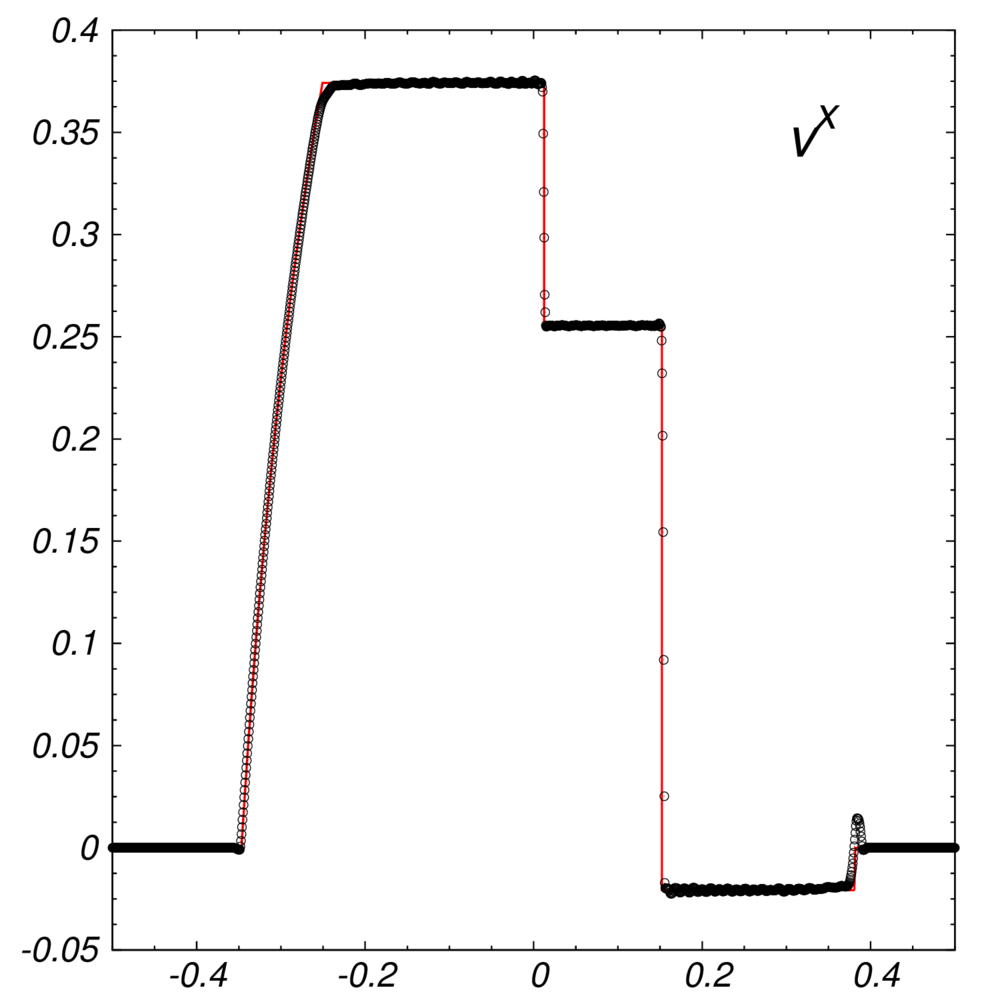}
\includegraphics[width=4.0cm]{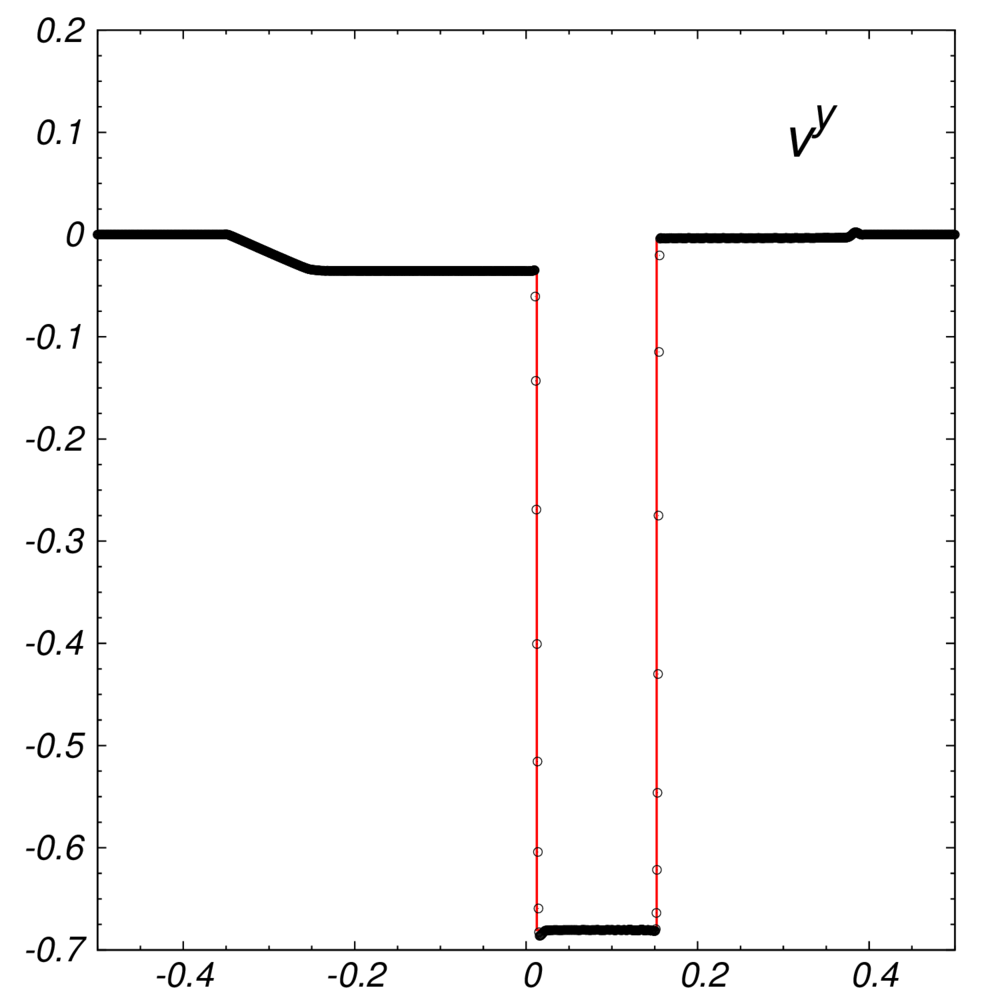}
\end{center}
\caption{\label{fig:Balsara1} Test 3: Balsara 1 test at time $t=0.4$. We use $\Delta x=1/1600$ in a domain $x\in[-0.5:0.5]$ and $CFL=0.1$.} 
\end{figure}

This RMHD test corresponds to the relativistic generalization of the classical Brio-Wu problem \citep{1988JCoPh..75..400B}. The results for the different variables and the various expected features are shown in Figure \ref{fig:Balsara1} at time $t=0.4$: a left fast rarefaction wave in the region $x \in [-3.6,-2.5]$, at $x \sim 0.01$ a slow compound wave, a contact discontinuity at $x \sim 0.1$, a slow shock at $x \sim 0.15$, and a right fast rarefaction wave in the region $x=[0.3,0.38]$.  The slow shock appears in a strongly magnetically dominated region. In this zone the magnetic energy is greater than the fluid rest mass energy or the fluid thermal energy. It is worth to mention that the slow compound wave appears only in the numerical solution; the exact solution omits this by construction \citep{2006JFM...562..223G}; however, the numerical solution is consistent with that of previous numerical RMHD codes.

\subsubsection{Test 4:  Balsara 2 test}

\begin{figure}
\begin{center}
\includegraphics[width=4.0cm]{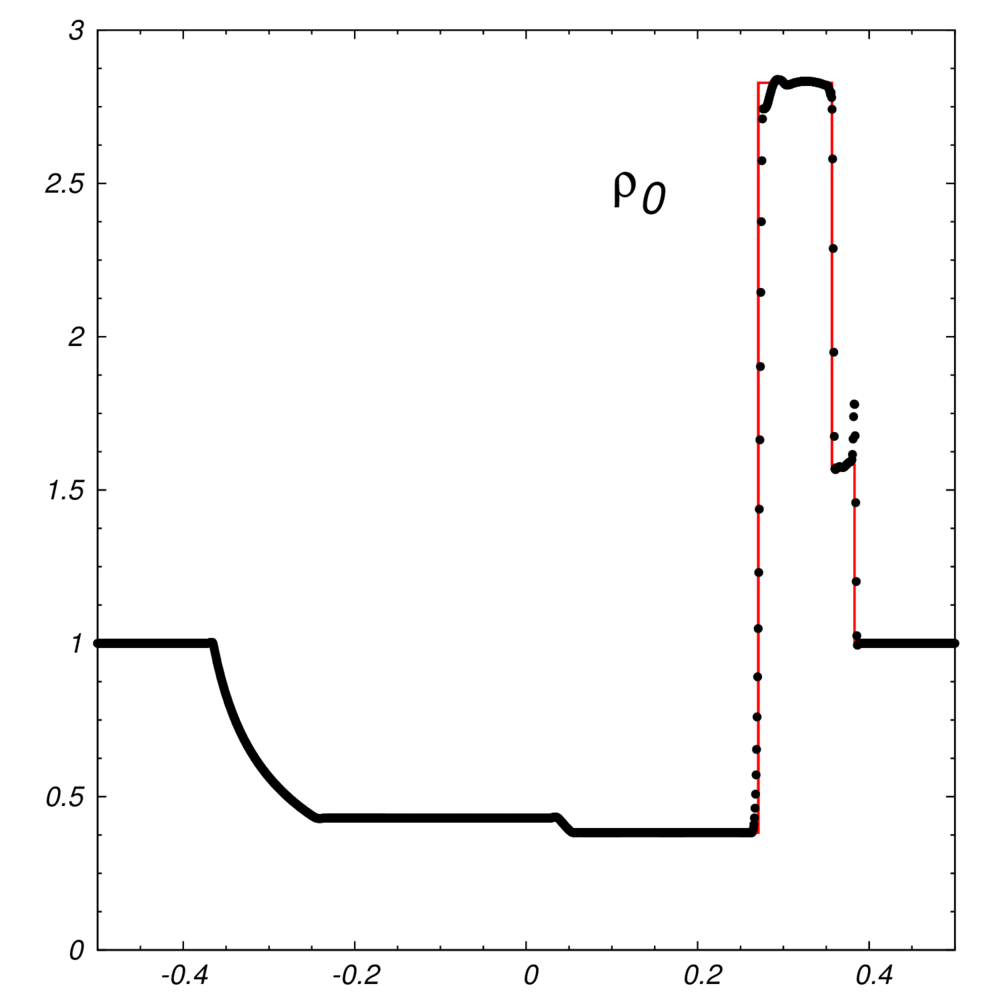}
\includegraphics[width=4.0cm]{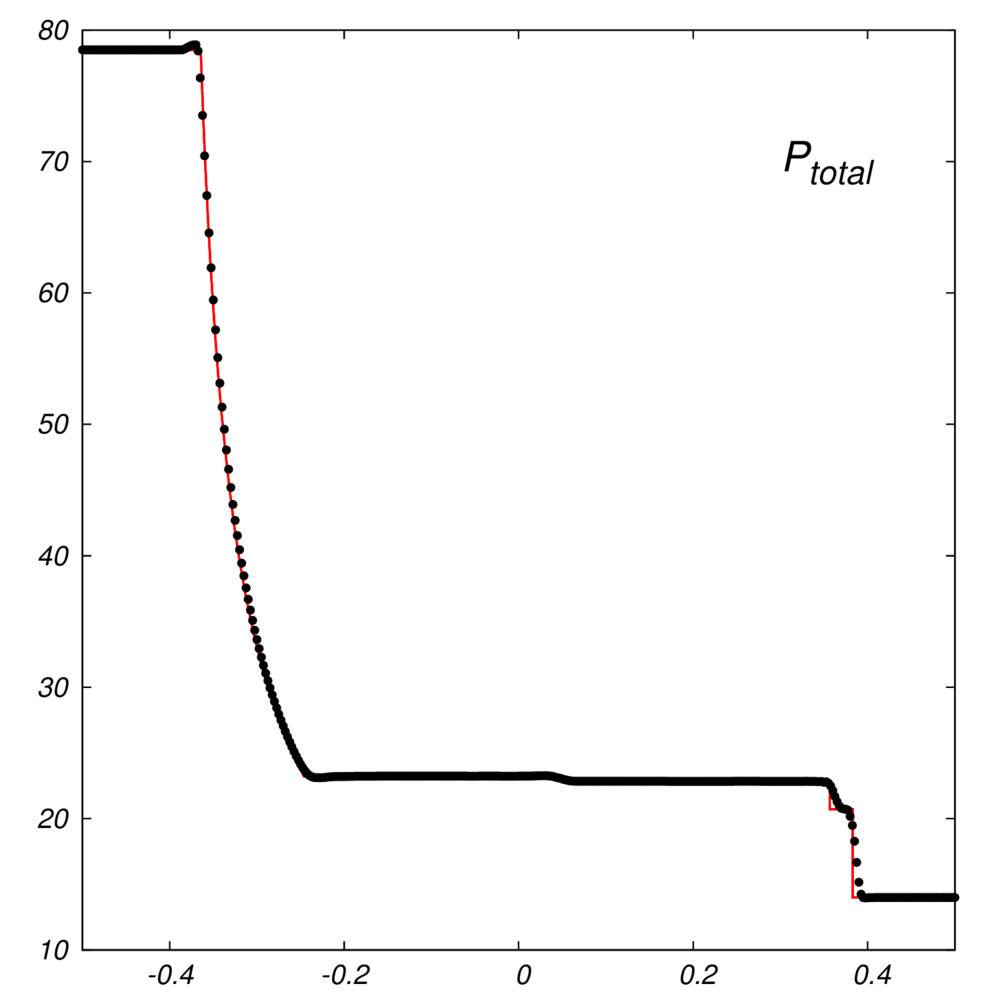}
\includegraphics[width=4.0cm]{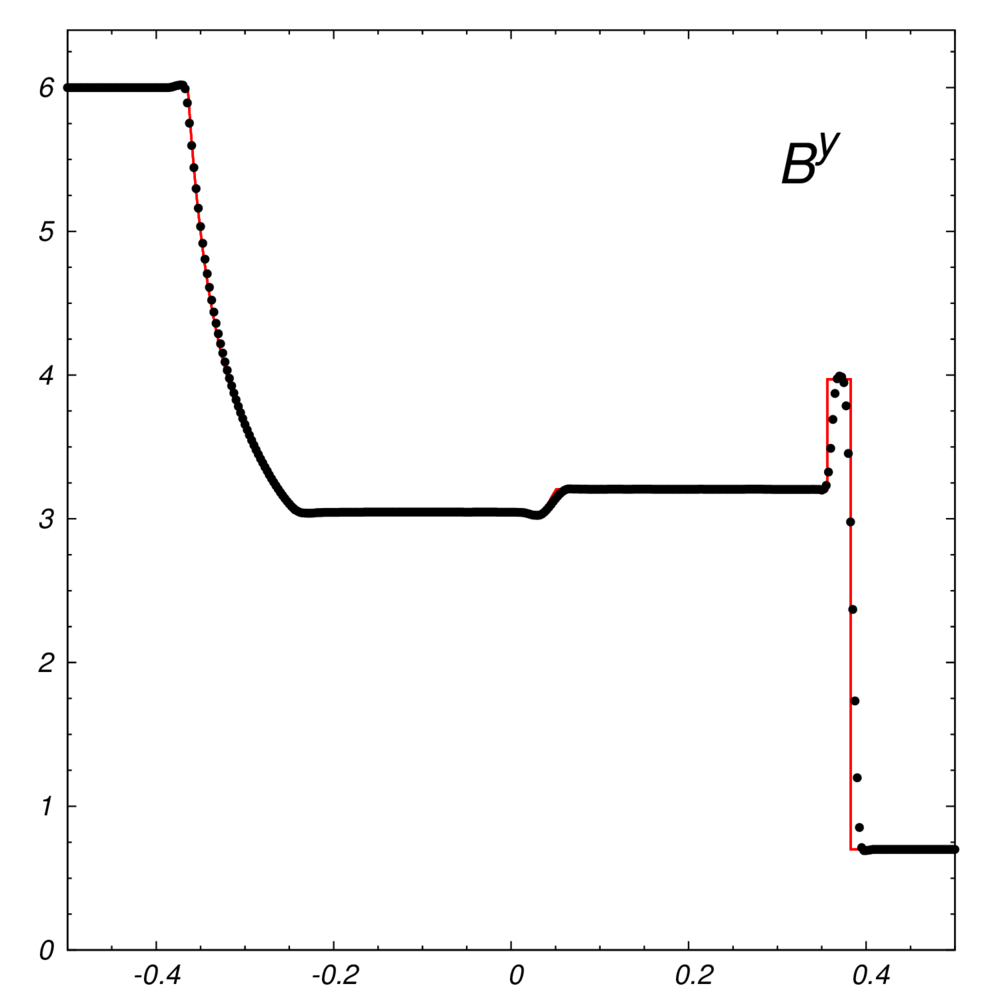}
\includegraphics[width=4.0cm]{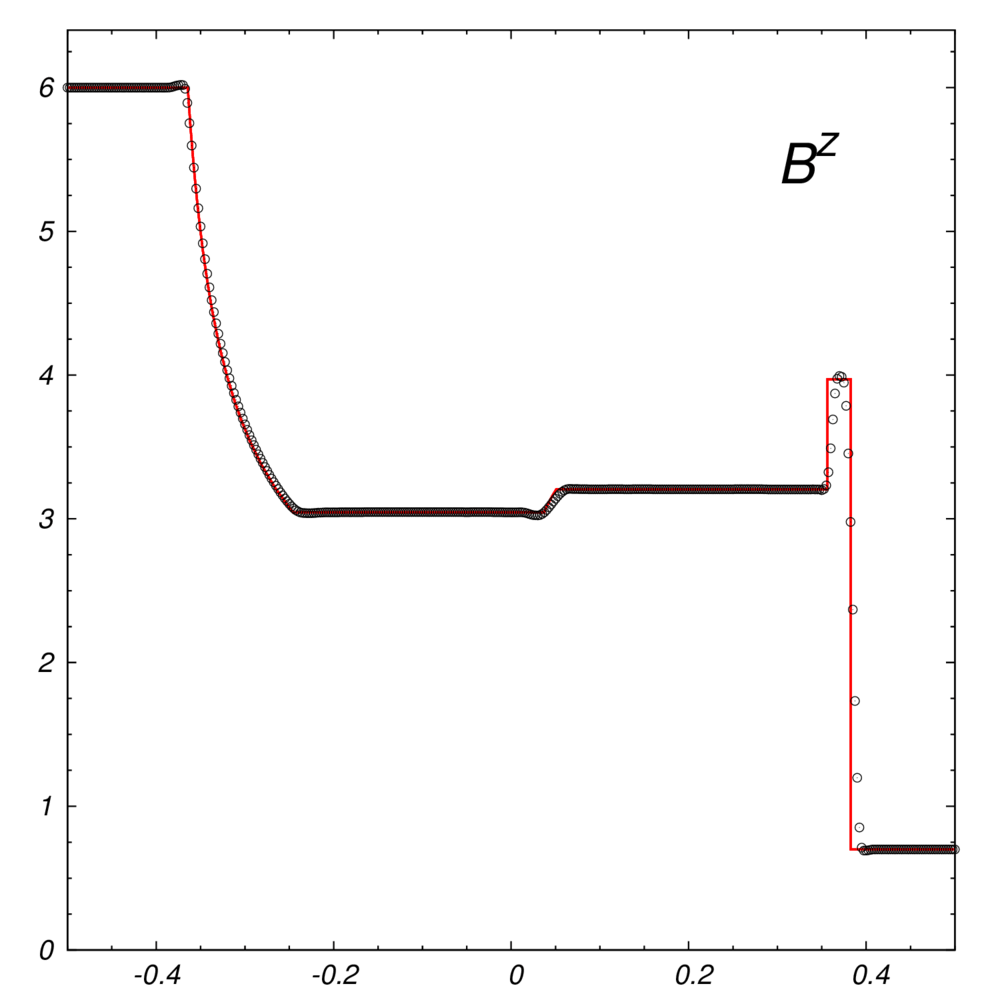}
\includegraphics[width=4.0cm]{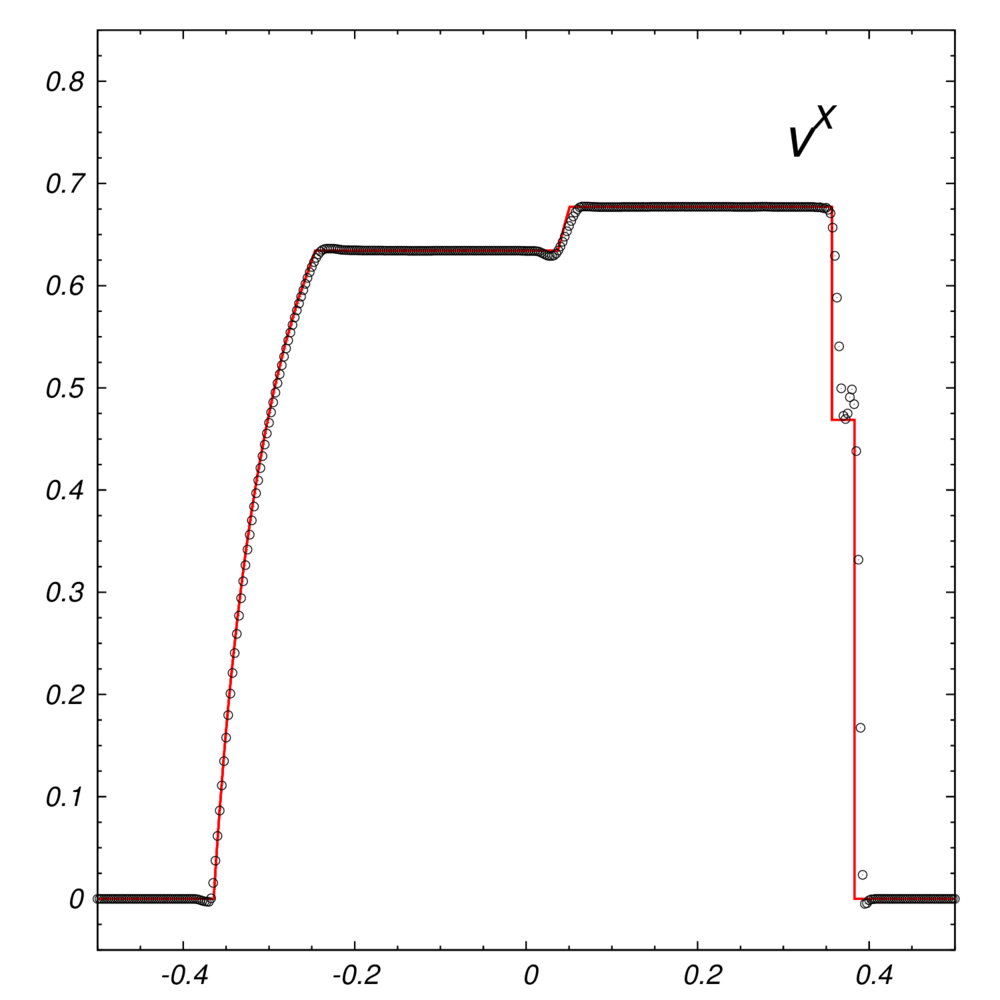}
\includegraphics[width=4.0cm]{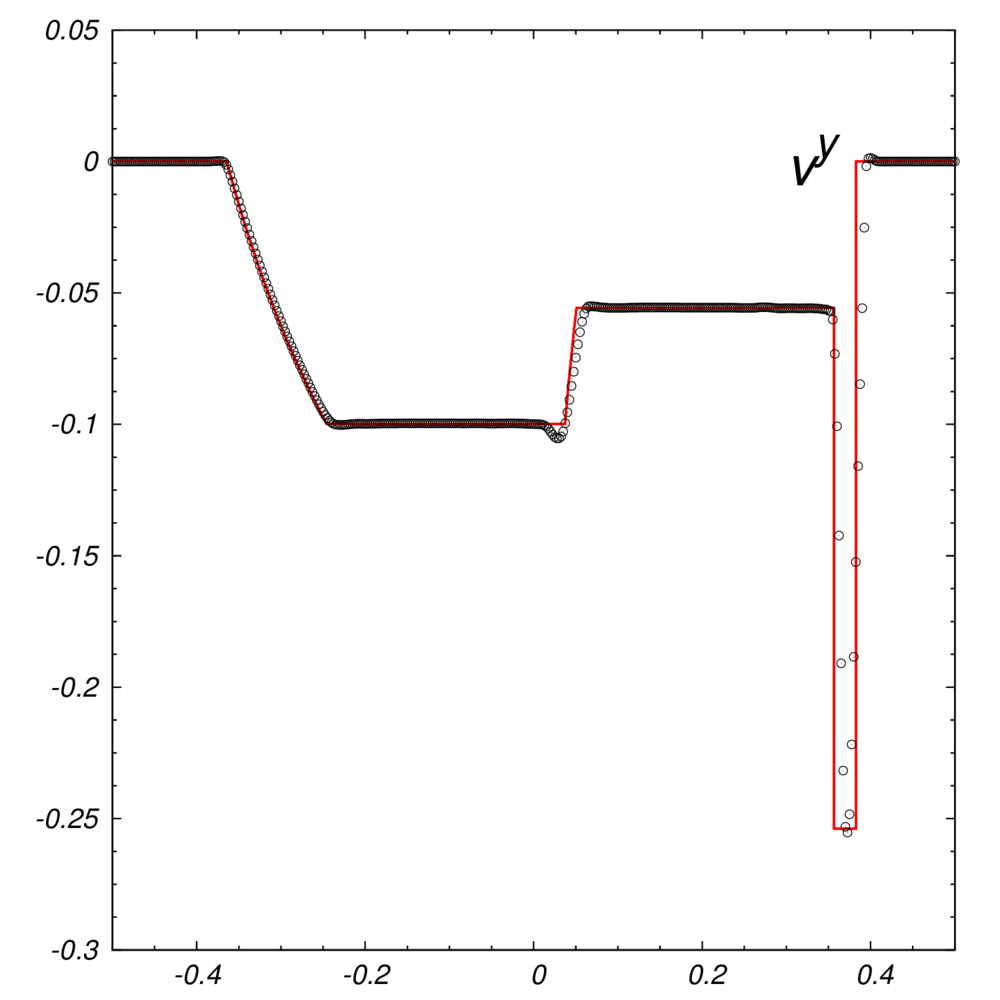}
\end{center}
\caption{\label{fig:Balsara2} Test 4: Balsara 2 blast wave at time $t=0.4$. We use $1600$ cells in the numerical domain $[-0.5,0.5]$ and use  $CFL=0.1$.}
\end{figure}

This corresponds to a weak blast wave, the initial configuration consists in a moderate initial discontinuity of the pressure and constant rest mass density. The ratio between pressures is $p_{L}/p_{R}= 30$.  In Figure \ref{fig:Balsara2} we show a snapshot at $t=0.4$. A slow shock wave is formed and propagates along the $x$ direction near the contact discontinuity, with maximum Lorentz factor $W=1.36$. In this case all the reconstructors and the constraint control methods produce well behaved results. The numerical domain $x\in[-0.5,0.5]$ is covered with resolution $\Delta x= 1/1600$.

\subsubsection{Test 5:  Balsara 3 test}

\begin{figure}[!h]
\begin{center}
\includegraphics[width=4.0cm]{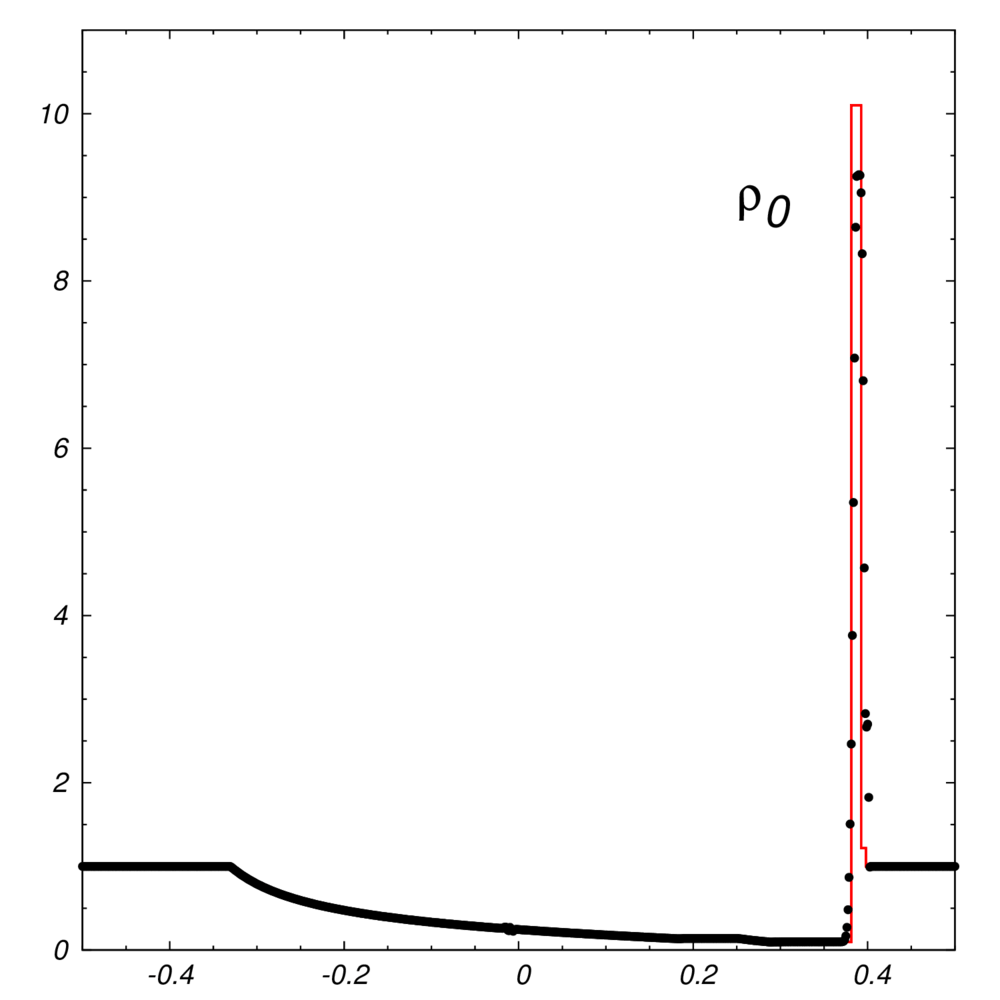}
\includegraphics[width=4.0cm]{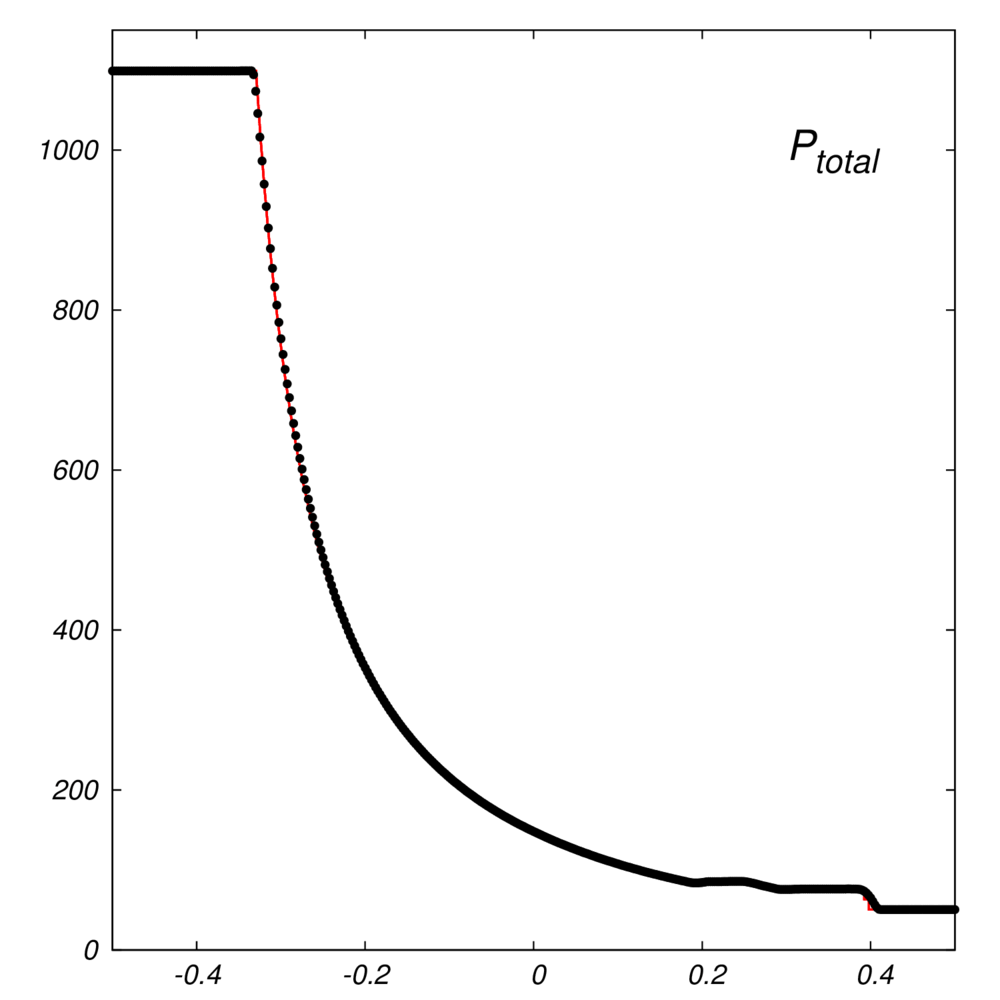}
\includegraphics[width=4.0cm]{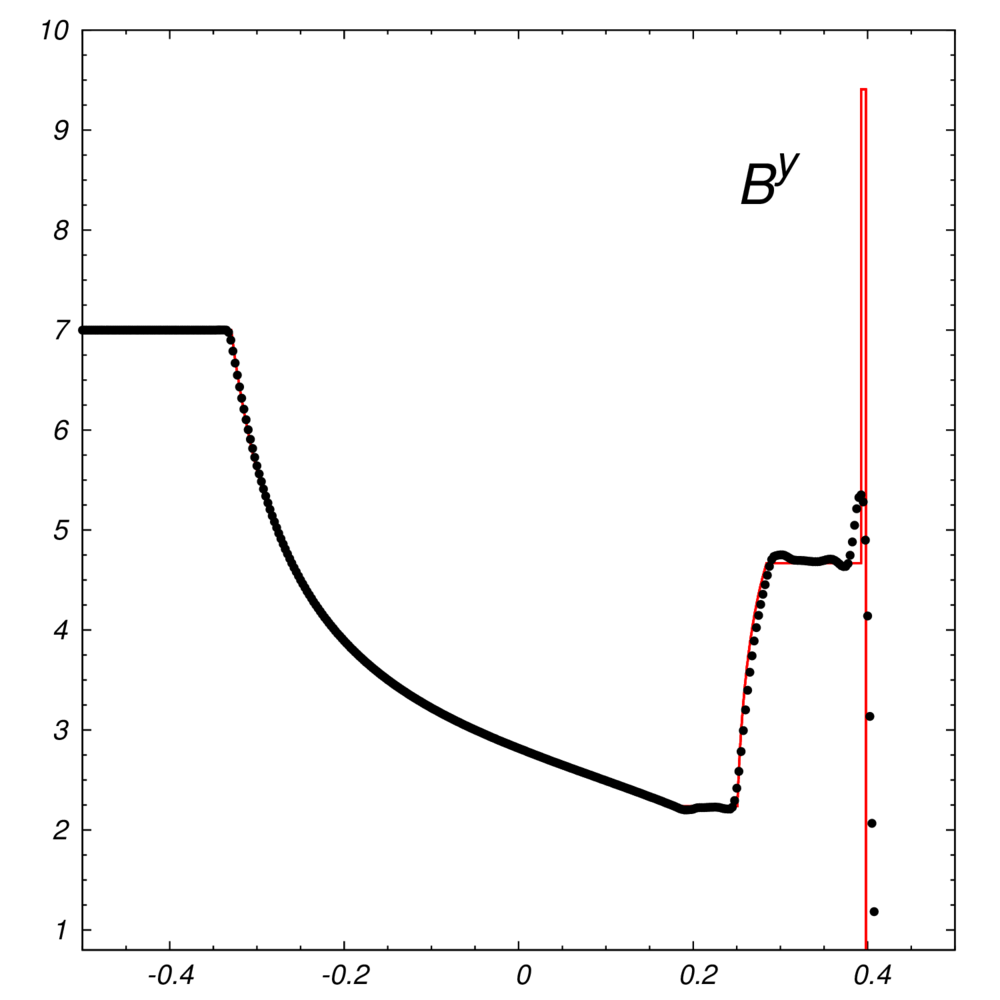}
\includegraphics[width=4.0cm]{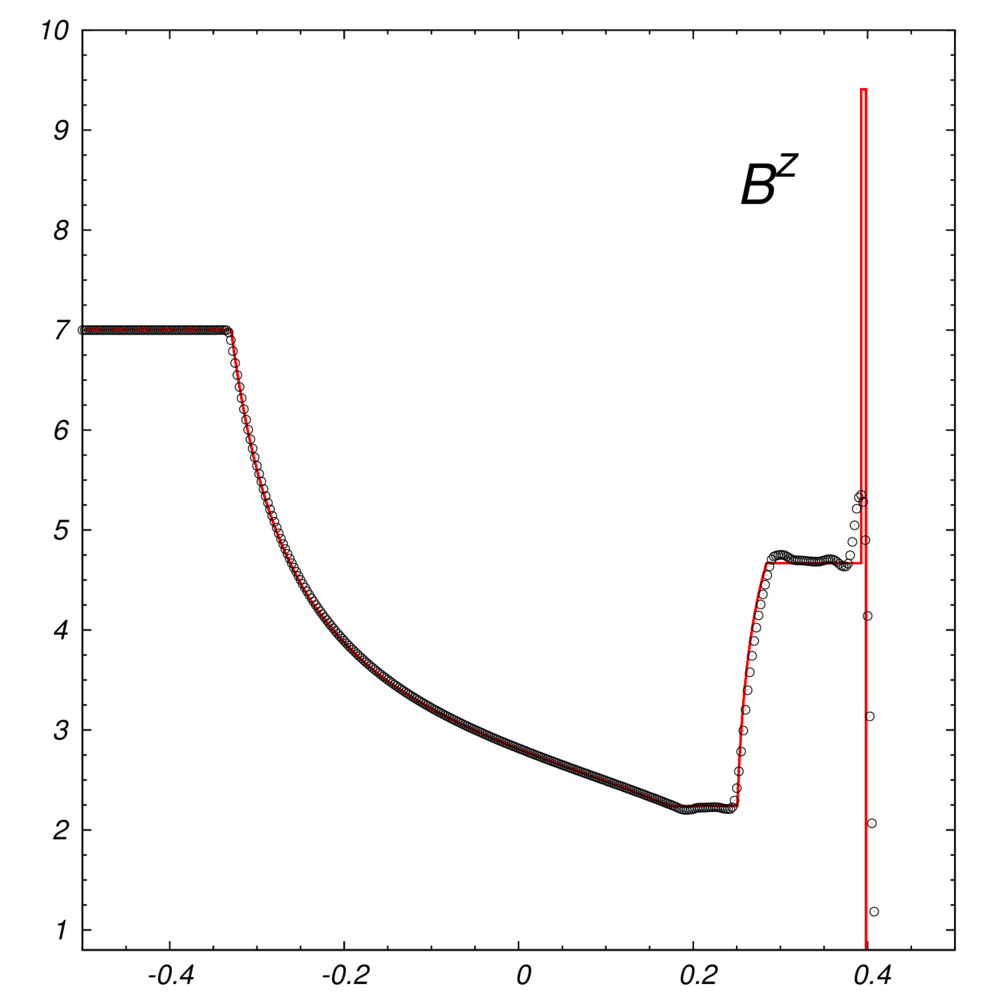}
\includegraphics[width=4.0cm]{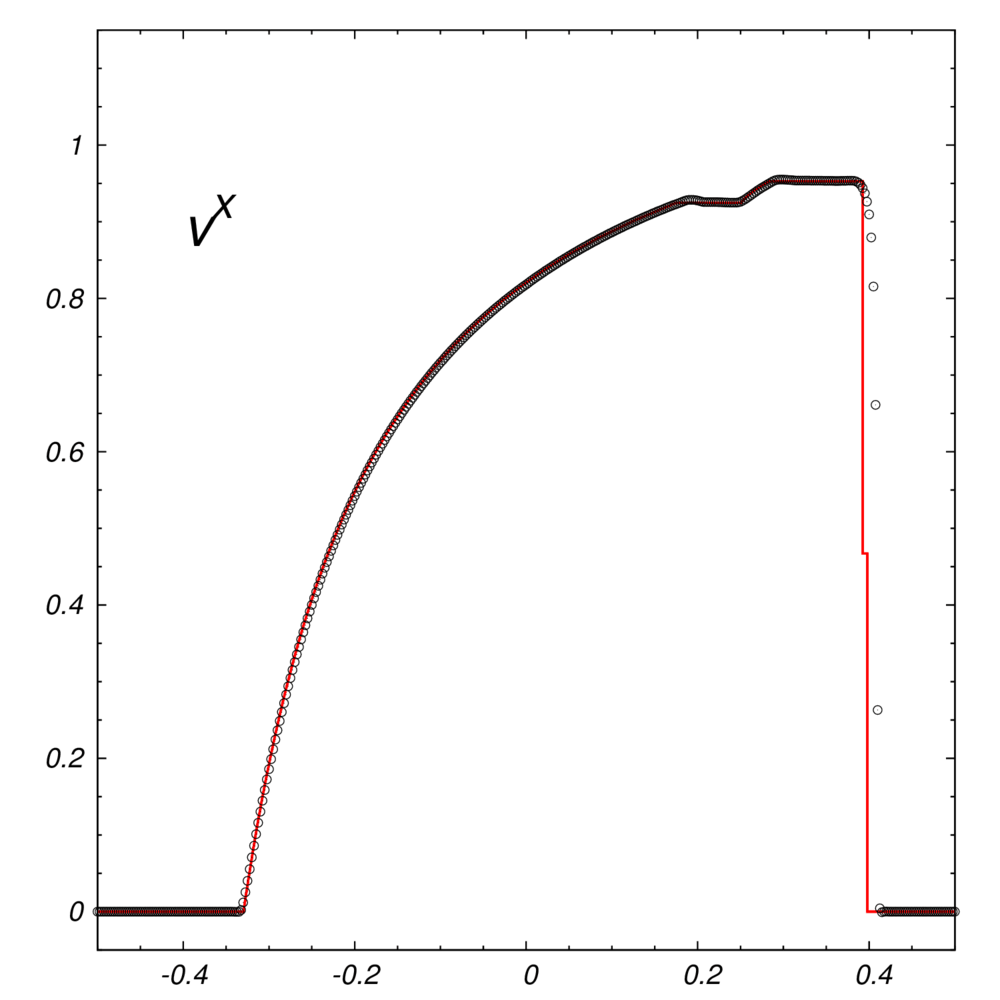}
\includegraphics[width=4.0cm]{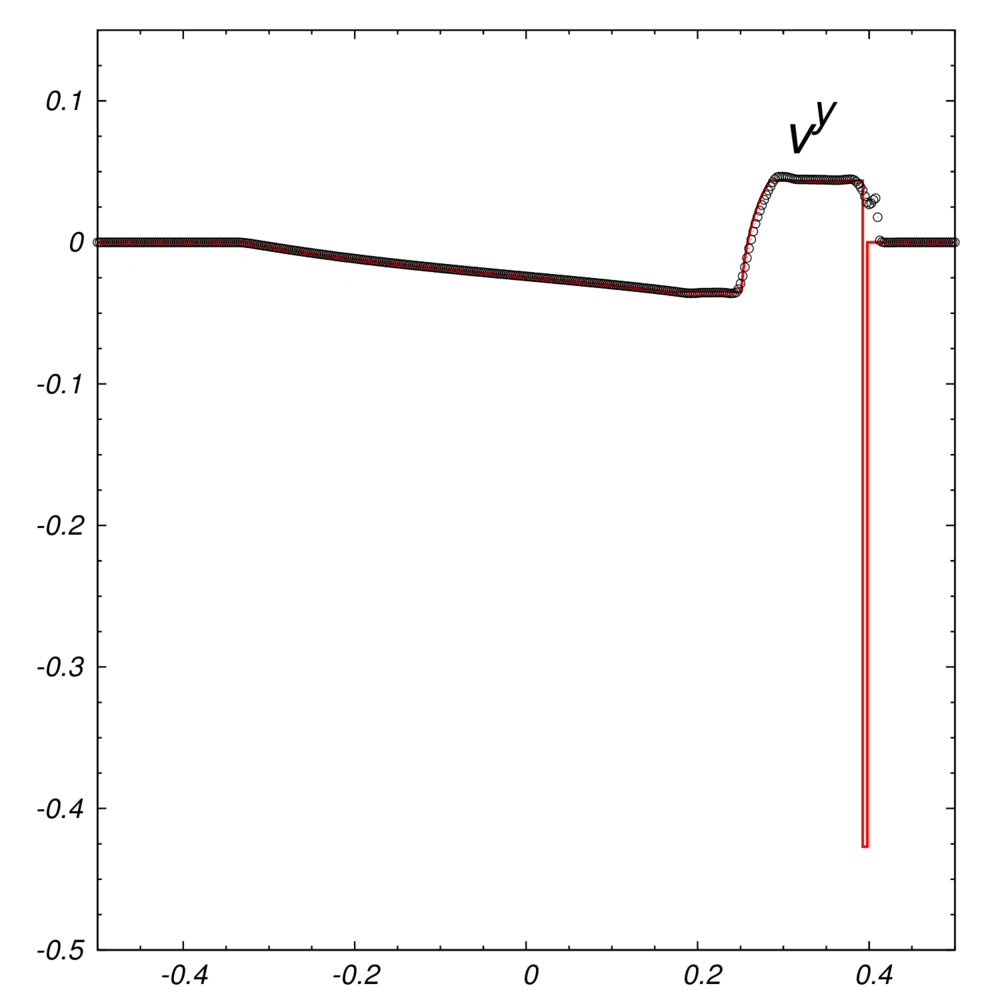}
\end{center}
\caption{\label{fig:Balsara3} Test 5: Balsara 3 blast wave along the $x$ direction at time $t=0.4$. We cover the numerical domain  $[-0.5,0.5]$ with 1600 cells and $CFL=0.1$.}
\end{figure}

This is a strong blast wave with high difference between the pressure in the initial discontinuity of nearly four orders of magnitude, constant density and zero velocities at initial time. In Figure \ref{fig:Balsara3} we show the snapshot  at $t=0.4$. We see the typical  peak of the density in the blast wave and the effects on the  velocity and magnetic field. The Lorentz factor reaches values of $W \sim 3.5$. The numerical solution is consistent with the exact solution. In the rest mass density we can observe a fast and a slow rarefaction zones moving to the left, a contact wave and two (fast and slow) shocks moving to the right. The presence of the magnetic field makes the slow and fast shocks propagate closely giving as result a thin density shell, which is difficult to capture with low resolution. However, in Tables \ref{tab:errors_rmhd} and \ref{tab:errors_rmhd2} we show that a more accurate result  is also obtained using MC or WENO5.

\subsubsection{Test 6:  Balsara 4 test}

\begin{figure}
\begin{center}
\includegraphics[width=4.0cm]{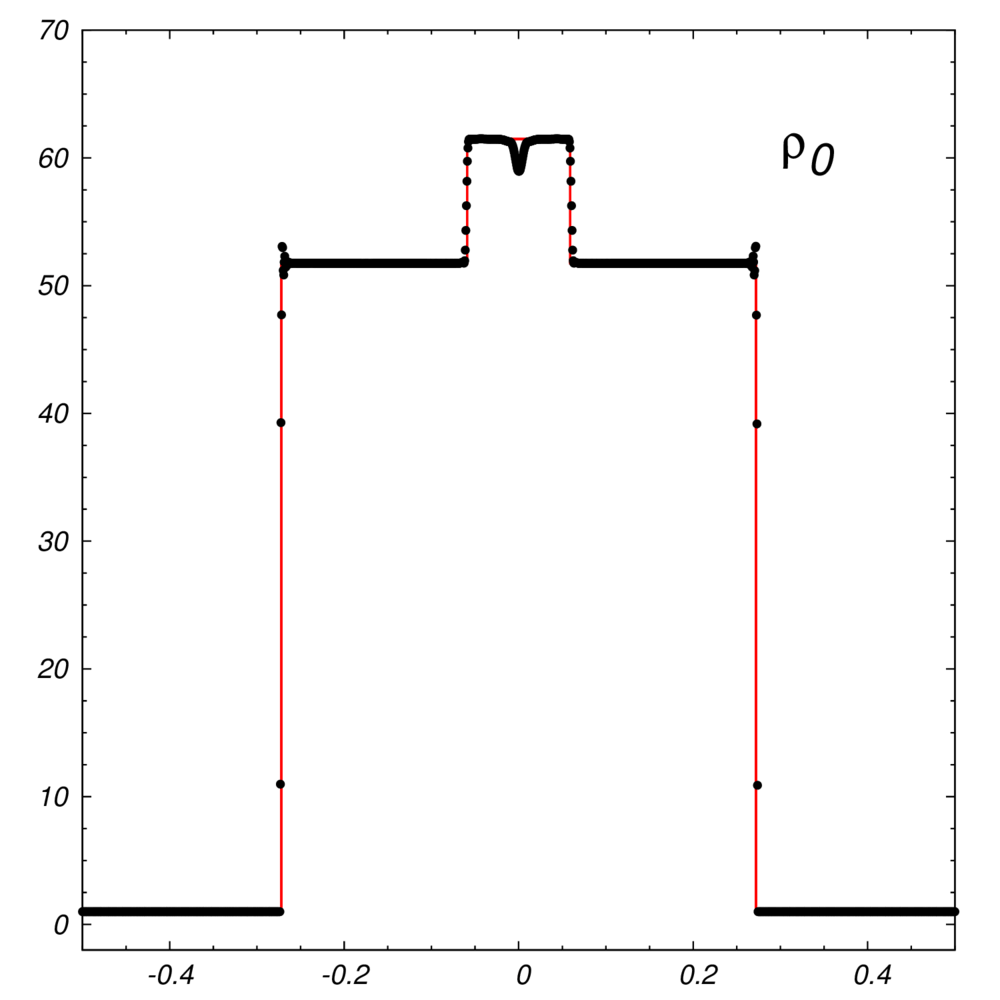}
\includegraphics[width=4.0cm]{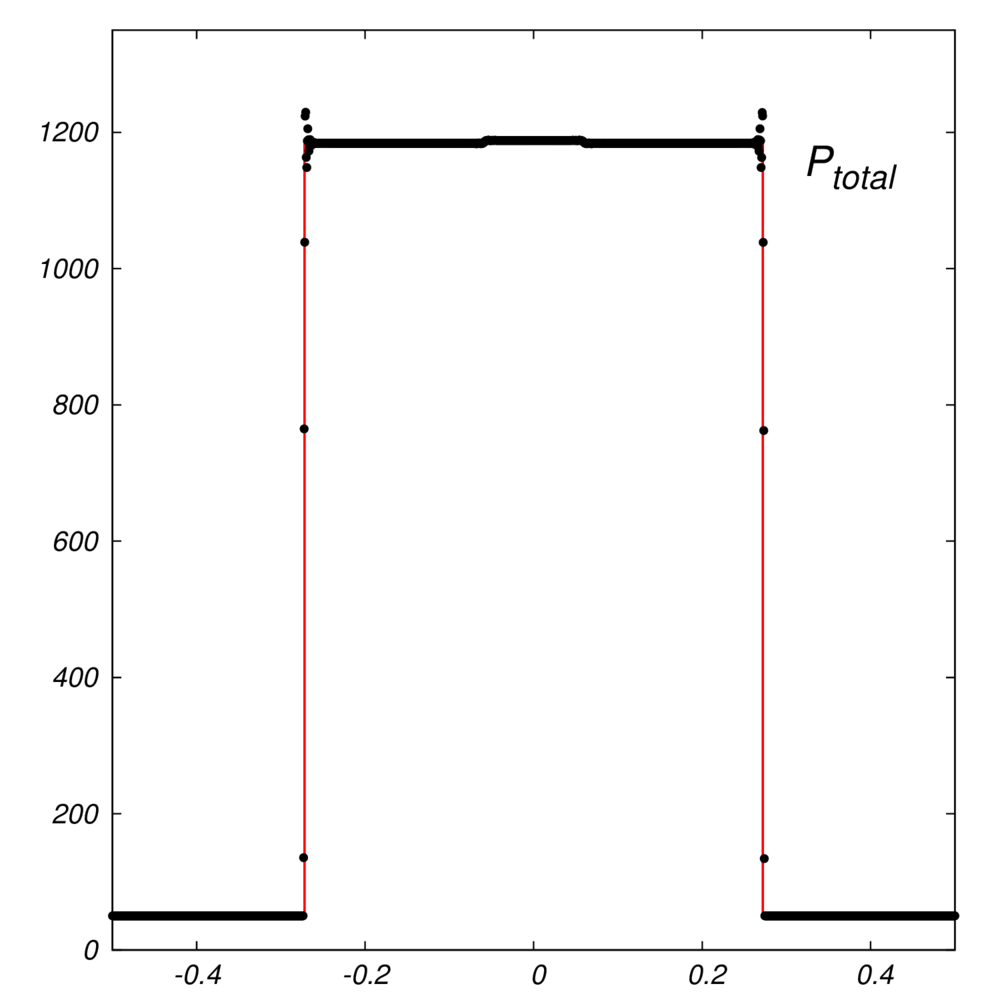}
\includegraphics[width=4.0cm]{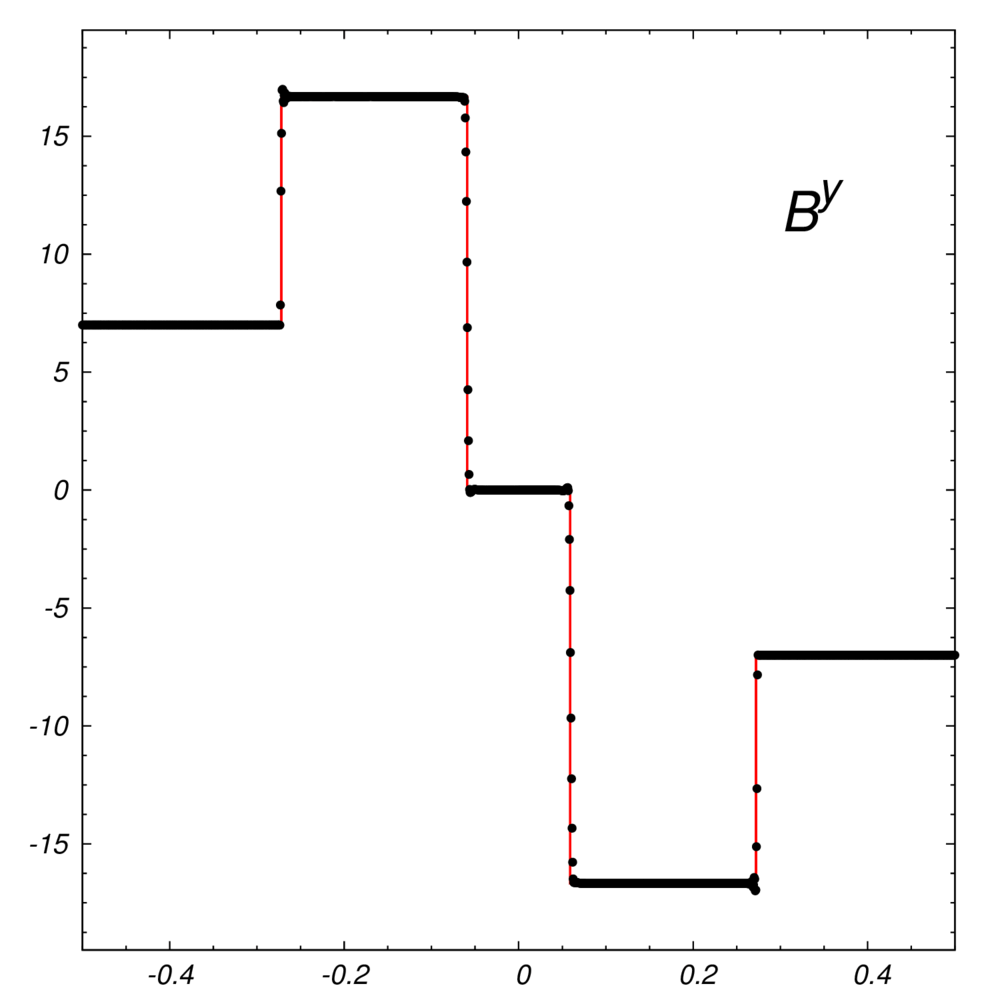}
\includegraphics[width=4.0cm]{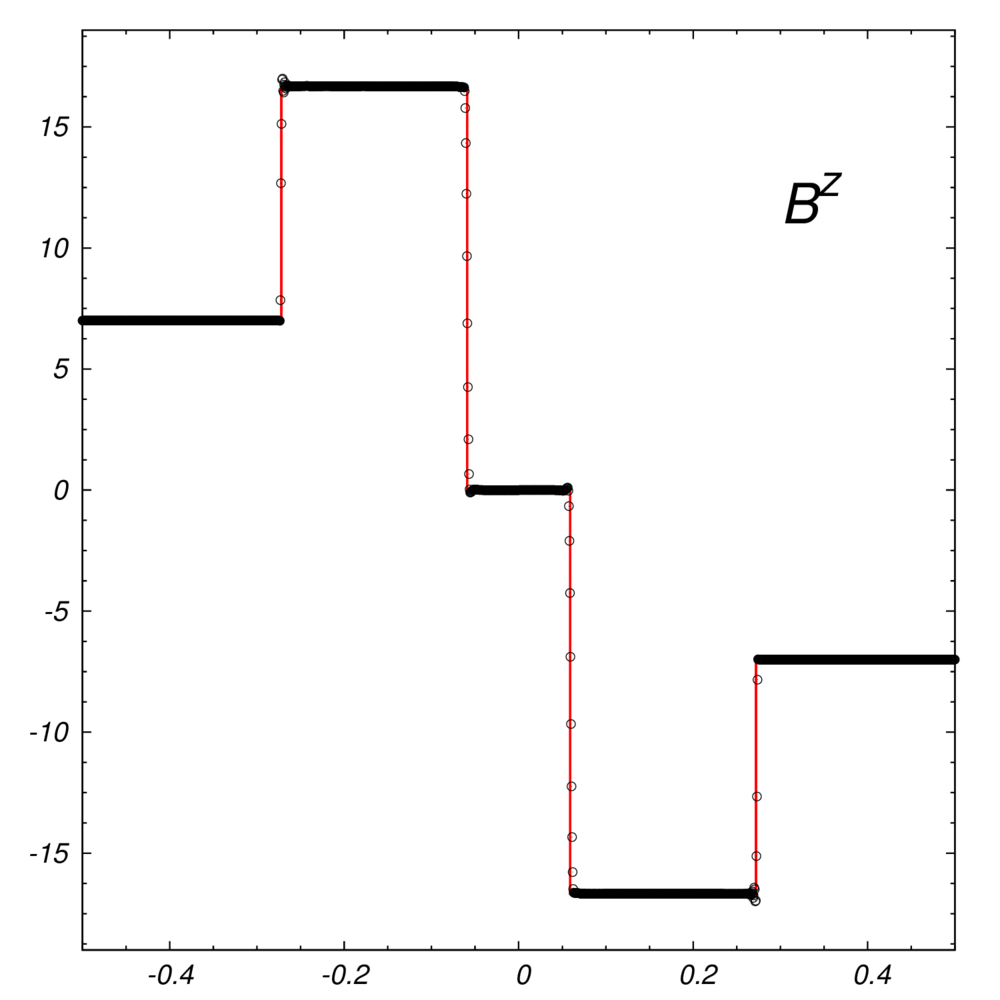}
\includegraphics[width=4.0cm]{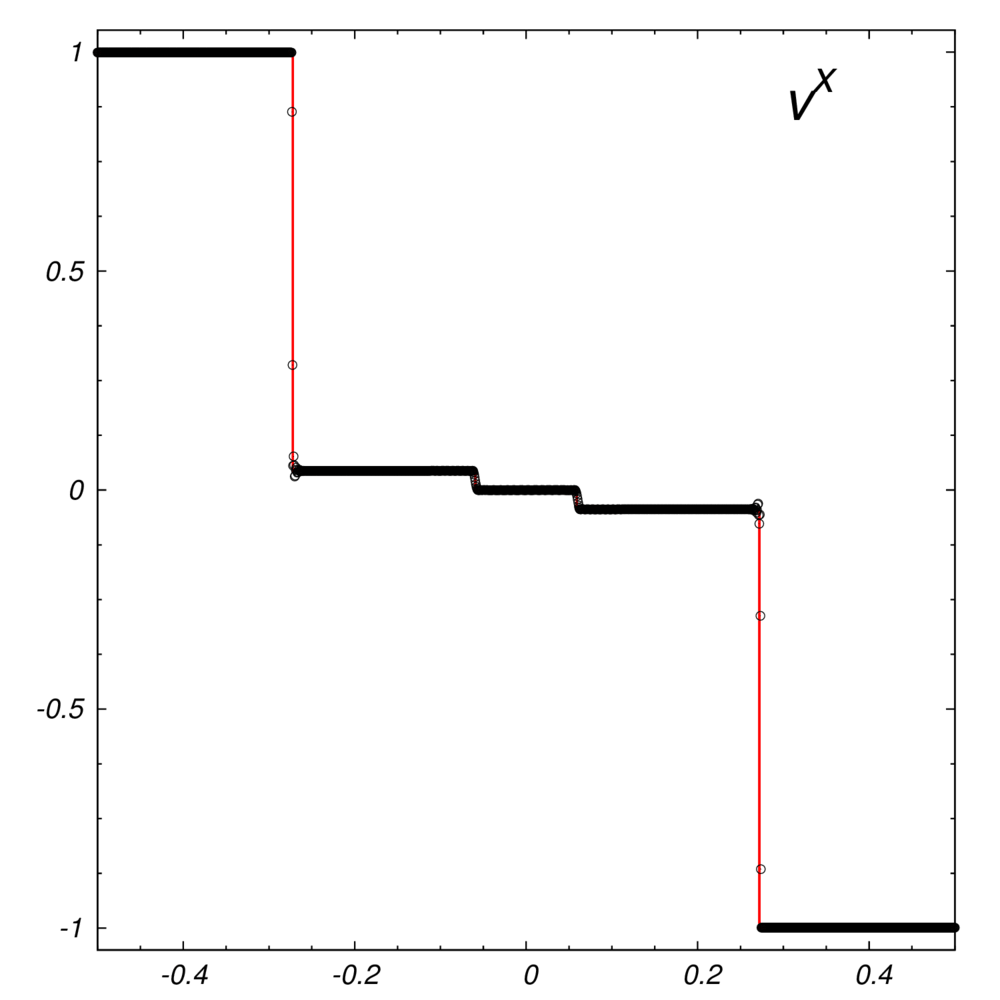}
\includegraphics[width=4.0cm]{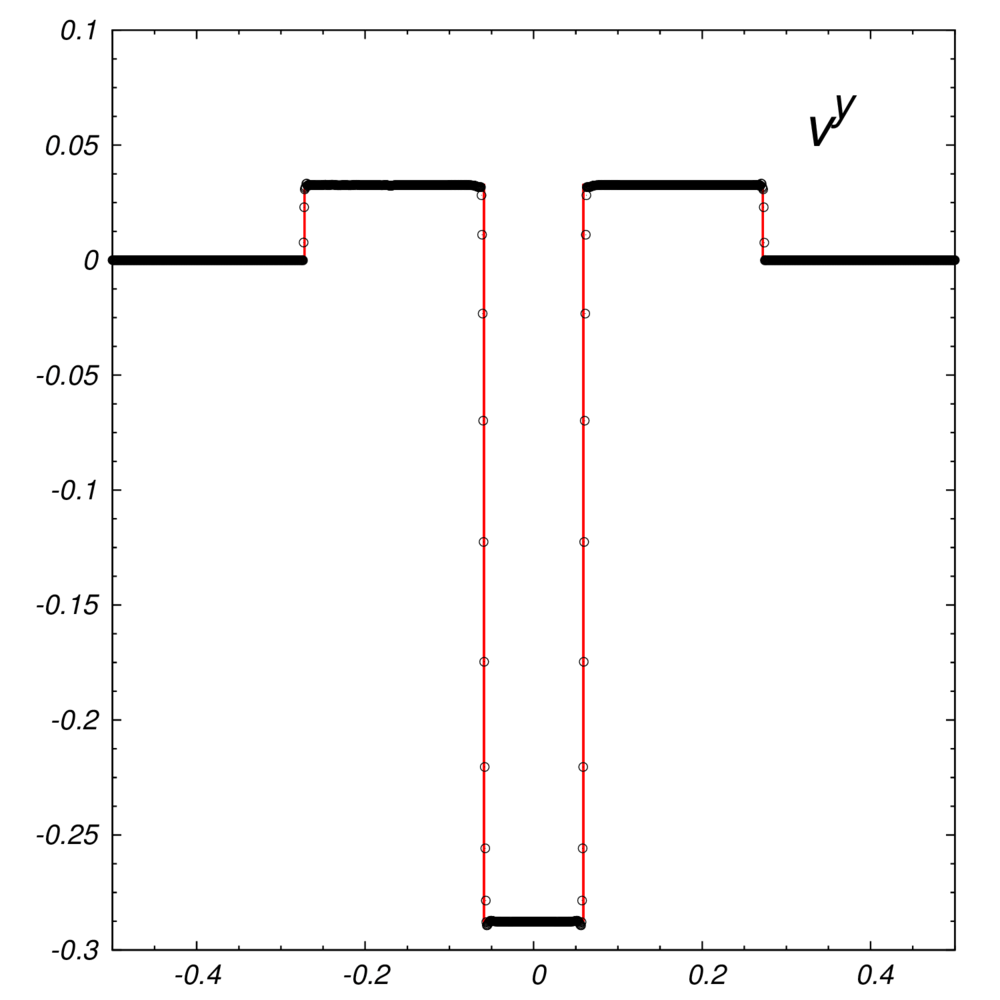}
\end{center}
\caption{\label{fig:Balsara4} Test 6: Balsara 4. We show the snapshot at $t=0.4$ where the expected shocks have been formed. We use $1600$ cells to cover the domain $[-0.5,0,5]$ and Courant Factor of $CFL=0.1$.}
\end{figure}

This is again the case of head-on collision of streams, however unlike the Komissarov collision test, in this case the transversal components of the magnetic field are non zero and the velocities are higher. The problem starts with two relativistic streams moving in opposite directions at nearly the speed of light, with initially constant pressure and rest mass density. In Figure \ref{fig:Balsara4}, we show a snapshot at $t=0.4$. In this particular case, in the plots some signals appear beyond $x>0.4$, however these effects are due to numerical diffusion.  The initial Lorentz factor is $W \sim 22.366$ and the initial pressure includes high values $p \sim 1200$. We can also see that two slow waves are moving in opposite directions. On the other hand,  in the strong shocks, spurious oscillations appear when the less dissipative reconstructors like MC and WENO5 are used. 

\subsubsection{Test 7:  Balsara 5 test}

\begin{figure}[!h]
\begin{center}
\includegraphics[width=4.0cm]{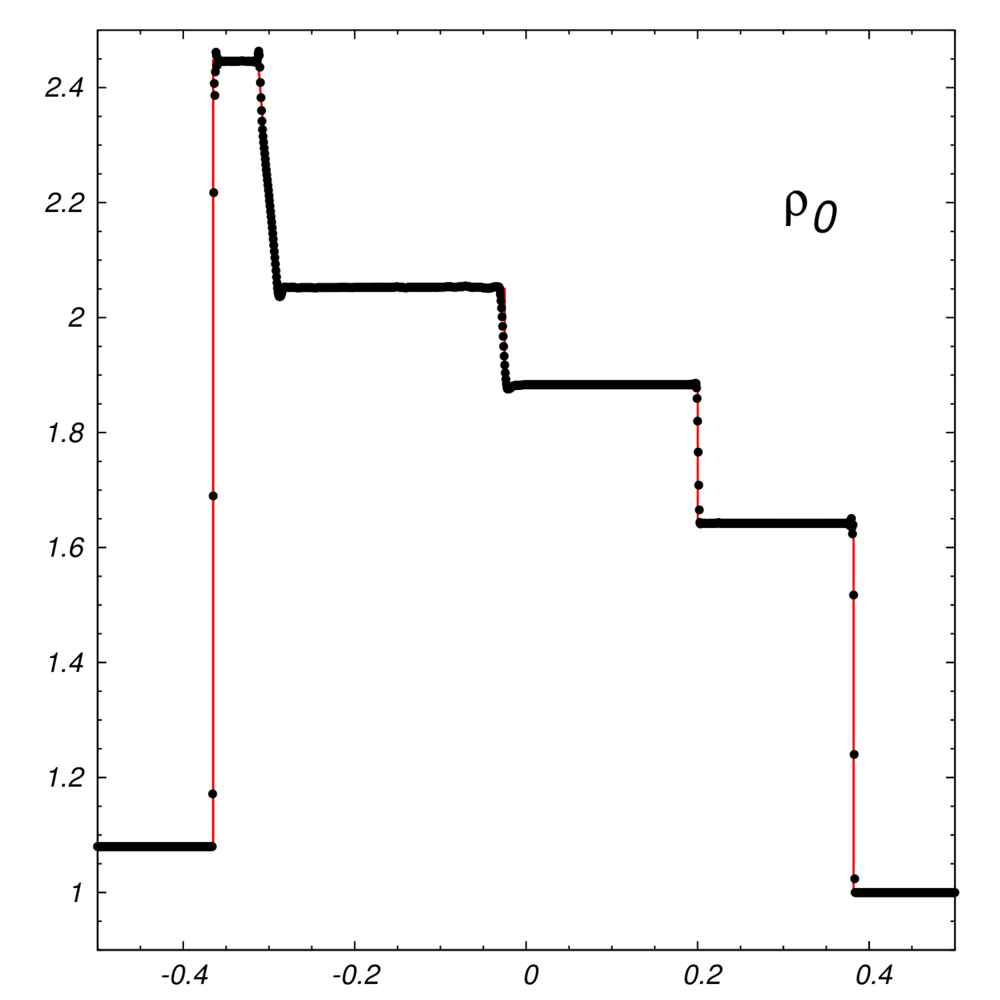}
\includegraphics[width=4.0cm]{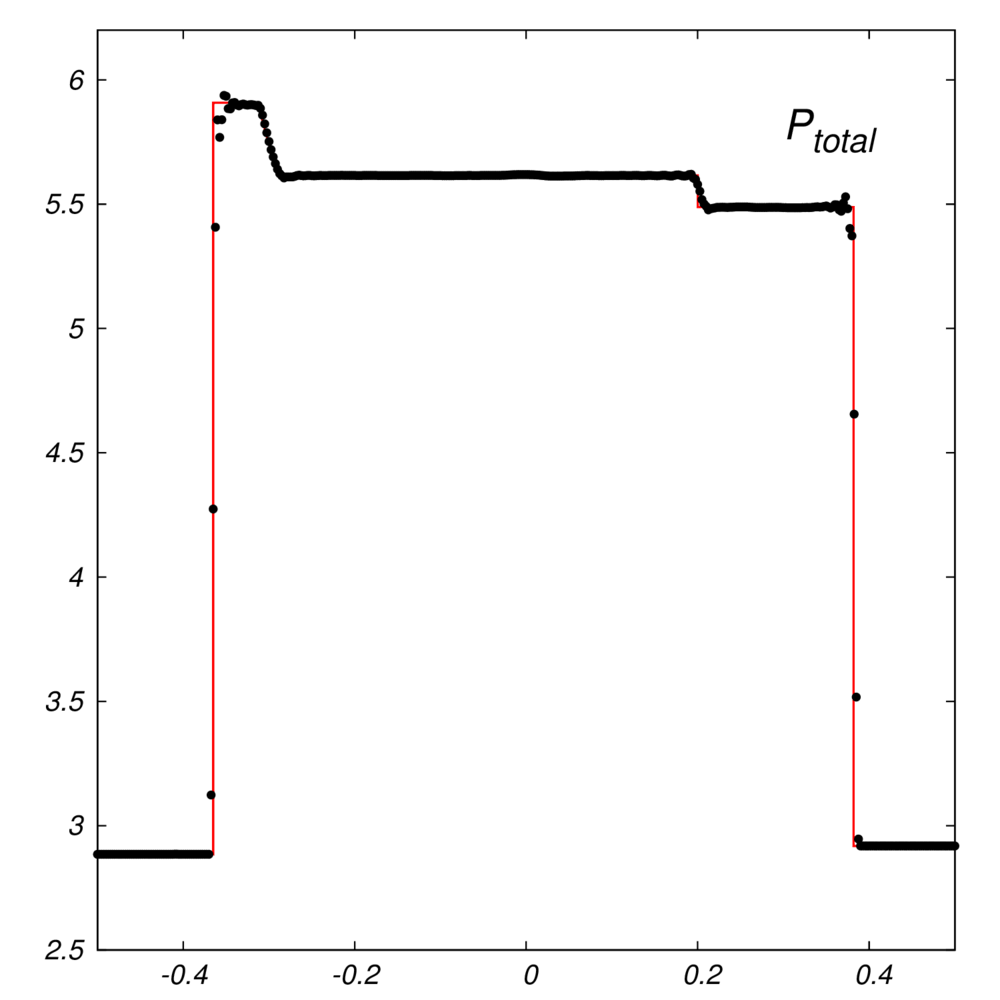}
\includegraphics[width=4.0cm]{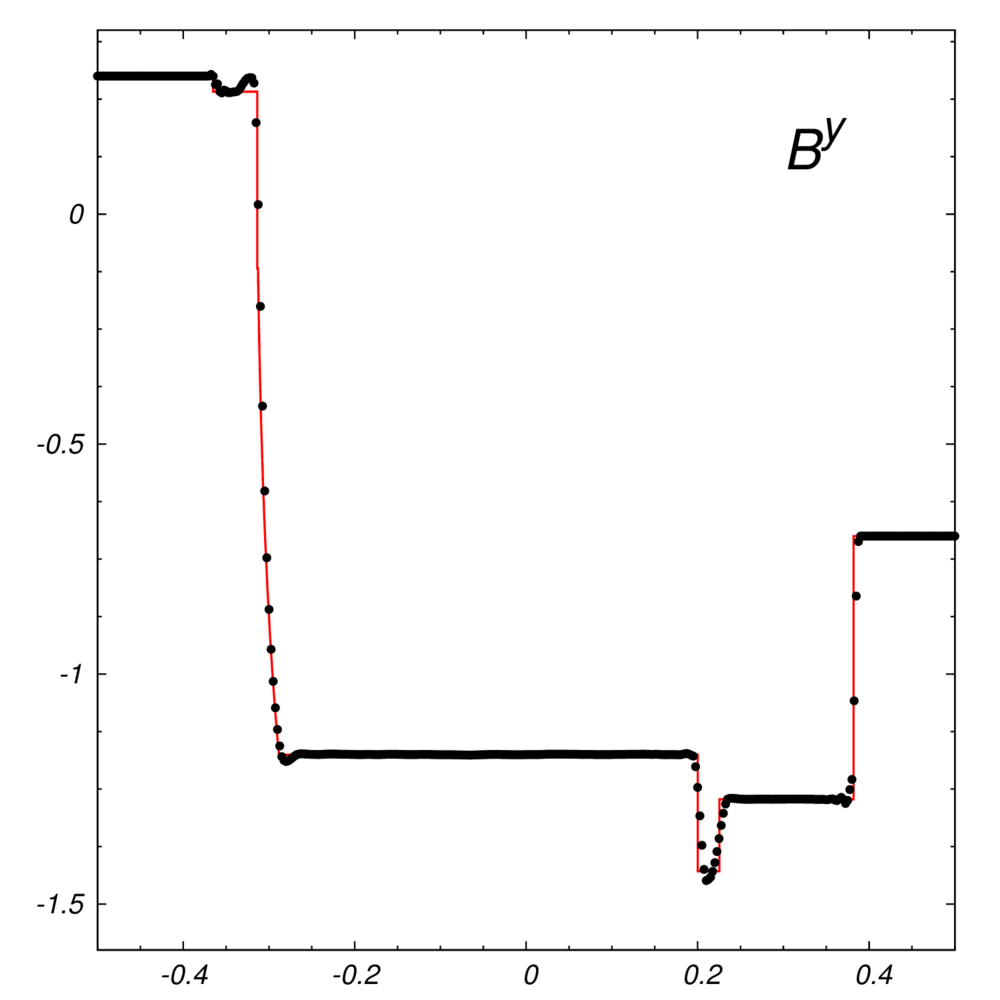}
\includegraphics[width=4.0cm]{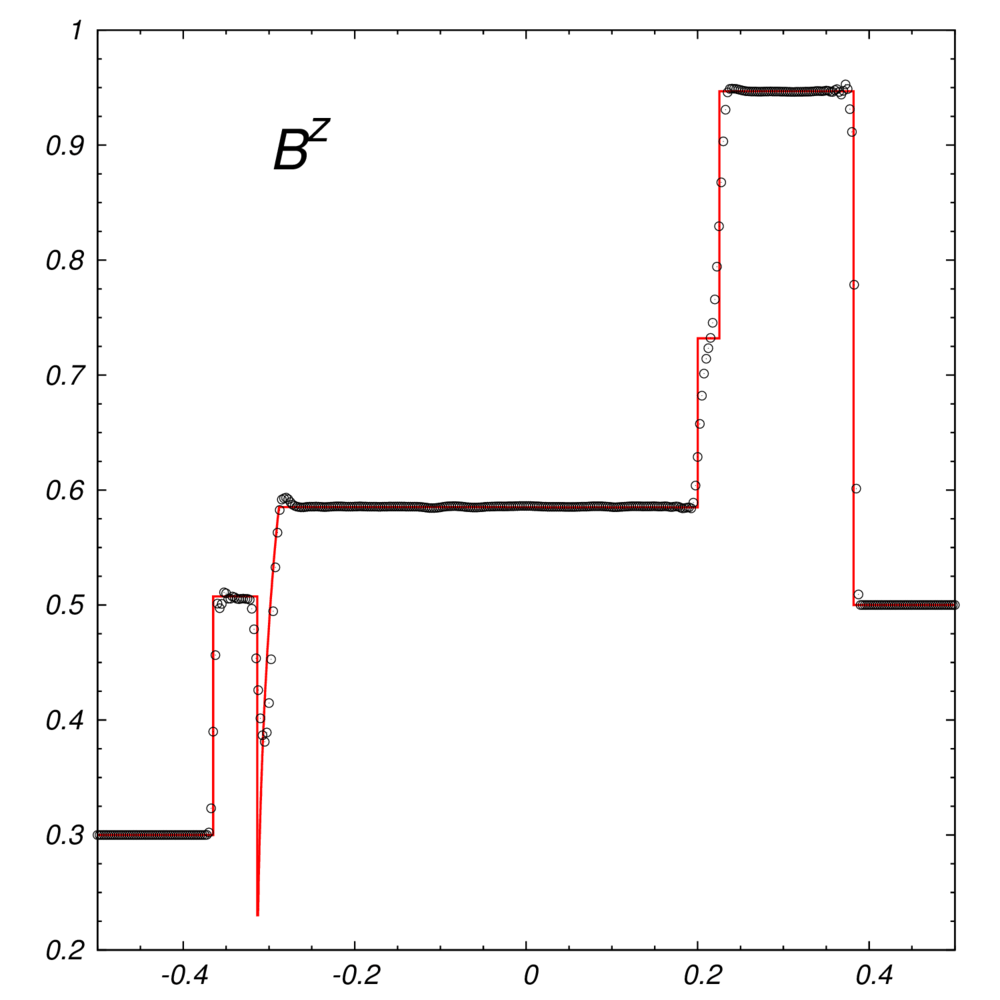}
\includegraphics[width=4.0cm]{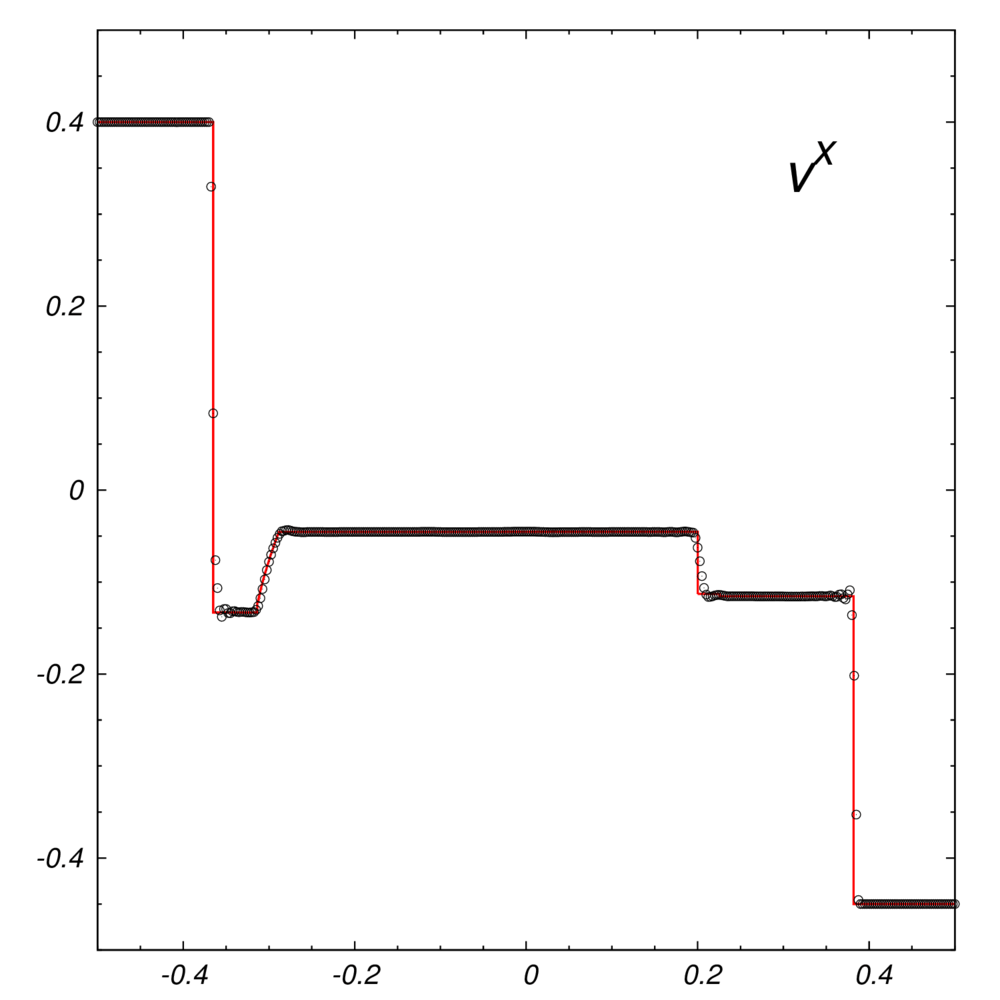}
\includegraphics[width=4.0cm]{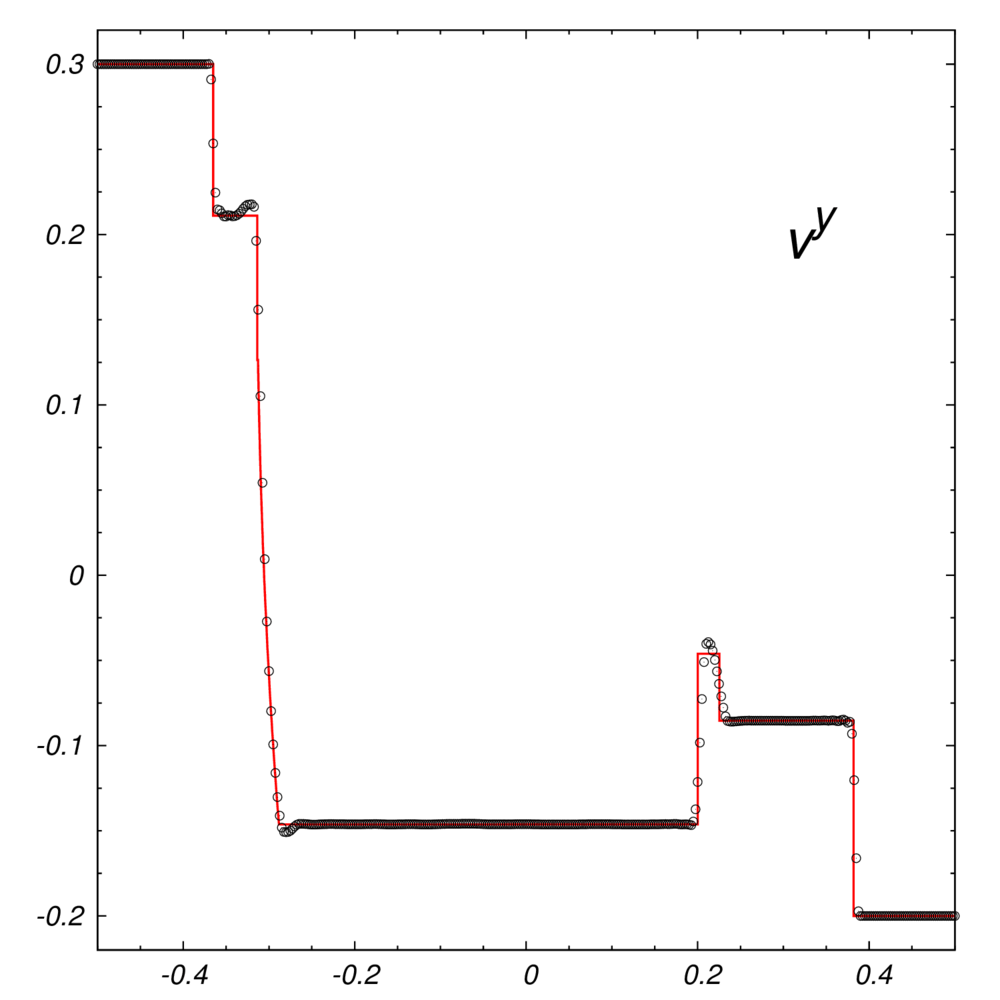}
\end{center}
\caption{\label{fig:Balsara5} Test 7: Balsara 5. In this test we cover the numerical domain $[-0.5,0.5]$ with 1600 cells and use $CFL=0.1$. We show a snapshot at $t=0.55$.}
\end{figure}

This test includes non zero transversal  and discontinuous components of the velocity and magnetic field. In Figure \ref{fig:Balsara5}, we show the snapshot at $t=0.55$. The Lorentz factor is rather small, of the order of $W \sim 1.86$. We can see also an Alfv\'en wave moving to the left and another one moving to the right. In this test we obtain similar errors when using the MINMOD, MC, PPM and WENO5 reconstructors.

\subsubsection{Test 8: Alfv\'en test}

\begin{figure}[!h]
\begin{center}
\includegraphics[width=4.0cm]{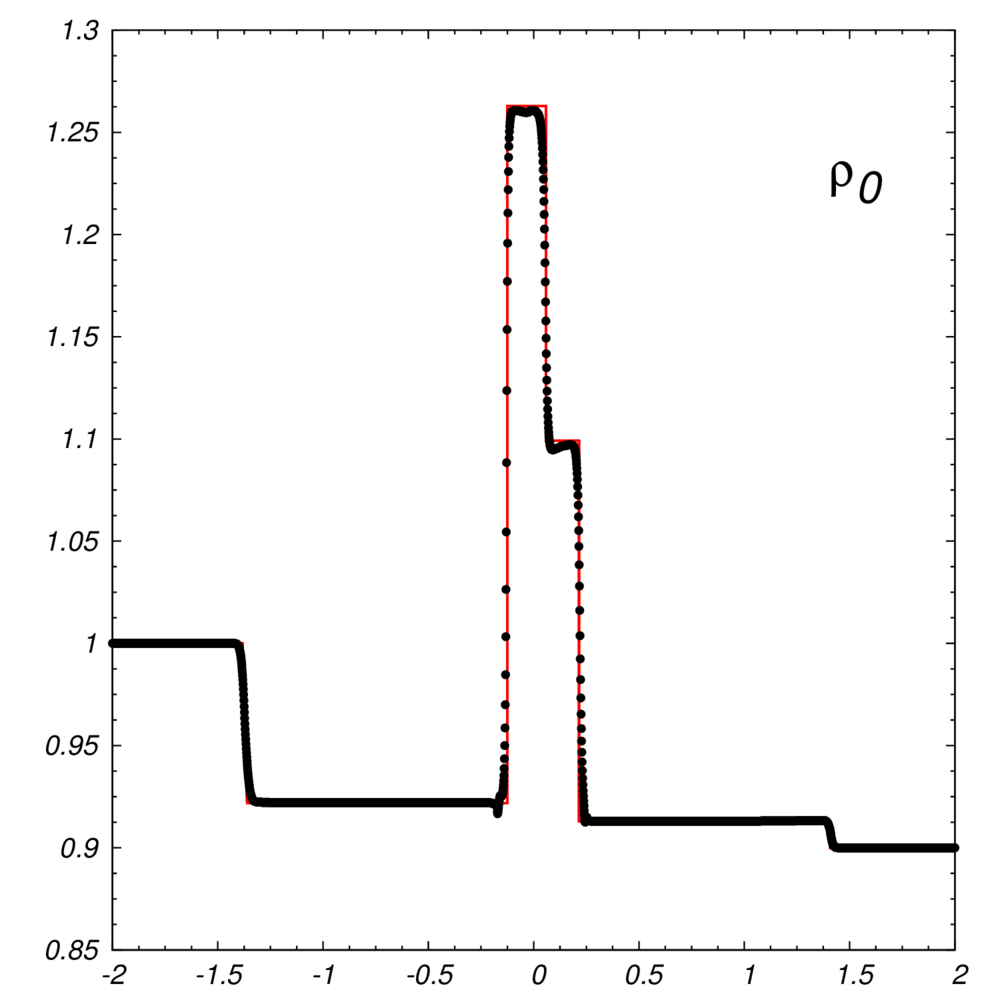}
\includegraphics[width=4.0cm]{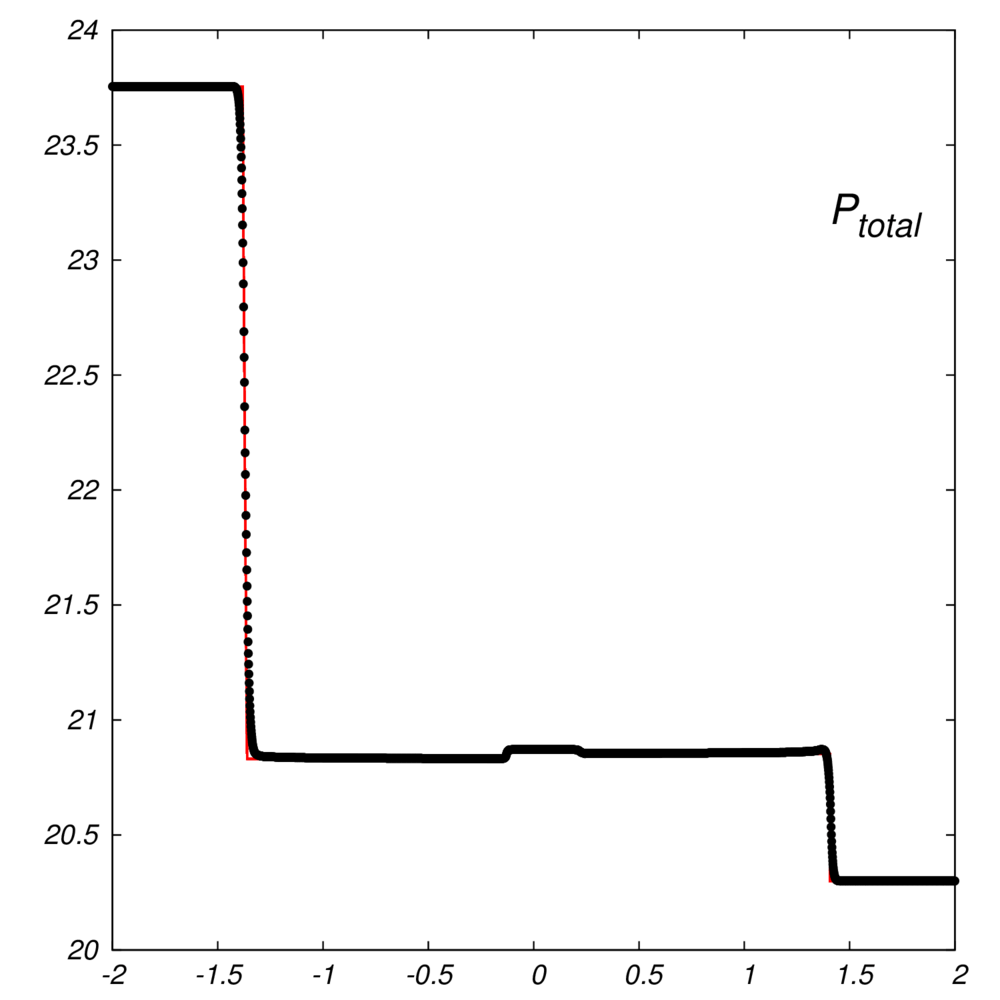}
\includegraphics[width=4.0cm]{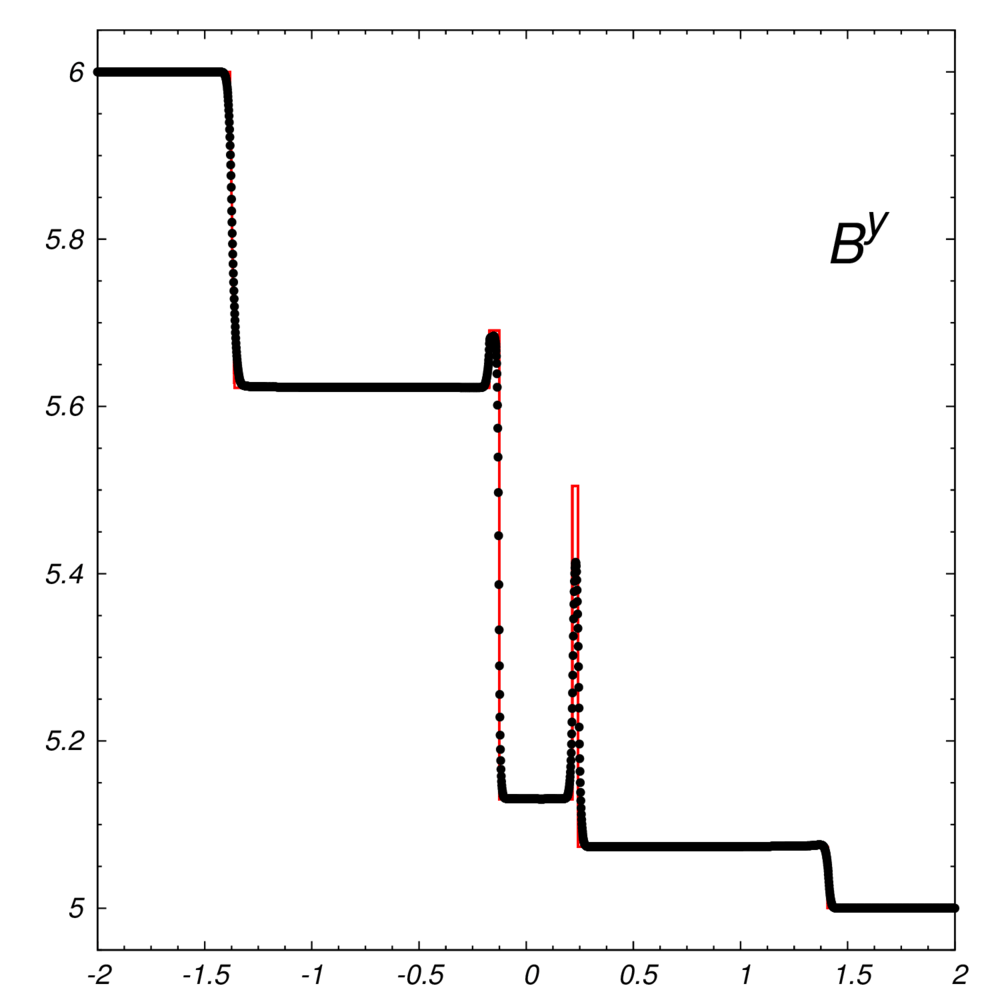}
\includegraphics[width=4.0cm]{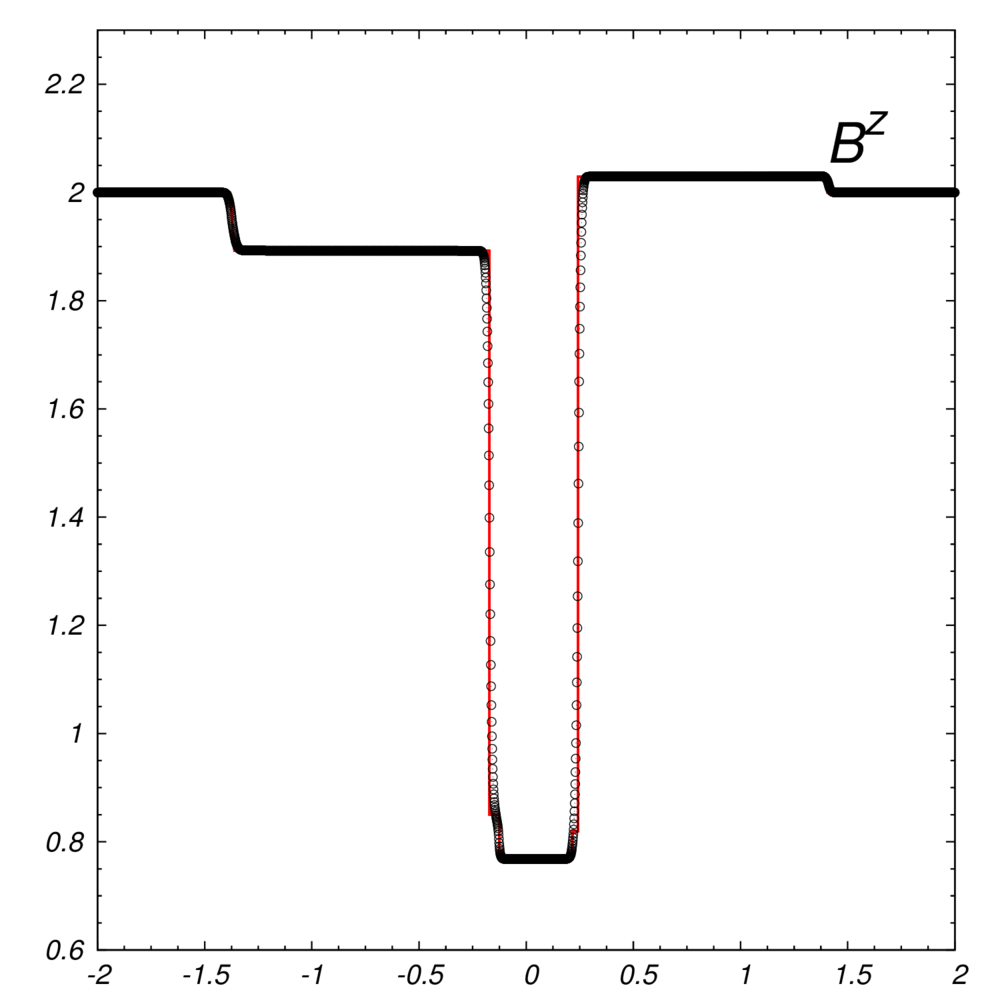}
\includegraphics[width=4.0cm]{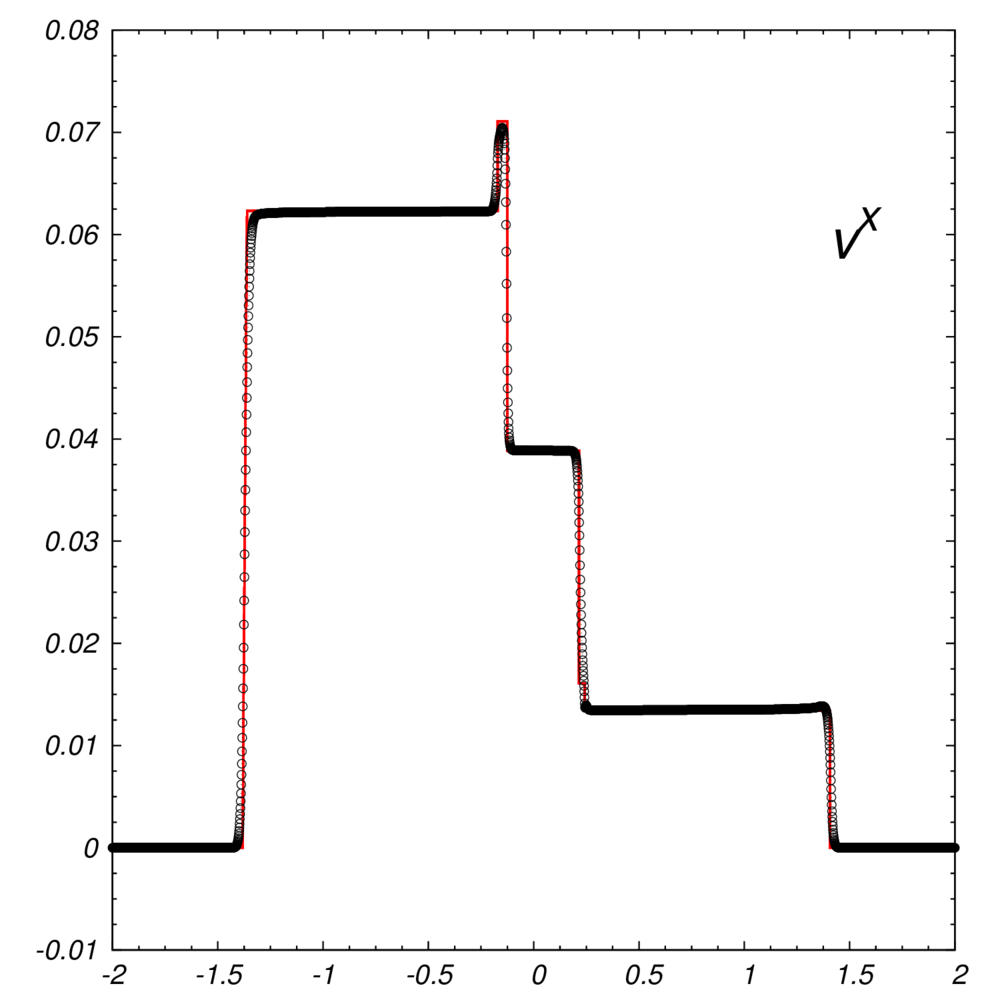}
\includegraphics[width=4.0cm]{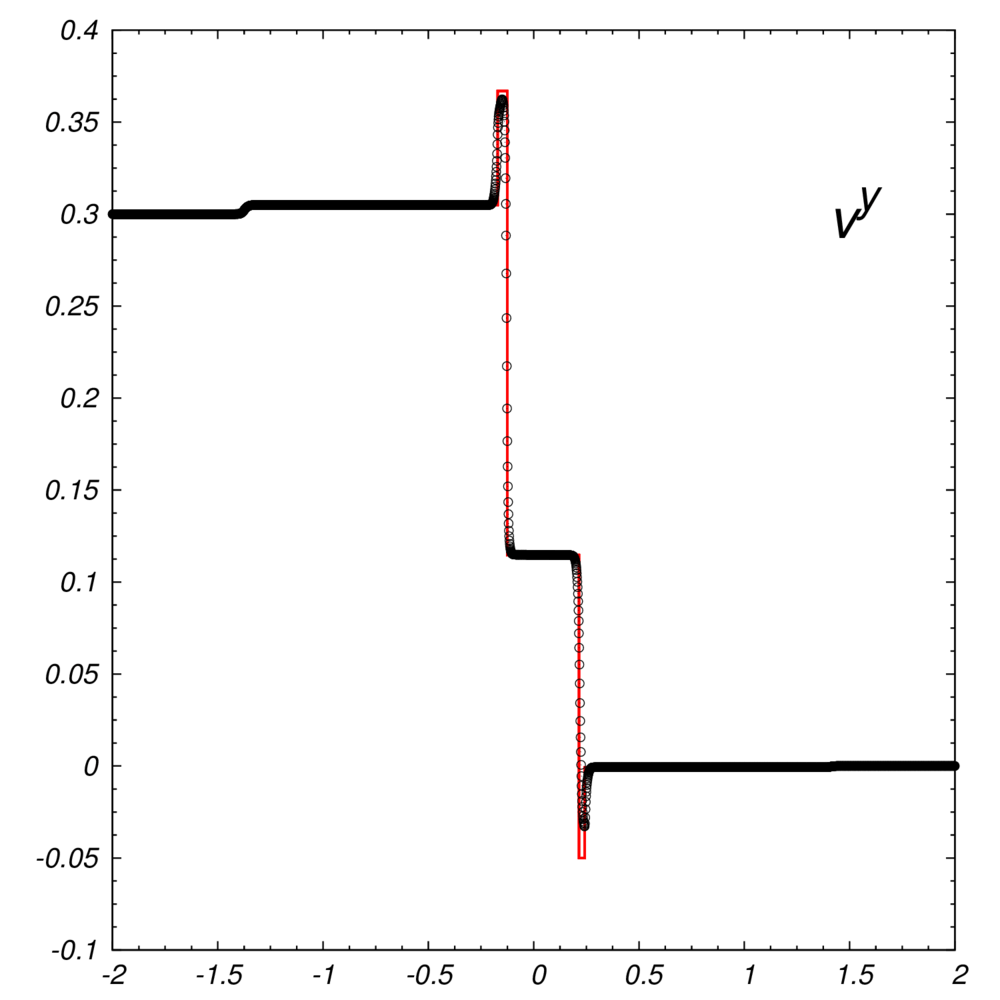}
\end{center}
\caption{\label{fig:Alfventest} Test 8: generic Alfv\'en test at $t=1.5$. We use $3200$ cells to cover the domain $[-2,2]$. We use $CFL=0.1$.}
\end{figure}

The last RMHD 1D test is the generic Alfv\'en wave. In Figure \ref{fig:Alfventest} we show the numerical results at $t=1.5$. During the evolution different regions are formed: a fast rarefaction region, an Alfv\'en wave and a slow shock  moving to the left, a contact wave and two (slow and fast) shocks moving to the right. We reproduce similar results with different  reconstructors, all of which are able to capture the thin shell formed in $B^{y}$ with a few cells. 

 \subsubsection{Error estimates for the 1D RMHD tests}

As in the RHD case, the numerical solution using the various limiters is consistent. We calculate the $L_1$ norm of the error of the 1D tests and the results are shown in Tables \ref{tab:errors_rmhd} and  \ref{tab:errors_rmhd2}. Convergence is considerably more difficult to achieve than in the RHD. For the tests 1 to 8 we obtain nearly first order convergence for the constraint control methods and various  reconstructors, as expected for initial data containing strong discontinuities and our RK3 integrator.

\begin{table*}
\begin{center}
\begin{tabular}{c||c|c|c|c||c|c|c|c} \hline \hline
 ${\it Resolution}$ & MM & MC & WENO5 & PPM & MM & MC & WENO5 & PPM \\ \hline \hline
\multicolumn{1}{|c}{} & \multicolumn{4}{c|}{Flux-CT Error} & \multicolumn{4}{|c|}{Order of convergence} \\ \hline \hline
\multicolumn{1}{|c}{} & \multicolumn{8}{c|}{${\bf Test ~ 1}$} \\ \hline \hline
$\Delta x_1$   & 1.18e-1& 1.10e-1 & 1.14e-1 & 1.18e-1 & .... & .... & .... & .... \\ \hline \hline
$\Delta x_1$   & 8.09e-2 & 6.97e-2& 7.14e-2 & 1.02e-1 & 0.54 & 0.65 & 0.67 & 0.21 \\ \hline \hline
$\Delta x_2$   & 5.20e-2 & 3.96e-2& 3.68e-2 & 8.21e-2 & 0.63 & 0.81 & 0.95 & 0.31 \\ \hline \hline
$\Delta x_3$   & 3.05e-2 &2.15e-2 & 1.95e-2 & 6.39e-2 & 0.77 & 0.88 & 0.91 & 0.36 \\ \hline \hline
$\Delta x_4$   & 1.65e-2 & 1.12e-3& 9.99e-3 & 4.95e-2 & 0.88 & 0.94 & 0.96 & 0.37 \\ \hline \hline
\multicolumn{1}{|c}{} & \multicolumn{8}{c|}{${\bf Test ~ 2}$} \\ \hline \hline
$\Delta x_1$   & 3.39e-1 & 3.16e-1 & 3.15e-1 & 3.45e-1 & .... & .... & .... & .... \\ \hline \hline
$\Delta x_1$   & 2.25e-1 & 2.50e-1 & 2.30e-1 & 2.35e-1 & 0.59 & 0.62 & 0.65 & 0.55 \\ \hline \hline
$\Delta x_2$   & 1.47e-1 & 1.57e-1 & 1.42e-2 & 1.53e-1 & 0.63 & 0.67 & 0.69 & 0.62 \\ \hline \hline
$\Delta x_3$   & 9.41e-2 & 9.80e-2 & 8.60e-2 & 9.87e-2 & 0.64 & 0.68 & 0.72 & 0.63 \\ \hline \hline
$\Delta x_4$   & 5.91e-2 & 6.04e-2 & 4.97e-2 & 6.24e-2 & 0.67 & 0.69 & 0.79 & 0.66 \\ \hline \hline
\multicolumn{1}{|c}{} & \multicolumn{8}{c|}{${\bf Test ~ 3}$} \\ \hline \hline
$\Delta x_1$   & 2.10e-2 & 1.56e-2 & 1.59e-2 & 2.09e-2 & .... & .... & .... & .... \\ \hline \hline
$\Delta x_2$   & 1.33e-2 & 8.86e-3 & 9.36e-3 & 1.33e-2 & 0.65 & 0.81 & 0.76 & 0.65 \\ \hline \hline
$\Delta x_3$   & 8.04e-3 & 4.76e-3 & 5.05e-3 & 8.00e-3 & 0.72 & 0.90 & 0.89 & 0.73 \\ \hline \hline
$\Delta x_4$   & 4.83e-3 & 2.48e-3 & 2.68e-3 & 4.83e-3 & 0.73 & 0.89 & 0.91 & 0.73 \\ \hline \hline
$\Delta x_5$   & 2.89e-3 & 1.39e-3 & 1.40e-3 & 2.89e-3 & 0.74 & 0.89 & 0.93 & 0.74 \\ \hline \hline
\multicolumn{1}{|c}{} & \multicolumn{8}{c|}{${\bf Test ~ 4}$} \\ \hline \hline
$\Delta x_1$   & 1.65e-1 & 1.51e-1 & 1.49e-1 & 1.60e-1 & .... & .... & .... & .... \\ \hline \hline
$\Delta x_2$   & 1.25e-1 & 1.14e-1 & 1.12e-1 & 1.22e-1 & 0.40 & 0.40 & 0.41 & 0.39 \\ \hline \hline
$\Delta x_3$   & 7.98e-2 & 7.11e-2 & 6.76e-1 & 7.86e-1 & 0.64 & 0.68 & 0.72 & 0.63 \\ \hline \hline
$\Delta x_4$   & 5.05e-2 & 4.37e-2 & 3.98e-2 & 5.00e-2 & 0.66 & 0.70 & 0.76 & 0.65 \\ \hline \hline
$\Delta x_5$   & 3.15e-2 & 2.57e-2 & 2.23e-2 & 3.13e-2 & 0.68 & 0.76 & 0.83 & 0.67 \\ \hline \hline
\multicolumn{1}{|c}{} & \multicolumn{8}{c|}{${\bf Test ~ 5}$} \\ \hline \hline
$\Delta x_1$   & 2.26e-1 & 2.24e-1 & 2.24e-1 & 2.94e-1 & .... & .... & .... & .... \\ \hline \hline
$\Delta x_2$   & 1.84e-1 & 1.58e-1 & 1.48e-1 & 2.28e-1 & 0.47 & 0.50 & 0.59 & 0.37 \\ \hline \hline
$\Delta x_3$   & 1.28e-1 & 1.04e-1 & 9.39e-2 & 1.63e-1 & 0.52 & 0.60 & 0.65 & 0.44 \\ \hline \hline
$\Delta x_4$   & 8.70e-2 & 6.62e-2 & 5.74e-2 & 1.12e-1 & 0.55 & 0.65 & 0.71 & 0.54 \\ \hline \hline
$\Delta x_5$   & 5.41e-2 & 3.96e-2 & 3.36e-2 & 0.68e-2 & 0.68 & 0.74 & 0.77 & 0.71 \\ \hline \hline
\multicolumn{1}{|c}{} & \multicolumn{8}{c|}{${\bf Test ~ 6}$} \\ \hline \hline
$\Delta x_1$   & 2.31e0  & 2.15e0  & 2.24e0 & 2.31e0 & .... & .... & .... & ....\\ \hline \hline
$\Delta x_2$   & 1.56e0  & 1.41e0  & 1.46e0 & 1.58e0 & 0.56 & 0.60 & 0.61 & 0.54 \\ \hline \hline
$\Delta x_3$   & 1.05e0  & 8.89e-1 & 9.08e-1& 1.07e0 & 0.57 & 0.66 & 0.68 & 0.56\\ \hline \hline
$\Delta x_4$   & 6.68e-1 & 5.33e-1 & 5.30e-1& 6.89e-1& 0.65 & 0.73 & 0.77 & 0.63 \\ \hline \hline
$\Delta x_5$   & 4.13e-1 & 3.11e-1 & 2.99e-1& 5.50e-1& 0.71 & 0.77 & 0.82 & 0.69 \\ \hline \hline
\multicolumn{1}{|c}{} & \multicolumn{8}{c|}{${\bf Test ~ 7}$} \\ \hline \hline
$\Delta x_1$   & 1.66e-1 & 1.64e-1 & 1.67e-1 & 1.75e-1 & .... & .... & .... & .... \\ \hline \hline
$\Delta x_1$   & 1.14e-1 & 1.12e-1 & 1.13e-1 & 1.22e-1 & 0.54 & 0.55 & 0.56 & 0.52 \\ \hline \hline
$\Delta x_2$   & 7.32e-2 & 6.96e-2 & 7.01e-2 & 7.94e-2 & 0.63 & 0.68 & 0.69 & 0.62 \\ \hline \hline
$\Delta x_3$   & 4.47e-2 & 4.16e-2 & 4.01e-2 & 4.88e-2 & 0.71 & 0.74 & 0.80 & 0.70\\ \hline \hline
$\Delta x_4$   & 2.58e-2 & 2.36e-2 & 2.24e-2 & 2.89e-3 & 0.79 & 0.81 & 0.84 & 0.75 \\ \hline \hline
\multicolumn{1}{|c}{} & \multicolumn{8}{c|}{${\bf Test ~ 8}$} \\ \hline \hline
$\Delta x_1$   & 1.85e-1 & 1.94e-2 & 1.96e-1 & 1.80e-1 & .... & .... & .... & .... \\ \hline \hline
$\Delta x_1$   & 1.20e-1 & 1.24e-1 & 1.23e-1 & 1.16e-1 & 0.62 & 0.64 & 0.67 & 0.63 \\ \hline \hline
$\Delta x_2$   & 7.21e-2 & 7.41e-2 & 7.22e-2 & 7.11e-2 & 0.73 & 0.74 & 0.76 & 0.70 \\ \hline \hline
$\Delta x_3$   & 4.11e-2 & 4.12e-2 & 4.00e-2 & 4.19e-2 & 0.82 & 0.84 & 0.85 & 0.76 \\ \hline \hline
$\Delta x_4$   & 2.23e-2 & 2.21e-2 & 2.11e-2 & 2.29e-2 & 0.88 & 0.90 & 0.92 & 0.87 \\ \hline \hline
\end{tabular} 
\end{center}
\caption{\label{tab:errors_rmhd}. $L_1$ norm of the error in density for each reconstructor and the Flux-CT method to control the magnetic field divergence free constrain. We use resolutions $\Delta x_1 = 1/400 $, $\Delta x_2 = 1/800$, $\Delta x_3 = 1/1600$, $\Delta x_4 = 1/3200$, and $\Delta x_4 = 1/6400$. A dash indicates that the reconstructor was unable to carry out the simulation.}
\end{table*}


\begin{table*}
\begin{center}
\begin{tabular}{c||c|c|c|c||c|c|c|c} \hline \hline
 ${\it Resolution}$ & MM & MC & WENO5 & PPM & MM & MC & WENO5 & PPM \\ \hline \hline
\multicolumn{1}{|c}{} & \multicolumn{4}{c|}{Divergence cleaning error} & \multicolumn{4}{|c|}{Order of convergence} \\ \hline \hline
\multicolumn{1}{|c}{} & \multicolumn{8}{c|}{${\bf Test ~ 1}$} \\ \hline \hline
$\Delta x_1$   & 1.18e-1 & 1.10e-1 & 1.14e-1 & 1.38e-1 & .... & .... & .... &      \\ \hline \hline
$\Delta x_1$   & 8.10e-2 & 6.98e-2 & 7.15e-2 & 1.03e-1 & 0.54 & 0.65 & 0.67 & 0.42     \\ \hline \hline
$\Delta x_2$   & 5.20e-2 & 3.96e-2 & 3.68e-2 & 5.50e-2 & 0.63 & 0.81 & 0.95 & 0.85 \\ \hline \hline
$\Delta x_3$   & 3.05e-2 & 2.15e-2 & 1.95e-2 & 3.00e-2 & 0.76 & 0.88 & 0.92 & 0.87 \\ \hline \hline
$\Delta x_4$   & 1.65e-2 & 1.16e-2 & 1.01e-2 & 1.63e-2 & 0.88 & 0.89 & 0.94 & 0.88 \\ \hline \hline
\multicolumn{1}{|c}{} & \multicolumn{8}{c|}{${\bf Test ~ 2}$} \\ \hline \hline
$\Delta x_1$   & 5.79e+1 & 3.18e+1 & 3.14e+1 & 1.48e+1 & .... & .... & .... & .... \\ \hline \hline
$\Delta x_2$   & 3.98e+1 & 1.86e+1 & 1.76e+1 & 1.11e+1 & 0.54 & 0.77 & 0.83 & 0.41 \\ \hline \hline
$\Delta x_3$   & 2.32e+1 & 1.03e+1 & 9.65e0  & 6.80e0  & 0.77 & 0.85 & 0.86 & 0.70 \\ \hline \hline
$\Delta x_4$   & 1.35e+1 & 5.67e0  & 4.89e0  & 3.98e0  & 0.80 & 0.86 & 0.95 & 0.77 \\ \hline \hline
$\Delta x_5$   & 7.51e0  & 3.03e0  & 2.51e0  & 2.24e0  & 0.84 & 0.90 & 0.96 & 0.82 \\ \hline \hline
\multicolumn{1}{|c}{} & \multicolumn{8}{c|}{${\bf Test ~ 3}$} \\ \hline \hline
$\Delta x_1$   & 2.16e-2 & 1.59e-3 & 1.61e-3 & 1.91e-2 & .... & .... & .... & .... \\ \hline \hline
$\Delta x_1$   & 1.39e-2 & 9.25e-3 & 9.55e-3 & 1.38e-2 & 0.63 & 0.78 & 0.75 & 0.46 \\ \hline \hline
$\Delta x_2$   & 8.45e-3 & 4.84e-3 & 5.02e-3 & 8.42e-3 & 0.71 & 0.93 & 0.92 & 0.71 \\ \hline \hline
$\Delta x_3$   & 5.11e-3 & 2.57e-3 & 2.71e-3 & 5.10e-3 & 0.72 & 0.91 & 0.89 & 0.72 \\ \hline \hline
$\Delta x_4$   & 3.06ee3 & 1.41e-3 & 1.40e-4 & 3.08e-3 & 0.73 & 0.86 & 0.95 & 0.72 \\ \hline \hline
\multicolumn{1}{|c}{} & \multicolumn{8}{c|}{${\bf Test ~ 4}$} \\ \hline \hline
$\Delta x_1$   & 1.65e-1 & 1.52e-1 & 3.02e-1 & 1.66e-1 & .... & .... & .... & .... \\ \hline \hline
$\Delta x_1$   & 1.26e-1 & 1.15e-1 & 1.53e-1 & 1.27e-1 & 0.38 & 0.40 & 0.98 & 0.38 \\ \hline \hline
$\Delta x_2$   & 8.29e-2 & 7.49e-2 & 9.31e-2 & 9.38e-2 & 0.61 & 0.61 & 0.71 & 0.43 \\ \hline \hline
$\Delta x_3$   & 5.23e-2 & 4.62e-2 & 5.01e-2 & 6.42e-2 & 0.66 & 0.69 & 0.90 & 0.61 \\ \hline \hline
$\Delta x_4$   & 3.03e-2 & 2.65e-2 & 2.55e-2 & 3.91e-2 & 0.78 & 0.80 & 0.97 & 0.71 \\ \hline \hline
\multicolumn{1}{|c}{} & \multicolumn{8}{c|}{${\bf Test ~ 5}$} \\ \hline \hline
$\Delta x_1$   & 2.56e-1 & 2.57e-1 & 2.55e-1 & 2.46e-1 & .... & .... & .... & .... \\ \hline \hline
$\Delta x_1$   & 2.28e-1 & 2.04e-1 & 2.01e-1 & 2.00e-1 & 0.16 & 0.33 & 0.34 & 0.30 \\ \hline \hline
$\Delta x_2$   & 1.74e-1 & 1.29e-1 & 1.19e-1 & 1.29e-1 & 0.38 & 0.66 & 0.75 & 0.63 \\ \hline \hline
$\Delta x_3$   & 1.03e-1 & 7.42e-2 & 6.62e-2 & 8.04e-2 & 0.75 & 0.79 & 0.84 & 0.68 \\ \hline \hline
$\Delta x_4$   & 6.13e-2 & 4.05e-2 & 3.58e-2 & 4.91e-2 & 0.75 & 0.87 & 0.88 & 0.71 \\ \hline \hline
\multicolumn{1}{|c}{} & \multicolumn{8}{c|}{${\bf Test ~ 6}$} \\ \hline \hline
$\Delta x_1$   & 2.32e-1 & 2.15e-1 & 2.23e-1 & 2.33e-1 & .... & .... & .... & .... \\ \hline \hline
$\Delta x_1$   & 1.58e-1 & 1.42e-1 & 1.47e-1 & 1.72e-1 & 0.55 & 0.59 & 0.60 & 0.43 \\ \hline \hline
$\Delta x_2$   & 1.08e-1 & 9.36e-2 & 9.39e-2 & 1.21e-1 & 0.54 & 0.60 & 0.64 & 0.50 \\ \hline \hline
$\Delta x_3$   & 7.21e-2 & 6.00e-2 & 5.89e-2 & 8.08e-2 & 0.58 & 0.64 & 0.74 & 0.58 \\ \hline \hline
$\Delta x_4$   & 4.42e-2 & 3.63e-2 & 3.39e-2 & 5.22e-2 & 0.70 & 0.72 & 0.79 & 0.63 \\ \hline \hline
\multicolumn{1}{|c}{} & \multicolumn{8}{c|}{${\bf Test ~ 7}$} \\ \hline \hline
$\Delta x_1$   & 1.14e-1 & 6.33e-2 & 6.61e-2 & 1.62e-1 & .... & .... & .... & .... \\ \hline \hline
$\Delta x_1$   & 8.21e-2 & 4.39e-2 & 4.13e-2 & 1.14e-1 & 0.47 & 0.52 & 0.65 & 0.50 \\ \hline \hline
$\Delta x_2$   & 5.21e-2 & 2.78e-2 & 2.50e-2 & 7.35e-2 & 0.65 & 0.66 & 0.72 & 0.63 \\ \hline \hline
$\Delta x_3$   & 3.24e-2 & 1.72e-2 & 1.44e-2 & 4.78e-2 & 0.68 & 0.69 & 0.79 & 0.62 \\ \hline \hline
$\Delta x_4$   & 1.98e-2 & 1.02e-2 & 8.16e-3 & 2.96e-2 & 0.71 & 0.75 & 0.82 & 0.69 \\ \hline \hline
\multicolumn{1}{|c}{} & \multicolumn{8}{c|}{${\bf Test ~ 8}$} \\ \hline \hline
$\Delta x_1$   & 1.69e-1 & 1.70e-1 & 1.68e-1 & 1.70e-1 & .... & .... & .... & .... \\ \hline \hline
$\Delta x_1$   & 1.19e-1 & 1.14e-1 & 1.13e-1 & 1.15e-1 & 0.50 & 0.57 & 0.57 & 0.56 \\ \hline \hline
$\Delta x_2$   & 7.86e-2 & 7.36e-2 & 7.14e-2 & 7.73e-2 & 0.60 & 0.63 & 0.66 & 0.57 \\ \hline \hline
$\Delta x_3$   & 4.85e-2 & 4.26e-2 & 4.11e-2 & 4.91e-2 & 0.70 & 0.78 & 0.79 & 0.65 \\ \hline \hline
$\Delta x_4$   & 2.85e-2 & 2.42e-2 & 2.28e-3 & 2.89e-2 & 0.76 & 0.81 & 0.85 & 0.76 \\ \hline \hline
\end{tabular} 
\end{center}
\caption{\label{tab:errors_rmhd2}. $L_1$ norm of the error in density for each reconstructor and the Divergence Cleaning method to control the magnetic field divergence free constrain. We use the resolutions $\Delta x_1 = 1/400 $, $\Delta x_2 = 1/800$, $\Delta x_3 = 1/1600$, $\Delta x_4 = 1/3200$, and $\Delta x_4 = 1/6400$. A dash indicates that the reconstructor was unable to carry out the simulation.}
\end{table*}

\subsubsection{Magnetic Rotor Test}

\begin{figure*}
\begin{center} 
\includegraphics[width=8.0cm]{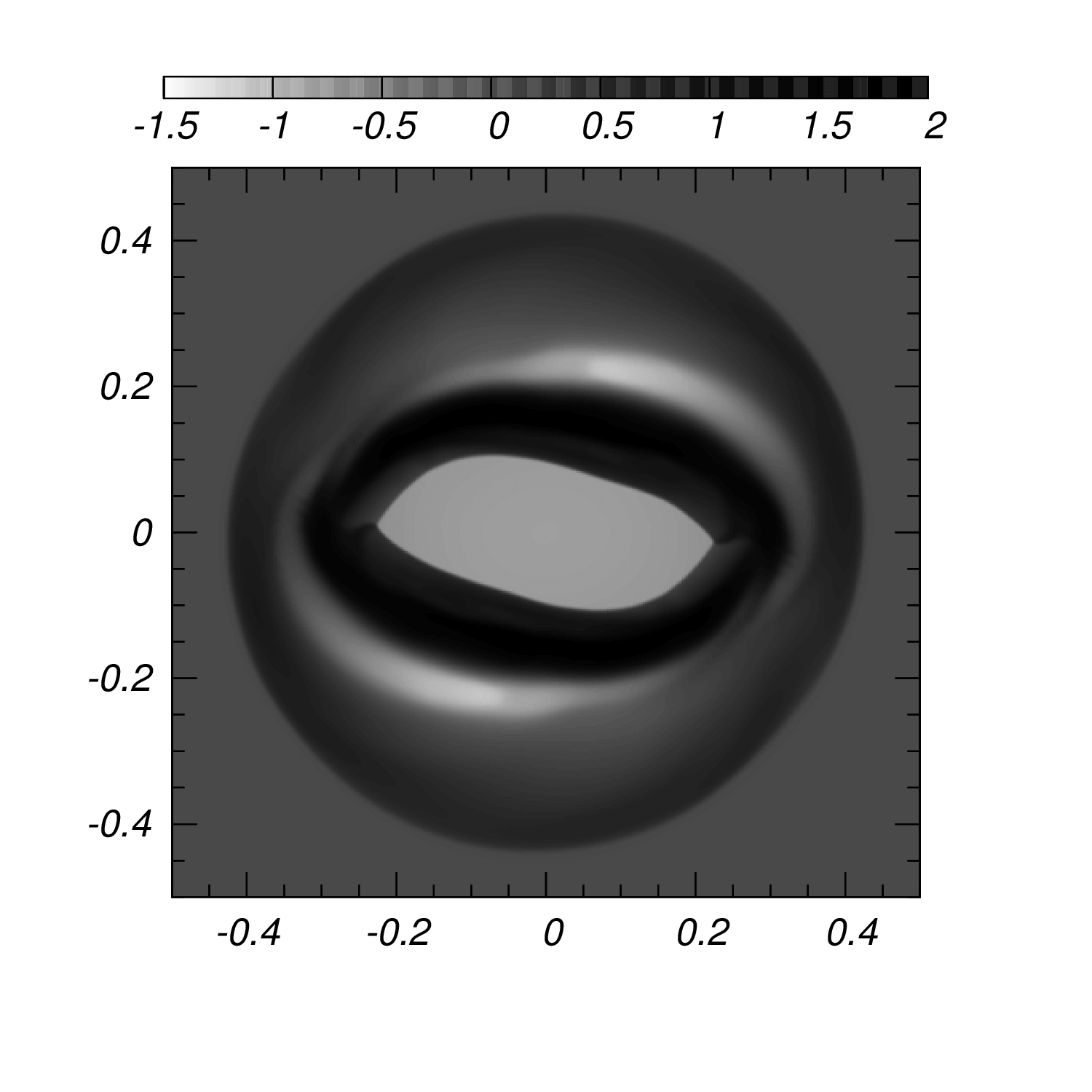}
\includegraphics[width=8.0cm]{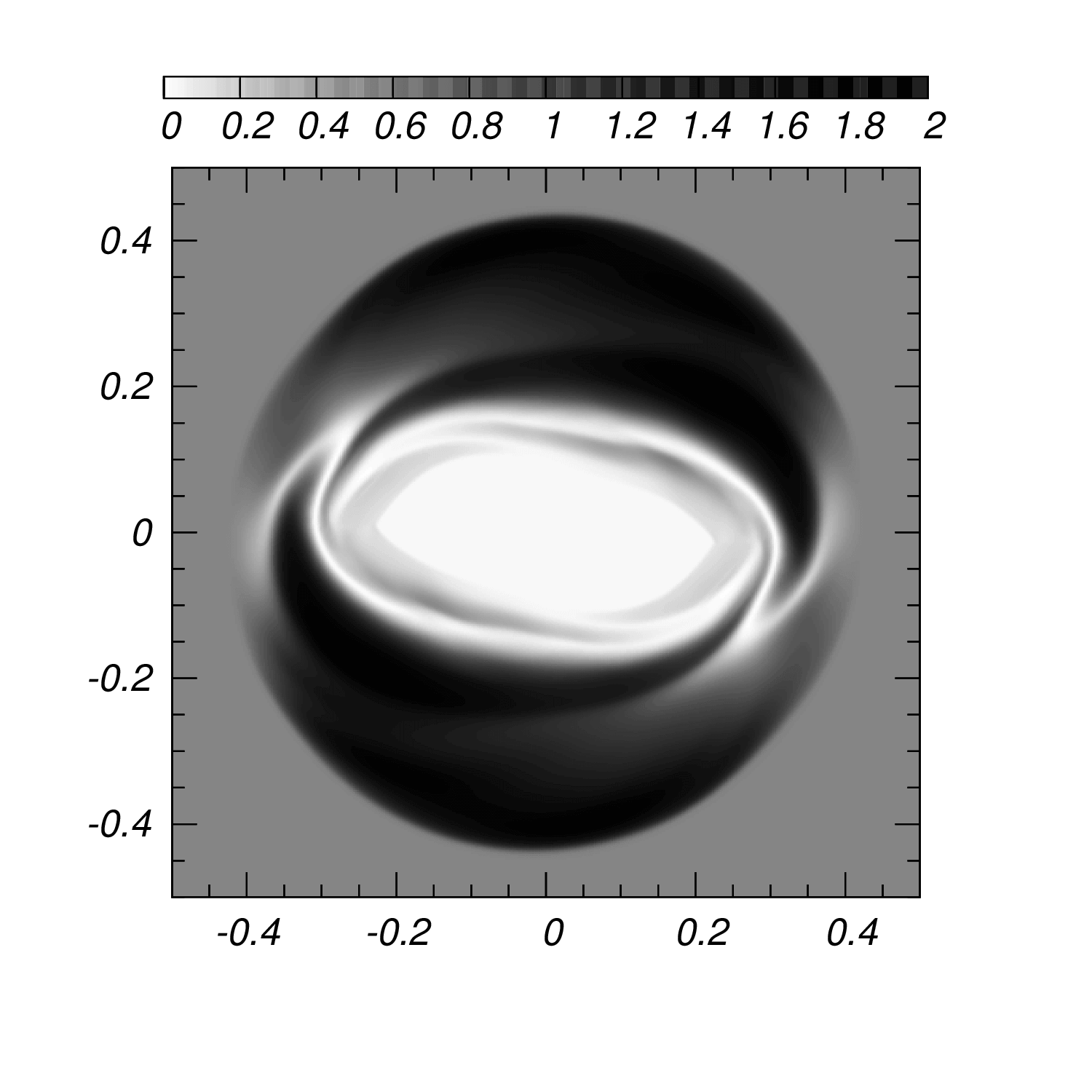}
\includegraphics[width=8.0cm]{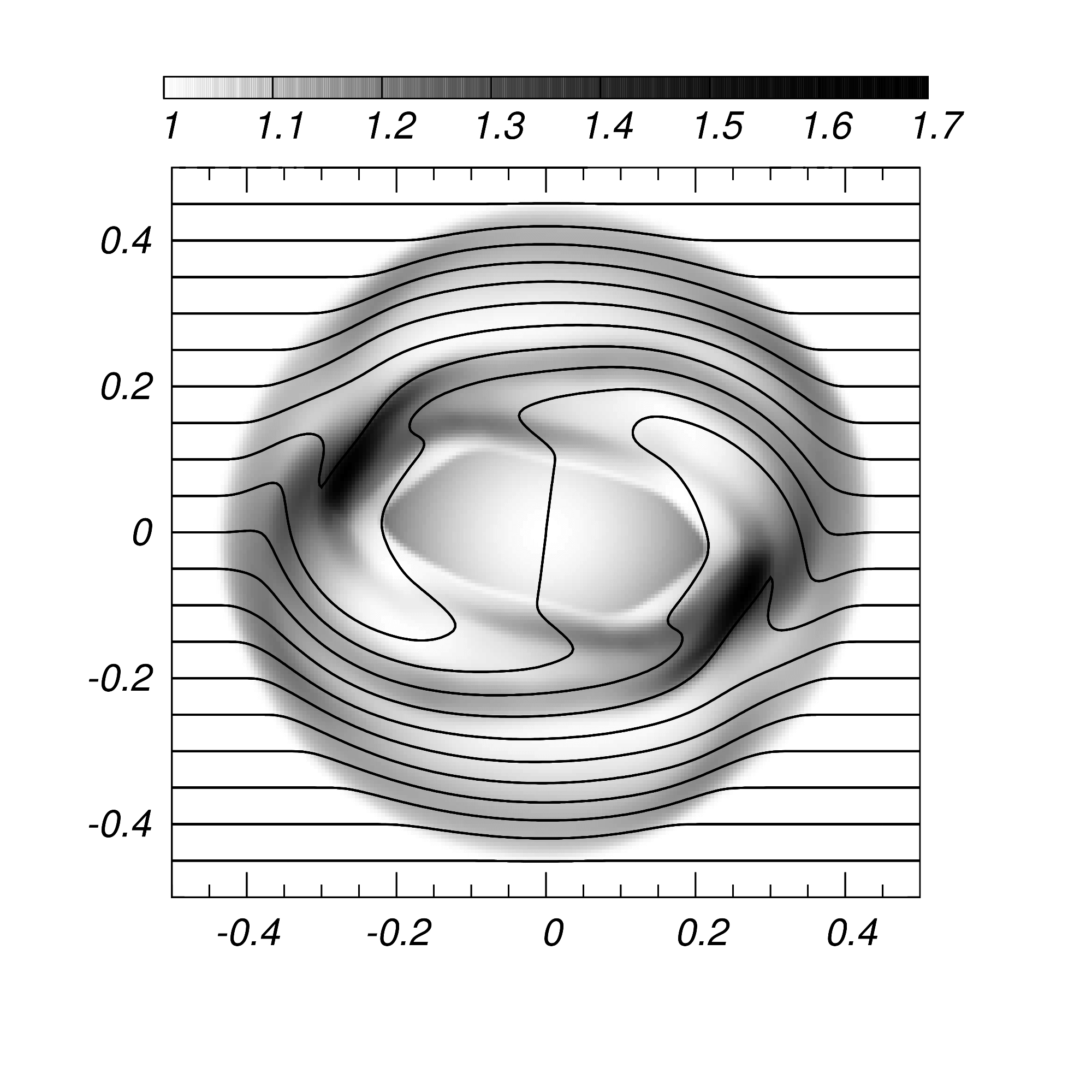}
\includegraphics[width=8.0cm]{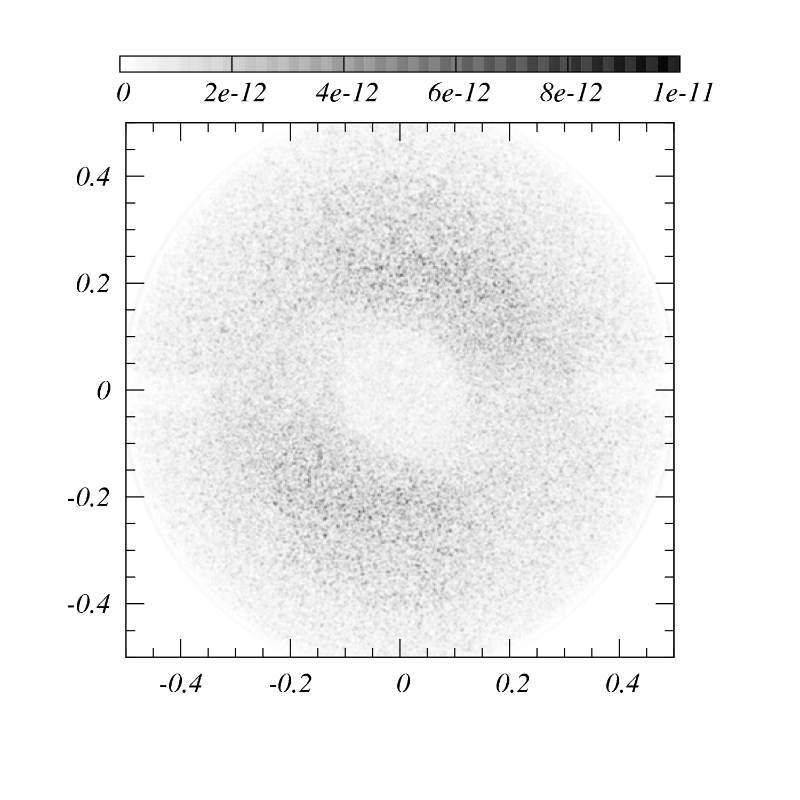}
\end{center}
\caption{\label{fig:RMRotor2D} 2D Magnetic Rotor. We show a snapshot at $t=0.4$ of the logarithm of the proper rest mass density $\rho_0$ (top-left), the magnetic pressure $p$ (top-right), Lorentz factor and the magnetic field lines (bottom-left) and the divergence of the magnetic field $\nabla \cdot \vec{B}$ (bottom-right). We use MINMOD and $200 \times 200$ cells to cover the domain $[-0.5,0.5] \times [-0.5,0.5]$, and use $CFL=0.25$. In all faces outflow boudary conditions were used.} 
\end{figure*}

\begin{figure*}
\begin{center}
\includegraphics[width=10.0cm]{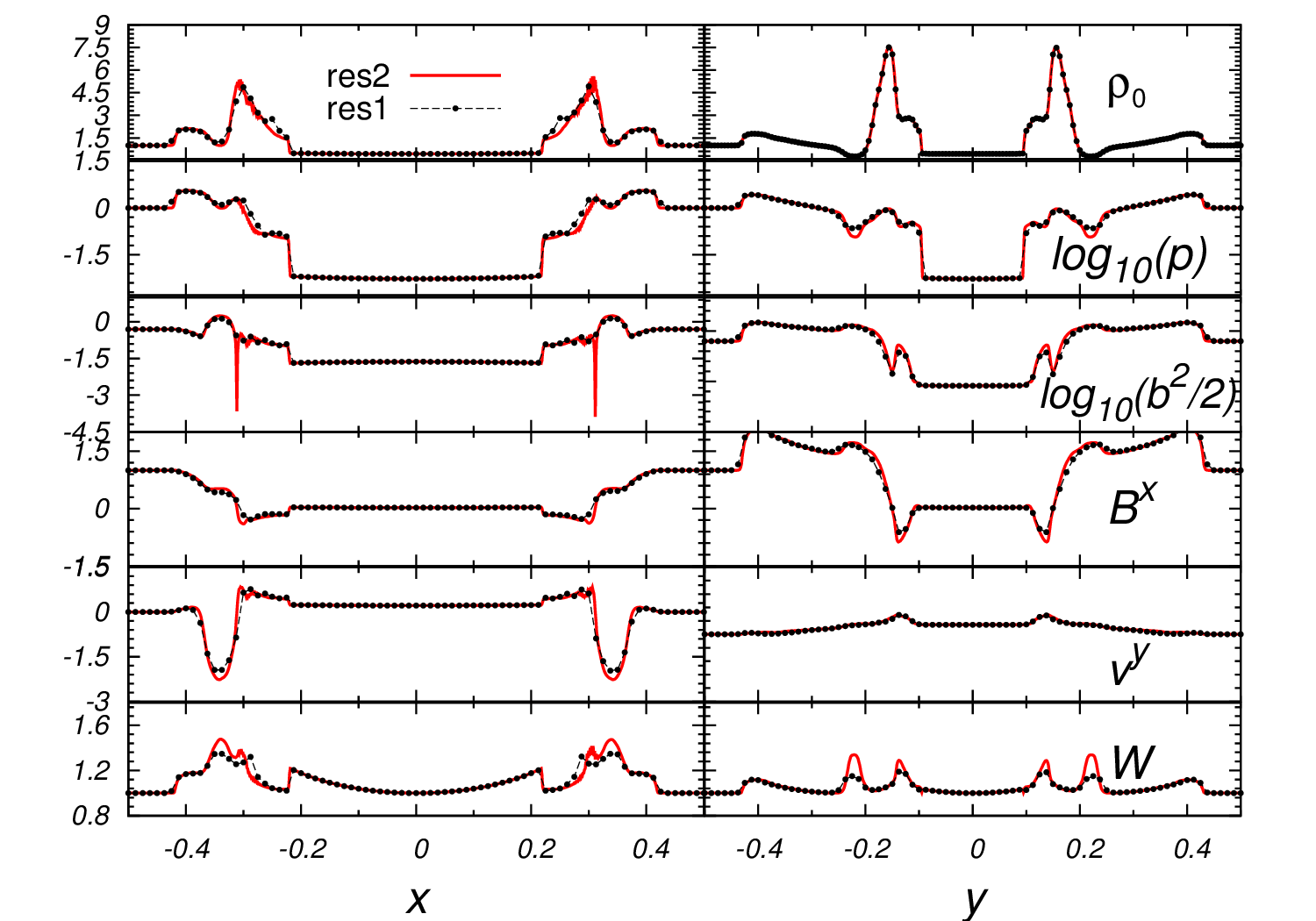}
\end{center}
\caption{\label{fig:RMRotor1D} Variables of the 1D magnetic rotor test along the $x$
and $y$ axes using two resolutions. On the left we show the variables along $x$ and on the right the
variables along $y$. The two resolutions we use are $200\times 200$ and $400\times 400$ cells that cover the numerical domain $[-0.5, 0.5]\times [-0.5,0.5]$. }
\end{figure*}

The first  2D test is a magnetic rotor defined on the $xy$ plane. The initial density within a cylinder of  radius $r_{in} = 0.1$ is $\rho_{in} = 10$, and has angular velocity $\omega_{z} =9.55$. The initial pressure is constant in the whole domain $p=1$ and the magnetic field has components $B_{x}=1$, $B_y=0$.  The components of the initial velocity inside the cylinder are defined by $v_{in}^{x}= -\omega_{z} y$ and $v_{in}^{y}=\omega_{z} x$. In the exterior region ($r> r_{in}$), the fluid density is $\rho=1$ and the velocity is zero. The results of the evolution for $\Gamma=5/3$ are shown in Figure \ref{fig:RMRotor2D} at $t=0.4$,  where the different shocks and the rotational Alfv\'en waves can be observed and reproduce the morphology in \citep{2003A&A...400..397D}. We show additionally the violation of the divergence of the magnetic field constraint calculated with the flux-CT method. In this case the violation is bounded to be $\nabla \cdot \vec{B} \sim 10^{-11}$ for MINMOD. As expected, the magnetic field slows down the rotor velocity, reducing the Lorentz factor from $W\sim 10$ at initial time to $W\sim 4$ at time $t=0.4$ and all the variables are affected by the dragging of the initial angular velocity. In Figure \ref{fig:RMRotor1D}, we show 1D slices for two different resolutions, showing the variable profiles along the $x$ and $y$ axes. The variable profiles reproduce those in \citep{2014CQGra..31a5005M}. 

\subsubsection{Cylindrical Explosion Test}

\begin{figure*}
\begin{center}
\includegraphics[width=5.5cm]{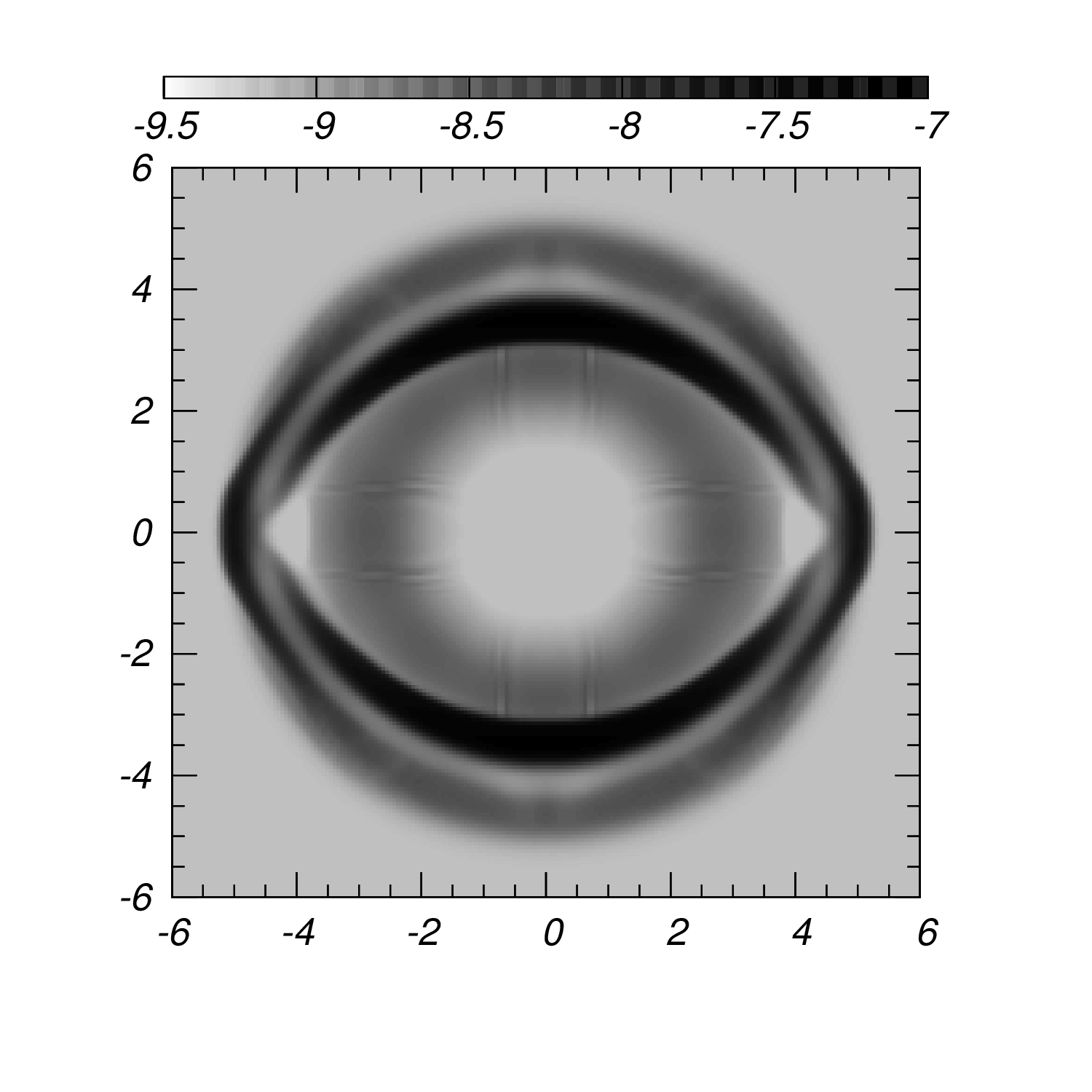}
\includegraphics[width=5.5cm]{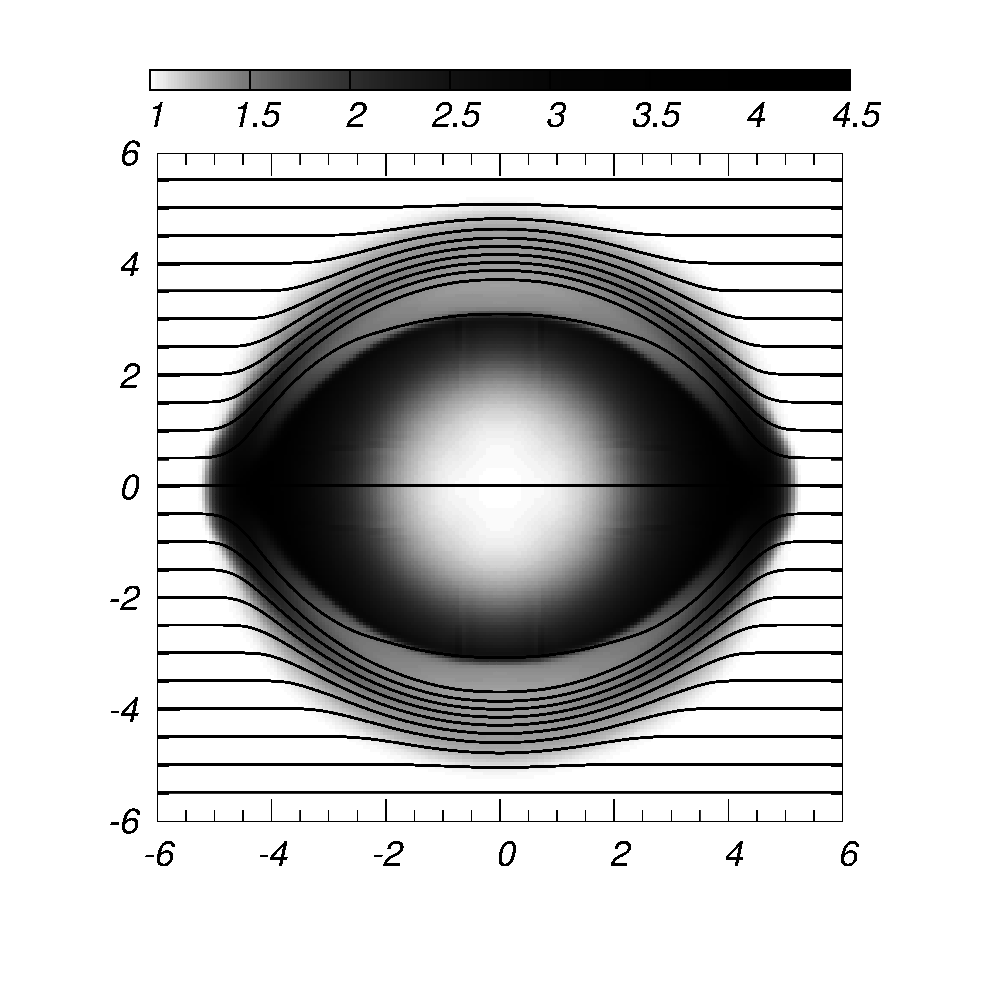}
\includegraphics[width=5.5cm]{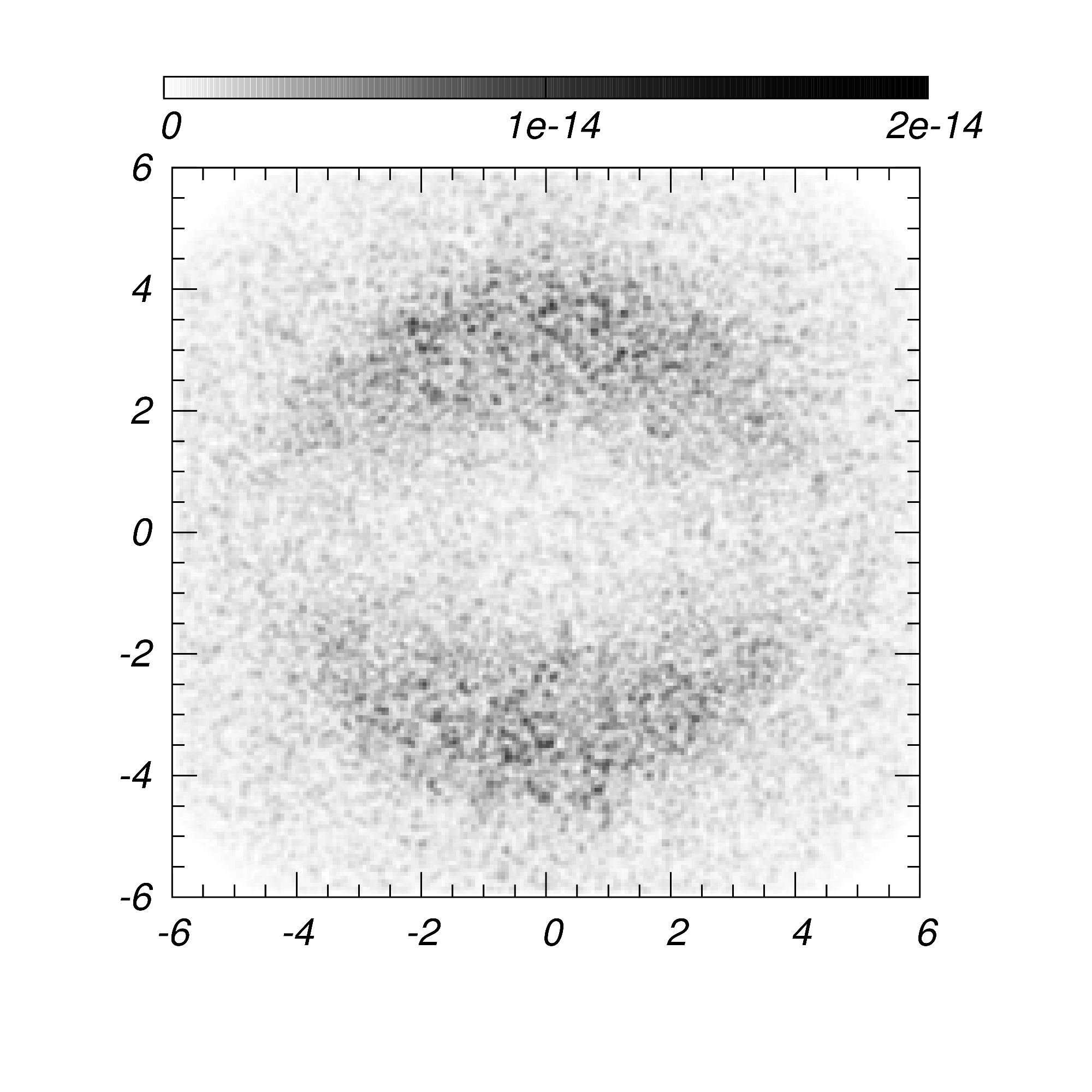}
\includegraphics[width=5.5cm]{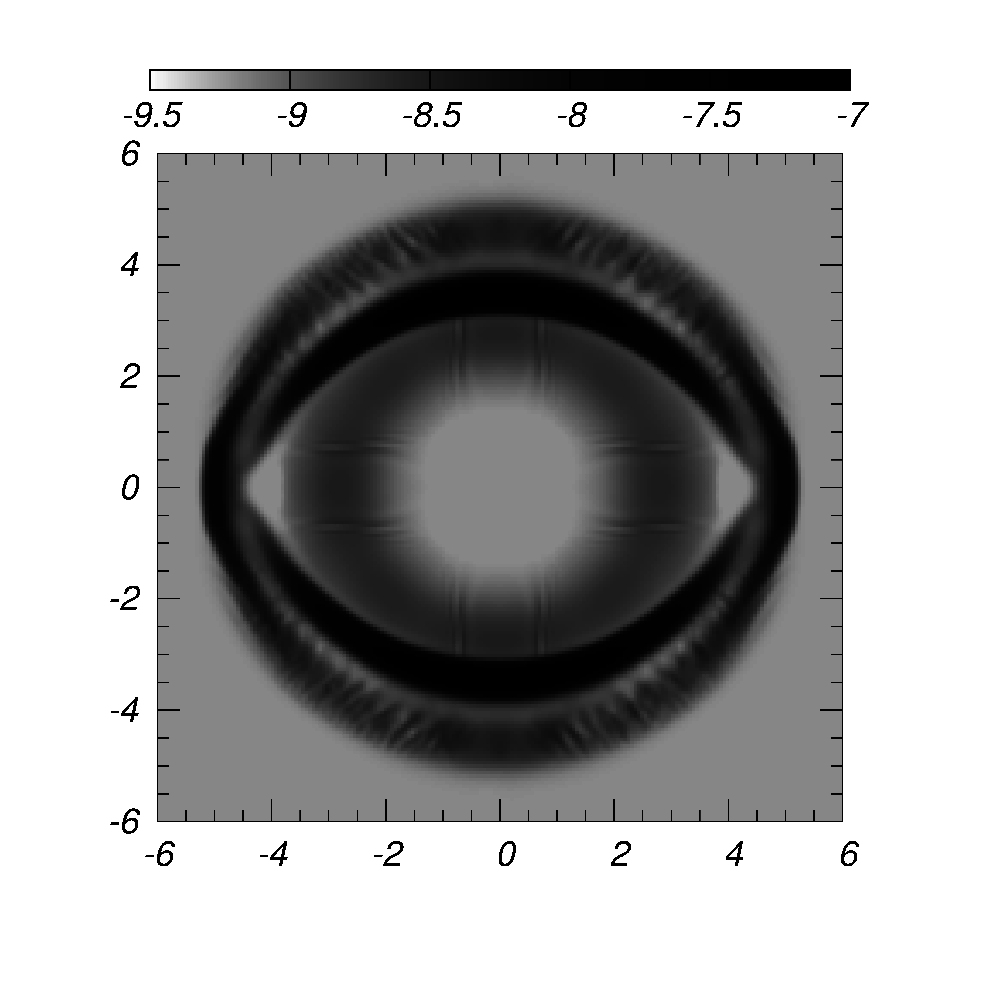}
\includegraphics[width=5.5cm]{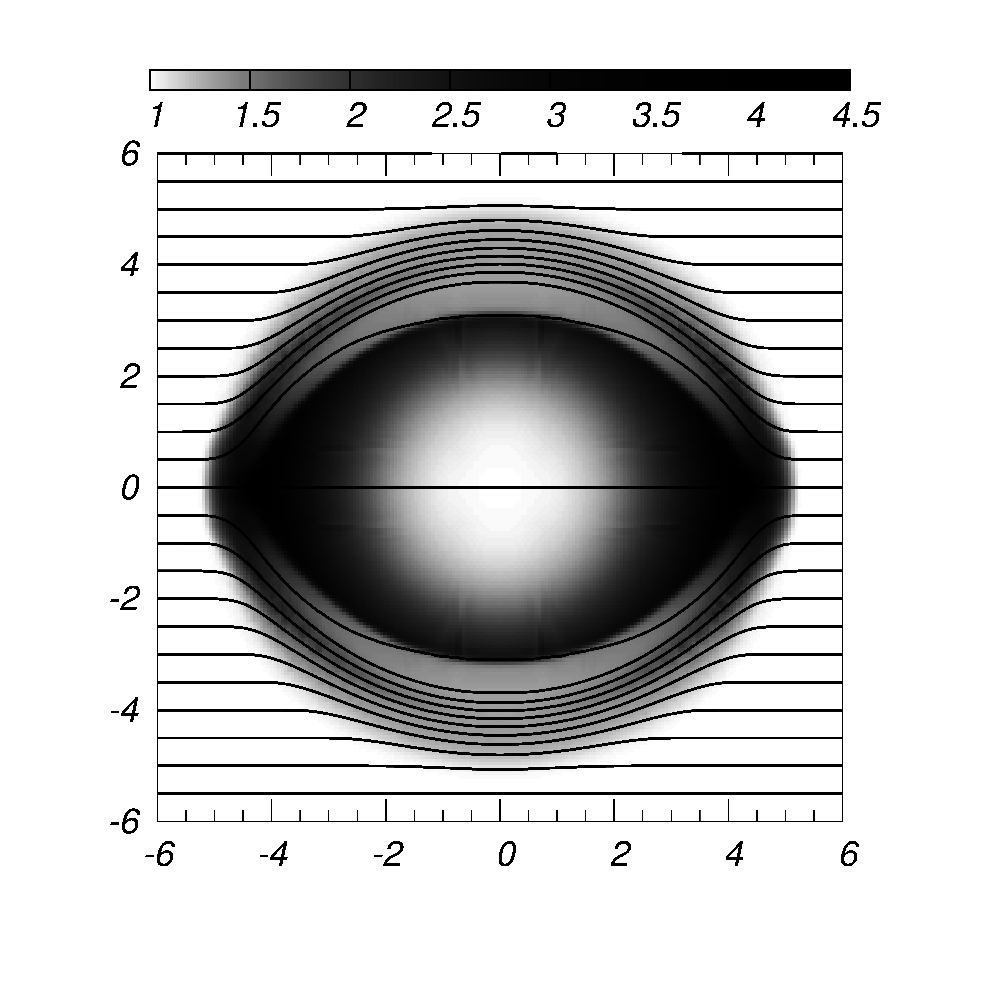}
\includegraphics[width=5.5cm]{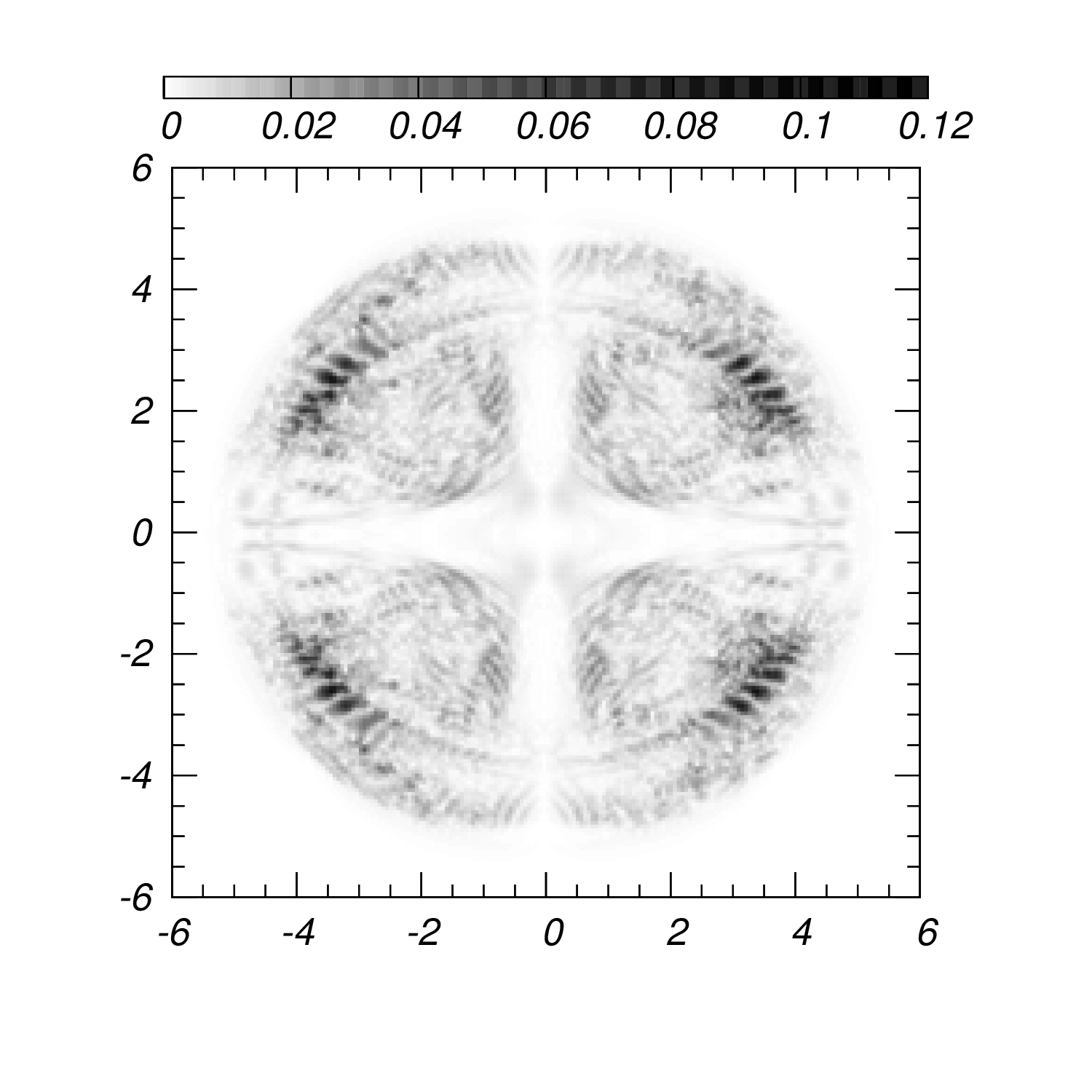}
\end{center}
\caption{\label{fig:Cylindrical2D} 2D Cylindrical Explosion Test. We show a snapshot at $t=4$ for the logarithm of the proper rest mass density $\rho_0$ (left), Lorentz factor and the magnetic field lines (middle) and the divergence of the magnetic field $\nabla \cdot \vec{B}$ (right). The plots on the top row are calculated using the flux-CT method whereas those on the bottom are computed with the divergence cleaning method. We use $200 \times 200$ cells to cover the domain $[-6,6]\times[-6,6]$ and use $CFL = 0.25$. We use outflow boundary conditions.} 
\end{figure*}

\begin{figure*}
\begin{center}
\includegraphics[width=10.0cm]{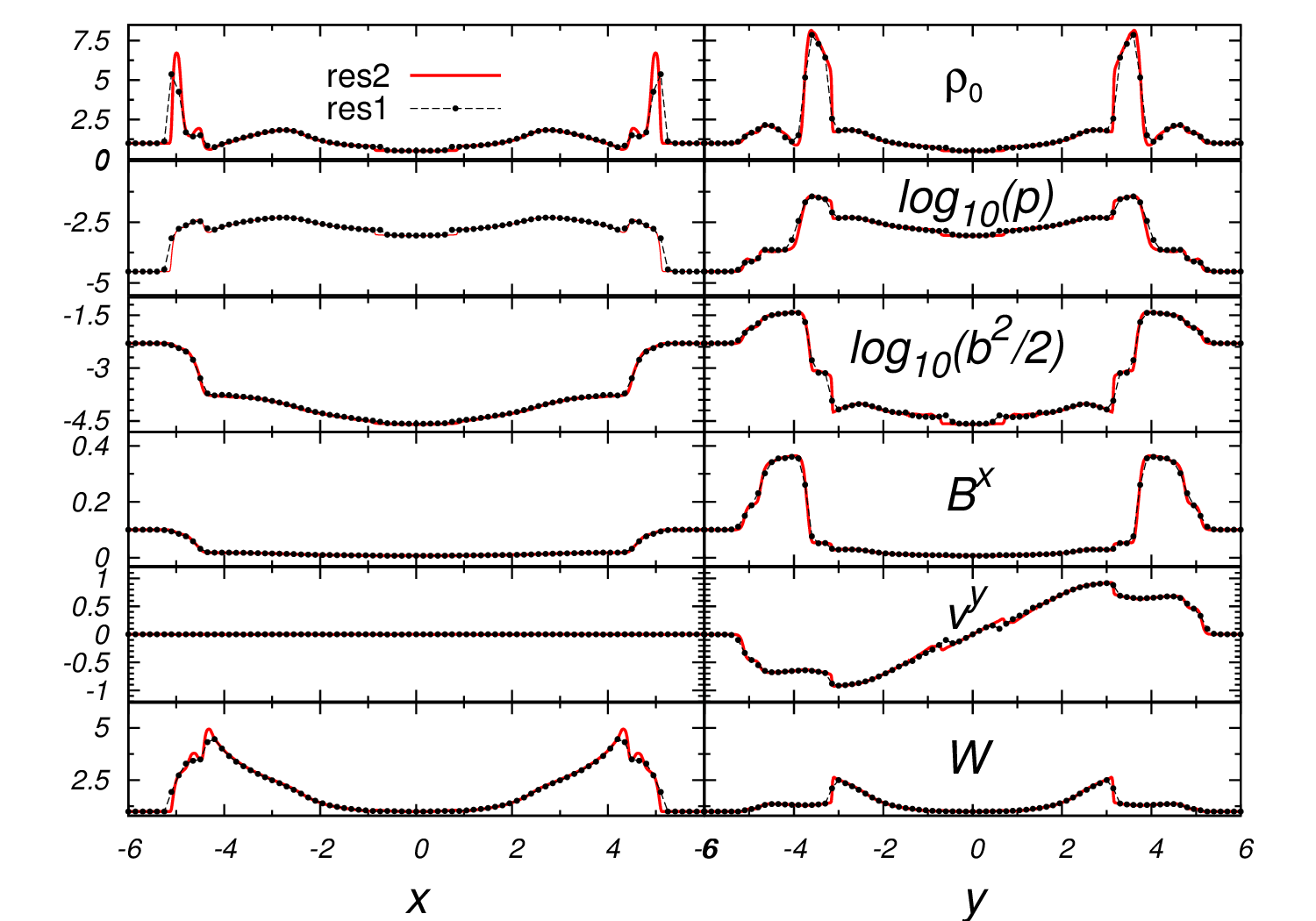}
\end{center}
\caption{\label{fig:Cylindrical1D} The variables of the explosion projected along the $x$ and $y$ axes. The three resolutions correspond to $200\times 200$ and $400\times 400$ cells used to cover the numerical domain $[-6, 6]\times[-6,6]$. }
\end{figure*}

The second 2D test is the cylindrical explosion, starting the evolution with a cylindrical inner region with radius $r_{in} = 0.8$, where the rest mass density is $\rho_{in} = 10^{-2}$ and the pressure is $p_{in} = 1$. In the outer region $r>r_{out} = 1.0$ the variables are $\rho_{out}= 10^{-4}$ and $p_{out} = 3\times 10^{-5}$. The  magnetic field is uniform in the whole domain initially  $B_{x}=0.1$ and $B_y=0$, whereas the adiabatic index is $\Gamma =4/3$. In this configuration the fluid is initially at rest. We also use a smoothing function for the density and pressure for $r_{in}<r<r_{out}$ as in \citep{2014CQGra..31a5005M}.

\begin{eqnarray}
\rho_{0} =\left\{ 
\begin{tabular}{ccc}
$\rho_{in}$                     && $r \leq r_{in}$ \\
&&\\
$e^{\big\{\frac{(r_{out}-r)ln(\rho_{out})+(r-r_{in})ln(\rho_{in})}{r_{out}-r_{in}}\big\}}$ & & $r_{in}<r<r_{out}$\\
&&\\
$\rho_{out}$                    &       & $r \geq r_{out}$,
\end{tabular}
\right.
\end{eqnarray}

\noindent where $r=\sqrt{x^{2}+y^{2}}$. A similar smoothing function is used for the pressure. We  again practice the test in the $xy$ plane. The results of the evolution are shown in Figure \ref{fig:Cylindrical2D}, and reproduce those in \citep{2003A&A...400..397D}. In this test, in order to compare the two different methods implemented to prevent the growth of the constraint violation of the magnetic field, we present the numerical calculations with flux-CT and divergence cleaning methods. In the first case, the divergence of the magnetic field remains of the order $\nabla \cdot \vec{B} \sim 10^{-14}$ whereas in the second one the violation is of the order $\sim 10^{-2}$ in some regions, which is comparable  with previous analyses \citep{Neilsen:2005rq}. Additionally, like in the relativistic rigid rotor test, in Figure \ref{fig:Cylindrical1D} we show 1D slices for two different resolutions, showing the variable profiles along the $x$ and $y$ axes. The variable profiles are better resolved with higher resolution as in \citep{2014CQGra..31a5005M}. The evolution shows an exterior shock wave expanding radially at nearly the speed of light with a very small amplitude. 

\begin{figure*}
\begin{center}
\includegraphics[width=8.0cm]{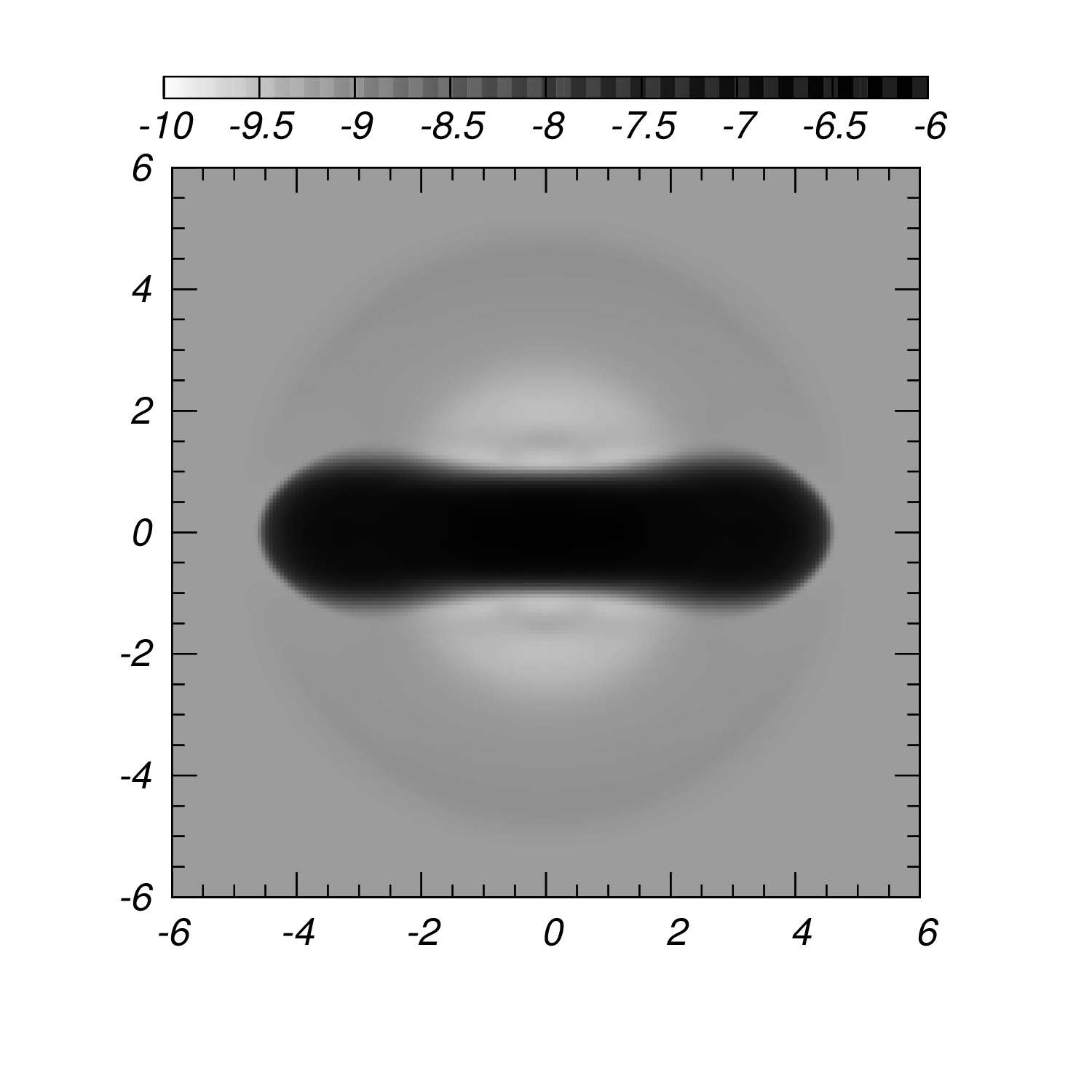}
\includegraphics[width=8.0cm]{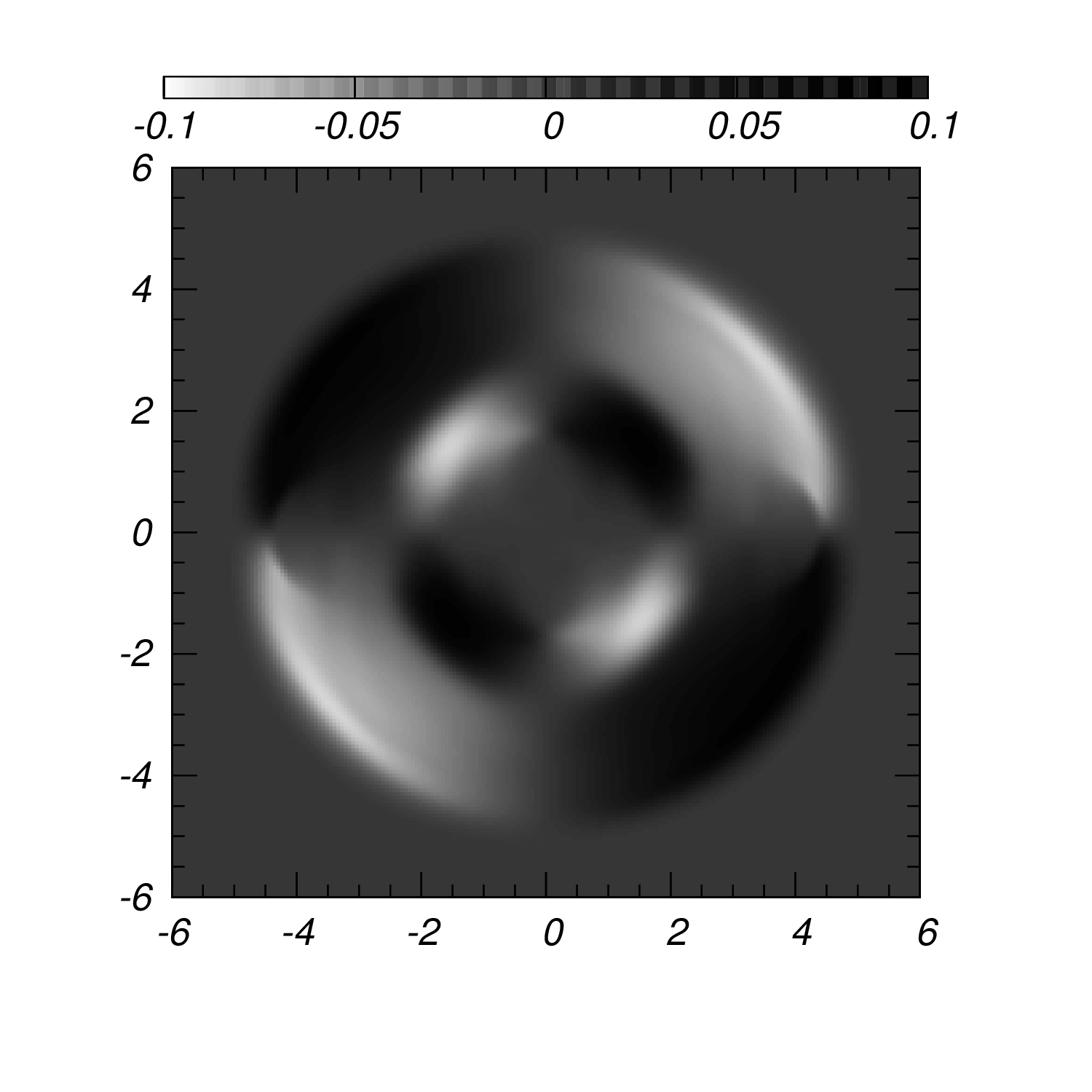}
\includegraphics[width=8.0cm]{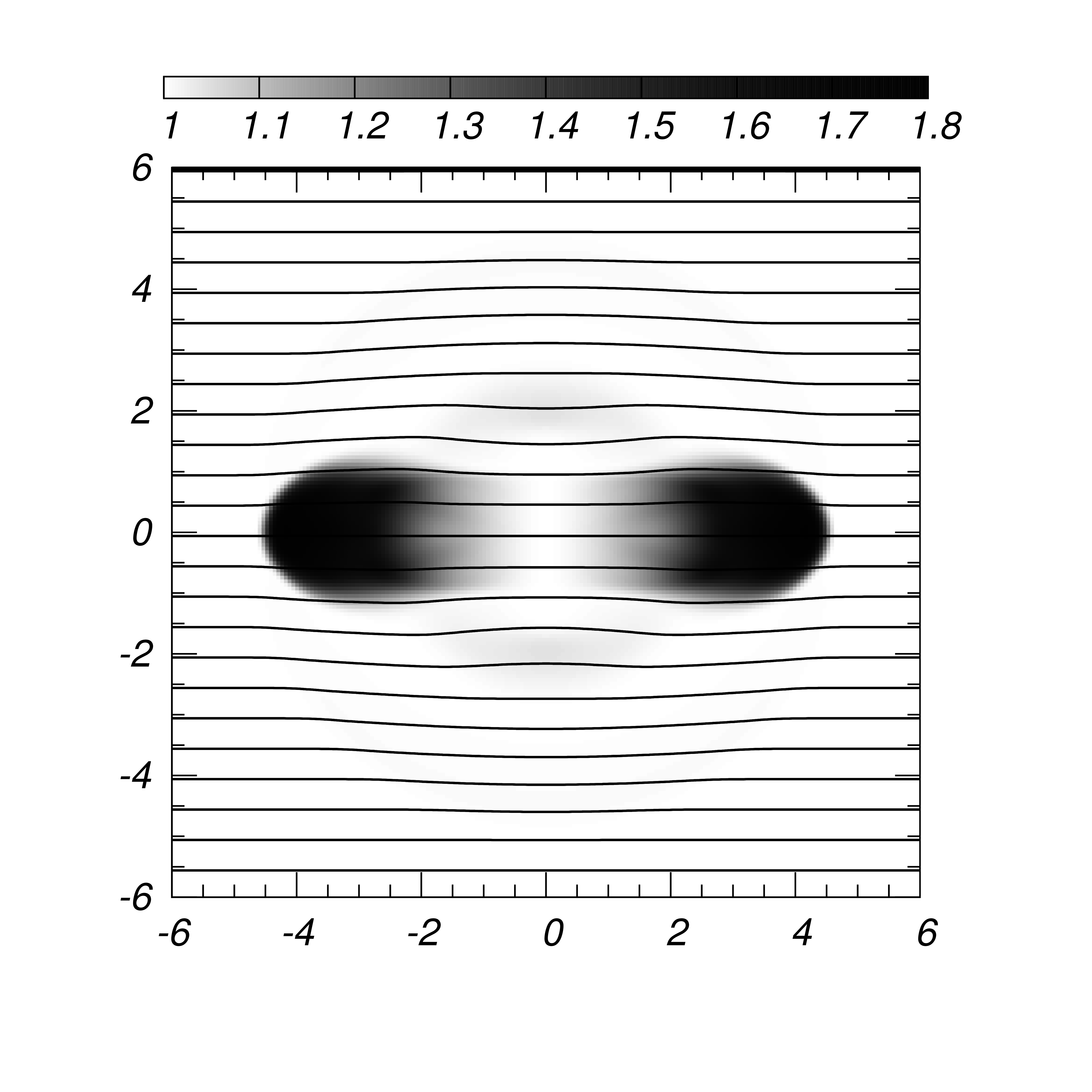}
\includegraphics[width=8.0cm]{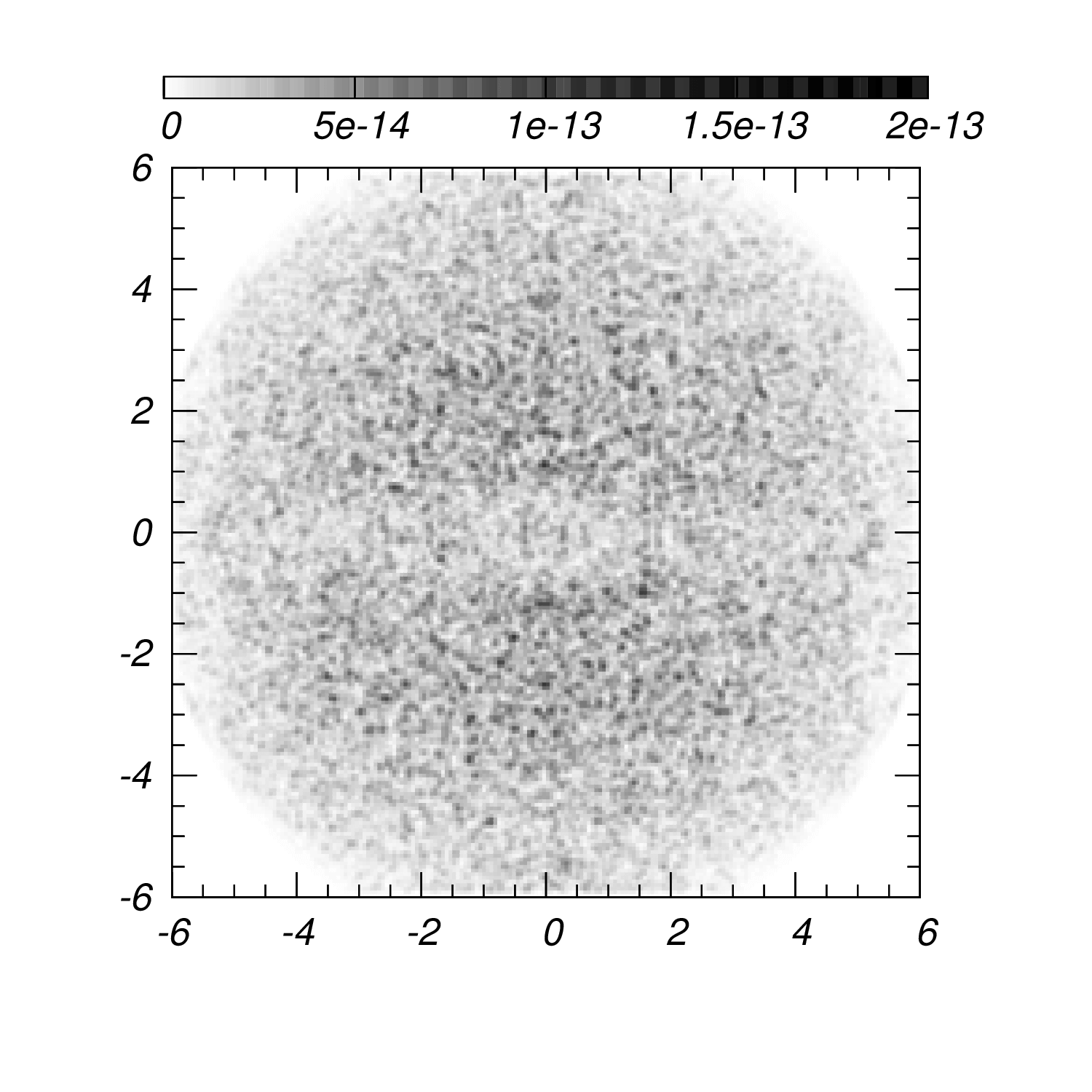}
\end{center}
\caption{\label{fig:Cylindrical2Dstrong} 2D Cylindrical Explosion Test with a strong magnetic field $B_x=1.0$ at $t=4$. We show the logarithm of the proper rest mass density $\rho_0$ (top-left), $B_y$ (top-right), Lorentz factor with the field lines (bottom-left) and the divergence of the magnetic field $\nabla \cdot \vec{B}$ (bottom-right). We use the flux-CT method and $200 \times 200$ cells to cover the domain $[-6,6]\times[-6,6]$ with $CFL = 0.1$. } 
\end{figure*}

It is also possible to handle stronger magnetic fields, for instance, we evolved the cylindrical explosion test with $B_x=1$ using the MC reconstructor, the flux-CT method to control the constraint and a Courant factor 0.1, however using a higher external pressure $5\times 10^{-3}$. In Figure \ref{fig:Cylindrical2Dstrong} we show the results, which are consistent with those in \citep{2007A&A...473..11D} and \citep{2011ApJS..193....6B} for such strong field and our code tolerates magnetic filed strength up to $B_x=1.5$. Without increasing the external pressure it has been possible to carry out this test, however using HLLC fluxes \citep{2006MNRAS.368.1040M}.

\subsubsection{Magnetized Kelvin Helmholtz Test}

\begin{figure*}
\begin{center}
\includegraphics[width=8.0cm]{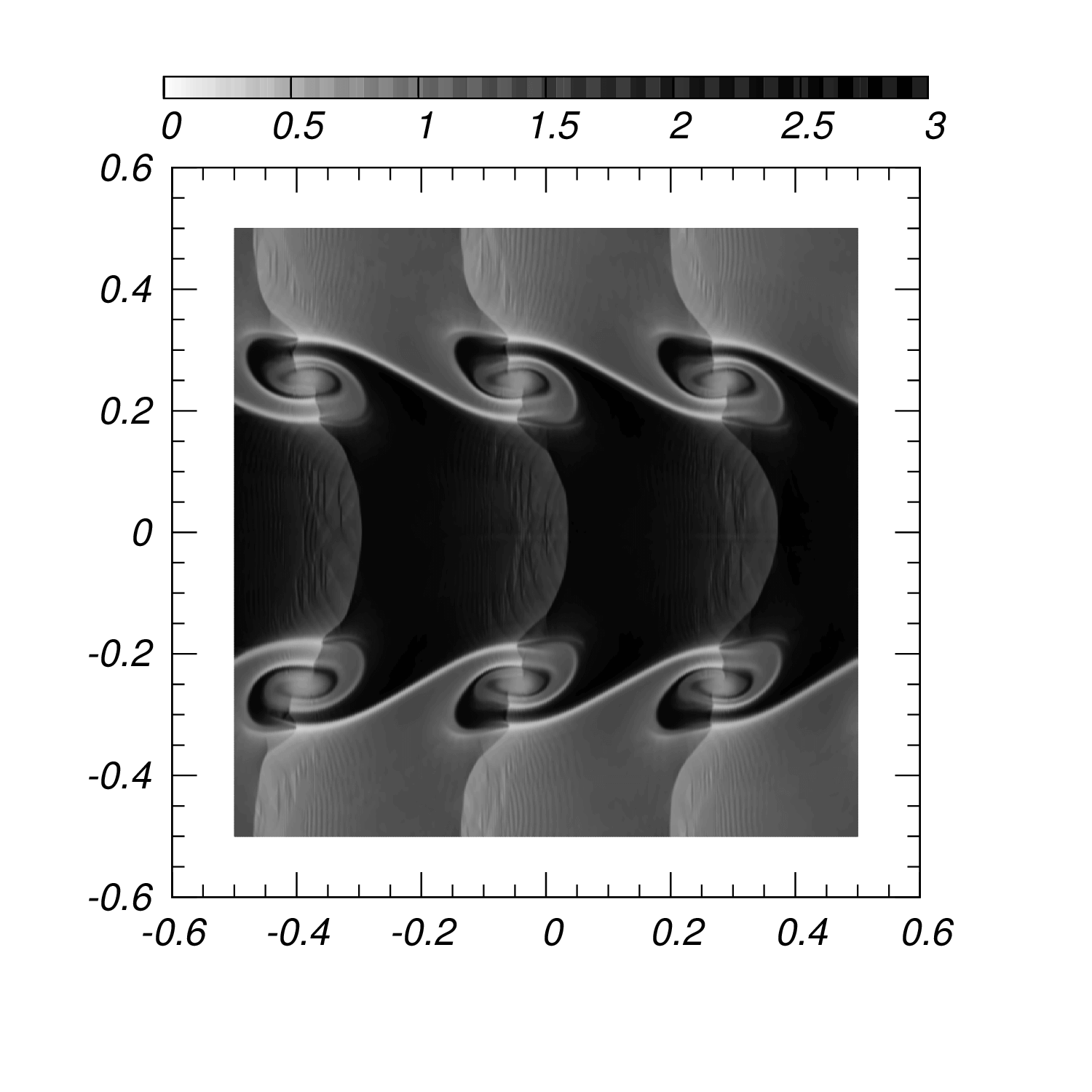}
\includegraphics[width=8.0cm]{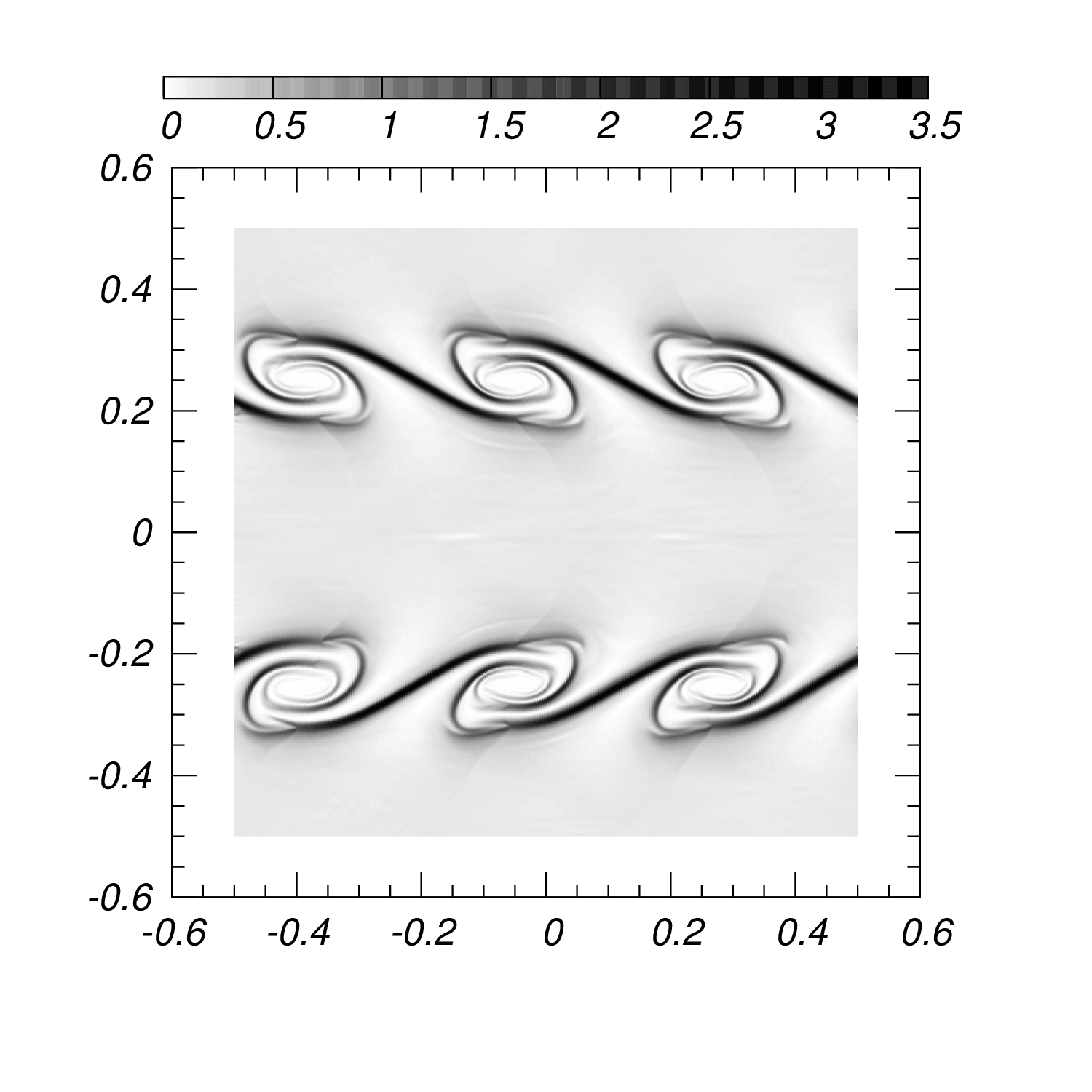}
\includegraphics[width=8.0cm]{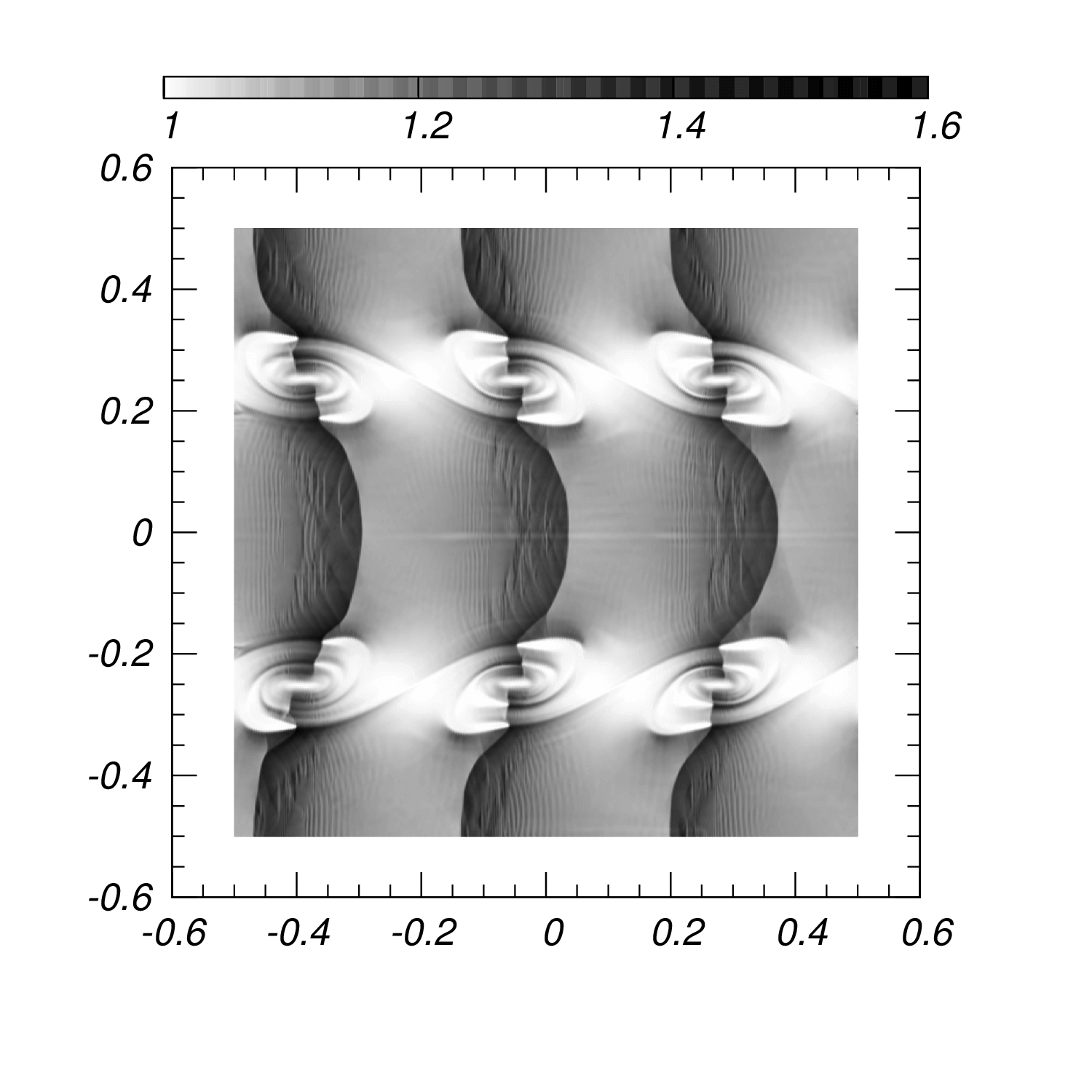}
\includegraphics[width=8.0cm]{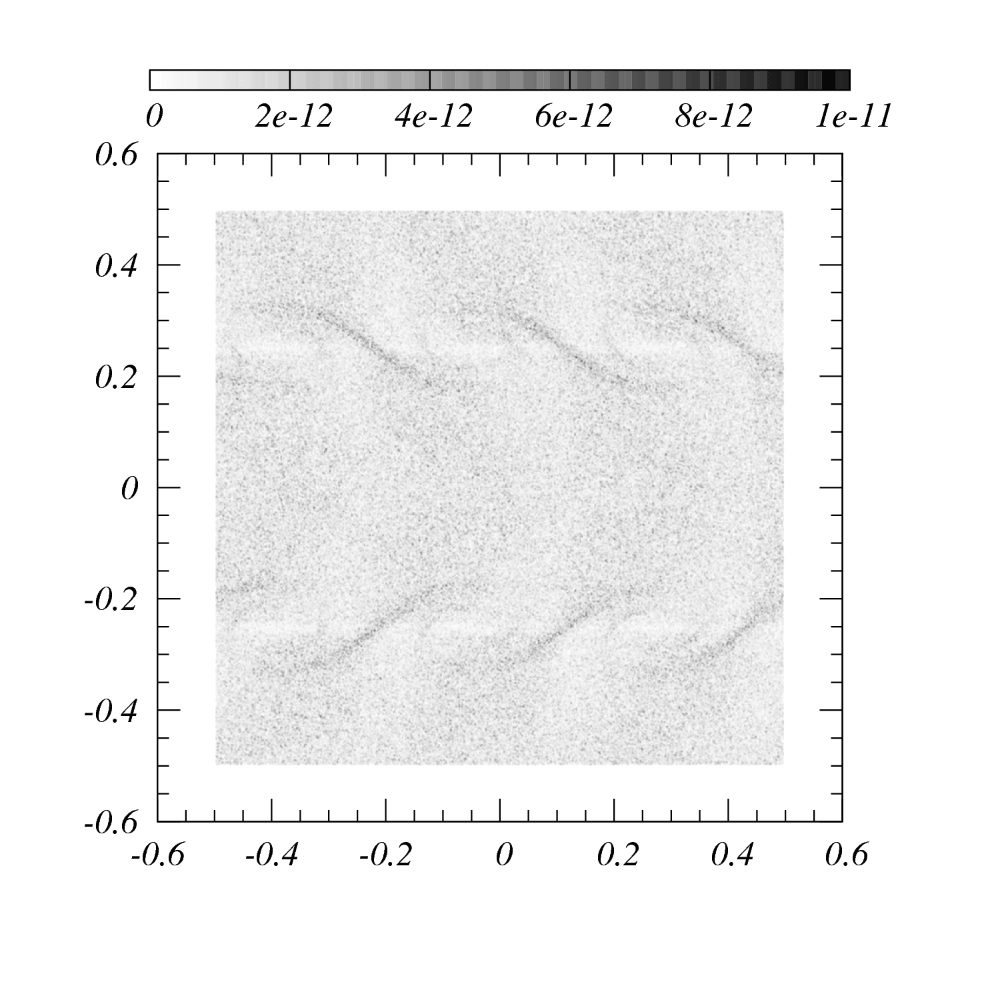}
\end{center}
\caption{\label{fig:MKelvin2D} Magnetized Kelvin Helmholtz Test. We show at $t=2$, the proper rest mass density $\rho_0$ (top-left), the magnetic pressure $p_{mag}$ (top-right), the Lorentz factor $W$ (bottom-left) and the divergence of the magnetic field $\nabla \cdot \vec{B}$ (bottom-right)  calculated with WENO5. The simulations were calculated for $l=3$ on the domain $[-0.5, 0.5] \times [-0.5, 0.5]$ covered with $600 \times 600$ cells, using a Courant factor $CFL=0.25$. }    
\end{figure*}

Another standard 2D test is the magnetized Kelvin-Helmholtz (MKH) instability. The initial condition sets the fluid  in three separate regions. In one of them it moves in one direction and in the other two in the opposite direction. Along the layers dividing the three regions the velocity is perturbed. We study this test in the $xy$ plane covering the domain $[-0.5, 0.5] \times [-0.5, 0.5]$, using $600 \times 600$ cells. The fluid in the strip $|y|\leqslant 0.25$  moves along the $x$ direction with velocity $v^{x}=0.5$, and density $\rho=1$. The fluid outside this area moves in the opposite direction with a speed  $v_{0}^{x}=-0.5$ and density  $\rho=2$. The initial pressure is constant throughout the domain, the adiabatic index is $\Gamma=1.4$, and the magnetic field is uniform along the $x$ direction, $B^{x}=0.5$. The $y$ component of the velocity is also perturbed, and the velocity field including the perturbation at initial time is \\

\begin{eqnarray}
v^{x}&=&v^{x}_{0}(1+ 0.01 \cos(2\pi y l)  \sin(2\pi x l) ),\\
v^{y}&=& 0.01 \cos(2\pi y l)  \sin(2\pi x l),
\end{eqnarray} 

\noindent where $l=3$ is the number of nodes of the perturbation along the domain. The perturbation triggers the instabilities shown in Figure \ref{fig:MKelvin2D}, where we show the proper rest mass density,  magnetic pressure, Lorentz factor and the constraint violation at $t=2$ using WENO5. As we can see, the constraint violation is of the order of $10^{-12}$ when using the flux-CT method. \\

Like in the non-magnetized case, it would also be interesting to estimate the saturation time and to follow the evolution until the turbulent regime as in \citep{2006A&A...454..393B}, however this would deserve a separate space elsewhere.

\subsubsection{Relativistic Magnetic Field Loop advection 3D Test}

This is a test modeling a loop of magnetic field that is being advected similar to that in \citep{2011ApJS..193....6B}. The initial pressure gradients are zero and the dynamics is ruled by the velocity field that carries the magnetic field with it. We set the initial constant density and pressure to $\rho=p=1.0$ with adiabatic index $\Gamma=4/3$.

The magnetic field is initialized using a vector potential defined by $A_3 = MAX([A ( R_0 - r )],0)$ with $r$ the 2D cylinder-type radius measured from an axis parallel to $A_3$. In order to have an oblique advection we choose the vector potential such that its only component lies along the diagonal of the $xz$ plane. Since we choose the numerical domain to be $[-0.5, 0.5] \times [-0.5, 0.5] \times [-1,1]$, the rotation $(x_1,x_2,x_3)=((2x+z)/\sqrt{5},y,-x+2z)/\sqrt{5})$ makes the vector potential to have the single component $A_3$ along such diagonal. The 2D radius perpendicular to this diagonal is thus $r=\sqrt{x_{1}^2+x_{2}^2}=\sqrt{(2x+z)^2/5 + y^2}$. We choose the amplitude $A$ to be small so that the field is weak compared to the gas pressure and thus maintain magnetostatic equilibrium. We use $A = 10^{-3}$ and the loop radius $R_0 = 0.3$. Face-centered magnetic fields are computed using finite differences to calculate ${\bf B} = {\bf \nabla} \times {\bf A}$ to set $ {\bf \nabla} \cdot {\bf B}= 0$ initially up to numerical error. 

\begin{figure*}
\begin{center}
\includegraphics[width=8.0cm]{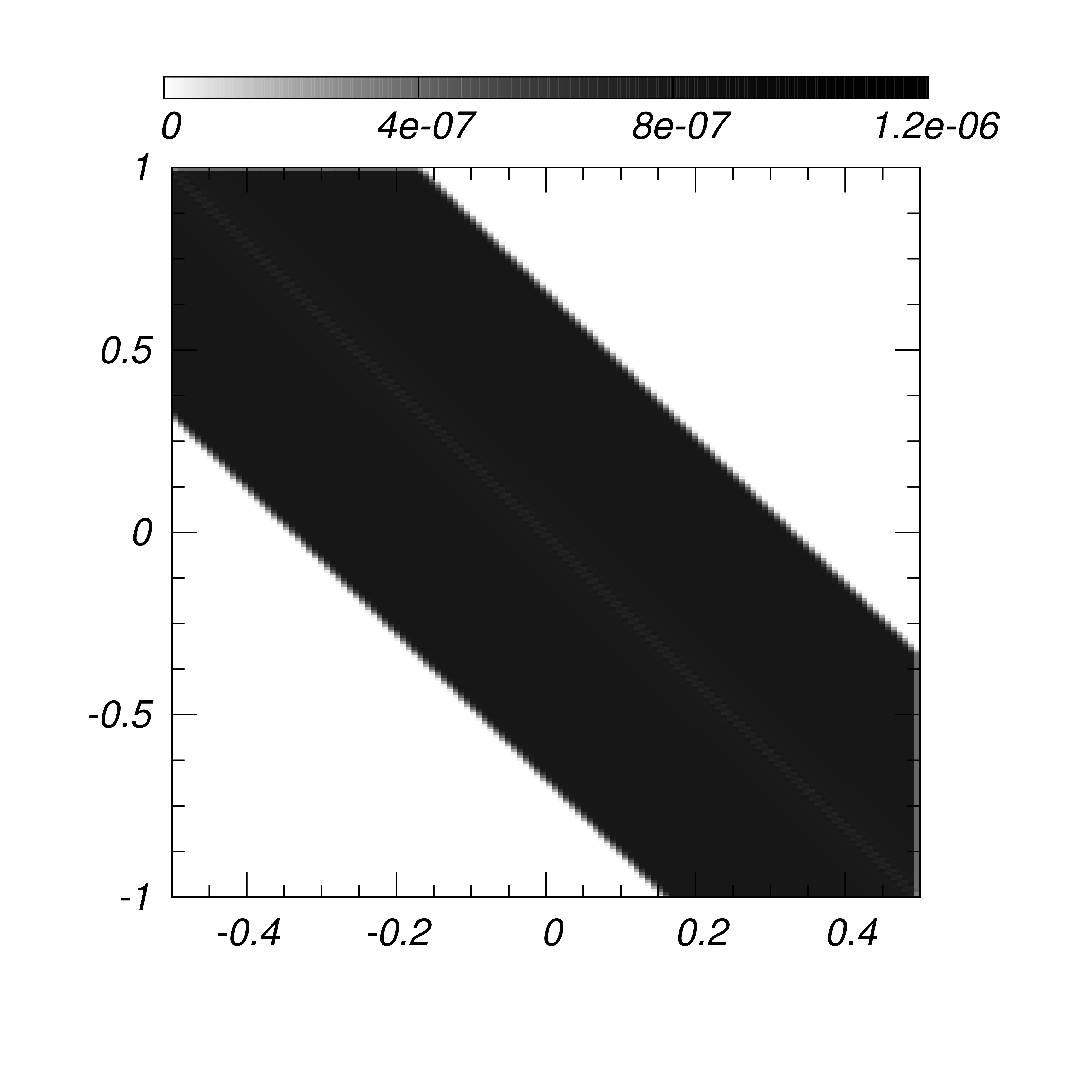}
\includegraphics[width=8.0cm]{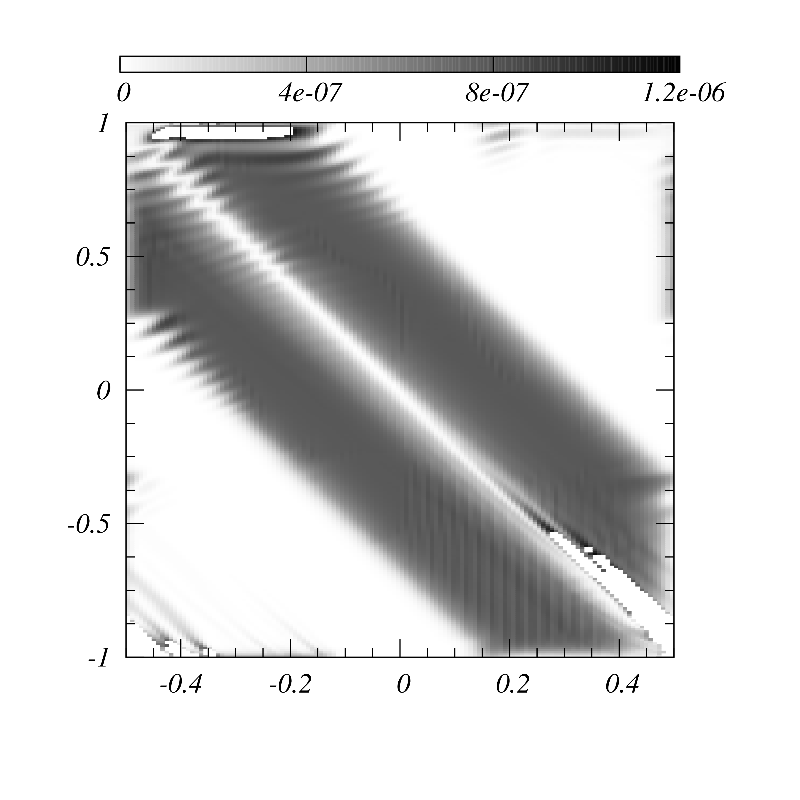}
\includegraphics[width=8.0cm]{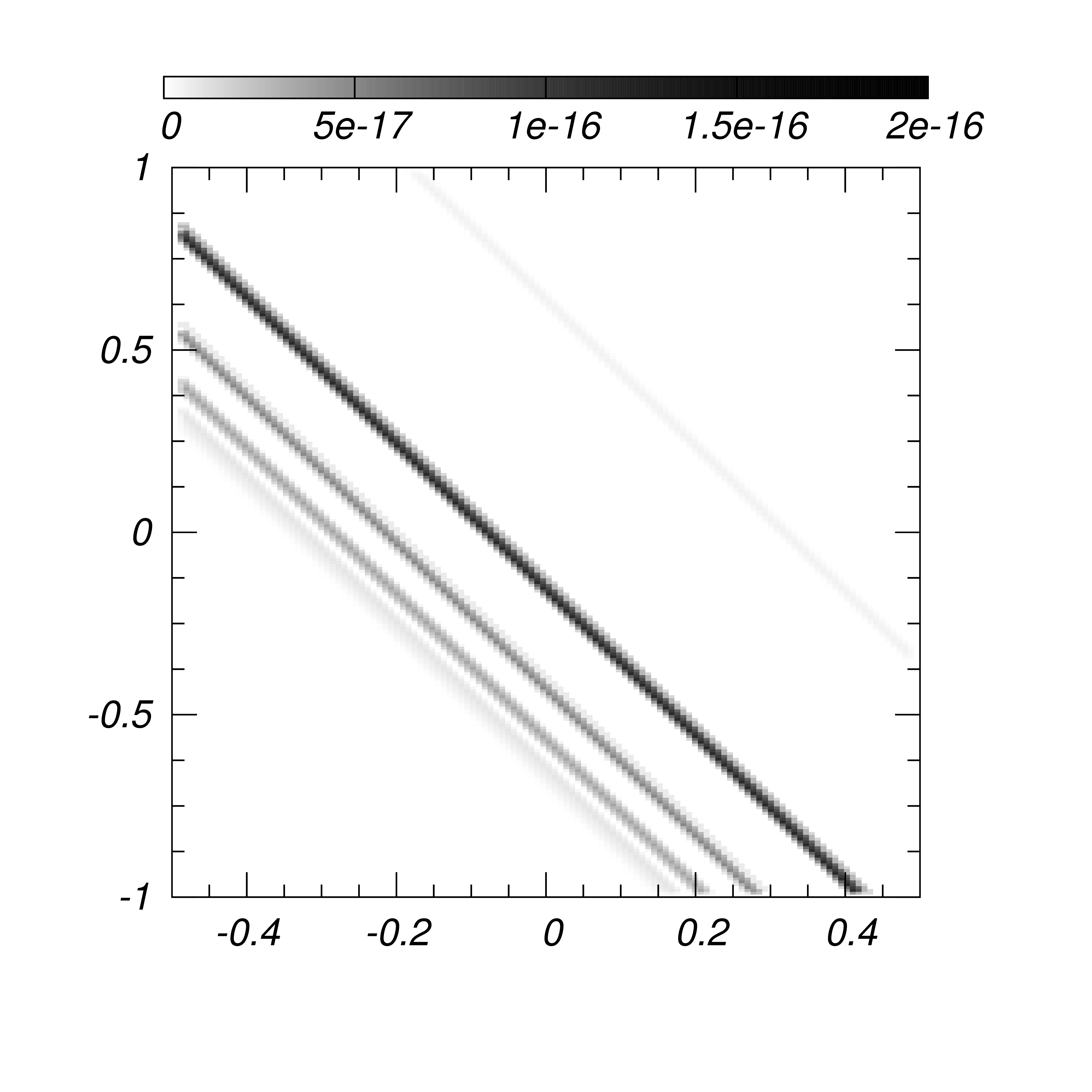}
\includegraphics[width=8.0cm]{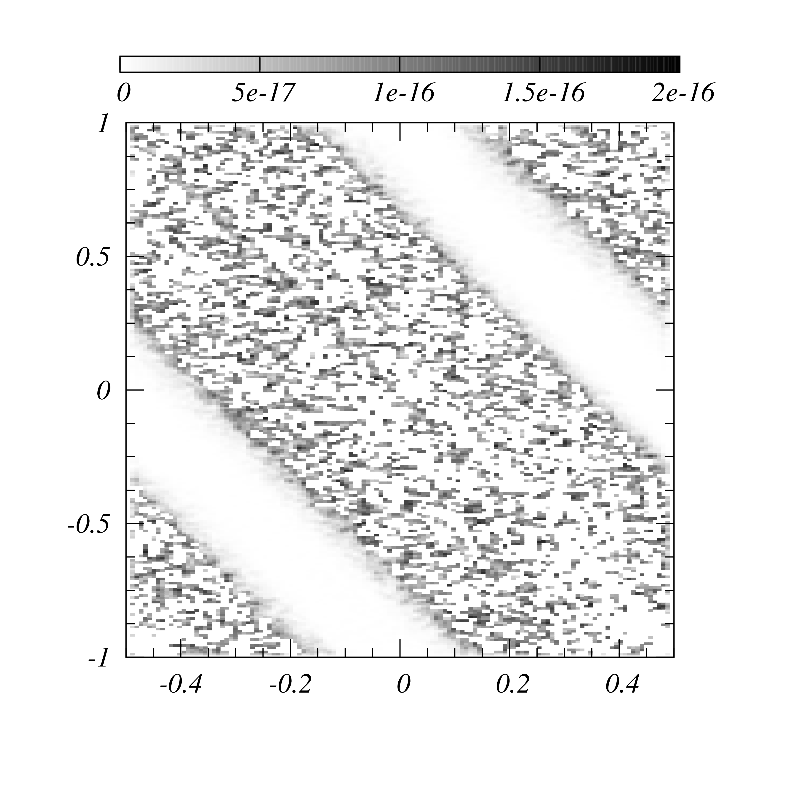}
\end{center}
\caption{\label{fig:MHDloop} Oblique magnetic loop field advection test. We show snapshots on the $xz$ plane of the magnetic field squared magnitude $\| {\bf B} \|^2$ (top row) and Magnetic Field Divergence (bottom row), measured by an Eulerian observer: from left to right at $t=0$ and $t=3.36$ respectively.}    
\end{figure*}

The velocity field is defined by $v_x=-0.3$, $v_y=0.0$ and $v_z=0.6$, such that the loop propagates along the diagonal of the $xz$ plane. In order to preserve the magnetic field constraint we use the Flux constraint transport method. The 3D simulation was calculated  using $128 \times 128 \times 256$ cells, a Courant factor $CFL=0.25$ and periodic boundary conditions. 
In Figure \ref{fig:MHDloop} we present snapshots of the squared magnitude of the magnetic field. This shows the magnetic field is being advected across the domain. It is also shown that the magnetic field constraint is kept under control during the evolution.

\section{Discussion and conclusions}
\label{sec:conclusions}

We have presented a new 3D code designed to solve the Relativistic ideal Magnetohydrodynamics equations and shown that it passes the RHD and RMHD standard tests.

Among the various combinations between methods in a HRSC implementation, we have shown the tests only for a reduced set of linear reconstructors (MINMOD, MC), parabolic PPM and the fifth order WENO5 limiter, all of them combined with the HLLE flux formula and the RK3 time integrator.

We have also presented an error analysis for the RHD and RMHD 1D tests. In all the cases the numerical solutions are consistent. Furthermore, the convergence achieved is the expected nearly first order for initial data with shocks and second for smooth data, at least within a resolution regime and a particular linear reconstructor. In the 2D and 3D cases of RMHD, we have shown the ability of our code to keep the violation of the constraint under control, by using either the flux-CT and cleaning methods. In our case, in unigrid mode both methods are comparably easy to implement, however the flux-CT method may offer extra complication when implemented on adaptive mesh refinement, unlike the cleaning method, which is implemented as an extra evolution equation. Also the flux-CT shows the advantage that the errors in the constraint are kept very low of the order of round off error, whereas the divergence cleaning method is easy to implement but the errors are not as low, which on the other hand is consistent with previous experience.

Finally we want to confirm some of the general limitations of the methods used here that have been described in the past. We confirm that MC captures the shocks better than MINMOD in most of the cases, however, depending on the strength of the shocks, it introduces high frequency noise on discontinuities. 
On the other hand, even though the PPM is a third order method, the parameters we used for this reconstructor show similar errors and convergence rates as those for the linear reconstructors. Finally, the WENO5 is a fifth order reconstructor, which is expensive but the reconstruction is in most cases free of unphysical oscilations.

\section*{Acknowledgments}

We are thankful to the anonymous referee, whose feedback has been essential. We also thank Ian Hawke and Luciano Rezzolla for reading the manuscript and providing important criticism and suggestions. This research is partly supported by grants CIC-UMSNH-4.9 and CONACyT 106466. F.S.G. acknowledges support from the CONACyT program for sabbatical visits in foreign countries. F.D.L-C gratefully acknowledges DGAPA postdoctoral grant to Universidad Nacional Aut\'onoma de M\'exico (UNAM) and finantial support from CONACyT 57585.

\begin{appendix}

\section{SRMHD in cylindrical Coordinates}
\label{appendix}


In the particular test corresponding to the jets our numerical code requieres the implementation of the SRMHD equaitons in cylindrical coordinates $(r,~\phi,~z)$. In this case, the Minkowski metric becomes $\eta_{\alpha \beta}=diag(-1,1,r,1)$. Then the SRMHD equations  can be written in a conservative form as follows 

\begin{equation}
\frac{\partial {\bf f}^{(0)}}{\partial t}  + \frac{1}{r}\frac{\partial (r{\bf f}^{r}) }{\partial r}+ \frac{1}{r}\frac{\partial {\bf f}^{\phi} }{\partial \phi}+ \frac{\partial {\bf f}^{z} }{\partial z} = {\bf s}, 
\end{equation}

\noindent where the conservative variables, the fluxes in each direction and the source vector are: 

\begin{equation}
{\bf f}^{(0)} = \left[ D,M_{r}, M_{\phi},M_{z},\tau,B^r, B^\phi,B^z \right],
\end{equation}

\begin{eqnarray} 
{\bf f}^{(r)}= 
 \left[ \begin{array}{l}
 D v^{r}                              \\
 M_{r}v^{r} + p^* - b_{r} B^{r}/W    \\
 M_{\phi}v^{r} - b_{\phi} B^{r}/W   \\
 M_{z}v^{r} - b_{z} B^r/W         \\
 \tau v^{r} + p^*v^{r} - b^{0} B^{r}/W\\
 0\\
 v^{r} B^\phi - v^\phi B^{r} \\
 v^{r} B^{z} - v^{z} B^{r} 
 \end{array} \right],~~~~
{\bf f}^{(\phi)}= 
 \left[ \begin{array}{l}
 D v^{\phi}                              \\
 M_{r}v^{\phi} - b_{r} B^{\phi}/W    \\
 M_{\phi}v^{\phi} + p^*  - b_{\phi} B^{\phi}/W   \\
 M_{z}v^{\phi} - b_{z} B^{\phi}/W         \\
 \tau v^{\phi} + p^*v^{\phi} - b^{0} B^{\phi}/W\\
 v^{\phi} B^r - v^r B^{\phi} \\
 0               \\
 v^{\phi} B^{z} - v^{z} B^{\phi} 
 \end{array} \right],~~~~
{\bf f}^{(z)}= 
 \left[ \begin{array}{l}
 D v^{z}                         \\
 M_{r}v^{z} - b_{r} B^{z}/W      \\
 M_{\phi}v^{z} - b_{\phi} B^{z}/W   \\
 M_{z}v^{z}  + p^* - b_{z} B^{z}/W         \\
 \tau v^{z} + p^*v^{z} - b^{0} B^{z}/W\\
 v^{z} B^r - v^r B^{z} \\
 v^{z} B^{\phi} - v^{\phi} B^{z}\\ 
 0         
 \end{array} \right],
\end{eqnarray}

\begin{eqnarray}
{\bf s} = \left[0, \frac{\rho h^{*}W^{2}v^{\phi}v^{\phi}+ p^{*}-b^{\phi}b^{\phi}}{r},0,\frac{\rho h^{*}W^{2}v^{r}v^{\phi}-b^{r}b^{\phi}}{r}, 0,0,0,\frac{B^{\phi}v^{r}-B^{r}v^{\phi}}{r} \right]^{T}. 
\end{eqnarray}

\noindent The above equations in cylindrical coordinates can be written in a semi-discrete form as follow

\begin{eqnarray}
\frac{d }{d t} {\bf f}^{(0)}_{(i,j,k)} =  &-& \frac{\left(r_{(i+1/2,j,k)}{\bf f}^{(r)}_{(i+1/2,j,k)} -r_{(i-1/2,j,k)}{\bf f}^{(r)}_{(i-1/2,j,k)} \right)}{r_{(i,j,k)}\Delta r}- \frac{ \left({\bf f}^{(\phi)}_{(i,j+1/2,k)} -{\bf f}^{(\phi)}_{(i,j-1/2,k)} \right)}{r_{(i,j,k)}\Delta \phi}\\
&-& \frac{\left( {\bf f}^{(z)}_{i,j,k+1/2}- {\bf f}^{(z)}_{i,j,k-1/2}\right)}{\Delta z} + {\bf s}_{(i,j,k)} \nonumber,
\end{eqnarray}

\noindent where the numerical fluxes are computed at each respective intercell.

\end{appendix}



\end{document}